\UseRawInputEncoding
\documentclass[reprint,
superscriptaddress,
showpacs,
nofootinbib,
twocolumn,
longbibliography,
prc,
aps,
]{revtex4-1}

\usepackage{amsmath,amssymb,amsfonts,amsthm}
\usepackage[english]{babel}
\usepackage{url}
\usepackage{bm}
\usepackage{graphicx,rotating}
\usepackage{csvsimple}
\usepackage{longtable}
\usepackage{booktabs}
\usepackage{multirow, array}
\usepackage{epsfig}
\usepackage{ulem}

\usepackage{dcolumn}
\usepackage{verbatim}

\usepackage[colorlinks=true,linkcolor=black, citecolor=blue]{hyperref}

\allowdisplaybreaks[1]

\newcommand{\BR}{B\!R}

\newcommand{\add}[1]{{\color{blue} #1}}

\begin{document}

\title{$\mathcal{F}t$ values of the mirror $\beta$ transitions and the weak magnetism induced current in allowed nuclear $\beta$ decay}

\author{N. Severijns}%
\email{nathal.severijns@kuleuven.be}
\affiliation{KU Leuven University, Instituut voor Kern- en Stralingsfysica, Celestijnenlaan 200D, B-3001 Leuven, Belgium}%
\author{L. Hayen}%
\affiliation{KU Leuven University, Instituut voor Kern- en Stralingsfysica, Celestijnenlaan 200D, B-3001 Leuven, Belgium}%
\affiliation{Department of Physics, North Carolina State University, Raleigh, North Carolina 27695, USA}
\affiliation{Triangle Universities Nuclear Laboratory, Durham, North Carolina 27708, USA}
\author{V. De Leebeeck}%
\author{S. Vanlangendonck}%
\affiliation{KU Leuven University, Instituut voor Kern- en Stralingsfysica, Celestijnenlaan 200D, B-3001 Leuven, Belgium}%
\author{K. Bodek}%
\author{D. Rozpedzik}%
\affiliation{Institute of Physics, Jagiellonian University, 30348 Cracow, Poland}%
\author{I.S. Towner}%
\affiliation{Cyclotron Institute, Texas A \& M University, College Station,
Texas 77843, USA}%

\begin{abstract}
In recent years a number of correlation measurements in nuclear $\beta$ decay have been performed reaching a  precision of the order of 1\% and below and it is expected that even higher precision will be reached in the near future. At these levels of precision higher-order corrections due to e.g. recoil terms induced by the strong interaction and radiative corrections cannot necessarily be neglected anymore when interpreting these results in terms of new physics or extracting a value for the $V_{ud}$ quark-mixing matrix element. We provide here an update of the $\mathcal{F} t$ values of the $T=1/2$ mirror $\beta$ decays as well as an overview of current experimental and theoretical knowledge of the most important recoil term, weak magnetism, for both the $T=1/2$ mirror $\beta$ transitions and a large set of $\beta$ decays in higher isospin multiplets. The matrix elements determining weak magnetism were calculated in the nuclear shell model and cross-checked against experimental data, showing overall good agreement. Additionally, we show that further insight can be obtained from properly deformed nuclear potentials, in particular for mirror $T=1/2$ decays. The results provide new insights in the size of weak magnetism, extending the available information to $\beta$ transitions of nuclei with masses up to $A =$ 75. This provides important guidance for the planning and interpretation of ongoing and new precise correlation measurements in nuclear $\beta$ decay searching for new physics or to extract the $V_{ud}$ quark-mixing matrix element in mirror $\beta$ decays. This more detailed knowledge of weak magnetism can also be of interest for further theoretical work related to the reactor neutrino problem.
\end{abstract}

\maketitle

\add{\tableofcontents}

\section{Introduction}
In the past, so-called induced or recoil terms in nuclear $\beta$ decay have received considerable attention, both theoretically and experimentally (e.g. \cite{Holstein1974, Holstein1974err, Azuelos1975, Calaprice1976, Calaprice1977, Behrens1978}). Their presence originates from QCD effects in the weak interaction of a bound quark, and folds in nuclear structure effects in heavier nuclei. Its study was motivated mainly in relation to experimental tests of the Conserved Vector Current (CVC) hypothesis \cite{Feynman1958} (see e.g. \cite{Tribble1974, Kaina1977, Wu1977, Elmbt1987, Calaprice1978, Camp1990, Grenacs1985, Minamisono2011a}) and searches for second class currents \cite{Grenacs1985} (e.g. \cite{Wilkinson1970, Holstein1971, Holstein1974, Holstein1976, Tribble1975, Calaprice1975Ne, Calaprice1976, Calaprice1977, Fifield1977, Behrens1978, DupuisRolin1978, Tribble1978, Tribble1981, Grenacs1985, Rosa1988, Wilkinson2000, Minamisono1998, Minamisono2001, Smirnova2003, Minamisono2008, Minamisono2011}). The most important of these induced currents, the weak magnetism term, has recently received new attention for several reasons. First, measurements of correlations between the spins and momenta of the particles involved in nuclear $\beta$ decay, such as e.g. $\beta-\nu$ correlation and $\beta$-asymmetry parameter measurements \cite{Jackson1957a}, have now reached the level of precision where recoil terms cannot be neglected anymore (e.g. \cite{Pitcairn2009, Wauters2009a, Wauters2010, Soti2014, Fenker2018, Markisch2019}). In such measurements one is searching either for new physics beyond the standard model (SM), such as e.g. scalar and tensor weak currents \cite{Severijns2011, Naviliat2013, Holstein2014, Vos2015, Gonzalez2019}, or aiming at the determination of the $V_{ud}$ quark-mixing matrix element in the $\beta$ decay of mirror isotopes \cite{Naviliat2009, Lienard2015, Brodeur2016b, Beck2019}. Secondly, weak magnetism is also important with respect to the so-called reactor antineutrino anomaly \cite{Mention2011, Mueller2011, Huber2011, Hayes2016, Hayes2017, Hayes2018, Hayen2019, Hayen2019b}, i.e., the fact that the ratio of the observed antineutrino event rate at nuclear reactors to the predicted rate is lower than unity. Each of these respective topics is briefly commented upon here below.

\subsection{Searches for exotic scalar and tensor charged weak currents}

In $\beta$ decay strong limits on the existence of charged weak scalar currents have been obtained from the corrected $\mathcal{F} t$ values of the superallowed $0^+ \rightarrow 0^+$ pure Fermi $\beta$ transitions \cite{Hardy2015}. Complimentary information as well as limits on tensor currents are traditionally obtained from accurate measurements of correlations between the spins and momenta of the particles involved in $\beta$ decay (reviews are given in e.g. \cite{Severijns2011, Naviliat2013, Holstein2014, Vos2015, Gonzalez2019, Falkowski2021}). The best-suited correlations are the $\beta$-$\nu$ angular correlation and the $\beta$ asymmetry with respect to the polarization axis of oriented parent nuclei \cite{Jackson1957a, Severijns2006}. In recent years the precision of such measurements has improved significantly and several results with a precision of about 1\% or better have been reported already (see e.g. for the $\beta$-$\nu$ correlation \cite{Adelberger1999,Gorelov2005,Vetter2008,Flechard2011,Sternberg2015,Beck2020}, for the $\beta$-asymmetry parameter \cite{Wauters2009a, Wauters2010, Soti2014, Fenker2018, Brown2018, Markisch2019, Combs2020}, and for other correlations \cite{Pitcairn2009, Melconian2007, Beck2011, Kozela2012, Chupp2012, Sun2020, Saul2020}). Many new experiments in search of exotic currents are ongoing or planned. 
An overview can be found in Table 3 in Ref. \cite{Gonzalez2019}. Most of these are aiming at a precision down to the few per-mille level. At such a precision, searches for physics beyond the Standard Model in $\beta$ decay remain competitive to direct searches for new bosons in collider experiments such as at the Large Hadron Collider (e.g. \cite{Bhattacharya2012, Naviliat2013, Cirigliano2013, Cirigliano2013a, Falkowski2021}). However, as experiments reach ever higher precision and dive below the 1\% level, additional theoretical corrections arise \cite{Hayen2020a}. The most important of these is the so-called weak magnetism correction which is nuclear structure dependent. Its importance and possible limitations related to its exact knowledge were already demonstrated in the $\beta$-decay correlation measurements with e.g. $^{37}$K \cite{Fenker2018}, $^{60}$Co \cite{Wauters2010}, $^{67}$Cu \cite{Soti2014}, $^{80}$Rb \cite{Pitcairn2009} and $^{114}$In \cite{Wauters2009a}, and discussed recently in \cite{Hayen2020a, Vanlangendonck2021}.

\subsection{Determining the $V_{ud}$ quark-mixing matrix element from mirror beta transitions}

Apart from searches for new physics, correlation measurements in nuclear $\beta$ decay can also be used to determine the $V_{ud}$ quark-mixing matrix element or perform an independent test of the CVC hypothesis, i.e. by checking the constancy of the $V_{ud}$ values for the individual transitions, as was shown for isospin $T$~=~1/2 mirror $\beta$ transitions \cite{Naviliat2009}. 
Such measurements can also help to better understand the isospin-symmetry breaking correction (noted as $\delta_c$) that is required to obtain high-precision corrected ${\mathcal F} t$ values for superallowed $\beta$ transitions (e.g.~\cite{Towner2010, Hardy2015}). Recently, very precise results were reported for the $\beta$ asymmetry parameter in de mirror beta decay of $^{37}$K \cite{Fenker2018} and $^{19}$Ne \cite{Combs2020}. In addition, a double-counting instance in the extraction of $V_{ud}$ from the mirror $\beta$ transitions was corrected in Ref.~\cite{Hayen2021}. The current value for $V_{ud}$ from mirror $\beta$ transitions, i.e. $|V_{ud}|$~=~0.9739(10) \cite{Hayen2021} (see also \cite{Falkowski2021}), depends primarily on the availability of precise experimental $\mathcal{F}t$ values \cite{Severijns2008, Hayen2019a} and existing experimental results, mainly for the $\beta$-asymmetry parameter, $A$ \cite{Naviliat2009, Falkowski2021}.

In Refs.~\cite{Naviliat2009traps, Severijns2013, Hayen2020a} prospects for a more precise determination of $V_{ud}$ from these $T = 1/2$ mirror $\beta$ transitions are given. The mirror $\beta$ decays of $^{3}$H and $^{19}$Ne appear to be the most sensitive in case of $\beta$-$\nu$ correlation measurements, with all other mirror transitions up to $^{41}$Sc still having a quite good sensitivity \cite{Severijns2013, Hayen2020a, Vanlangendonck2021}. Such measurements have recently been performed with $^{19}$Ne and $^{35}$Ar \cite{Fabian2014, Fabian2018, Combs2020} and new ones are planned \cite{Lienard2014a, Ron2014, Brodeur2016}. For the $\beta$-asymmetry parameter measurements in the mirror $\beta$ decays of $^{3}$H, $^{19}$Ne, $^{33}$Cl and $^{35}$Ar provide the best sensitivity \cite{Severijns2013, Combs2020, Hayen2020a, Vanlangendonck2021}, with a measurement of the $\beta$-asymmetry parameter in the decay of polarized $^{35}$Ar recently being attempted \cite{Velten2014, Gins2019, Gins2019a}. Further improvement on the $\beta$ asymmetry parameter in the mirror beta decay of $^{37}$K is anticipated as well \cite{Fenker2018}. In order to obtain sufficiently precise values for $V_{ud}$ from such measurements a relative precision of typically 0.5\% or better is needed \cite{Severijns2013}, again requiring that higher-order effects are duly taken into account \cite{Hayen2020a, Vanlangendonck2021}.

\subsection{Reactor neutrino anomaly}

As several modern reactor neutrino experiments have come online in the search for neutrino oscillations, comparisons to theoretical predictions have resulted in large discrepancies. Using two independent analysis methods, the ratio of experimentally detected versus theoretically predicted antineutrinos is found to be $R=0.943 \pm 0.023$ \cite{Mention2011}. If genuine, this could be related to the existence of a sterile neutrino with $\Delta m^2 \geq$~1~eV$^2$ \cite{Mention2011, Kopp2011}. Over the past few years, apart from other work, the treatment of forbidden transitions \cite{Dwyer2015, Hayes2014, Hayen2019, Martinez-Pinedo2019, Hayen2019b, Estienne2019} and the evaluation of weak magnetism \cite{Hayes2016, Hayes2017} has undergone some scrutiny. Recently, it was suggested \cite{Hayen2019b, Martinez-Pinedo2019} that a more proper treatment of forbidden transitions and inclusion of more pandemonium-free data in calculating the reactor antineutrino energy spectrum \cite{Estienne2019} could possibly explain the flux problem as well as the so-called spectral shoulder with respect to the theoretical spectrum. As to weak magnetism, all calculations of the expected antineutrino flux at reactors assume the weak magnetism contribution to depend solely on the difference between the proton and neutron magnetic moments and the $\beta$-particle energy. However, as will be shown explicitly later in this document, its precise value depends strongly on the isotope at hand. Using the same value for all transitions is then a substantial oversimplification, with up to now poorly understood consequences. As current experimental information on weak magnetism is limited to nuclei with $A \leq 43$, a study including also heavier masses, and especially measurements of weak magnetism for nuclei in the fission region, would be most helpful. At present several precision $\beta$-spectrum shape measurements to provide direct experimental information on weak magnetism are ongoing, e.g. with the isotopes $^6$He \cite{Huyan2016, Huyan2018, Huyan2018a, Mukul2018}, $^{20}$F \cite{Naviliat2021, Hughes2018}, and $^{114}$In \cite{DeKeukeleere2021, Vanlangendonck2021a}, the latter effectively being in the fission isotope region. Here we present and analyse the currently existing experimental data allowing to determine weak magnetism for $\beta$ transitions up to mass $A$ = 75, and provide evidence for a nuclear structure and mass dependence of the size of the weak magnetism term.

\subsection{Motivation and outline of paper}

Based on the above formulated arguments, a more detailed knowledge of the weak magnetism form factor in $\beta$ decay would clearly be of great value. Previously, the weak magnetism term was evaluated for superallowed mixed Fermi/Gamow-Teller (F/GT) mirror $\beta$ transitions between isospin $T=1/2$ doublets in mirror nuclei up to $^{43}$Ti \cite{Calaprice1976, Azuelos1977}, and for $T$~=~1 to $T$~=~0 Gamow-Teller decays in nuclei up to mass $A$~=~32 \cite{Calaprice1976, Huber2011}. 

In this paper, the currently available experimental data leading to the weak magnetism form factor are analysed for both the isospin T$ = 1/2$ mixed Fermi/Gamow-Teller (F/GT) $\beta$ transitions and Gamow-Teller transitions in $T = 1, 3/2$ and $2$ multiplets, based on the CVC hypothesis. For the mirror $\beta$ transitions, the input data for the corrected $\mathcal{F} t$ values for these transitions are reviewed and updated $\mathcal{F} t$ values are provided. Combining these with the nuclear magnetic moments of the mirror transitions' initial and final states allows determining values for the weak magnetism form factor for these transitions up to $^{75}$Sr. For the Gamow-Teller transitions, the data base of the Brookhaven Laboratory National Nuclear Data Center database \cite{ENSDF} was scanned for $\beta$ transitions for which analog (i.e. within the same isospin multiplet) gamma transitions are known and with sufficient information being available for both transitions in order to determine the weak magnetism form factor for the $\beta$ transition, again using the CVC hypothesis.

In Sec.~\ref{Ft-values} we first review the currently available input data (the literature cut-off was April 2021) leading to the corrected ${\mathcal F}t$ values for the mirror $\beta$ transitions with $A = 3-75$, thereby updating the previously published $\mathcal{F} t$ values for the mirror $\beta$ transitions up to $A = 45$ \cite{Severijns2008} and extending them further to $A = 75$. 

Sec.~\ref{sec:weak_magnetism} focuses on the weak magnetism form factors for both the mirror $\beta$ transitions between $T=1/2$ states, and the Gamow-Teller transitions in higher-$T$ multiplets. In Sec.~\ref{sec:weak_magnetism_theory}, the general formalism with respect to the induced form factors and the weak magnetism term is introduced. Then, in Sec.~\ref{wm-mirror-nuclei}, the nuclear magnetic moments of the  mirror $\beta$ transitions' mother and daughter states are reviewed and the weak magnetism term is extracted for the transitions up to $A = 75$, using the CVC hypothesis. Combining these with the $\mathcal{F} t$ values allows extracting both the Gamow-Teller and orbital angular momentum matrix elements, $M_{GT}$ and $M_L$, respectively. These are compared to shell model calculations in Sec.~\ref{sec:Theoretical form factor comparison}. 

In Sec.~\ref{triplet-states} the analysis of weak magnetism for Gamow-Teller $\beta$ decays of states belonging to $T = 1$, $T = 3/2$, and $T = 2$ isospin multiplets up to mass $A=53$ is presented, again using the CVC hypothesis. The weak magnetism term for the respective $\beta$ transitions is now obtained from the width of the analog $E1$ gamma transition. Also for these transitions the $ft$ values are calculated in order to extract the Gamow-Teller and orbital angular momentum matrix elements, with these again being compared to shell model calculations (for a subset of the transitions considered).

Finally, the nuclear structure and mass dependence of the weak magnetism term and its relevance to the reactor anomaly in face of these new data, are discussed.

\section{${\mathcal F}t$ values for mirror $\beta$ transitions} \label{Ft-values}

In Ref.~\cite{Severijns2008} the corrected ${\mathcal F}t$ values for mirror $\beta$ transitions in nuclei with mass $A$ up to 45 were presented for the first time. Since then, many new experimental results for the half-life, branching ratio and decay energies for these transitions, as well as for mirror transitions in nuclei with higher masses have been reported. First, an update of the relevant definitions and formalism will be given (Sec.~\ref{definition-Ft}). Thereafter, the updated data set is presented and evaluated, together with new theoretical calculations of nuclear structure and radiative corrections (Sec.~\ref{data-Ft-mirrors}). Finally, the current values for the ${\mathcal F}t$ values for the mirror $\beta$ transitions are reported (Sec.~\ref{Ft-values mirrors}).

\subsection{Formalism} \label{definition-Ft}

For a general $\beta$ transition, the uncorrected partial half-life can be written as \cite{Severijns2008}
\begin{equation} \label{partial-t-1}
    t = \frac{K}{G_F^2V_{ud}^2}\frac{1}{\xi f}
\end{equation}
where $G_F$ is the Fermi coupling constant, $V_{ud}$ is the $\it{up}$-$\it{down}$ matrix element of the CKM matrix, 
\begin{align}
    K/(\hbar c)^6 &= \frac{2\pi^3\ln 2\hbar}{(m_ec^2)^5} \\
    &= (8120.27648\pm 0.00026) \cdot 10^{-10} \text{GeV}^{-4}s ~ , \nonumber \\
    \xi &= g_V^2M_F^2 + g_A^2M_{GT}^2
\end{align}
with $g_V$ and $g_A$ the strength of the vector and axial-vector interactions (in units of $G_F$) as defined in the Hamiltonian of Jackson, Treiman and Wyld \cite{Jackson1957a}. Here, $M_{F, GT}$ is the main Fermi or Gamow-Teller matrix element, respectively. Further, $f$ is the statistical rate function defined as the integral over the normalized spectrum shape
\begin{align} 
    f = \int_1^{W_0}&pW(W_0-W)^2F(Z, W) \nonumber \\
    &\times C(Z, W)K(Z, W)dW ~ ,
    \label{eq:f_def_integral}
\end{align}
where $W$ is the total $\beta$ energy in units of its rest mass, i.e. $W = E_\mathrm{kin}/m_ec^2 + 1$, and $W_0$ the spectral endpoint. Here $F(Z, W)$ is the well-known Fermi function, $C(Z, W)$ the nuclear-structure dependent shape factor, and $K(Z, W)$ all higher-order corrections \cite{Hayen2018}.

The product $ft$ is then independent of the kinematics and produces the main matrix elements. Its experimental value for a specific $\beta$ transition depends on three quantities: ($i$) the total decay transition energy, $Q_{EC}$, ($ii$) the half-life, $t_{1/2}$, of the decaying state and ($iii$) the branching ratio, $BR$, of the particular transition of interest. The $Q_{EC}$ value determines the statistical rate function, $f$. The half-life and the branching ratio, together with the electron-capture fraction, $P_{EC}$, determine the partial half-life, $t$, of the transition of interest, the latter of which is defined as
\begin{equation}     \label{eq:partial-t}
    t = t_{1/2} \biggl( \frac{ 1 + P_{EC} } {BR} \biggr) ~ .
\end{equation}
The superallowed mirror $\beta$ transitions between the analog $T = 1/2$ states situated at both sides of the $N = Z$ line are mixed transitions with both Fermi and Gamow-Teller components. Despite this increased complexity, they harbour a particular advantage. As the decay occurs within the same isospin multiplet - and taking the isospin limit in QCD - the Fermi matrix element is determined simply by using the Wigner-Eckart theorem. In the case of mirror $T=1/2$ transitions this is particularly simple and one finds $M_F^0 = 1$, with the superscript denoting the isospin-conserving limit. In nuclei, however, the presence of the Coulomb interaction between nucleons breaks the purported isospin symmetry, albeit at only the few percent level. Small corrections must therefore be introduced to the single particle matrix element
\begin{equation}
    M_F = M_F^0(1-\delta_C)^{1/2},
\end{equation}
with $\delta_C$ a correction arising from the imperfect overlap of proton and neutron radial wave functions due to their slightly different nuclear potentials \cite{Towner2010}. This must be provided by nuclear theory and its estimate is of particular importance in, e.g., the $V_{ud}$ extraction in superallowed $0^+ \to 0^+$ Fermi decays and mirror decays, as discussed below.

The weak axial-vector current, occurring in Gamow-Teller decays, is not conserved and is renormalized through QCD effects. This prevents one, in theory, from writing down an analogous isospin-invariant GT matrix element, $M_{GT}^0$, although this has historically been commonplace \cite{Severijns2008, Naviliat2009}. With previous nuclear many-body calculations, dominated by the nuclear shell model, this is at first glance merely a notational matter as $M_{GT}$ could typically not be calculated to high enough precision for the distinction to be relevant. With the developing maturity of nuclear \textit{ab initio} methods (see e.g. \cite{Pastore2018, Cirigliano2019}), however, the distinction is much more clearly defined and of interest as they enter the $A=10$-$19$ mass range and beyond.

Besides strong interaction effects, additional electroweak corrections arise both at the nucleon and nucleus level. Traditionally these have been separated into a QCD-invariant, infrared-singular part (the \textit{outer} correction, $\delta_R'$) and an energy-independent renormalization of the weak coupling constants from the ultraviolet (the \textit{inner} correction, $\Delta_R^{V, A}$) \cite{Sirlin1967}. Both contribute roughly at the few percent level. The latter is typically calculated for a single nucleon, i.e. in a loop calculation the nucleon interacting with both weak gauge bosons is one and the same. In a nucleus, however, this is not necessarily the case and additional nuclear-structure dependent corrections arise, noted as $\delta_{NS}$ \cite{Towner1992a}. This once again requires nuclear theory input, and has garnered significant attention following recent progress \cite{Seng2018, Seng2019, Czarnecki2019}. 

In summary, Eq.~(\ref{partial-t-1}) can be written as \cite{Severijns2008, Towner2010}
\begin{align} \label{eq:ft_general}
    1/t &= \frac{G_F^2 ~ V_{ud}^2}{K} ~ \left( 1 + \delta_R^{\prime} \right)   \nonumber \\
&\times \biggl( f_V |M_F^0|^2 \left( 1 + \delta_{NS}^V - \delta_C^V \right) g_V^2 \left( 1 + \Delta_R^V \right)   \nonumber \\
&+ f_A |M_{GT}|^2 \left( 1 + \delta_{NS}^A\right) g_A^2  \left( 1 + \Delta_R^A \right) \biggr),
\end{align}
where $f_V$ and $f_A$ are the statistical rate functions for, respectively, the vector (Fermi) and axial-vector (Gamow-Teller) parts of the transitions, and we write $M_{GT}$ as noted above. As denoted in Eq. (\ref{eq:f_def_integral}), these take into account higher-order nuclear corrections besides the main F/GT matrix elements in the shape factor, $C(Z, W)$. Traditionally, these corrections have been calculated in the elementary particle approach - i.e. a form factor decomposition encoding the full nuclear response - and a matrix element reduction using the impulse approximation (IA) which treats the nucleus as a sum of independent nucleons in a mean-field potential. While the results presented below are calculated in the Behrens-B\"uhring formalism \cite{Behrens1982}, we write the results in the Holstein formalism \cite{Holstein1974} for convenience. For comparisons between them we refer the reader to Refs. \cite{Behrens1978, Hayen2018, Hayen2020a}.

We can absorb all constant corrections and define the Gamow-Teller/Fermi mixing ratio as
\begin{align} \label{eq:rho}
\rho \equiv \frac{c}{a} &\stackrel{\text{IA}}{=} \frac{g_A\mathcal{F}_{GT}(0)}{g_VM_F^0} \left[ \frac{1+\Delta_R^A-\Delta_R^V}{1+\delta_{NS}^V-\delta_C^V}\right]^{1/2} \\
&=\frac{g_A M_{GT}^0}{g_V M_F^0} \left[ \frac{(1 +
	\delta_{NS}^A-\delta_C^A)(1 + \Delta_R^A)} {(1 + \delta_{NS}^V -
	\delta_C^V)(1 + \Delta_R^V)} \right]^{1/2} ~ . 
\end{align}
\noindent with $a$ and $c$, respectively, the Fermi and Gamow-Teller form factors. The first line contains a definition analogous to Ref. \cite{Hayen2020a}, with $\mathcal{F}_{GT}(0)$ the form factor evaluated at zero momentum, while the second has the more traditional notation. The benefit lies in the generality of Eq. (\ref{eq:rho}), meaning experimental analysis extracts the same quantity, whether it comes from a $\beta$-asymmetry, a $\beta$-$\nu$ correlation or another correlation. One can define the ``corrected" $ft$ value for the mixed Gamow-Teller/Fermi $T=1/2$ mirror $\beta$ transitions to be \cite{Severijns2008, Towner2010}
\begin{align} \label{master}
\mathcal{F} t^\mathrm{mirror} &\equiv f_V t (1 + \delta_R^{\prime}) (1+
\delta_{NS}^V - \delta_C^V ) \nonumber \\
&= \frac{2 \mathcal{F} t^{0^+
\rightarrow 0^+}} {|M_0^F|^2 \left( 1 + \frac{f_A}{f_V} \rho^2 \right)} ~ ,
\end{align}

where

\begin{align} \label{Fermi}
    {\mathcal F}t^{0^+ \rightarrow 0^+} &= \frac{K}{2G_F^2V_{ud}^2g_V^2(1+\Delta_R^V)}\nonumber \\
    &= 3072.24(185) s ~ ,
\end{align}
is the corrected ${\cal F}t$ value for the superallowed $0^+ \rightarrow 0^+$ pure Fermi transitions \cite{Hardy2020}.


The Gamow-Teller/Fermi mixing ratio, $\rho$, can to good approximation be written as \cite{Severijns2008}:
\begin{equation} \label{eq:rho1}
\rho  \cong \frac{g_A M_{GT}^0}{g_V M_F^0}  ~ , 
\end{equation}
\noindent where we use the fact that the small radiative and isospin symmetry-breaking corrections do not differ much for the vector and axial-vector parts of the transition \cite{Severijns2008}. The radiative corrections $\Delta_R^V$ and $\Delta_R^A$ were recently shown to indeed differ only at the $10^{-4}$ level or even less \cite{Hayen2021, Gorchtein2021}. As the vector coupling constant $g_V = 1$ and $|M_F^0|^2 = 1$ for these $T = 1/2$ mirror $\beta$ transitions, one has
\begin{equation} \label{eq:rho2}
\rho \simeq g_A M_{GT}^0 ~ . 
\end{equation}

Before moving on, we comment on the evaluation of $f_A/f_V$ values in relation to recent work \cite{Hayen2019c, HayenGTRC}. It was found that a double-counting instance occurs when combining results into Eq. (\ref{master}) which were typically calculated in different formalisms. Specifically, the expressions used for $f_A$ and $f_V$ were obtained in the Behrens-B\"uhring formalism, which was observed to contain parts of the Gamow-Teller-specific inner radiative correction $\Delta_R^A$ (and missing cancellations from the full $\mathcal{O}(\alpha)$ calculation) \cite{HayenGTRC, Hayen2021}. Experimental results for $\rho$, on the other hand, were typically obtained using (simplified) results in the Holstein formalism \cite{Holstein1974}, which do not contain this contribution and so return a fully renormalized mixing ratio. When combining it into Eq. (\ref{master}) then, the electroweak renormalization was partially double-counted. Removing this issue brings significantly improved internal consistency in the $V_{ud}$ mirror data set and additional lowered uncertainty. We will come back to this point below when reporting on $f_A/f_V$ for the data set presented here.

In the next section, a detailed survey of the data leading to the ${\cal F}t$ values for the $T = 1/2$ mirror $\beta$ transitions is made. In addition to bringing the previously published results for the mirror transitions up to $^{45}$V \cite{Severijns2008} up to date, we are now also incorporating existing data on the heavier $T = 1/2$ mirror $\beta$ transitions, leading to the ${\cal F}t$ values for the transitions up to $^{75}$Sr.

\subsection{Experimental data} \label{data-Ft-mirrors}

\subsubsection{Data selection}

We considered all measurements published up to April 2021 as well as any additional final results (available in preprint or as a PhD thesis) we were aware of. Since the previous survey \cite{Severijns2008} about 45 new half-life measurements and four new branching ratios have been reported, for almost 30 different isotopes. As in our previous survey \cite{Severijns2008} the $Q_{EC}$ values have again been taken from the most recent Atomic Mass Evaluation (AME2020) \cite{AME2020, AME2020-1, AME2020-2}. Since the Atomic Mass Evaluation 2003 (AME2003) \cite{Audi2003}, which was used in \cite{Severijns2008}, the precision with which the $Q_{EC}$ value is known has been (significantly) improved for 11 out of the 19 mirror $\beta$ transitions discussed in \cite{Severijns2008}.

In the tables and throughout this work, ``error bars" and ``uncertainties" always refer to plus/minus one standard deviation (68\% confidence level). In analyzing the tabulated data we have followed the statistical procedures used by the Particle Data Group (e.g. ref.~\cite{Olive2014}) that were adopted also for the surveys of the superallowed $0^+ \rightarrow 0^+$ $\beta$ decays (e.g. \cite{Hardy2005, Hardy2015}) as well as for the previous survey \cite{Severijns2008}. Thus, the one standard deviation error bars on weighted averages were increased by a scaling factor 
\begin{equation} \label{eq:scale}
S = \left[ \chi^2 / (N-1) \right]^{1/2} ~ ,
\end{equation}
with $\chi^2/(N-1)$ the reduced $\chi^2$. Further, data that were clearly inconsistent and results with an uncertainty that is a factor 10 or more larger than that of the most precise measurement for each quantity were rejected. Also results that deviate 4 or more standard deviations from the weighted average of mutually consistent results from other measurements (which are in most cases more recent and also more precise), were not considered. For the latter cases, the average values that would be obtained when the deviating results are not omitted, are given as well. These all agree within error bars with the finally adopted value, but have of course larger error bars and also a larger scaling factor, $S$, than the adopted value. 

All data that are now rejected on top of the ones that had already been rejected in the initial survey \cite{Severijns2008} are listed in Tables~\ref{unusedrefs} and \ref{unusedrefsBR}. As in e.g. Ref.~\cite{Hardy2005} an alphanumeric code comprising the initial two letters of the first author's name and the two last digits of the publication date is used to link each datum appearing in the tables to its original journal reference. The reference key linking these alphanumeric codes with the actual reference numbers is given in Table~\ref{reference-key-2}.

Table \ref{table:halflives_mirrors} shows the collected half-lives of the considered isotopes. A breakdown of the branching ratios is collected in Table \ref{table:branchings_mirrors}.  In comparison to the previous compilation \cite{Severijns2008} a more detailed analysis of the branching ratios was now performed. It includes all known branchings to excited states in the daughter isotopes of the mirror nuclei. Because of this, some branching ratios for the mirror $\beta$ transitions have slightly shifted in central value or some error bars have slightly increased. Of course, the more complete analysis that was performed renders the newly adopted values more reliable.

\newpage

{ \footnotesize

\begin{longtable*}{ l | r@{ }c@{ }l @{$\,$} l @{  } r@{ }c@{ }l @{$\,$} l   r@{ }c@{ }l @{$\,$} l   | r@{ }c@{ }l  c }

\caption{Half-lives, $t_{1/2}$, of the mirror nuclei from $^3$H to $^{39}$Ca, expressed in seconds unless specified differently under the name of the parent nucleus (days (d), minutes (min)). References to data listed in this table are given in Table~\ref{reference-key-2}. References to data that were not used are listed in Table~\ref{unusedrefs} together with the reason for their rejection. The scale factor $S$ listed in the last column is defined in Eq.~\ref{eq:scale}.}
\label{table:halflives_mirrors}\\

\hline Parent &    \multicolumn{12}{c|}{Measured half-lives, $t_{1/2}$ (s) } &  \multicolumn{3}{c}{Average half-life}   & scale        \\
\cline{2-17}
Nucleus&   \multicolumn{4}{c}{1}   &      \multicolumn{4}{c}{2}  &  \multicolumn{4}{c|}{3}      &     \multicolumn{3}{c}{$t_{1/2}$ (s)} &  $S$            \\
\hline
\endfirsthead

\multicolumn{17}{c}%
{ \tablename\ \thetable{}. (\textit{Continued})} \\
\hline Parent &    \multicolumn{12}{c|}{Measured half-lives, $t_{1/2}$ (s) } &  \multicolumn{3}{c}{Average half-life}   & scale        \\
\cline{2-17}
Nucleus&   \multicolumn{4}{c}{1}   &      \multicolumn{4}{c}{2}  &  \multicolumn{4}{c|}{3}      &     \multicolumn{3}{c}{$t_{1/2}$ (s)} &  $S$            \\
\hline
\endhead

\hline \hline
\endfoot

\hline \hline
\endlastfoot


   $^{3}$H &     $4419$ &      $\pm$ &    $183$   &     [No47] &     $4551$ &      $\pm$ &     $54$   &     [Je50] &     $4530$ &      $\pm$ &     $27$   &     [Jo51] &            &            &            &            \\
   (d)     &     $4479$ &      $\pm$ &     $11$   &     [Jo55] &     $4596$ &      $\pm$ &     $66$   &     [Po58] &     $4496$ &      $\pm$ &     $16$   &     [Me66] &     &       &     &         \\
      &  $4474$ &      $\pm$ &     $11$   &     [Jo67] &     $4501$ &      $\pm$ &      $9$   &     [Ru77] &   $4498$ &      $\pm$ &     $11$   &     [Si87] &           &            &            &            \\
       &  $4521$ &      $\pm$ &     $11$   &     [Ol87] &     $4485$ &      $\pm$ &     $12$   &     [Ak88] &     $4497$ &      $\pm$ &     $11$   &     [Bu91] &            &            &            &            \\
       &  $4504$ &      $\pm$ &      $9$   &     [Un00] &     $4500$ &      $\pm$ &      $8$   &     [Lu00] &     $4479$ &      $\pm$ &      $7$   &     [Ak04] &          &            &            &            \\
    &  $4497$ &      $\pm$ &      $4$   &     [Ma06] &     &       &    &    &      &       &      &     &  $4497$ &      $\pm$ & $4$ d\footnote{We did not perform the analysis of the tritium half-lives ourselves, but rather used the value (and the references) from [Ma06]. An interesting effect is mentioned in [Ak04]; the half-life of molecular and atomic $^{3}$H would differ by about 9 days. Due to a lack of additional information on this effect we have not included it in the present compilation. All measurements, except for [Ak04], have been performed on molecular tritium.} &         \\ 
\hline 
 
  $^{11}$C &  $20.34$ &      $\pm$ &   $0.04$   &    [Ka64a] &    $20.40$ &      $\pm$ &   $0.04$   &     [Aw69] &  $20.38$ &      $\pm$ &   $0.02$   &     [Az75] &           &            &            &            \\
  (min) &   $20.334$ &      $\pm$ &  $0.024$   &     [Wo02] &    $20.3378$ &    $\pm$ &   $0.0043$   &    [Va18] &     &       &      &      &    $20.3401$ &      $\pm$ &   $0.0053$ min &        1.3 \\
\hline
 
   $^{13}$N &     $9.96$ &      $\pm$ &   $0.03$   &     [Ar58] &    $9.965$ &      $\pm$ &  $0.005$   &     [Ja60] &     $9.93$ &      $\pm$ &   $0.05$   &     [Ki60] &          &            &            &            \\
     (min)   &   $10.05$ &      $\pm$ &   $0.05$   &     [Bo65] &     $9.96$ &      $\pm$ &   $0.02$   &     [Eb65] &    $9.963$ &      $\pm$ &  $0.009$   &     [Ri68] &          &            &            &            \\
     &  $9.965$ &      $\pm$ &  $0.010$   &     [Az77] &            &            &            &            &           &            &            &            &   $9.9647$ &      $\pm$ &  $0.0039$ min &          1 \\
\hline
 
  $^{15}$O &   $122.1$ &      $\pm$ &       $0.1$ &     [Ja60] &   $122.23$ &      $\pm$  &     $0.23$ &     [Az77]  &   
 $122.308$ &   $\pm$   &   $0.049$   &   [Bu20]    &    $122.27$ &      $\pm$ &      $0.06$ &        1.3 \\
\hline
 
  $^{17}$F &     $64.50$ &      $\pm$ &     $0.25$ &     [Al72] &    $64.31$ &      $\pm$ &     $0.09$ &     [Az77] &    $64.347$ &      $\pm$ &    $0.035$ &     [Gr15a]   &            &            &            &            \\
     &  $64.402$ &      $\pm$ &  $0.042$   &     [Br16] &            &            &            &            &           &            &            &            &   $64.366$  &   $\pm$   &   $0.026$\footnote{If the results from [Wo69] and [Al77] (see Table~\ref{unusedrefs}) are included one has $t_{1/2} = 64.411 \pm 0.068$ s (S = 2.8).}  &    1   \\
\hline
 
$^{19}$Ne &   $17.219$ &      $\pm$ &    $0.017$ &     [Az75] &   $17.237$ &      $\pm$ &    $0.014$ &     [Pi85] &   $17.262$ &    $\pm$   &   $0.007$  &  [Tr12]  &            &            &           \\    &   $17.254$ &   $\pm$    &   $0.005$ &     [Uj13]   &   $17.2832$  &    $\pm$   &   $0.0083$ &   [Br14] &   $17.2569$  &    $\pm$   &   $0.0021$ &   [Fo17] &    $17.2573$ &      $\pm$ &     $0.0034$\footnote{Note that $t_{1/2} = 17.2561 \pm 0.0025$ s (S = 1.4) when excluding [Br14] (which differs by 3.1$\sigma$ from the weighted average), while $t_{1/2} = 17.2569 \pm 0.0019$ s (S = 1) when considering only the three most precise results, i.e. [Tr12], [Uj13], and [Fo17].}  &   1.9 \\
\hline

 $^{21}$Na &   $22.422$   &   $\pm$     &            $0.010$   &    [Gr15]  &    $22.4506$ &      $\pm$ &     $0.0033$ &   [Fi17] &    $22.4615$ &      $\pm$ &     $0.0042$ &   [Sh18] &  $22.4527$ &      $\pm$ &     $0.0067$ &          2.7 \\
\hline

$^{23}$Mg &     $11.327$ &    $\pm$ &    $0.014$ &     [Az75] &   $11.317$ &      $\pm$ &    $0.011$ &     [Az77] &  $11.3027$  &  $\pm$  &  $0.0033$  & [Ma17] &    $11.3050$ &      $\pm$ & $0.0044$ &        1.4 \\
\hline

 $^{25}$Al &     $7.23$ &      $\pm$ &     $0.02$ &     [Ju71] &    $7.177$ &      $\pm$ &    $0.023$ &     [Ta73] &   $7.174$ &      $\pm$ &    $0.007$ &     [Az75] &         &            &            &            \\
    &   $7.1657$ &      $\pm$ &    $0.0024$ &     [Lo17] &   &      &     &   &    &    &    &    &   $7.1674$ &      $\pm$ &    $0.0044$ &        1.9 \\
\hline

 $^{27}$Si &     $4.109$ &      $\pm$ &    $0.004$ &     [Az75]  &  $4.1117$ &      $\pm$ &     $0.0020$ &     [Ma17]  &    &    &      &    &      $4.1112$ &      $\pm$ &     $0.0018$  &    1 \\
\hline

  $^{29}$P &     $4.15$ &      $\pm$ &     $0.03$ &     [Sc70] &     $4.083$ &      $\pm$ &     $0.012$ &     [Az75] &    $4.084$ &      $\pm$ &    $0.022$ &     [Wi80] &        &            &            &            \\
    &   $4.1055$ &      $\pm$ &     $0.0044$ &     [Lo20] &          &          &         &         &       &            &            &            &     $4.1031$ &      $\pm$ &     $0.0058$ &        1.4 \\
\hline

 $^{31}$S  &     $2.57$ &      $\pm$ &     $0.01$ &     [Ja60] &    $2.543$ &      $\pm$ &    $0.008$ &     [Az77] &    $2.562$ &      $\pm$ &    $0.007$ &     [Wi80]  &       &            &            &            \\
     &  $2.5534$  &    $\pm$  &   $0.0018$ & [Ba12]  &      &     &   &    &    &    &    &  &  $2.5539$ &      $\pm$ & $0.0023$\footnote{If the result from [Al74] (see Table~\ref{unusedrefs}) is not omitted the weighted average is $t_{1/2} = 2.5549 \pm 0.0040$ s (S = 2.4).}  &    1.4\\
\hline

$^{33}$Cl &     $2.53$ &      $\pm$ &     $0.02$ &     [Mu58] &     $2.51$ &      $\pm$ &     $0.02$ &     [Ja60] &     $2.47$ &      $\pm$ &     $0.02$ &     [Sc70] &        &            &            &            \\
   &   $2.513$ &      $\pm$ &    $0.004$ &     [Ta73] &    $2.507$ &      $\pm$ &    $0.008$ &     [Az77] &   $2.5038$   &  $\pm$   &  $0.0022$  &   [Gr15a]   &   $2.5059$ &      $\pm$ &    $0.0025$ &        1.3 \\
\hline

 $^{35}$Ar &     $1.79$ &      $\pm$ &     $0.01$ &     [Ja60] &    $1.770$ &      $\pm$ &    $0.006$ &     [Wi69] &    $1.774$ &      $\pm$ &    $0.003$ &     [Az77] &      &            &            &            \\
    &  $1.7754$ &      $\pm$ &   $0.0011$ &     [Ia06] &     &     &   &    &    &    &    &  &   $1.7752$ &      $\pm$ &   $0.0010$ &       1 \\
\hline

   $^{37}$K &      $1.223$ &      $\pm$ &    $0.008$ &     [Az77] &  $1.23651$ &  $\pm$     &    $0.00094$ &   [Sh14]   & $1.23635$ &  $\pm$     &    $0.00088$ &   [Ku17]   & $1.23634$ &      $\pm$ &    $0.00076$ &   1.2   \\
\hline

 $^{39}$Ca &    $0.860$ &      $\pm$ &    $0.005$ &     [Mi58] &    $0.873$ &      $\pm$ &    $0.008$ &     [Li60] &  $0.8604$ &      $\pm$ &   $0.0030$ &     [Al73] &     &            &            &            \\
   &   $0.8594$ &      $\pm$ &   $0.0016$ &     [Az77]  &   $0.8607$ &  $\pm$   &    $0.0010$   &   [Bl10]  &   &   &   &   &   $0.86046$ &      $\pm$ &    $0.00080$\footnote{If the result from [Kl54] (see Table~\ref{unusedrefs}) is not omitted the weighted average is $t_{1/2} = 0.861 \pm 0.049$ s (S = 1.9).}  &   1 \\
\hline

 $^{41}$Sc &    $0.628$ &      $\pm$ &    $0.014$ &     [Ja60] &    $0.596$ &      $\pm$ &    $0.006$ &     [Yo65] &   $0.5963$ &      $\pm$ &   $0.0017$ &     [Al73] &           &            &            &            \\
   &   $0.591$ &      $\pm$ &    $0.005$ &     [Ta73] &     &      &     &   &     &      &     &   &    $0.5962$ &      $\pm$ &   $0.0022$ &          1.4 \\
\hline

 $^{43}$Ti &     $0.528$ &      $\pm$ &    $0.003$ &     [Ja60] &     $0.56$ &      $\pm$ &     $0.02$ &     [Ja61] &  $0.50$ &      $\pm$ &     $0.02$ &     [Pl62] &          &            &            &            \\
   &   $0.49$ &      $\pm$ &     $0.01$ &     [Al67] &     $0.54$ &      $\pm$ &     $0.01$ &     [Va69] &    $0.509$ &      $\pm$ &    $0.005$ &     [Ho87] &  $0.5223$ &      $\pm$ & $0.0057$ &        2.4 \\
\hline

  $^{45}$V &    $0.539$ &      $\pm$ &    $0.018$ &     [Ho82] &   $0.5472$ &      $\pm$ &   $0.0053$ &     [Ha87] &            &            &            &            &  $0.5465$ &      $\pm$ &    $0.0051$ &          1 \\
\hline

 $^{47}$Cr &   $0.4600$ &      $\pm$ &   $0.0015$ &     [Ed77] &    $0.4720$ &      $\pm$ &   $0.0063$ &     [Ha87] & &  &  &  &   $0.4606$ &      $\pm$ &    $0.0027$\footnote{If the result from [Bu85] (see Table~\ref{unusedrefs}) is not omitted the weighted average is $t_{1/2} = 0.4616 \pm 0.0051$ s (S = 3.6).} &        1.9 \\
\hline

 $^{49}$Mn &    $0.384$ &      $\pm$ &    $0.017$ &     [Ha80] &   $0.3817$ &      $\pm$ &   $0.0074$ &     [Ha87] &  $0.380$ &      $\pm$ &   $0.030$ &     [Ku17a]  &  $0.3820$ &      $\pm$ &    $0.0066$ &          1 \\
\hline

  $^{51}$Fe &    $0.310$ &      $\pm$ &    $0.005$ &     [Ay84] &   $0.3050$ &      $\pm$ &   $0.0043$ & [Ha87] &  $0.298$  &  $\pm$  &   $0.014$ &  [Ho89]  &  &  &  &  \\ 
  &  $0.301$  &  $\pm$  &   $0.004$ &  [Su13]  &  $0.308$ &      $\pm$ &    $0.005$ &     [Sh15] &  $0.288$ &      $\pm$ &   $0.006$ &     [Ku17a]  &   $0.3031$ &      $\pm$ &    $0.0029$ &    1.4 \\
\hline

 $^{53}$Co &  $0.240$ &      $\pm$ &    $0.020$ &     [Ho89] &   $0.240$ &  $\pm$ & $0.009$ & [Lo02]\footnote[7]{See also [Bl02].} &   $0.230$ &      $\pm$ &     $0.017$ &     [Su13] & & & &  \\
    &   $0.245$ &   $\pm$ &   $0.003$ & [Ku17a] &     &      &     &   &     &      &     &   &    $0.2440$ &      $\pm$ &   $0.0028$ &   1 \\
\hline

 $^{55}$Ni &    $0.189$ &    $\pm$ &    $0.005$ &     [Ho77] &    $0.208$ &      $\pm$ &    $0.005$ &     [Ay84] &   $0.2121$ &      $\pm$ &   $0.0038$ &     [Ha87] &        &            &            &             \\
   &   $0.204$ &      $\pm$ &    $0.003$ &     [Re99] &  $0.196$ &    $\pm$ &   $0.005$ &   [Lo02]\footnotemark[7] &  $0.203$ &    $\pm$ &    $0.002$ &     [Ku17a] &       $0.2032$ &      $\pm$ &    $0.0025$ &     1.8  \\
\hline

$^{57}$Cu &   $0.1994$ &    $\pm$ &   $0.0032$ &     [Sh89] &   $0.1963$ &      $\pm$ &   $0.0007$ &     [Se96] &   $0.195$ &      $\pm$ &   $0.004$ &     [Ku17a]      &   $0.19640$ &      $\pm$ &   $0.00067$ &          1 \\
\hline

  $^{59}$Zn &   $0.1820$ &      $\pm$ &   $0.0018$ &     [Ar84] &    $0.173$ &      $\pm$ &    $0.014$ &  [Lo02]\footnotemark[7] &  $0.174$ &      $\pm$ &   $0.002$ &     [Ku17a] &     $0.1784$ &      $\pm$ &   $0.0028$ &     2.1  \\
\hline

 $^{61}$Ga &     $0.150$ &      $\pm$ &     $0.030$ &     [Wi93] &    $0.168$ &      $\pm$ &    $0.003$ &     [We02] &    $0.148$ &      $\pm$ &    $0.019$ &  [Lo02]\footnotemark[7] &      &            &            &             \\
   &   $0.162$  &    $\pm$  &  $0.010$  & [Ro14]  &   $0.163$  &    $\pm$  &  $0.005$  & [Ku17a]   &    &      &  &    &  $0.1660$ &   $\pm$ &    $0.0025$ &          1 \\
\hline

 $^{63}$Ge &    $0.095$ &     $^+_-$ & $^{0.023}_{0.020}$ &     [Wi93]\footnote[8]{These asymmetric errors have been symmetrized for the analysis by using standard recommendations of the Particle Data Group.} &   $0.150$ &      $\pm$ &    $0.009$ & [Lo02]\footnotemark[7] &   $0.149$    &   $\pm$    &   $0.004$ &    [Ro14]  &  &     &  &   \\
 &    $0.156$ &     $\pm$ & $0.011$ &     [Ku17a] &    &     &  &   &    &     &  &    & $0.1485$ &      $\pm$ &    $0.0051$ &        1.5 \\
\hline

 $^{65}$As &     $0.126$ &      $\pm$ &    $0.016$ & [Lo02]\footnotemark[7] &   $0.126$  & $\pm$    &    $0.007$   &   [Ro14]  &  &  &   &   &  $0.1260$ &   $\pm$ &   $0.0064$  &   1 \\
\hline

 $^{67}$Se &    $0.107$ &      $\pm$ &    $0.035$ &     [Ba94] &    $0.136$ &      $\pm$ &    $0.012$ &  [Lo02]\footnotemark[7] &  $0.133$   &  $\pm$     &  $0.004$   &   [Ro14]   &  $0.1330$ &      $\pm$ &    $0.0038$ &        1 \\
\hline

 $^{69}$Br &   $<100$ ns &        &      &     [Bl95] & $<24$ ns  &        &      &     [Pf96]  & $<24$ ns  &        &      &     [Ja99] &    $<24$ ns\footnote[9]{Proton unbound.}  &       &     &           \\
 \hline

 $^{71}$Kr &   $0.100$ &      $\pm$ &    $0.003$ &     [Oi97] & $0.0988$ &    $\pm$ &   $0.0003$ &   [Si19] & &    &     &    &  $0.0988$ &     $\pm$ &    $0.0003$ &  1   \\
\hline

 $^{73}$Rb &   $<30$ ns  &        &      &     [Pf96]  &   $<30$ ns  &        &      &     [Ja99] &   $<81$ ns &        &      &     [Su17]   &    $<30$ ns\footnotemark[9]  &       &     &           \\
\hline

 $^{75}$Sr &   $ 0.088$ &      $\pm$ &    $0.003$ &     [Hu03] &    $ 0.0817$ &      $\pm$ &    $0.0034$ &     [Si19]   &            &            &            &            &    $0.0852$ &      $\pm$ &    $0.0031$ &     1.4      \\
\hline

  $^{77}$Y &    $0.057$ &     $^+_-$ & $^{0.022}_{0.012}$ &     [Ki01]\footnotemark[8],\footnote[10]{Also reported in [Fa02].} &         &         &         &       &        &            &            &            &    $0.065$ &      $\pm$ &    $0.017$ &           \\
\hline

 $^{79}$Zr &   $ 0.056$ &      $\pm$ &    $0.030$ &     [Bl99] &        &            &            &            &            &            &            &            &    $0.056$ &      $\pm$ &    $0.030$ &         \\
\hline

  $^{81}$Nb &   $<80$ ns  &        &      &     [Ja99] &   $<200$ ns &        &      &     [Ki01] &   $<38$ ns &        &      &     [Su13a] &    &    &     &   \\
 &    $<40$ ns &        &      &     [Su17a] &    &     &  &   &    &     &  &    & $<38$ ns\footnotemark[9] &    &    &    \\
 \hline
 
$^{83}$Mo &    $0.006$ &     $^+_-$ & $^{0.030}_{0.003}$ &     [Ki01]\footnotemark[8] &         &      &       &        &            &            &            &            &    $0.028$ &      $\pm$ &    $0.019$ &           \\
\hline

 $^{85}$Tc &   $<100$ ns  &        &      &     [Ja99] &   $<43$ ns &        &      &     [Ja99] &   $<42$ ns &        &      &     [Su13a] &    &    &     &   \\
 &    $<43$ ns &        &      &     [Su17a] &    &     &  &   &    &     &  &    & $<42$ ns\footnotemark[9] &    &    &    \\
\hline

 $^{89}$Rh &   $>500$ ns  &      &      &     [Ja99] &   $<120$ ns  &        &      &     [Ce16] &            &            &            &            &    $<120$ ns\footnotemark[9]  &       &     &           \\
\hline

 $^{93}$Ag &   $228$ & $\pm$ & $16$ ns&   [Ce16] &     &      &      &      &      &         &            &            &   $228$ & $\pm$ & $16$ ns&           \\
\hline

 $^{97}$In &   $>3~\mu$s  &        &      &     [Ce16] &      &        &      &      &            &            &            &            &    $>3~\mu$s  &       &     &           \\


\end{longtable*}

}



{ \footnotesize

\begin{longtable*}{ l | l@{ } l@{ }    l@{ } l@{ }   l@{ } l@{ }   l@{ } l@{ }   l@{ } l@{ }    | c }

\caption{Branching ratios, $\BR$, for the $T = 1/2$  mirror $\beta$ transitions up to A = 75. References to data listed here are given in Table \ref{reference-key-2}. References to rejected data are listed in Table \ref{unusedrefs}. The scale factor, $S$, is defined in the caption of Table~\ref{table:halflives_mirrors}. If branching ratios in literature were quoted relative to values that are to date known more accurately because of later measurements, the published values have been updated. When weighted averages of several input values are made the scale factor $S$ in Eq.~\ref{eq:scale} is also given.}
\label{table:branchings_mirrors}\\

\hline    & \multicolumn{10}{c|}{Measured branching ratios to excited states (\%)} &  {Ground state}   \\
\cline{2-11}
          & \multicolumn{2}{c}{1} & \multicolumn{2}{c}{2} & \multicolumn{2}{c}{3} & \multicolumn{2}{c}{4} & \multicolumn{2}{c|}{5} & {branching ratio}   \\
          &          &            &          &            &          &            &          &            &          &             & {$BR$ $(\%)$}   \\

\hline
\endfirsthead

\multicolumn{12}{c}%
{ \tablename\ \thetable{}. (\textit{Continued})} \\
\hline   & \multicolumn{10}{c|}{Measured branching ratios to excited states (\%)} &  {Ground state}   \\
\cline{2-11}
         & \multicolumn{2}{c}{1} & \multicolumn{2}{c}{2} & \multicolumn{2}{c}{3} & \multicolumn{2}{c}{4} & \multicolumn{2}{c|}{5} & {branching ratio}   \\
         &          &            &          &            &          &            &          &            &          &             & {$BR$ $(\%)$}   \\

\hline
\endhead

\hline \hline
\endfoot

\hline \hline
\endlastfoot

   $^{3}$H &         &        &        &          &         &          &          &          &          &         &      $100$ [Ti87] \\
\hline
  $^{11}$C &         &        &        &          &         &          &          &          &          &         &      $100$ [Aj75] \\
\hline
  $^{13}$N &         &        &        &          &         &          &          &          &          &         &      $100$ [Aj70] \\
\hline
  $^{15}$O &         &        &        &          &         &          &          &          &          &         &      $100$ [Aj70] \\
\hline
  $^{17}$F &         &        &        &          &         &          &          &          &          &         &      $100$ [Aj82] \\
\hline
 $^{19}$Ne & \textbf{1.55 MeV}   &            & \textbf{0.11 MeV}   &        &         &          &          &          &         &          &      \\
           & $0.0021(3)$& [Al76] & $0.012(2)$  & [Ad81] &          &          &          &          &         &         &       \\
           & $0.0023(3)$& [Ad83] & $0.0099(7)$ & [Re19] &          &          &          &          &         &         &    \\           
           & $0.0017(5)$& [Re19] &             &        &          &          &          &          &         &         &    \\
\cline{2-2} \cline{4-4}
           & \textbf{0.00212(20)}& (S=1) & \textbf{0.0101(7)} & (S=1) &          &          &          &          &         &         & $\textbf{99.9878(7)}$\\
\hline
 $^{21}$Na & \textbf{0.35 MeV} &        &    &         &         &          &          &         &         &         &       \\
           & $5.1(2)$& [Al74] &        &    &         &         &          &          &         &         &      \\
           & $4.2(2)$& [Az77] &        &    &         &         &          &          &         &         &        \\
           & $4.97(16)$& [Wi80] &        &    &         &         &          &          &         &         &        \\
           & $4.74(4)$& [Ia06] &        &    &         &         &          &          &         &         &      \\
\cline{2-2}
           & \textbf{4.746(77)}& (S=2.1)   &      &      &          &          &          &          &         &         & $\textbf{95.254(77)}$  \\
\hline
 $^{23}$Mg &\textbf{2.39 MeV}&  & \textbf{0.44 MeV}   &    &      &        &         &          &          &          &         \\
           & 0.0064(9) & [Ma74] & $9.1(5)$& [Ta60] &    &      &        &          &          &          &        \\
           &           &        & $8.6(3)$& [Go68a] &    &      &        &          &          &          &       \\
           &           &        & $9.1(4)$& [Al74] &    &      &        &          &          &          &        \\
           &           &        & $8.1(4)$& [Ma74] &    &      &        &          &          &          &          \\
           &           &        & $7.79(15)$& [Az77] &    &      &        &          &          &          &        \\
           &           &        & $7.805(81)$& [Ma17] &    &      &        &          &          &          &        \\
\cline{2-2} \cline{4-4}
           & \textbf{0.0064(9)} & & \textbf{7.91(14)} & (S=2.1) &  &      &          &          &          &          &  \textbf{92.08(14)}\footnote{A small correction because the 440 keV level is also fed in the decay of the 2390 keV level with a branching ratio of 0.0025(3)\%, lowering the g.s. branching ratio by this amount, is negligible at the present level of precision.}\\
\hline
 $^{25}$Al &  \textbf{1.611 MeV}  &    &   \textbf{0.975 MeV}     &         &         &        &        &       &    &    & \\
           &  $0.9(2)$  & [Ma69] & $0.036(17)$ & [Mo71] &          &          &         &         &   &    &       \\
           &  $0.7(2)$  & [Ba70] & $0.044(7)$ & [Ma76] &          &          &         &         &   &    &    \\
           &  $0.84(7)$ & [Ju71] & $0.040(14)$ & [Az77] &          &          &         &         &   &    &    \\
           &  $0.794(35)$ & [Az77] &          &          &          &          &         &         &   &    &    \\
           &  $0.776(28)$ & [Wi80] &          &          &          &          &         &         &    &    &   \\
\cline{2-2} \cline{4-4}
           &  \textbf{0.788(21)} & (S=1)  &   \textbf{0.042(6)}   &  (S=1)  &    &    &    &      &    &     &   $\textbf{99.170(22)}$ \footnote{A branching ratio of 0.006(4)\% to the level at 1.965 MeV is reported in [Az77] but is not retained in the ENSDF database.}\\
\hline
 $^{27}$Si & \textbf{2.98 MeV} &  & \textbf{2.73 MeV} &  & \textbf{2.21 MeV} &  & \textbf{1.01 MeV} &  &   &  &    \\
           & $0.0249(40)$ &[De71] & $0.0155(17)$ &[Ma74] & $0.10(2)$ &[Go64] & $0.0024(8)$ &[De71] &       &       &        \\
           & $0.0259(24)$& [Ma74] & $0.0142(10)$ &[Da85] & $0.18(5)$ &[Be71] & $0.0049(24)$ &[Ma74]&       &       &        \\
           & $0.0214(14)$& [Da85] &              &      & $0.15(7)$  &[De71] & $0.0050(8)$ &[Da85]  &       &       &       \\
           &              &      &              &      & $0.181(14)$ &[Ma74] &              &       &       &       &      \\
           &              &      &              &      & $0.148(7)$ &[Az77]  &              &       &       &       &         \\
           &              &      &              &      & $0.164(28)$ &[Ma17] &              &      &       &       &         \\
           &              &      &              &      &           &       &                &      &       &       &         \\
\cline{2-2} \cline{4-4} \cline{6-6} \cline{8-8}
           & \textbf{0.0227(14)}& (S=1.2) & \textbf{0.0145(9)}& (S=1) &\textbf{0.151(9)}& (S=1.5) &\textbf{0.0038(9)} & (S=1.7) &  &  & \textbf{99.808(9)}    \\
\hline
 $^{29}$P & \textbf{2.43 MeV}   &            & \textbf{1.27 MeV}   &        &         &          &          &          &         &          &      \\
           & $0.453(16)$  & [Wi80] & $0.8(2)$ & [Ro55] &          &          &          &          &         &         &       \\
           & $0.4(1)$     & [Lo02] & $1.2(2)$ & [Lo62] &          &          &          &          &         &         &    \\
           &              &        & $1.255(22)$ & [Wi80] &          &          &          &          &         &         &    \\
\cline{2-2} \cline{4-4}
           & $\textbf{0.455(14)}$& (S=1)  & $\textbf{1.249(35)}$ & (S=1.6)  &          &          &          &          &         &         & $\textbf{98.296(38)}$\\
\hline
 $^{31}$S  & \textbf{3.51 MeV} &  & \textbf{3.13 MeV} &  & \textbf{1.266 MeV} &  &         &          &         &         &     \\
           & $0.011(3)$   & [De71] & $0.032(5)$   & [De71] & $1.1(1)$  & [Ta60] &          &          &         &         &       \\
           & $0.0121(10)$ & [Wi80] & $0.0326(16)$ & [Wi80] & $0.8(4)$  & [De71] &          &          &         &         &    \\
           &              &        &              &        & $1.25(6)$  & [Al74] &          &          &         &         &       \\
           &              &        &              &        & $0.98(20)$  & [Az77] &          &          &         &         &    \\
           &              &        &              &        & $1.097(33)$  & [Wi80] &          &          &         &         &    \\
\cline{2-2} \cline{4-4} \cline{6-6}
           & $\textbf{0.0120(9)}$& (S=1) & $\textbf{0.0325(15)}$ & (S=1) & $\textbf{1.126(34)}$  & (S=1.3) &     &    &    &    & $\textbf{98.830(34)}$\\
\hline
 $^{33}$Cl  & \textbf{4.75 MeV} &  & \textbf{4.05 MeV} &  & \textbf{2.87 MeV} &  & \textbf{2.31 MeV} &  & \textbf{1.97 MeV} &  &     \\
           & $\textbf{0.00041(21)}$&[Wi80]& $\textbf{0.00047(15)}$&[Wi80]&$\textbf{0.443(58)}$ &[Wi80]&$\textbf{0.0353(48)}$ &[Wi80]&$\textbf{0.460(60)}$ &[Wi80]&          \\
\cline{2-11}
           & \textbf{0.84 MeV}   &      &        &    &         &         &          &          &         &         &      \\
           & $\textbf{0.479(64)}$ &[Wi80]&        &    &         &         &          &          &         &         & $\textbf{98.58(11)}$   \\

\hline
 $^{35}$Ar & \textbf{3.97 MeV} &      & \textbf{3.92 MeV} &      & \textbf{3.00 MeV} &        & \textbf{2.69 MeV} &      & \textbf{1.76 MeV} &      &    \\
           & $0.00744(87)$     &[Wi80]& $0.00790(52)$     &[Wi80]&$0.07(2)$          &[De71]  &$0.19(3)$          &[De71]&$0.25(7)$          &[Wi69]&        \\
           & $0.00763(74)$     &[Ad84]& $0.0104(42)$      &[Ad84]&$0.11(3)$          &[Ge71]  &$0.16(4)$          &[Ge71]&$0.234(13)$        &[Wi69]&        \\
           &                   &      &                   &      &$0.0901(33)$       &[Wi80]  &$0.1606(65)$       &[Wi80]&$0.22(3)$          &[De71]&       \\
           &              &      &              &      &$0.0932(56)$ &[Ad84]&$0.171(11)$&[Ad84]&$0.22(5)$ &[Ge71]&      \\
           &              &      &              &      &$0.084(6)$ &[Da85] &$0.160(9)$  &[Da85]&$0.272(10)$ &[Wi80]&         \\
           &              &      &              &      &           &       &            &      &$0.250(16)$ &[Ad84]&         \\
           &              &      &              &      &           &       &            &      &$0.260(11)$ &[Da85]&         \\
\cline{2-2} \cline{4-4} \cline{6-6} \cline{8-8} \cline{10-10}
           & $\textbf{0.00755(56)}$& (S=1) & $\textbf{0.00794(52)}$& (S=1) &$\textbf{0.0895(25)}$& (S=1) &$\textbf{0.1630(47)}$ & (S=1) &$\textbf{0.2571(68)}$ & (S=1.2) \\
\cline{2-11}
           & \textbf{1.22 MeV}   &      &        &    &         &         &          &          &         &         &          \\
           & $1.04(24)$ &[Wi69]&        &    &         &         &          &          &         &         &        \\
           & $1.223(46)$ &[Wi69]&        &    &         &         &          &          &         &         &             \\
           & $1.22(20)$ &[Ge71]&        &    &         &         &          &          &         &         &          \\
           & $1.228(30)$ &[Wi80]&        &    &         &         &          &          &         &         &            \\
           & $1.23(7)$ &[Ad84]&        &    &         &         &          &          &         &         &           \\
\cline{2-2}
           & $\textbf{1.225(9)}$ & (S=1) &        &    &         &         &          &          &         &         & $\textbf{98.250(12)}$  \\
\hline
 $^{37}$K  & \textbf{3.94 MeV} &  & \textbf{3.60 MeV} &  & \textbf{3.17 MeV} &  & \textbf{2.80 MeV} &  & \textbf{2.49 MeV} &  &     \\
           & $\textbf{0.00116(13)}$& [Ha97] & $\textbf{0.0224(12)}$& [Ha97]& $\textbf{0.0027(2)}$ & [Ha97]& $2.0(4) $ & [Ka64] & $\textbf{0.0029(4)}$ & [Ha97]&          \\
           &              &        &              &      &           &      & $2.22(21)$ & [Az77] &           &      &          \\
           &              &        &              &      &           &      & $2.07(11) $ & [Ha97] &           &      &          \\
           &              &        &              &      &           &      & $2.20(17)$ & [Ku17]\footnote{See also [Se17]} &           &      &          \\
\cline{8-8}
           &              &        &              &      &           &      & $\textbf{2.121(70)}$ & (S=1) &    &      &          \\
\cline{2-11}
           & \textbf{1.61 MeV} &       & \textbf{1.41 MeV} &       &         &         &         &          &          &         &   \\
           & \textbf{0.0025(20)}   & [Ha97]& \textbf{0.00422(75)}   & [Ha97]&         &         &         &          &          &         & $\textbf{97.843(70)}$   \\
\hline
 $^{39}$Ca & \textbf{0.35 MeV} &        &    &         &         &          &          &         &         &         &       \\
           & $0.00226(58)$& [Ad84] &        &    &         &         &          &          &         &         &      \\
           & $0.00250(27)$& [Ha94] &        &    &         &         &          &          &         &         &        \\
 \cline{2-2}
           & $\textbf{0.00246(24)}$& (S=1) &      &      &          &          &          &          &         &         & $\textbf{99.9975(32)}$\footnote{The error is obtained from the quadratic sum of the error bar on the BR to the state at 2.52 MeV and the upper limits for the BR to states at higher excitation energies reported in [Ha94].}  \\
\hline
 $^{41}$Sc & \textbf{2.96 MeV} &   & \textbf{2.58 MeV}   &        &         &          &          &          &         &          &      \\
           & \textbf{0.0139(14)} & [Wi80] & \textbf{0.0232(29)} & [Wi80] &          &          &          &          &         &         &  $\textbf{99.9629(52)}$\footnote{The error is obtained from the quadratic sum of the error bars on the BR to the states at 2.58 and 2.96 MeV and the upper limits for the BR to states at higher excitation energies reported in [Wi80].}\\
\hline
 $^{43}$Ti  & \textbf{3.63 MeV} &  & \textbf{3.26 MeV} &  & \textbf{2.76 MeV} &  & \textbf{2.46 MeV} &  & \textbf{2.34 MeV} &  &     \\
           & \textbf{0.016(4)}&[Ho87]& \textbf{0.011(3)}&[Ho87]&\textbf{0.20(3)} &[Ho87]&\textbf{0.92(13))} &[Ho87]&\textbf{0.39(6)} &[Ho87]&          \\
\cline{2-11}
           & \textbf{2.29 MeV} &  & \textbf{1.96 MeV} &  & \textbf{1.88 MeV} &  & \textbf{1.41 MeV} &  & \textbf{0.85 MeV} &  &     \\
           & \textbf{4.7(7)}&[Ho87]& \textbf{0.024(10)}&[Ho87]&\textbf{0.20(4)} &[Ho87]&\textbf{0.68(10)} &[Ho87]&\textbf{2.7(8)} &[Ho87]&  \textbf{90.2(11)}        \\
\hline
 $^{45}$V & \textbf{0.040 MeV} &        &    &         &         &          &          &         &         &         &       \\
           & \textbf{4.3(15)}\footnote{This value includes a correction for the internal conversion coefficient $\alpha$ = 0.223(10) for the 40 keV $\gamma$ ray to the ground state [Bu08].}& [Ho82] &        &    &         &         &          &          &         &         &   \textbf{95.7(15)}  \\
\hline
 $^{47}$Cr & \textbf{0.087 MeV} &        &    &         &         &          &          &         &         &         &       \\
           & \textbf{3.9(12)}\footnote{This value includes a correction for the internal conversion coefficient $\alpha$ = 0.041(4) for the 87 keV $\gamma$ ray to the ground state [Bu07].} & [Bu85] &        &    &         &         &          &          &         &         &   \textbf{96.1(12)}  \\
\hline
 $^{49}$Mn & \textbf{2.50 MeV}   &            & \textbf{0.27 MeV}   &        &         &          &          &          &         &          &      \\
           & \textbf{2.3(9)} & [Ho89] & $6.4(26)$ & [Ha80] &          &          &          &          &         &         &       \\
           &          &        & $5.8(26)$  & [Ho89] &          &          &          &          &         &         &      \\
 \cline{4-4}
           &          &       & \textbf{6.1(18)} & (S=1) &          &          &          &          &         &         & \textbf{91.6(20)}\\
\hline
 $^{51}$Fe  & \textbf{3.56 MeV} &  & \textbf{3.43 MeV} &  & \textbf{2.91 MeV} &  & \textbf{2.14 MeV} &  & \textbf{1.82 MeV} &  &     \\
           & \textbf{0.16(5)}&[Ho89]& \textbf{0.20(6)}&[Ho89]&\textbf{0.10(3)} &[Ho89]&\textbf{0.24(7)} &[Ho89]&\textbf{0.49(14)} &[Ho89]&          \\
\cline{2-11}
           & \textbf{0.237 MeV}   &      &        &    &         &         &          &          &         &         &      \\
           & \textbf{5.0(13)} &[Ay84]&        &    &         &         &          &          &         &         & \textbf{93.8(13)}   \\
\hline
 $^{53}$Co & \textbf{1.33 MeV} &        &    &         &         &          &          &         &         &         &       \\
           & \textbf{5.6(17)}& [Ho89] &        &    &         &         &          &          &         &         &   \textbf{94.4(17)}\\
\hline
 $^{55}$Ni  &         &        &        &    &         &         &          &          &         &         &   \textbf{100.0(20)}\footnote{Based on the fact that the first excited state which may be populated by allowed $\beta$ decay is near 2.5 MeV, making 100\% g.s. feeding very probable [Ju08]. A 2\% error bar was assumed.}  \\
\hline
 $^{57}$Cu & \textbf{3.01 MeV} &  & \textbf{2.44 MeV} &  & \textbf{1.11 MeV} &  & \textbf{0.77 MeV} &       &         &         &       \\
           & \textbf{0.35(4)}& [Se96] & \textbf{0.17(3)}& [Se96] & \textbf{8.60(60)}& [Se96] & \textbf{0.94(9)}& [Se96] &         &         &   \textbf{89.9(8)}  \\
\hline
 $^{59}$Zn & \textbf{high-E}   &    & \textbf{0.914 MeV} &    & \textbf{0.491 MeV} &          &          &          &         &          &      \\
           & $0.09(2)$ & [Ar84] & $4.8(6)$ & [Ar84] & $1.1(2)$ &  [Ar84] &          &          &         &         &       \\
           & $0.23(8)$ & [Ho81] & $5(2)$   & [Ho81] & $1.6(8)$ &  [Ho81] &          &          &         &         &    \\
\cline{2-2} \cline{4-4} \cline{6-6}
           & \textbf{0.098(33)}& (S=1.2) & \textbf{4.82(57)} & (S=1) & \textbf{1.13(19)} & (S=1) &    &      &     &    & \textbf{93.95(60)}\\
\hline
$^{61}$Ga & \textbf{0.94 MeV} &  & \textbf{0.76 MeV} &  & \textbf{0.42 MeV} &  & \textbf{0.12 MeV} &  &\textbf{0.088 MeV}  &          &       \\
& \textbf{0.68(18)} & [Zu15] & \textbf{0.86(24)} & [Zu15] & \textbf{1.3(4)} & [Zu15] & \textbf{1.2(4)} & [Zu15] & \textbf{1.9(6)} &  [Zu15]   &  \textbf{94.0(10)}\footnote{See also [We02].}  \\
\hline
$^{67}$Se & \textbf{high-E}\footnote[16]{$\beta$ decay to proton-decaying level(s).} &        &    &         &         &          &          &         &         &         &       \\
& \textbf{0.5(1)} & [Bl95] &        &    &         &         &          &          &         &         &  \textbf{99.5(1)}\footnote[17]{A level assumed to be at 0.35 MeV with a supposed 8\% branching ratio reported in [Ba94] was not observed in (HI,xn$\gamma$) reactions [Ju05].}\\
\hline

$^{71}$Kr\footnote[18]{See [Ab11] for more details} & \textbf{$\approx$ 4.25 MeV}\footnotemark[16]  &  & \textbf{0.407$?$ MeV}\footnote[19]{Level uncertain (see [Ab10])} &  & \textbf{0.207 MeV} &  &            &  &              &          &   \\
& $2.1(7)$\footnote[20]{The very different value of 5.2(6)\% reported in [Bl95] might be due to the presence of an isomer at $\approx 10$~keV [Fi05, Ab11].}    &      [Oi97] &          &           &  $15.8(14)$  &  [Oi97]   &    &     &      &        &  \textbf{82.1(16)\footnote[21]{Value for [Oi97].}}  \\
& $\approx 2$ &      [Fi05]\footnote[22]{No error bars given. Results in [Fi05] are based on a re-interpretation of the data in [Oi97].} &  $ \approx 15$  &  [Fi05]   &  $\approx 15$  &  [Fi05]   &    &     &      &        &  to \textbf{$\approx$ 68}\footnote[23]{Value for [Fi05]. See also [Ab11] and Sec.~\ref{Note on A>63}. Note that a minimum value of $68.1$\% is required in order to avoid a negative value for $\rho^2$ when solving Eq.~\ref{master}.} \\
\cline{2-2} \cline{4-4} \cline{6-6}
& $\textbf{2.1(7)}$ & & $\textbf{$\approx$ 15}$  &      & $\textbf{15.8(14)}$  &      &     &     &      &      &   \\

\hline

$^{75}$Sr & $\textbf{5.50 MeV}$\footnotemark[16]  &  & $\textbf{0.144 MeV}$ &    &    &          &          &         &         &         &       \\
& $6.5(33)$ & [Bl95] &  $5.2(11)$\footnote[24]{This value includes a correction for the internal conversion coefficient $\alpha = 0.15(10)$ for the 144 keV $\gamma$ ray to the ground state [Ne13].} & [Hu03] &         &         &          &          &         &         &   \\
& $5.2(9)$ & [Hu03] &        &    &         &         &          &          &         &         &   \\
\cline{2-2} \cline{4-4}
& $\textbf{5.3(9)}$& (S=1) & $\textbf{5.2(11)}$ &      &          &          &          &          &         &         & $\textbf{89.5(14)}$\\



\end{longtable*}

}


\begin{table*}
\caption{References from which results for half-lives have been rejected because the error bars are a factor 10 or more larger than that of the most precise measurement (top part) or other reasons (lower part). The correlation between the alphabetical reference code used here and the actual reference numbers is listed in Table~\ref{reference-key-2}.}
\label{unusedrefs}
\begin{ruledtabular}

\begin{tabular}{p{4cm} p{13cm} }

 & \\
 &  {\bf{ $<$Error bar more than 10 times larger$>$ } }   \\
 Parent nucleus   &    Value~[Reference]  \\
 \hline

 & \\


$^{11}$C~(min)  & $20.35(8)$~[Sm41]; $20.0(1)$~[Di51]; $20.74(10)$~[Ku53]; $20.26(10)$~[Ba55]; $20.8(2)$~[Pr57]; $20.11(13)$~[Ar58]; $20.32(12)$~[Be75] \\
$^{15}$O~(s)  & $123.95(50)$~[Pe57]; $124.1(5)$~[Ki59]; $122.6(10)$~[Ne63] \\
$^{19}$Ne~(s) & $17.43(6)$~[Ea62]; $17.36(6)$~[Go68]; $17.7(1)$~[Pe57]; $17.36(6)$~[Wi74] \\
$^{21}$Na~(s) & $23.0(2)$~[Ar58]; $22.55(10)$~[Al74]; $22.48(4)$~[Az75, Az77]\footnote{The value $t_{1/2} = 22.47(3)$~s reported in [Az75] was reanalyzed with the updated value being reported in [Az77].} \\
$^{23}$Mg~(s) & $11.36(4)$~[Al74]; $11.26(8)$~[Az74]; $11.41(5)$~[Go68]; $12.1(1)$~[Mi58] \\
$^{25}$Al~(s) & $7.24(3)$~[Mu58] \\
$^{27}$Si~(s) & $4.21(3)$~[Gr71]; $4.14(3)$~[Mi58]; $4.16(3)$~[Su62]; $4.19(2)$~[Bl68]; $4.09(2)$~[Ba77] \\
$^{31}$S~(s)  & $2.80(5)$~[Cl58]; $2.66(3)$~[Ha52]; $2.40(7)$~[Hu54]; $2.61(5)$~[Li60]; $2.72(2)$~[Mi58]; $2.58(6)$~[Wa60] \\
$^{37}$K~(s)  & $1.25(4)$~[Ka64]; $1.23(2)$~[Sc58]\\
$^{39}$Ca~(s) & $0.876(12)$~[Cl58]; $0.865^{+0.007}_{-0.017}$~[Ka68] \\
$^{43}$Ti~(s) & $0.58(4)$~[Sc48]; $0.40(5)$~[Va63] \\
$^{47}$Cr~(s) & $0.460(80)$~[Ku17a] \\
$^{53}$Co~(s) & $0.267(109)$~[Ha87] \\
$^{57}$Cu~(s) & $0.1994(32)$~[Sh84]; $0.183(17)$~[Lo02,Bl02] \\
$^{59}$Zn~(s) & $0.213(34)$~[Ro14] \\
$^{65}$As~(s) & $0.190(11)$~[Mo95]; $0.19^{+0.11}_{-0.07}$~[Wi93] \\
$^{71}$Kr~(s) & $0.097(9)$~[Ew81]; $0.100(3)$~[Oi97]; $0.092(9)$~[Ro14]; $0.083(48)$~[Lo02,Bl02]\\
$^{75}$Sr~(s) & $0.080^{+0.400}_{-0.040}$~[Ki01]\footnote{Also reported in [Fa02]}; $0.071^{+0.071}_{-0.024}$~[Bl95]  \\

\hline
\hline
& \\
    &   {\bf { $<$Other reasons for exclusion$>$ } }   \\
Parent nucleus [Refs.]  &   Explanation  \\
\hline

& \\

$^{17}$F [Al77], [Wo69]    &  The quoted half-lives (resp. 64.80(9) s and 65.2(2) s) are significantly higher (by resp. 4.8~$\sigma$ and 4.2~$\sigma$) than the average of the other, more recent and also more precise results. \\


$^{27}$Si[Ge76]    &  The quoted half-life (i.e. 4.206(8) s) was measured with a spectrometer and is significantly higher (by 12~$\sigma$) than the average of the other, mutually consistent results, as is also the case for the other half-life (for $^{24}$Na) that was obtained with the same apparatus. \\

$^{27}$Si[Go68]    &  The quoted half-life (i.e. 4.17(1) s) is significantly higher (by 5.9$\sigma$) than the average of the other, mostly more recent and also more precise and mutually consistent results. \\

$^{29}$P [Ja60], [Ta73]    &  Both quoted half-lives (resp. 4.19(02) s and 4.149(5) s) are significantly higher (by resp. 4.3~$\sigma$ and 9.2~$\sigma$) than the weighted average of the other results. \\

$^{31}$S [Al74]       &  The quoted half-life (i.e. 2.605(12) s) is significantly higher (by 4.3~$\sigma$) than the average of the other, mostly more recent and also mutually consistent results. \\

$^{39}$Ca [Kl54]      &  The larger value reported (i.e.~0.90(1)~s) is most probably due to a small contamination from $^{38}$K produced in the $^{40}$Ca($\gamma$,d)$^{38}$K reaction with a threshold below the maximum energy used for the $^{40}$Ca($\gamma$,n)$^{39}$Ca reaction. \\

$^{47}$Cr [Bu85]      &  The quoted half-life (i.e.~0.508(10) s) is significantly higher (by 4.7~$\sigma$) than the weighted average of the two other more precise, and mutually consistent results. \\

$^{53}$Co [Ko73]      &  The quoted half-life (i.e.~0.262(25) s) was obtained by a $\beta$-ray measurement possibly including the positrons from $^{53m}$Co [Ha87]. \\
$^{53}$Co, $^{55}$Ni, and $^{59}$Cu [Ru14]      &  The quoted half-lives of 0.2460(18) s, 0.2018(19) s, and 0.1733(33) s, respectively, were later replaced by the new values of 0.245(3) s, 0.203(2) s, and 0.174(2) s respectively [Ku17a], following a re-analysis of the same data \cite{Rubio2021}. \\

$^{63}$Ga [Sh93]      &  The half-lives from [Wi93] and [Sh93] listed in \cite{Severijns2008} were obtained from the same set of data. \\

$^{67}$Se and $^{71}$Kr [Bl95]      &  The quoted half-lives (i.e. $0.060^{+0.017}_{-0.011}$ s, resp. $0.064^{+0.008}_{-0.005}$ s) were  obtained with very low statistics. They are a factor of respectively about 2 and 1.5 lower than the other results and also differ by more than 5~$\sigma$ from the weighted average of these. \\


\end{tabular}
\end{ruledtabular}

\end{table*}

\begin{table*}
	\caption{References from which results for branching ratios have been rejected. The correlation between the alphabetical reference code used here and the actual reference numbers is listed in Table~\ref{reference-key-2}.}
	\label{unusedrefsBR}
	\begin{ruledtabular}
		
		\begin{tabular}{p{5cm} p{12cm} }
			
			& \\
			Parent nucleus [Reference(s)]   &    Reason for exclusion \\

			\hline

			& \\			
			{\bf $<$Branching ratios$>$}  &  \\
			
		    $^{19}$Ne [Sa93], $^{23}$Mg [Da85],    &  Error bars are a factor 10 or more larger than that of the most precise measurement. \\
		    $^{27}$Si [Go64], $^{33}$Cl [Ba70]   & \\
		    		
			$^{21}$Na [Ac07], [Ac10]   & The result from the priv. comm. [Ac07] in \cite{Severijns2008} has in the mean time been published as [Ac10] but it is not included here as the error bar is more than 10 times larger than the most precise measurement.  \\
			
			$^{29}$P [Az77]       & The branching ratios for $^{29}$P in [Az77] have an error bar that is more than 10 times larger than the one of the most precise result (level at 1.27 MeV), or deviates by more than 12~$\sigma$ from the weighted average of the other results in the literature (level at 2.43 MeV).  \\
			 					 & The branching ratio for $^{29}$P to the level at 2.43 MeV in $^{29}$Si reported in [Ro55] has an error bar that is 10 times larger than the one of the most precise result.  \\
			
			
			$^{35}$Ar [Az77], [Ha79]  & The branching ratios from [Az77] and [Ha79] were not considered as only three, resp. one, branch(es) were observed and the reported branching ratios differ significantly from the ones reported in other experiments. \\
			
			$^{37}$K [Ma76]  & The branching ratios from [Ma76] to the levels at 2.80 and 3.60 MeV were not considered as they are both about 4.2~$\sigma$ lower than the weighted average of other results in the literature. \\
			
			$^{57}$Cu [Sh84]  & The branching ratio result from [Sh84] was not considered as only the most intensive $\beta$ branch could be observed due to interference of the $\beta$ decays of $^{54g}$Co and $^{58}$Cu while, in addition, the lifetime for $^{57}$Cu reported in [Sh84] is significantly longer than more recent values in the literature. \\
			
			$^{61}$Ga [Oi99]  & The results reported in [Oi99] were not considered as this experiment suffered from strong $^{61}$Zn and $^{61}$Cu isobaric contaminations, resulting in a limited accuracy for the measurement of the $^{61}$Ga branching ratios [We02]. \\
			
			$^{67}$Se [Ba84]  & The results reported in [Ba84] were not considered as no error bars are given. \\
			
			$^{71}$Kr [Fi05]       & The branching ratios to the different levels in the daughter isotope $^{71}$Br as shown in Fig. 10 of [Fi05] are only estimated values based on the $\gamma$ spectra shown in [Oi97].  \\
			
			
		\end{tabular}
	\end{ruledtabular}
	
\end{table*}

\begin{table*}
\caption{Adopted values for the half-lives, $t_{1/2}$, the branching ratios, $\BR$, and the decay transition energies, $Q_{EC}$ for the $T = 1/2$
mirror $\beta$ transitions with $A \leq 83$ (i.e. $s$, $p$, $sd$, and $fp$ shells). Isotopes with half-lives in the range of 
$\mu s$ and below are not included here. The $Q_{EC}$ values are from the 2020 Atomic Mass Evaluation \cite{AME2020, AME2020-1, AME2020-2} 
unless otherwise indicated. Details on the calculation of the vector statistical rate factor, $f_V$, and
the electron-capture fraction, $P_{EC}$, are given in the text.} \label{adopted-t12-BR-QEC}
\begin{ruledtabular}

\begin{tabular}{ l   r@{ }c@{ }l@{ }     r@{ }c@{ }l@{ }    r@{ }c@{ }l@{ }  r@{ }c@{ }l@{ }   c@{\extracolsep{\fill} }    r@{ }c@{ }l@{ } }



Parent  &  \multicolumn{3}{c}{Half-life} & \multicolumn{3}{c}{Branching} & \multicolumn{3}{c}{$Q_{EC}$} & \multicolumn{3}{c}{$f_V$} &  {$P_{EC}$}  &  \multicolumn{3}{c}{Partial half-life}        \\
nucleus  &    \multicolumn{3}{c}{$t_{1/2}$(s)} &    \multicolumn{3}{c}{$BR(\%)$} &  \multicolumn{3}{c}{(keV)}   & \multicolumn{3}{c}{}  &   {$(\%)$}   & \multicolumn{3}{c}{$t (s)$}         \\

\hline
   &    &    &    &    &    &    &    &    &    &   &   &   &   &   &   &   \\

   $^{3}$H &  $38854(35)$ &  $\times$  &  $10^{4}$ &       \  &      $100$ &        \ &  $18.59202$ & $\pm$ & $0.00006$\footnote{From [My15] which is 17, respectively 24 times more precise than the two most precise previous results, i.e [Na06] resp. [Va93].} &   $2.864604(26)$  &  $\times$ &  $10^{-6}$ & N/A  &  $(38854 $ &   $\pm$ & $35 ) \times 10^4$ \\

  $^{11}$C &   $1220.41$ &      $\pm$ &   $0.32$  &     \  &    $100$ &      \ &     $1981.689$ &   $\pm$ &   $0.061$ &  $3.18289$ &   $\pm$ &    $0.00082$ &   $0.232$ & $1223.24$ &   $\pm$ &   $0.32$ \\

  $^{13}$N &   $597.88$ &      $\pm$ &    $0.23$  &        \ &   $100$ &     \ &    $2220.47$ &      $\pm$ &     $0.27$ &    $7.7143$ &  $\pm$ &  $0.0072$ &   $0.196$ & $599.05$ &   $\pm$ & $0.23$\\

  $^{15}$O &    $122.27$ &      $\pm$ &      $0.06$ &    \ &     $100$ &   \ &     $2754.18$ &      $\pm$ &      $0.49$  &   $35.496$ &  $\pm$ &  $0.043$ &  $0.100$ &  $122.392$ & $\pm$ & $0.060$  \\
 
  $^{17}$F &     $64.366$ &      $\pm$ &      $0.026$ & \ &   $100$ & \ &     $2760.47$ &      $\pm$ &     $0.25$  &   $35.208$ &  $\pm$ &    $0.022$ &  $0.146$ &  $64.460$ &   $\pm$ &   $0.026$ \\

 $^{19}$Ne &    $17.2573$ &      $\pm$ &     $0.0034$ &   $99.9878$   & $\pm$ &  $0.0007$  &  $3239.50$ &  $\pm$ & $0.16$  &   $98.650$ & $\pm$ &  $0.031$ &  $0.100$ &  $17.2767$ & $\pm$ & $0.0034$\\

 $^{21}$Na &    $22.4527$ &      $\pm$ &    $0.0067$ &    $95.254$   &      $\pm$ &   $0.077$  &   $3546.919$ &    $\pm$ &    $0.018$ &   $170.7580$ &  $\pm$ &  $0.0054$ &  $0.094$ &   $23.594$ &    $\pm$ &   $0.020$ \\

 $^{23}$Mg &   $11.3050$ &      $\pm$ &    $0.0044$ &   $92.08$   &      $\pm$ &   $0.14$  &   $4056.179$ &      $\pm$ &    $0.032$ &   $378.621$ &      $\pm$ &     $0.018$ &   $0.073$ &    $12.286$ &      $\pm$ &     $0.019$\\

 $^{25}$Al &    $7.1674$ &      $\pm$ &    $0.0044$ &   $99.170$  &  $\pm$ &   $0.022$  &  $4276.808$ &  $\pm$ &  $0.045$  &   $508.522$ & $\pm$ &  $0.032$ &   $0.079$ &  $7.2331$ &    $\pm$ &    $0.0047$ \\

 $^{27}$Si &     $4.1112$ &  $\pm$ & $0.0018$ &  $99.808$   & $\pm$ &  $0.009$  &  $4812.358$ &  $\pm$ &  $0.096$  &   $993.52$ &  $\pm$ & $0.11$ &  $0.065$  & $4.1218$ &  $\pm$ &  $0.0018$ \\

  $^{29}$P &     $4.1031$ &      $\pm$ &     $0.0058$ &    $98.296$   &      $\pm$ & $0.038$  &    $4942.23$ &      $\pm$ &      $0.36$ &   $1136.37$ &      $\pm$ &      $0.48$ &   $0.075$ &   $4.1774$ &   $\pm$ &     $0.0061$\\

  $^{31}$S &     $2.5539$ &      $\pm$ &  $0.0023$ &  $98.830$   & $\pm$ & $0.034$  & $5398.01$ & $\pm$ &  $0.23$  &   $1844.74$ &  $\pm$ &  $0.45$ &  $0.068$ & $2.5859$ &  $\pm$ & $0.0025$\\

 $^{33}$Cl &    $2.5059$ &  $\pm$ &    $0.0025$ &   $98.58$   &  $\pm$ & $0.11$ &   $5582.52$ & $\pm$ & $0.39$  &   $2189.71$ & $\pm$ & $0.86$ &   $0.075$ &  $2.5439$ &  $\pm$ &   $0.0038$ \\

 $^{35}$Ar &   $1.7752$ &   $\pm$ &   $0.0010$ &   $98.250$   &  $\pm$ & $0.012$  &   $5966.24$ &  $\pm$ &  $0.68$  &   $3122.0$ &  $\pm$ & $2.0$ &   $0.072$ &   $1.8081$ &  $\pm$ &   $0.0010$ \\

  $^{37}$K &    $1.23634$ &      $\pm$ &    $0.00076$ &   $97.843$   & $\pm$ & $0.070$  &   $6147.48$ &  $\pm$ &$0.23$  &   $3623.73$ &  $\pm$ &  $0.75$ &   $0.079$ & $1.2646$ &   $\pm$ &  $0.0012$\\

 $^{39}$Ca &    $0.86046$ &    $\pm$ &    $0.00080$ &  $99.9975$   & $\pm$ & $0.0032$  & $6524.49$ &  $\pm$ &  $0.60$  &  $4951.5$ & $\pm$ & $2.5$ &   $0.077$ &  $0.86114$ & $\pm$ &   $0.00080$ \\

 $^{41}$Sc &   $0.5962$ &   $\pm$ &   $0.0022$ &  $99.9629$ & $\pm$ & $0.0052$  &  $6495.55$ &  $\pm$ &  $0.16$  & $4745.45$ &  $\pm$ &  $0.65 $ &  $0.094$ &  $0.5970$ &  $\pm$ &   $0.0022$\\

 $^{43}$Ti &    $0.5223$ &      $\pm$ &    $0.0057$ & $90.2$   &  $\pm$ & $1.1$  &   $6872.6$ &  $\pm$ & $6.0$  &  $6364$ &   $\pm$ &  $31$ &   $0.092$ &   $0.5796$ & $\pm$ &   $0.0095$\\

  $^{45}$V &    $0.5465$ &      $\pm$ &    $0.0051$ &    $95.7$   &      $\pm$ &    $1.5$  &     $7123.82$ &      $\pm$ &       $0.21$ &   $7616.8$ &      $\pm$ &       $1.2$  & $0.096$ &   $0.572$ &  $\pm$ &    $0.010$  \\

  $^{47}$Cr &    $0.4606$ &      $\pm$ &      $0.0027$ &  $96.1$ &     $\pm$ & $1.2$ &     $7444.0$ &   $\pm$ &   $5.2$  &   $9521$ &   $\pm$ &   $36$ &   $0.097$ &  $0.4798$ & $\pm$ & $0.0066$  \\

  $^{49}$Mn &     $0.3820$ &      $\pm$ &      $0.0066$ & $91.6$ &   $\pm$ & $2.0$ &    $7712.43$ &   $\pm$ &   $0.23$  &   $11353.3$ &   $\pm$ &   $1.8$ &  $0.101$ &  $0.417$ &  $\pm$ &  $0.012$ \\

 $^{51}$Fe &    $0.3031$ &      $\pm$ &     $0.0029$ &   $93.8$ &   $\pm$ & $1.3$  &  $8054.0$ &  $\pm$ & $1.4$  &   $14123$ &  $\pm$ & $13$ &  $0.101$ &  $0.3235$ & $\pm$ & $0.0054$\\

 $^{53}$Co &    $0.2440$ &      $\pm$ &     $0.0028$ &    $94.4$ &     $\pm$ & $1.7$  &  $8288.11$ &   $\pm$ &   $0.44$ &   $16222.0$ &  $\pm$ &   $4.6$ &  $0.107$ &   $0.2588$ &  $\pm$ &  $0.0055$ \\

 $^{55}$Ni &   $0.2032$ &      $\pm$ &    $0.0025$ &   $100.0$ &  $\pm$ & $2.0$  &   $8694.04$ &  $\pm$ &   $0.58$ & $20642.3$ &   $\pm$ &   $7.4$ &  $0.105$ &   $0.2034$ &    $\pm$ &   $0.0048$\\

 $^{57}$Cu &    $0.19640$ &    $\pm$ &   $0.00067$ &  $89.9$ &   $\pm$ & $0.8$  & $8774.95$ &  $\pm$ &  $0.44$  &   $21374.2$ &  $\pm$ &  $5.7$ &   $0.117$ &  $0.2187$ &    $\pm$ &    $0.0021$ \\

 $^{59}$Zn &     $0.1784$ &  $\pm$ & $0.0028$ &  $93.95$ &     $\pm$ & $0.60$  &  $9142.78$ &  $\pm$ &  $0.60$  &   $26205.3$ &  $\pm$ &  $9.2$ &  $0.117$  & $0.1901$ &  $\pm$ &  $0.0032$ \\

 $^{61}$Ga &     $0.1660$ &   $\pm$ &   $0.0025$ &  $94.0$ &   $\pm$ & $1.0$  &   $9214$ &   $\pm$ &   $38$ &   $26914$ &   $\pm$ &  $597$ &   $0.131$ &  $0.1768$ &   $\pm$ &   $0.0033$\\

 $^{63}$Ge &     $0.1485$ &      $\pm$ &  $0.0051$ &  \ &   $-$ & \  & $9626$ & $\pm$ &  $37$  &   $33452$ & $\pm$ &  $688$ &  $0.129$ &  \ &   $-$ & \   \\

 $^{65}$As &    $0.1260$ &  $\pm$ &    $0.0064$ &   \ &   $-$ & \ &   $9541$ & $\pm$ & $85$  &   $31458$ & $\pm$ & $1517$ &   $0.151$ &    \ &   $-$ & \  \\

 $^{67}$Se &   $0.1330$ &   $\pm$ &   $0.0038$ &   $99.5$\footnote[2]{Upper limit. See Sec.~\ref{Note on A>63}.} &   $\pm$ &  $0.1$  &   $10007$ &  $\pm$ &  $67$  &   $39923$ &  $\pm$ &  $1435$ &   $0.146$ &   $0.1339$ &   $\pm$ &  $0.0038$\\

           &            &         &          &   $90.0$\footnote[3]{Assumed lower limit. See Sec.~\ref{Note on A>63}.} &   $\pm$ &  $1.5$  &             &        &        &        &   &    &           &   $0.1480$ &   $\pm$ &  $0.0049$\\

 $^{71}$Kr &    $0.0988$ &      $\pm$ &    $0.0003$ &  $82.1$\footnotemark[2] &  $\pm$  &  $1.6$  &   $10175$ &  $\pm$ & $129$  &   $42371$ &  $\pm$ & $2916$ &   $0.177$ & $0.1206$ &   $\pm$ &  $0.0024$\\

           &            &         &          &   $70.0$\footnotemark[3] &   $\pm$ &  $1.5$  &             &        &        &        &   &    &           &   $0.1414$ &   $\pm$ &  $0.0031$\\

 $^{75}$Sr &    $0.0852$ &    $\pm$ &    $0.0031$ &  $89.5$\footnotemark[2] &   $\pm$   & $1.4$  & $10600$ &  $\pm$ &  $220$  &  $51147$ &  $\pm$ &  $5841$ &   $0.197$ &  $0.0954$ & $\pm$ &   $0.0038$ \\

           &            &         &          &   $80.0$\footnotemark[3] &   $\pm$ &  $1.5$  &             &        &        &        &   &    &           &   $0.1067$ &   $\pm$ &  $0.0044$\\

 $^{77}$Y &   $0.065$ &    $\pm$ &   $0.017$ &  \ &   $-$ & \  &  $11365$ &  $\pm$ &  $203$  & $71486$ &  $\pm$ &  $5734$ &  $0.179$ &    \ &   $-$ & \ \\

 $^{79}$Zr &    $0.056$ &      $\pm$ &    $0.030$ & \ &   $-$ & \  &   $11033$ &  $\pm$ & $310$  &  $61468$ &  $\pm$ & $10023$ &   $0.217$ &   \ &   $-$ & \ \\

  $^{83}$Mo &    $0.028$ &      $\pm$ &    $0.019$ &   \ &   $-$ & \  &  $11273$ &  $\pm$ &  $432$ &   $67103$ &  $\pm$ &  $14649$  & $0.253
  $ &   \ &   $-$ & \  \\


\end{tabular}

\end{ruledtabular}

\end{table*}


\subsubsection{Adopted input values for the ${\mathcal F}t^\mathrm{mirror}$ values}
\label{adopted input for mirror Ft}

The adopted values for the half-lives, $t_{1/2}$, the branching ratios, $BR$, and the decay transtion energies, $Q_{EC}$, are listed in Table~\ref{adopted-t12-BR-QEC}, together with the deduced statistical rate function, $f_V$, the electron-capture fraction, $P_{EC}$, and the resulting partial half-lives, $t$ (Eq.~\ref{eq:partial-t}). Note that the isotopes listed in Table~\ref{table:halflives_mirrors} that are unbound to proton decay are not further considered. The statistical rate functions, $f_V$, listed in column 5 of Table~\ref{adopted-t12-BR-QEC} were calculated using the same code as in the survey of the superallowed pure Fermi transitions \cite{Hardy2015}. The basic methodology for the calculation is described in the Appendix to Ref.~\cite{Hardy2005}. They were evaluated with the $Q_{EC}$ values and their uncertainties listed in column 4 of Table~\ref{adopted-t12-BR-QEC}. The $P_{EC}$ values, finally, were calculated as outlined in Ref.~\cite{Hardy2005} using the tables of Bambynek et al. \cite{Bambynek1977} and Firestone \cite{Firestone1996}. No errors were assigned to these $P_{EC}$ values as they are expected to be accurate to a few parts in 100 \cite{Bambynek1977, Hardy2005} such that they do not contribute perceptibly to the overall uncertainties. All quantities listed in Table~\ref{adopted-t12-BR-QEC} finally allow obtaining the partial half-life, $t$ (Eq.~\ref{eq:partial-t}), for each $\beta$ transition considered. These values are listed in the last column of Table~\ref{adopted-t12-BR-QEC}.

\subsection{The ${\mathcal F}t^\mathrm{mirror}$ values} \label{Ft-values mirrors}

Having surveyed the experimental data we can now turn to the determination of the ${\cal F}t^\mathrm{mirror}$ values according to Eq.~(\ref{master}). To do this we must, however, still deal with the different correction terms. The values for the nucleus-dependent radiative correction $\delta_R^\prime$ have been calculated using the $Q_{EC}$ values in Table~\ref{adopted-t12-BR-QEC} following the same procedures as adopted for the superallowed $0^+ \rightarrow 0^+$ $\beta$ decays \cite{Towner2002} and are listed in column 4 of Table~\ref{input-data-and-calculated-quantities}. 
As to the nuclear-structure-dependent corrections $\delta_C^V - \delta_{NS}^V$ (column 5 in Table~\ref{input-data-and-calculated-quantities}), for the mirror $\beta$ decays from $^3$H to $^{45}$V the values previously reported in \cite{Severijns2008} were used. For the mirror decays from $^{47}$Cr to $^{75}$Sr estimates based on the systematic of the superallowed $0^+ \rightarrow 0^+$ $\beta$ decays \cite{Hardy2020} and the lower-mass mirror $\beta$ decays \cite{Severijns2008} were used, with generous error bars attached to them. In none of these cases the uncertainty on $\delta_C^V - \delta_{NS}^V$ is the dominant contribution to the total uncertainty on the ${\cal F}t^\mathrm{mirror}$ value.

\begin{figure}
	\includegraphics[width=0.48\textwidth]{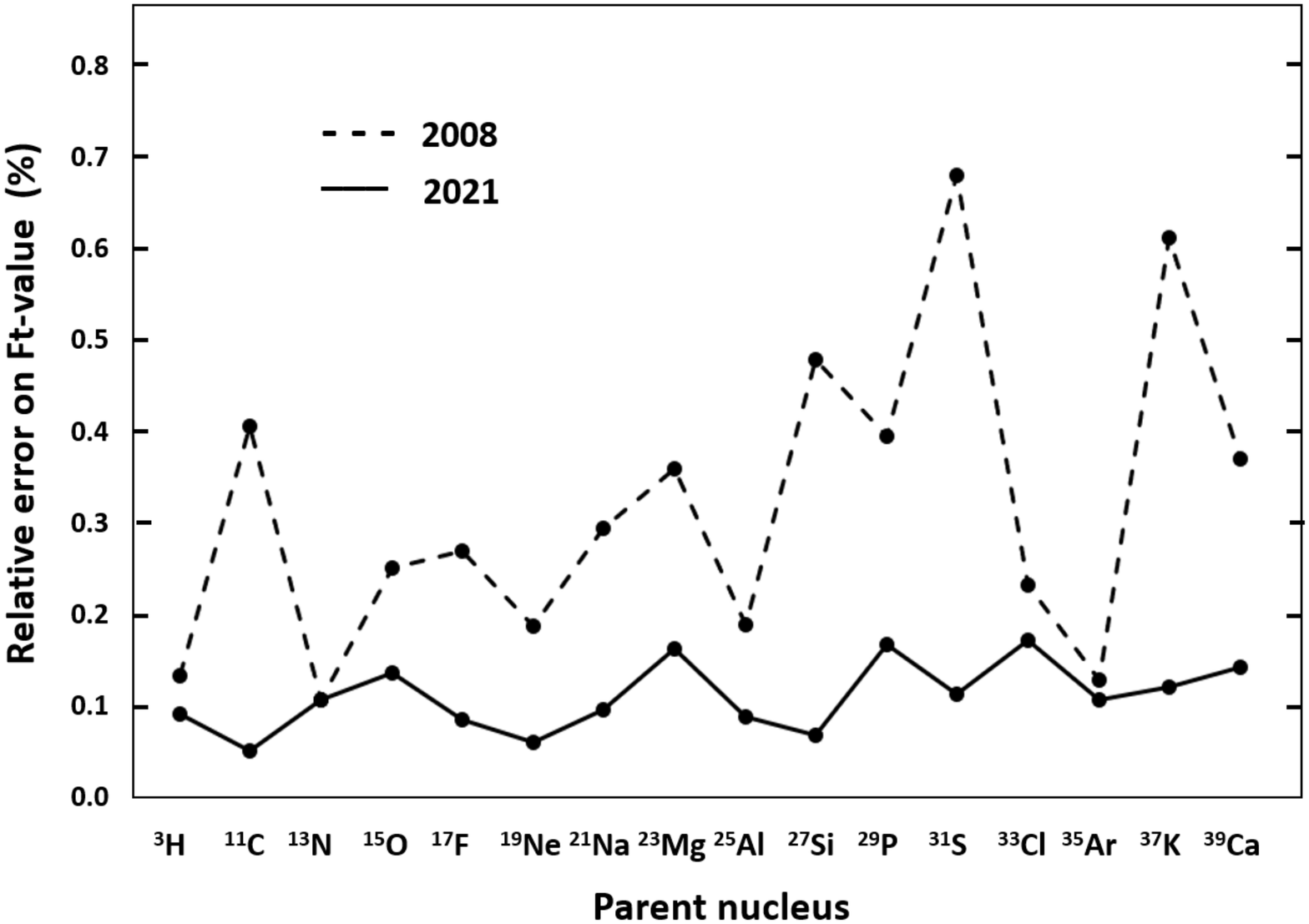}
	\caption{Progress in the relative uncertainties of the ${\mathcal F}t^\mathrm{mirror}$ values for the mirror $\beta$ transitions up to $A =39$ from 2008 \cite{Severijns2008} (dashed) till now (solid).}
	\label{fig:relative-uncertainty}
\end{figure}

\begin{figure}
	\centering
	\includegraphics[width=0.48\textwidth]{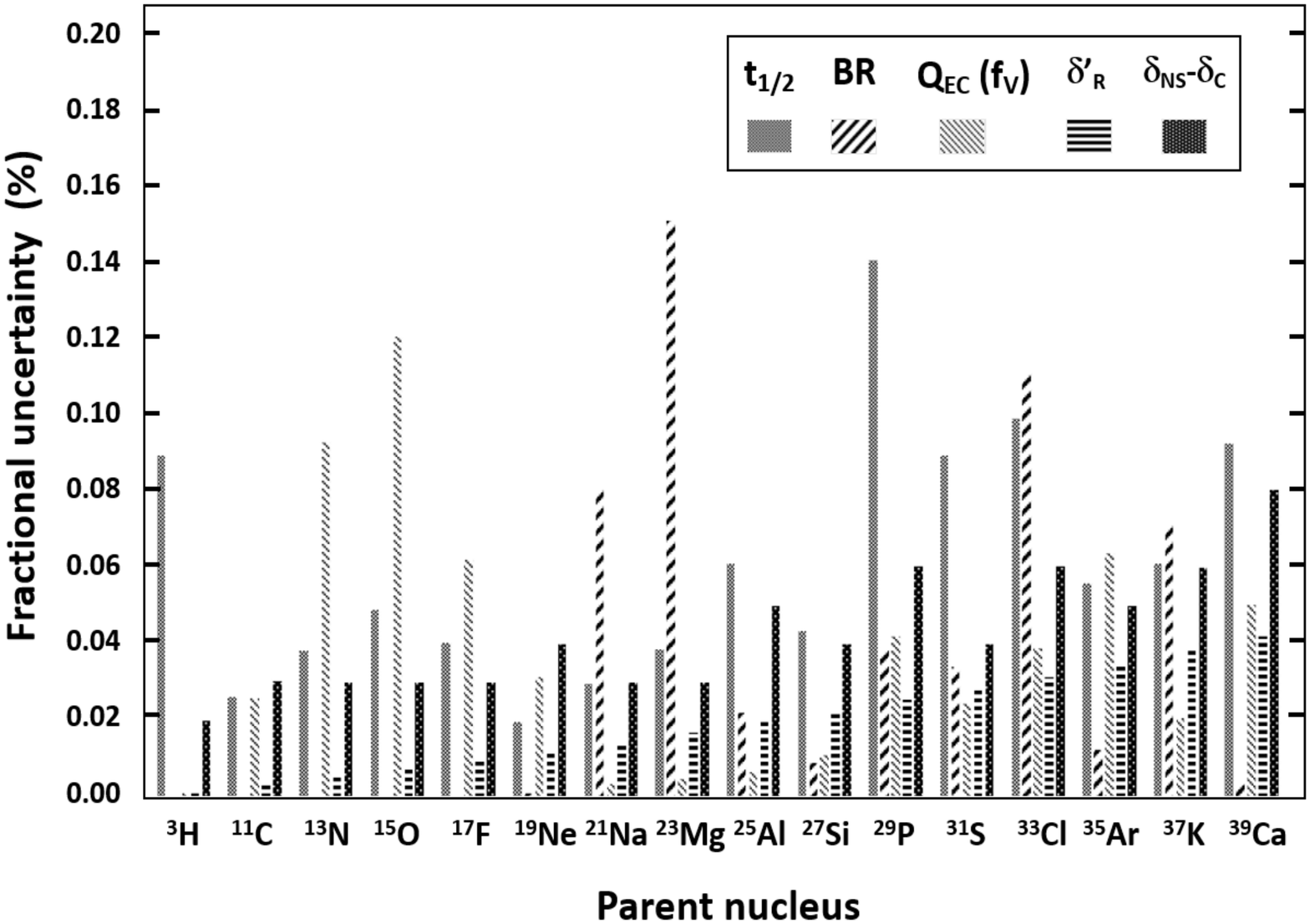}
	\caption{Fractional contribution of the experimental and theoretical input factors to the final ${\mathcal F}t^\mathrm{mirror}$ values for the transitions up to $A = 39$.}
	\label{fig:fractional-contribution-3-39}
\end{figure}

\begin{figure}
	\centering
	\includegraphics[width=0.48\textwidth]{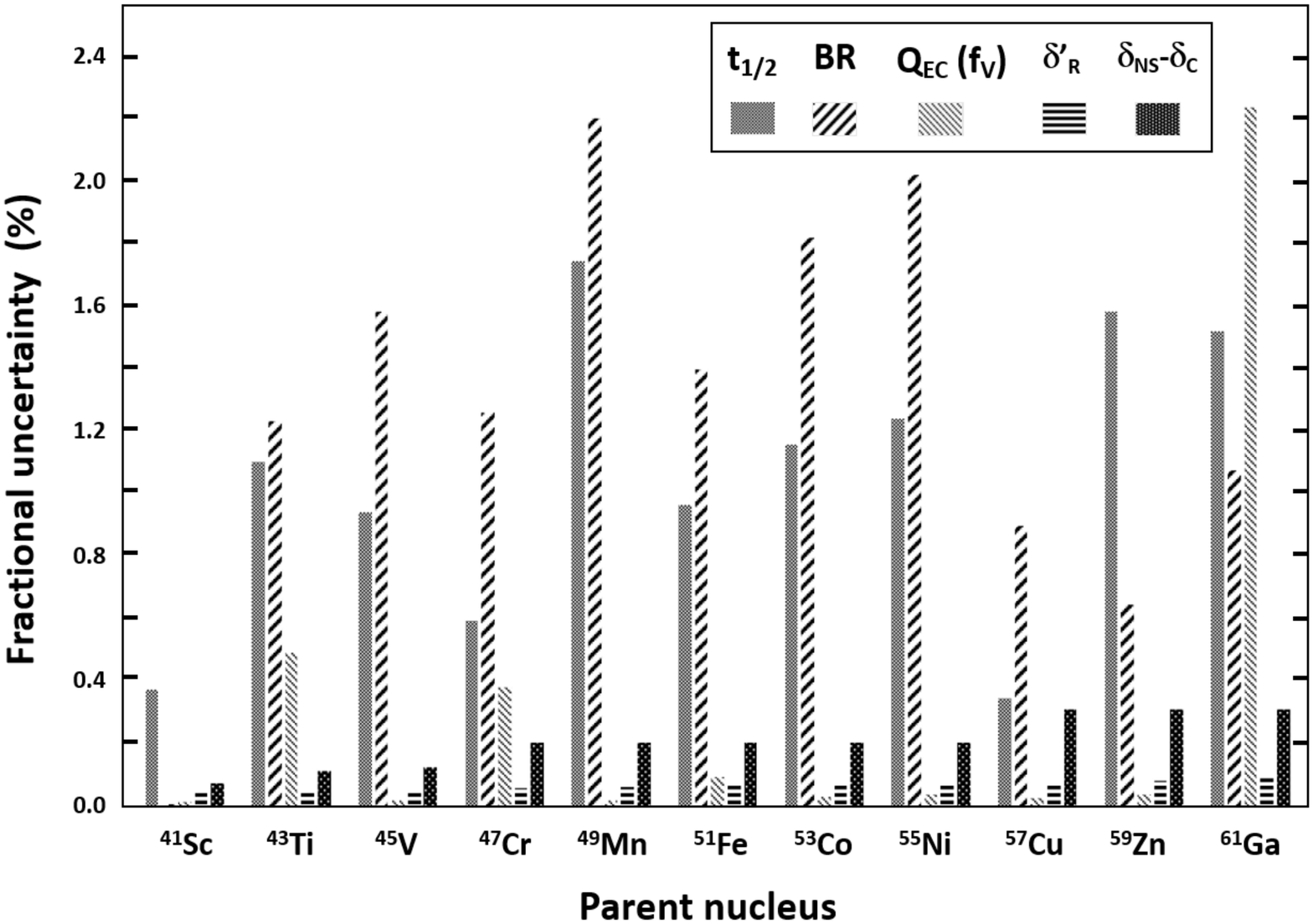}
	\caption{Fractional contribution of the experimental and theoretical input factors to the final ${\mathcal F}t^\mathrm{mirror}$ values for the transitions in the mass $A = 41-61$ range.}
	\label{fig:fractional-contribution-41-61}
\end{figure}

The resulting ${\cal F}t^\mathrm{mirror}$ values are listed in column 6 of Table~\ref{input-data-and-calculated-quantities}. Since the previous survey \cite{Severijns2008} the ${\mathcal F}t^\mathrm{mirror}$ values for all but three (i.e. $^{13}$N, $^{41}$Sc, and $^{43}$Ti) of the 19 mirror $\beta$ transitions with $A$ = 3 to 45 have been improved, with now a relative precision $\leq$ 0.2\% for all transitions (previously only 5) up to $^{39}$Ca, as can be seen in Fig.~\ref{fig:relative-uncertainty}, and precisions between 0.4\% and 2.9\% for the $fp$-shell mirror transitions up to $^{61}$Ga. Figures~\ref{fig:fractional-contribution-3-39} and \ref{fig:fractional-contribution-41-61} show the fractional uncertainties for each of the experimental and theoretical input factors to the ${\cal F}t^\mathrm{mirror}$ values for the $sp$-$sd$ shell transitions and the $fp$ shell transitions, respectively. Note the difference in the vertical scales between both figures.

\subsubsection{The Gamow-Teller/Fermi mixing ratio $\rho$}

From these $\mathcal{F}t^\mathrm{mirror}$ values the Gamow-Teller/Fermi mixing ratio, $\rho$, can be calculated (Eq.~\ref{master}). Comparing this with values obtained from e.g. beta-neutrino correlation or $\beta$ asymmetry parameter measurements allows testing the Conserved Vector Current hypothesis \cite{Naviliat2009} and extracting a value for the $V_{ud}$ quark-mixing matrix element \cite{Naviliat2009, Fenker2018, Karthein2019}. 

Extracting $\rho$ from Eq.~(\ref{master}) requires the ratio of statistical rate functions, $f_A/f_V$ to be known. The values for $f_A/f_V$ have been calculated earlier for the mirror $\beta$ decays from $^3$H to $^{45}$V \cite{Severijns2008}. However, it was recently pointed out \cite{Hayen2021, HayenGTRC} that a double-counting instance occurs when extracting the mixing ratio $\rho$ from experimental data and then using it in combination with Eq.~\ref{master}. 
The values for $f_A/f_V$ listed in Table~\ref{input-data-and-calculated-quantities} have therefore been calculated according to Ref.~\cite{Hayen2021} resulting in values that differ from the previous ones (e.g. \cite{Severijns2008}) by up to about 2.5\% for the mirror $\beta$ transitions in the $sd$-shell. The effect of this shift on the extraction of the mirror $\beta$ transition-independent $\mathcal{F}t_0$ value (see \cite{Severijns2008}) to test CVC and extract a value for the $V_{ud}$ quark-mixing matrix element is discussed in \cite{Hayen2021}. A shift of the central value towards the much more precise value obtained from the superallowed pure Fermi-transitions is observed, bringing the values obtained for both types of transitions in agreement with each other. 

To date only a limited number of mixing ratios for mirror $\beta$ transitions have been determined experimentally \cite{Calaprice1975, Garnett1988, Severijns1989, Masson1990, Converse1993, Melconian2007, Vetter2008, Fenker2018, Brown2018, Markisch2019, Combs2020}. With more measurements being planned and experimental precision further improving (a non-exhaustive list of ongoing and planned measurements with the precision that is aimed at is given in Table 3 of \cite{Gonzalez2019}) the value for the $V_{ud}$ quark-mixing matrix element extracted from the mirror $\beta$ transitions (including the neutron) provides a complementary approach to $V_{ud}$ compared to the superallowed pure Fermi transitions \cite{Hardy2020}.

\subsubsection{Note on the mirror transitions with $A >$61}
\label{Note on A>63}

For the mirror isotopes with $A > 61$ no detailed decay spectroscopy has been performed yet, as is indicated in Table~\ref{adopted-t12-BR-QEC} by a $"-"$ for several isotopes in the column listing the branching ratios, $BR$. For three isotopes, i.e. $^{67}$Se, $^{71}$Kr, and $^{75}$Sr, some information is available which is, however, likely incomplete as these isotopes are for now still difficult to produce copiously at radioactive beam facilities. For the mirror $\beta$ decays of these three isotopes we therefore list in Table~\ref{adopted-t12-BR-QEC} the upper limits on the branching ratios, $BR$, based on the currently available spectroscopic data as well as estimated values for the lower limits. 
Note that for $^{71}$Kr the lower limit of $BR = 68.1$\% is the minimum value that is required in order to avoid a negative value for $\rho^2$ when solving Eq.~(\ref{master}).

\begin{table*}
\caption{Calculated quantities and corrections needed to obtain the ${\cal F}t^{\mbox{mirror}}$ values (Eq.~(\ref{master})). Details are given in the text.}
\label{input-data-and-calculated-quantities}
\begin{ruledtabular}

\begin{tabular*}{0.99\textwidth}{ l@{ }   r@{ }  c@{ } l@{\extracolsep{\fill} }    c@{ \extracolsep{\fill}}   c@{ \extracolsep{\fill}}  c@{ \extracolsep{\fill}}   r@{}  c@{ } l@{ }   c@{ \extracolsep{\fill}}   l@{ \extracolsep{\fill}}  }
Parent & \multicolumn{3}{c}{$f_V t$} & $f_A/f_V$ & $\delta_R^\prime$ & $\delta^V_C - \delta^V_{NS}$  & \multicolumn{3}{c}{${\cal F}t^\mathrm{mirror}$} & $\delta({\cal F}t^\mathrm{mirror})$ & $\rho = c$ \\
nucleus  &   \multicolumn{3}{c}{$(s)$} &                    &    $(\%)$          &            $(\%)$             &  \multicolumn{3}{c}{$(s)$}   &    $\%$      & (Eq.~\ref{eq:rho2}) \\
\hline
   &     &     &     &      &      &      &      &      &   &    &    \\
   $^{3}$H &   $1113.0$ & $\pm$  &  $1.0$ & $1.00027$ &    $1.767(1)$ &    $0.16(2)$  &  $1130.9$ & $\pm$  &  $1.0$  &  0.09  &  $+2.1053(12)$  \\
   
  $^{11}$C &  $3893.4$  &  $\pm$  & $1.4$ & $0.99923$ &    $1.660(4)$ &    $1.04(3)$  &  $3916.9$ & $\pm$  &  $1.9$  &  0.05  &  $-0.75442(28)$  \\

  $^{13}$N &  $4621.3$  &  $\pm$  & $4.7$ &  $0.99802$ &    $1.635(6)$ &    $0.33(3)$  &  $4681.3$ & $\pm$  &  $4.9$   &  0.11  &  $-0.55962(34)$  \\

  $^{15}$O &  $4344.3$  &  $\pm$  & $5.7$ & $0.99637$ &    $1.555(8)$ &    $0.22(3)$  &  $4402.3$ & $\pm$  &  $5.9$   &  0.13  &  $+0.63023(46)$  \\

  $^{17}$F &  $2269.5$  &  $\pm$  & $1.7$ & $1.00196$ &    $1.587(10)$ &    $0.62(3)$  &  $2291.2$ & $\pm$  &  $1.9$   &  0.08  &  $+1.29555(65)$  \\

 $^{19}$Ne &  $1704.31$  &  $\pm$  & $0.63$ & $1.00110$ &    $1.533(12)$ &     $0.52(4)$  &  $1721.5$ & $\pm$  &  $1.0$   &  0.06  &  $-1.60203(65)$  \\

 $^{21}$Na &  $4028.8$  &  $\pm$  & $3.5$ & $1.00198$ &    $1.513(14)$ &    $0.41(3)$  &  $4073.0$ & $\pm$  &  $3.8$   &  0.09  &  $+0.71245(39)$   \\

 $^{23}$Mg &  $4651.9$  &  $\pm$  & $7.3$ & $0.99940$ &    $1.476(17)$ &    $0.40(3)$  &  $4701.6$ & $\pm$  &  $7.6$   &  0.16  &  $-0.55413(47)$   \\

 $^{25}$Al &  $3678.2$  &  $\pm$  & $2.4$ & $1.00193$ &    $1.475(20)$ &    $0.52(5)$  &  $3713.0$ & $\pm$  &  $3.2$   &  0.08  &  $+0.80844(42)$   \\

 $^{27}$Si &  $4095.1$  &  $\pm$  & $1.9$ & $1.00024$ &    $1.443(23)$ &    $0.42(4)$  &  $4136.7$ & $\pm$  &  $2.7$   &  0.07  &  $-0.69659(30)$   \\

  $^{29}$P &  $4747.0$  &  $\pm$  & $7.2$ & $1.00077$ &    $1.453(26)$ &    $1.07(6)$  &  $4764.5$ & $\pm$  &  $7.9$   &  0.17  &  $+0.53798(47)$   \\

  $^{31}$S &  $4770.3$  &  $\pm$  & $4.7$ & $0.99919$ &    $1.430(29)$ &     $0.79(4)$  &  $4800.3$ & $\pm$  &  $5.3$   &  0.11  &  $-0.52939(33)$   \\

 $^{33}$Cl &  $5570.4$  &  $\pm$  & $8.6$ & $0.98952$ &    $1.435(32)$ &    $0.93(6)$  &  $5597.8$ & $\pm$  &  $9.5$   &  0.17  &  $-0.31416(28)$   \\

 $^{35}$Ar &  $5645.0$  &  $\pm$  & $4.9$ & $0.99293$ &    $1.421(35)$ &    $0.53(5)$  &  $5694.8$ & $\pm$  &  $6.0$   &  0.11  &  $+0.28199(17)$  \\

  $^{37}$K &  $4582.5$  &  $\pm$  & $4.4$ & $0.99550$ &    $1.431(39)$ &    $0.79(6)$  &  $4611.4$ & $\pm$  &  $5.5$   &  0.12  &  $-0.57789(39)$   \\

 $^{39}$Ca &  $4264.0$  &  $\pm$  & $4.5$ & $0.99551$ &    $1.422(43)$ &     $0.95(8)$  &  $4283.5$ & $\pm$  &  $6.0$   &  0.14  &  $+0.66061(50)$   \\

 $^{41}$Sc &  $2833$  &  $\pm$  & $10$ & $1.00193$ &    $1.454(47)$ &    $0.86(7)$  &  $2849$ & $\pm$  &  $11$   &  0.38  &  $+1.0743(21)$   \\

 $^{43}$Ti &  $3688$  &  $\pm$  & $63$ & $0.99547$ &    $1.444(50)$ &     $0.63(11)$  &  $3718$ & $\pm$  &  $64$   &  1.7  &  $-0.8097(69)$   \\

  $^{45}$V &  $4354$  &  $\pm$  & $79$ & $1.00418$ &    $1.438(53)$ &     $0.93(12)$  &  $4375$ & $\pm$  &  $80$   &  1.8  &  $+0.6346(58)$   \\

  $^{47}$Cr &  $4568$  &  $\pm$  & $65$ & $1.00325$ &    $1.439(58)$ &    $0.8(2)$  &  $4596$ & $\pm$  &  $66$   &  1.4  &  $-0.5794(42)$  \\

  $^{49}$Mn &  $4739$  &  $\pm$  & $132$ & $0.99908$ &    $1.438(61)$ &    $0.8(2)$  &  $4769$ & $\pm$  &  $133$   &  2.8  &  $+0.5373(75)$  \\

  $^{51}$Fe &  $4568$  &  $\pm$  & $77$ & $0.99700$ &    $1.442(66)$ &    $0.8(2)$  &  $4597$ & $\pm$  &  $78$   &  1.7  &  $-0.5811(49)$  \\

 $^{53}$Co &  $4197$  &  $\pm$  & $90$ & $1.00385$ &    $1.443(70)$ &     $0.8(2)$  &  $4224$ & $\pm$  &  $91$   &  2.1  &  $+0.6730(72)$  \\

 $^{55}$Ni &  $4199$  &  $\pm$  & $99$ & $0.99650$ &    $1.433(73)$ &    $0.8(2)$  &  $4225$ & $\pm$  &  $100$   &  2.4  &  $-0.6752(80)$   \\

 $^{57}$Cu &  $4675$  &  $\pm$  & $45$ & $0.99118$ &    $1.455(79)$ &    $1.5(3)$  &  $4672$ & $\pm$  &  $47$   &  1.0  &  $+0.5639(28)$   \\

 $^{59}$Zn &  $4982$  &  $\pm$  & $84$ & $0.98563$ &    $1.440(81)$ &    $1.5(3)$  &  $4978$ & $\pm$  &  $86$   &  1.7  &  $-0.4876(42)$   \\

 $^{61}$Ga &  $4759$  &  $\pm$  & $137$ & $0.99331$ &    $1.461(87)$ &    $1.5(3)$  &  $4756$ & $\pm$  &  $138$   &  2.9 &  $+0.5421(79)$   \\

 $^{67}$Se\footnote[1]{Values for upper limit of BR; see also Table~\ref{adopted-t12-BR-QEC} and Sec.~\ref{Note on A>63}.} &  $5344$  &  $\pm$  & $245$ & $1.01842$ &    $1.461(99)$ &    $1.7(3)$  &  $5330$ & $\pm$  &  $245$   &  4.6 &  $-0.3873(89)$   \\

 $^{67}$Se\footnote[2]{Values for lower limit of BR; see also Table~\ref{adopted-t12-BR-QEC} and Sec.~\ref{Note on A>63}.} &  $5908$  &  $\pm$  & $289$ &           &                &              &  $5893$ & $\pm$  &  $288$   &  4.9 &  $-0.2048(50)$   \\

 $^{71}$Kr\footnotemark[1] &  $5108$  &  $\pm$  & $366$ & $0.99758$ &    $1.474(109)$ &    $1.7(3)$  &  $5095$ & $\pm$  &  $365$   &  7.2  &  $+0.454(16)$  \\

 $^{71}$Kr\footnotemark[2] &  $5991$  &  $\pm$  & $432$ &           &                &              &  $5976$ & $\pm$  &  $432$   &  7.2 &  $+0.1682(61)$   \\

 $^{75}$Sr\footnotemark[1] &  $4879$  &  $\pm$  & $590$ & $0.95210$ &    $1.484(118)$ &    $1.7(3)$  &  $4867$ & $\pm$  &  $588$   &  12  &  $+0.525(32)$   \\

 $^{75}$Sr\footnotemark[2] &  $5458$  &  $\pm$  & $662$ &           &                &              &  $5445$ & $\pm$  &  $661$   &  12 &  $+0.367(22)$   \\

\end{tabular*}
\end{ruledtabular}

\end{table*}


\begin{table*}
\caption{Reference key relating the reference codes used in Tables~\ref{table:halflives_mirrors} and \ref{adopted-t12-BR-QEC} to the actual reference numbers.}
\label{reference-key-2}
\begin{ruledtabular}

\begin{tabular}{l c l c l c l c l c l c}


Table & Reference & Table & Reference & Table & Reference & Table & Reference & Table & Reference & Table & Reference \\
code  &   No.     & code  &    No.    & code  &   No.     & code  &    No.    & code  &    No.    & code  &    No.    \\

\hline

 
Ab11 & \cite{Abusaleem2011} & Ac07 & \cite{Achouri2007} & Ac10 & \cite{Achouri2010} & Ad81 & \cite{Adelberger1981} & Ad83 & \cite{Adelberger1983} & Ad84 & \cite{Adelberger1984} \\
Aj70 & \cite{Ajzenberg1970} & Aj75 & \cite{Ajzenberg1975} & Aj82 & \cite{Ajzenberg1982} & Ak04 & \cite{Akulov2004} & Ak88 & \cite{Akulov1988} & Al67 & \cite{Aldridge1967} \\
Al72 & \cite{Alburger1972} & Al73 & \cite{Alburger1973} & Al74 & \cite{Alburger1974} & Al76 & \cite{Alburger1976} & Al77 & \cite{Alburger1977} & Ar58 & \cite{Arnell1958} \\
Ar84 & \cite{Arai1984} & Aw69 & \cite{Awschalom1969} & Ay84 & \cite{Aysto1984} & Az74 & \cite{Azuelos1974} & Az75 & \cite{Azuelos1975} & Az77 & \cite{Azuelos1977} \\
Ba12& \cite{Bacquias2012} & Ba55 & \cite{Bashkin1955} & Ba70 & \cite{Bardin1970} & Ba77 & \cite{Barker1977} & Ba94 & \cite{Baumann1994} & Be71 & \cite{Berenyi1971} \\
Be75 & \cite{Behrens1975} & Bl02 & \cite{Blank2002} &  Bl10 & \cite{Blank2010} & Bl68 & \cite{Black1968} & Bl95 & \cite{Blank1995a} & Bl99 & \cite{Blank1999} \\
Bo65 & \cite{Bormann1965} & Br14 & \cite{Broussard2014} & Br16 & \cite{Brodeur2016} & Bu07 & \cite{Burrows2007} & Bu08 & \cite{Burrows2008} & Bu20 & \cite{Burdette2020} \\
Bu85 & \cite{Burrows1985} & Bu91 & \cite{Budick1991} & Ce16 & \cite{Celikovic2016} & Cl58 & \cite{Cline1958} & Da85 & \cite{Daehnick1985} & De71 & \cite{Detraz1971} \\
Di51 & \cite{Dickson1951} & Ea62 & \cite{Earwaker1962} & Eb65 & \cite{Ebrey1965} & Ed77 & \cite{Edmiston1977} & Ew81 & \cite{Ewan1981} & Fa02 & \cite{Faestermann2002} \\
Fi05 & \cite{Fischer2005} & Fi17 & \cite{Finlay2017} & Fo17 & \cite{Fontbonne2017} & Ge71 & \cite{Geiger1971} & Ge76 & \cite{Genz1976} & Go64 & \cite{Gorodetzky1964} \\
Go68 & \cite{Goss1968} & Go68a & \cite{Gorodetzky1968} & Gr15 & \cite{Grinyer2015} & Gr15a & \cite{Grinyer2015a} & Gr71 & \cite{Grober1971} & Ha52 & \cite{Haslam1952} \\
Ha79 & \cite{Hagberg1979} & Ha80 & \cite{Hardy1980} & Ha87 & \cite{Hama1987} & Ha94 & \cite{Hagberg1994} & Ha97 & \cite{Hagberg1997} & Ho77 & \cite{Hornshoj1977} \\
Ho81 & \cite{Honkanen1981} & Ho82 & \cite{Hornshoj1982}  & Ho87 & \cite{Honkanen1987} & Ho89 & \cite{Honkanen1989} & Hu03 & \cite{Huikari2003} & Hu54 & \cite{Hunt1954} \\
Ia06 & \cite{Iacob2006} & Ja60 & \cite{Janecke1960}  & Ja61 & \cite{Janecke1961} & Ja99 & \cite{Janas1999} & Je50 & \cite{Jenks1950} & Jo51 & \cite{Jones1951} \\
Jo55 & \cite{Jones1955} & Jo67 & \cite{Jones1967} & Ju05 & \cite{Junde2005} & Ju08 & \cite{Junde2008} & Ju71 & \cite{Jundt1971} & Ka14 & \cite{Kankainen2014} \\
Ka64 & \cite{Kavanagh1964} &Ka64a & \cite{Kavanagh1964a} & Ka68 & \cite{Kadlecek1968} & Ki01 & \cite{Kienle2001} & Ki59 & \cite{Kistner1959} & Ki60 & \cite{King1960} \\ 
Kl54 & \cite{Kline1954} & Ko73 & \cite{Kochan1973} & Ku53 & \cite{Kundu1953} & Ku17 & \cite{Kurtukian2017} & Ku17a & \cite{Kucuk2017} & Li60 & \cite{Lindenberger1960} \\ 
Lo02 & \cite{Lopez2002} & Lo17 & \cite{Long2017} & Lo20 & \cite{Long2020} & Lo62 & \cite{Lonsjo1962} & Lu00 & \cite{Lucas2000} & Ma06 & \cite{MacMahon2006} \\
Ma17 & \cite{Magron2017} & Ma69 & \cite{Makela1969} & Ma74 & \cite{Mann1974} & Ma76 & \cite{Mann1976} & Me66 & \cite{Merritt1966} & Mi58 & \cite{Mihailovic1958}  \\
Mo71 & \cite{Moss1971} & Mo95 & \cite{Morrissey1995} & Mu58 & \cite{Muller1958} & Ne13 & \cite{Negret2013} & Ne63 & \cite{Nelson1963} & No47 & \cite{Novick1947} \\
Oi97 & \cite{Oinonen1997} & Oi99 & \cite{Oinonen1999} &  Ol87 & \cite{Oliver1987} & Pe57 & \cite{Penning1957} & Pf96 & \cite{Pfaff1996} & Pi85 & \cite{Piilonen1985} \\
Pl62 & \cite{Plendl1962} & Po58 & \cite{Povov1958} & Pr57 & \cite{Prokoshkin1957} & Re19 & \cite{Rebeiro2019} & Re99 & \cite{Reusen1999} & Ri68 & \cite{Ritchie1968}  \\
Ro14 & \cite{Rogers2014} & Ro55 & \cite{Roderick1955} & Ru14 & \cite{Rubio2014} & Ru77 & \cite{Rudy1977} & Sa93 & \cite{Saettler1993} & Sc48 & \cite{Schelberg1948} \\
Sc58 & \cite{Schweizer1958} & Sc70 & \cite{Scanlon1970} & Se17 & \cite{Severijns2017} & Se96 & \cite{Semon1996} & Sh14 & \cite{Shidling2014} & Sh15 & \cite{Shen2015} \\
Sh18 & \cite{Shidling2018} & Sh84 & \cite{Shinozuka1984} & Sh89 & \cite{Shinozuka1989} & Sh93 & \cite{Sherrill1993} & Si19 & \cite{Sinclair2019} & Si87 & \cite{Simpson1987} \\
Sm41 & \cite{Smith1941} & Su13 & \cite{Su2013} & Su13a & \cite{Suzuki2013a} & Su17 & \cite{Suzuki2017} & Su17a & \cite{Suzuki2017a} & Su62 & \cite{Sutton1962} \\
Ta60 & \cite{Talbert1960} & Ta73 & \cite{Tanihata1973} & Ti87 & \cite{Tilleya1987}  & Tr12 & \cite{Triambak2012} & Uj13 & \cite{Ujic2013} & Un00 & \cite{Unterweger2000} \\
Va18 & \cite{Valverde2018} & Va63 & \cite{Vasilev1963} & Va69 & \cite{Vasilev1969} & Wa60 & \cite{Wallace1960} & We02 & \cite{Weissmann2002} & Wi69 & \cite{Wick1969} \\
Wi74 & \cite{Wilkinson1974} & Wi80 & \cite{Wilson1980} & Wi93 & \cite{Winger1993} & Wo02 & \cite{Woods2002} & Wo69 & \cite{Wohlleben1969} & Yo65 & \cite{Youngblood1965} \\
Zu15 & \cite{Zuber2015} &  &  &  &  &  &  &  &  &  &  \\



\end{tabular}
\end{ruledtabular}

\end{table*}

\section{Weak Magnetism}
\label{sec:weak_magnetism}

The second part of this work focuses on the weak magnetism correction to the $\beta$ spectrum shape (and corresponding $\mathcal{F}t$ values) and on $\beta$-correlation measurements. Its importance has been hinted at throughout the previous sections, and cannot be understated as experimental precision reaches the sub-percent level. Its correct evaluation is an important factor in the extraction of new physics as well as in the extraction of $V_{ud}$ from precise $\beta$-correlation measurements, such that a thorough discussion of its origin and properties is worthwhile as the field continues and grows.

We give here a fairly in-depth presentation of the theoretical origin and state, followed by a discussion of experimental behaviour in mirror and other isospin multiplet $\beta$ decays. In both cases we compare against results of the nuclear shell model, assisted by simple single-particle results to aid in its interpretation. Based on these results, we discuss the ability of nuclear theory to correctly predict this quantity and the prospects in experimentally unexplored territories.

\subsection{Theoretical foundations}
\label{sec:weak_magnetism_theory}


Throughout the years the effects of weak magnetism have been studied by various groups of authors, using different formalisms (e.g. \cite{Holstein1974, Calaprice1976, Behrens1978, Behrens1982}). Its precise evaluation comes with several subtleties and a number of caveats. It is of interest then, to lay out the theoretical foundations underlying the evaluation. To ease the interpretation of the reader, we will present all final results in the formalism by Holstein \cite{Holstein1974}, although some intermediate results have been derived from other works.

We will first discuss the generalization of the Hamiltonian to include induced currents and the electromagnetic interaction. We follow with the treatment of the conserved vector current hypothesis and how it connects these two components. Finally, we discuss the evaluation of the matrix elements that appear when comparing to theoretical results.

\subsubsection{Generalized Hamiltonian}
\label{sec:general_hamiltonian}

\paragraph{Generalized nuclear $\beta$ decay Hamiltonian}

In the original Fermi approach, the $\beta$ decay Hamiltonian is constructed as a simple current-current interaction, analogous to the electromagnetic interaction
\begin{equation}
H_\beta(x) = \frac{G_F \cos \theta_C}{\sqrt{2}}[H_\mu(x) L^\mu(x) + \text{h.c.}] ~ ,
\label{eq:current_current_hamiltonian}
\end{equation}
where $G_F$ is the Fermi coupling constant obtained from muon decay, $\theta_C$ is the Cabibbo angle and $H_\mu(x)$ and $L^\mu(x)$ are the hadron and lepton currents, respectively. The Standard Model expressions for the latter is defined as
\begin{equation}
    L^\mu = \bar{u}_e\gamma^\mu(1-\gamma^5) v_\nu ~ ,
\end{equation}
with $u, v$ the lepton wave functions and $\gamma_i$ the Dirac $\gamma$ matrices, which couples only to left-handed particles. In the absence of any other forces, all relevant particle states are simple plane waves. The nuclear medium is, however, hardly a place devoid of additional forces. This is mainly due to the strong interaction, which renormalizes the weak vertex. Further, the presence of an electrostatic potential (i.e. QED) forces a slew of changes compared to the simple plane wave picture. Since $\beta$ decay occurs only at low momentum transfer, however, most of the SM intricacies are not kinematically visible and instead serve to renormalize the coupling constants. The renormalization of the weak vertex due to QCD effects results in two changes. The first is a modification of the axial vector operator, $\langle n | \bar{u} \gamma^\mu \gamma^5 d| p \rangle$, which can, e.g., be calculated using first principles on the lattice \cite{Chang2018, Gupta2018}. The vector part, $\langle n | \bar{u}\gamma^\mu d | p \rangle$, is protected through the Conserved Vector Current (CVC) hypothesis discussed in the following section and remains unchanged, even to higher order \cite{Ademollo1964}. While in the Standard Model the simple $V$-$A$ behaviour is a direct consequence of the electroweak interaction, the nuclear medium allows for some modification. Additional operators which still adhere to the required $V$-$A$ behaviour can be constructed through a combination with the nuclear momentum transfer $q = p_i-p_f$. In preparation we write the hadronic current of Eq. (\ref{eq:current_current_hamiltonian}) more generally
\begin{equation}
H(x) = \langle f | V_\mu(x) + A_\mu(x) | i \rangle ~ .
\end{equation}
The approach spearheaded by Holstein \cite{Holstein1974} and contemporaries \cite{Behrens1982, Armstrong1972} consists of treating the nucleus as an elementary particle, and says nothing about its constituents or internal turmoil. This allows one to encode all behaviour through form factors that depend only on $q^2$ as mentioned above. For the simple neutron this becomes
\begin{align}
V_\mu(x) &=  [g_V\gamma_{\mu}-\frac{g_M-g_V}{2M}\sigma_{\mu\nu}q^{\nu}+i\frac{g_S}{2M}q_{\mu}]\tau^+ \label{eq:V_mu_neutron} ~ , \\
A_\mu(x) &= [g_A\gamma_5\gamma_{\mu}-\frac{g_{II}}{2M}\sigma_{\mu\nu}\gamma_5q^{\nu}+\frac{g_P}{2M}\gamma_5q_{\mu}]\tau^+ ~ ,
\label{eq:A_mu_neutron}
\end{align}
where all $g_i$ form factors are a dimensionless function of $q^2$, and $\tau^+$ is the isospin ladder operator. The appearance of factors $\mathcal{O}(q/M)$ traditionally lends them the name "recoil" corrections. In order of appearance, these are called the vector (Fermi), weak magnetism, induced scalar, axial vector (Gamow-Teller), induced tensor and induced pseudoscalar terms.

The presence of the electromagnetic field now has two additional consequences for nuclear $\beta$ decay. The first entails a renormalization of the decay vertex, as we have done before with the strong interaction in Eqs. (\ref{eq:V_mu_neutron})-(\ref{eq:A_mu_neutron}). Up to $\mathcal{O}(\alpha)$, with $\alpha$ the fine-structure constant, these simply renormalize the leading order terms (i.e. $\Delta_R^{V, A}$ as in Eqs. (\ref{eq:ft_general}) and (\ref{eq:rho})) and do not introduce additional structure. The experimental definition of the mixing ratio (and $g_A$ in the case of the neutron) simply absorbs the difference of vector and axial vector inner radiative corrections, in addition to the other small corrections in Eq. (\ref{eq:rho}). The second consequence concerns the operators in Eqs. (\ref{eq:V_mu_neutron})-(\ref{eq:A_mu_neutron}), specifically those coupling to the momentum $q_\mu$. In the traditional quantum mechanics calculation, turning on the electromagnetic field requires the standard substitution $\partial_{\mu} \to \partial_{\mu} - ieA_\mu$, where $A_\mu = (i\phi, \bm{A})$ is the electromagnetic four vector. It is in part this effect which in the traditional $\beta$ decay calculations \cite{Behrens1982} causes a contribution to what is now understood to be part of the inner radiative correction \cite{Hayen2021}. We will come back to this below.

We have, up to now, only discussed the generalization to a spin-$1/2$ system, leading to Eqs. (\ref{eq:V_mu_neutron})-(\ref{eq:A_mu_neutron}). For general allowed $\beta$ transitions where $\Delta J = 0,1$, several additional terms appear. Using the notation of Holstein \cite{Holstein1974a}, the form factors are then written as $a$ (vector), $b$ (weak magnetism), $c$ (axial vector), $d$ (induced tensor), $e$ (induced scalar), $h$ (induced pseudoscalar), $f$, $g$, $j_2$ and $j_3$. Full formulae can be found in several publications \cite{Holstein1974, Calaprice1976}. Once again, all form factors are a function of $q^2$. Typically, however, only the dominant form factors are expanded as
\begin{subequations}
\begin{align}
a(q^2) &\approx a_1 + a_2 q^2 ~ , \\
c(q^2) &\approx c_1 + c_2 q^2 ~ , 
\label{eq:c_expansion}
\end{align}
\end{subequations}
while all others are approximated as their value for $q^2 = 0$. This is possible because the momentum transfer in $\beta$ decay is sufficiently small such that $qR \ll 1$, where $R$ is the nuclear radius.

\paragraph{Nuclear electromagnetic interaction}

The influence of the electromagnetic field on the weak vertex was summarized in the previous section. As a charged particle, the nucleus also directly couples to the electromagnetic field via

\begin{equation}
\mathcal{L}(x) = ie \mathcal{J}_\mu A^\mu(x) ~ , 
\end{equation}

\noindent where $\mathcal{J_\mu}$ contains both nucleon, $J_\mu$, and electron, $l_\mu$, parts. As before, we can treat the interaction of the latter in the elementary particle approach and write

\begin{equation}
\langle f | J_\mu(0) | i \rangle = i\langle f | F_1 \gamma_\mu +F_2 \sigma_{\mu\nu}q^\nu | i \rangle ~ ,
\label{eq:form_factor_em}
\end{equation}

\noindent where again all $F_i$ are a function of $q^2$.  The first of these represents the charge, while the second corresponds to the magnetic dipole interaction. These can be split up into an isoscalar and isovector part, writing $F_i = F_i^{(S)}+F_i^{(V)}\tau_3$. The absence of higher order isospin operators in the electromagnetic interaction has obvious consequences for selection rules between nuclear levels \cite{Weinberg1958, Morpurgo1959}, and allows one to write down the mass formula within isospin multiplets \cite{BlinStoyle1973}

\begin{equation}
M(A, T, T_3) = a(A, T) - b(A, T)T_3 + c(A, T)T_3^2
\label{eq:mass_isospin_multiplet}
\end{equation}

\noindent to only second order in $T_3$, where $A$ is the mass number, $T$ the isospin and $T_3=\frac{1}{2}(Z-N)$. Here the parameters $a, b$ and $c$ can be found in several text books \cite{Wilkinson1969, BlinStoyle1973} and in a recent evaluation of the relevant experimental data \cite{MacCormick2014}, and depend mainly on the different isomoments of the Coulomb displacement energy and the difference in proton and neutron mass.

\subsubsection{Conserved Vector Current}
\label{sec:cvc}

Several years before the Weinberg-Salam model of electroweak interactions was proposed, a more intimate connection between electromagnetic and weak interactions was put forward by Feynman and Gell-Mann \cite{Feynman1958} and Sudarshan and Marshak \cite{Sudarshan1958}. Called the Conserved Vector Current hypothesis, the name only covers the `weak' part of the conjecture, namely that

\begin{equation}
\partial^\mu V_{\mu} = 0 ~ ,
\label{eq:cvc_weak}
\end{equation}

\noindent analogous to the classical continuity equation. Equation (\ref{eq:cvc_weak}) strictly only holds in the absence of electromagnetism, as the inclusion of the latter requires the substitution $\partial^\mu \to \partial^\mu - ie A^\mu$ as before. Application of the weak CVC principle forces $g_V(q^2 = 0) = 1$ and $g_S(q^2 = 0) = 0$, and allows one to express the Fermi matrix element as \cite{Behrens1982}

\begin{align}
\langle &J_fM_fT_fT_{3f} | V_0(0) | J_iM_iT_iT_{3i} \rangle = \nonumber \\
 &\sqrt{(T_i\pm T_{3i})(T_i \mp T_{3i}+1)} \delta_{J_iJ_f}\delta_{M_iM_f}  ~  .
\label{eq:MF_isospin}
\end{align}

\noindent Evaluating this for $0^+\to 0^+$ superallowed decays, one finds $\sqrt{2}$ and this is implicitly present in the factor $2$ in Eq. (\ref{master}) \cite{Behrens1982}.
The strong CVC principle states that the vector current from $\beta$ decay forms an isospin triplet with the electromagnetic current \cite{GellMann1958}

\begin{equation}
V_\mu = \mp \left[\tau^\pm, J_\mu^\text{em} \right] ~ ,
\label{eq:cvc_strong}
\end{equation}

\noindent where $\tau^\pm$ is the isospin ladder operator. Clearly the weak principle directly follows from Eq. (\ref{eq:cvc_strong}), and further allows one to relate \textit{weak} matrix elements to their \textit{electromagnetic} analogues. This further implies that Eqs. (\ref{eq:V_mu_neutron}) and (\ref{eq:form_factor_em}) can be directly compared. We consider here two cases.

The first deals with $\beta$ decays within the same isospin multiplet

\begin{equation}
\langle T T_{3i}\pm 1 | V_\mu^\pm | T T_{3i} \rangle ~ ,
\end{equation}

\noindent for which we then find\footnote{The CVC hypothesis relates only the \textit{isovector} part of the electromagnetic interaction to the corresponding weak vector decay, while the magnetic moment has both an isoscalar and isovector part. When taking the difference within an isospin multiplet, the isoscalar component obviously drops out.}

\begin{align}
b(0) &= \pm a(0) A \sqrt{\frac{J+1}{J}} (\mu_f-\mu_i) ~ ,
\label{eq:weak_magnetism_mirror}
\end{align}

\noindent where the upper/lower sign is for $\beta^-$/$\beta^+$ decay, respectively, $A$ is mass number, and $\mu_{i,f}$ are the nuclear dipole moments of respectively the mother and daughter nuclear states. Evaluating this for the neutron we find

\begin{equation}
\frac{1}{\sqrt{3}}b(0) = g_M = \mu_p-\mu_n = 4.706 ~ ,
\end{equation}

\noindent where we used $a(0)=1$ for the neutron. The natural extension of neutron decay from an isospin point of view are the mirror $\beta$ transitions, where $T=1/2, T_{3}=\pm 1/2$ for initial and final states. When experimental nuclear moment data is available, Eq. (\ref{eq:weak_magnetism_mirror}) allows for an unambiguous determination of the weak magnetism contribution. For the higher multiplets, with $T =$1 and higher, analog decays are typically energetically forbidden or connecting unbound state due to the behaviour of Eq. (\ref{eq:mass_isospin_multiplet}), so that Eq.~\ref{eq:weak_magnetism_mirror} cannot be used anymore.

The second case is that of a decay where $\Delta T = 1, \Delta T_{3} = \pm 1$. The strong CVC conjecture then allows one to relate the weak magnetism form factor to the \textit{isovector} part of the decay width of the corresponding analog $M1$ $\gamma$ transition, $\Gamma_{M1}^{\text{iso}}$. In this case, one finds\footnote{Note that CVC only relates the square of $b(0)$, such that a sign ambiguity remains.}

\begin{equation} 
b^2(0) = \eta \frac{\Gamma_{M1}^{\text{iso}}6M^2}{\alpha E_\gamma^3} ~ ,
\label{eq:weak_magnetism_triplet}
\end{equation}

\noindent with $\alpha$ the fine structure constant and $\eta$ a constant. The factor $\eta$ is unity in case the final state is the same for $\beta$ and $\gamma$ analog transitions, while it is $\eta = (2J_i+1)/(2J_f+1)$ when initial and final states of the $\gamma$ transition (resp. $J_i$ and $J_f$) are reversed relative to the $\beta$ decay, to compensate for the degeneracy in the $\gamma$ transition phase space.

It is of interest to note that for $\beta$ transitions between members of a common isotopic multiplet (like the mirror $\beta$ transitions) CVC requires the form factors $e(q^2) = f(q^2) = 0$, while the $d$ form factor can only be non-zero if second-class currents would exist \cite{Holstein1974, Holstein1974a}. In addition, as many of the transitions that are considered here involve low-spin states, the triangle inequality satisfied by Clesh-Gordan coefficients requires the vanishing of one or more form factors, i.e. one has (with $J$ and $J^\prime$ being the spin of the initial and final states of the $\beta$ transition, respectively):
\begin{align}
&J = 1 , J^\prime = 0:  ~  ~  a = e = f = g = j_2 = j_3 = 0 \nonumber \\
&J = 0, J^\prime = 1:  ~  ~  a = e = f = g = j_2 = j_3 = 0 \nonumber \\
&J = J^\prime = 1:  ~  ~  j_3 = 0 \nonumber \\
&J = 1/2, J^\prime = 1/2:  ~  ~  f = g = j_2 = j_3 = 0 \nonumber \\
&J = 1/2, J^\prime = 3/2:  ~  ~  a = e = j_3 = 0 \nonumber \\
&J = 3/2, J^\prime = 1/2:  ~  ~  a = e = j_3 = 0   ~  .
\label{eq:clebsh}
\end{align}
Finally, the effect of the higher-order form factors $a_2$, $c_2$, and the induced pseudoscalar form factor, $h$, that were not yet mentioned here, but which appear in the $\beta$-spectrum shape (cf. Eq.~(B7) in Ref.~\cite{Holstein1974}) and so determine the $ft$ values of a transition, are all reduced by a factor $M$ (the average mass of the initial and final nuclei) with respect to the $a$, $c$, $b$ and $d$ form factors. They can therefore be neglected at the level of precision of the $ft$ values we are dealing with in this paper, which is of the order of per mil to several percent.

\subsubsection{Form factor evaluation}
\label{sec:form_factor_evaluation}

The weak magnetism correction shows up in a specific combination in the Gamow-Teller $\beta$ spectrum as
\begin{equation}
    \delta_{\text{WM}} = \frac{b}{Ac_1} ~ , 
    \label{eq:bAc}
\end{equation}

\noindent where $c_1$ is the first component of the Gamow-Teller form factor as defined in Eq. (\ref{eq:c_expansion}). Considering we have generalized the nuclear current using the elementary particle approach, and related the form factors to electromagnetic observables for specific cases, one could consider the work to be done and leave the determination of these form factors to experiment. Often, however, application of CVC is impossible and one would like input from nuclear models. We consider here the impulse approximation and discuss the quenching of the axial vector coupling constant.

\paragraph{Impulse Approximation}
The easiest way through which we can accomplish such a feat is by using the impulse approximation, where we consider the nuclear current a sum of independent single particle nucleon currents. In doing this, one neglects meson exchange, off-shell mass effects and other many-body effects. This is typically a reasonable assumption for allowed transitions. Indeed, discrepancies resulting from using the impulse approximation as compared to the elementary particle treatment have been discussed in the light of CVC and PCAC, and were found to be in good agreement for allowed transitions \cite{Armstrong1972, Hwang1980}.

The usual approach is performed by generalizing Eqs. (\ref{eq:V_mu_neutron})-(\ref{eq:A_mu_neutron}) to a system of $A$ nucleons, and performing a Foldy-Wouthuysen transformation to yield the non-relativistic limit \cite{Foldy1950}. Comparing terms with that of the elementary particle approach, one can write the corresponding reduction of the form factors to proper nuclear matrix elements. A summary of the relevant form factors is given in Table \ref{table:summary}, with the required nuclear matrix elements being listed in Table \ref{table:expressionsmatrixelements}. For clarity, these are only given to first order. The correction term of Eq. (\ref{eq:bAc}) then becomes
\begin{equation}
    \frac{b}{Ac_1} = \frac{1}{g_A}\left(g_M + g_V\frac{M_L}{M_{GT}} \right) ~ ,
    \label{eq:bAc_impulse_approx}
\end{equation}
such that we are left with the evaluation of two matrix elements. Note that the value of $b/Ac$ measures the relative strength of $M_L$ and $M_{GT}$. For a pure-spin transition, with $M_L \ll M_{GT}$ and $g_A =$ 1.00 (see Sec.~\ref{Evaluation-g_A}), one obtains $b/Ac =$ 4.7. As $g_V = 1$, values of $b/Ac >$ 5.7 thus indicate a dominance of the orbital contribution.
Finally, at the level of precision we will be dealing with in the evaluation of $b/Ac$ later in this paper, we can neglect the $q^2$ dependence of the $c$ form factor in Eq.~(\ref{eq:c_expansion}) and so replace $c_1$ by $c$, i.e. the total Gamow-Teller form factor as it is present in the $\mathcal{F}t^\mathrm{mirror}$ value (Eqs.~(\ref{eq:rho} and \ref{master})).

\begin{table*}
\caption{Summary of the $a$, $b$, $c$, and $d$ form factors and their relation to the nuclear matrix elements defined in Table~\ref{table:expressionsmatrixelements}. Here, the impulse approximation is given to first order only, and the relativistic matrix elements are neglected, as is also done in \cite{Holstein1974}. Further, $M = 1/2 (M_i + M_f)$, with $M_i$ and $M_f$ being the masses of the initial and final states of the decay, is the nuclear mass, and $\Delta = M_i - M_f$ is the nuclear energy release. Important to note is that the induced tensor form factor, $d$, contains both a first-class and second-class contribution. The former can be shown to disappear for decays between analog states \cite{Holstein1974}, while no second-class currents have been observed till now \cite{Grenacs1985, Wilkinson2000, Minamisono2011}.}
\label{table:summary}
\begin{ruledtabular}
\centering
\begin{tabular}{c|ll}
Form factor  & Formula Imp. App. & Remark \\
\hline
\textit{a}  & $a\cong g_VM_F$ & $g_V=1$ (CVC) \\
$b$   & $b\cong A(g_MM_{GT}+g_VM_L)$  & $g_M = 4.706$ \\
$c$  & $c\cong g_A M_{GT}$ & $g_A \rightarrow g_{A,eff}=1$ \cite{Towner1987}\\
$d$  & $d\cong A(g_A M_{\sigma L} \pm g_{II} M_{GT})$ & $g_{II} \sim g_T\cong0$\footnote{This assumes the absence of second-class currents.} \cite{Holstein1974} \\
$f$  & $f\cong (2/3)^{1/2} M \Delta g_V M_Q (\hbar c)^2$  &  $M = 1/2 (M_i + M_f) = \rm{nuclear ~ mass}$ \\
$g$  & $g\cong (-4/3) M^2 g_V M_Q (\hbar c)^2$  &  $\Delta = M_i - M_f = \rm{nuclear~  energy ~ release}$ \\
$h$  & $h\cong (-2/\sqrt(10)) M^2 g_A M_{1y} (\hbar c)^2 - A^2 g_P M_{GT}$  &  $g_P = -181.03$ \\
$j_k$ & $j_k\cong (-2/3) M^2 g_A M_{ky} (\hbar c)^2$ \\
\end{tabular}

\end{ruledtabular}
\end{table*}
\begin{table}
\caption{Definitions of the reduced nuclear matrix elements relevant for the $a$, $b$, $c$, and $d$ form factors, from Ref.~\cite{Calaprice1977}.}
\label{table:expressionsmatrixelements}
\begin{ruledtabular}
\begin{tabular}{c | c}

Matrix element & Operator form \\
\hline
$M_F$ & $\langle\psi_{f}\|\Sigma\tau^{\pm}_i\|\psi_{i}\rangle$\\
$M_{GT}$ & $\langle\psi_{f}\|\Sigma\tau^{\pm}_i\overrightarrow{\sigma}_i\|\psi_{i}\rangle$ \\
$M_L$ & $\langle\psi_{f}\|\Sigma\tau^{\pm}_i\overrightarrow{l}_i\|\psi_{i}\rangle$ \\
$M_{\sigma L}$ & $\langle\psi_{f}\|\Sigma\tau^{\pm}_i i \overrightarrow{\sigma}_i\times\overrightarrow{l}_i \|\psi_{i}\rangle$ \\
$M_Q$ & $(4\pi/5)^{1/2} \langle\psi_{f}\|\Sigma\tau^{\pm}_i r^2_i Y_2(r_i) \|\psi_{i}\rangle$ \\
$M_{ky}$ & $(16\pi/5)^{1/2} \langle\psi_{f}\|\Sigma\tau^{\pm}_i \sigma^2_i C^{n n^\prime k}_{1 2 k} Y^{n^\prime}_2(r_i)\sigma_{in}  \|\psi_{i}\rangle$ \\
\end{tabular}

\end{ruledtabular}
\end{table}

One can understand Eq.~(\ref{eq:weak_magnetism_triplet}) more intuitively by writing down the M1 matrix element in the long wavelength approximation, in which case it becomes proportional to
\begin{equation}
\langle f | \sum_i^A\bm{l}_i\left(\frac{1+\tau_3^{(i)}}{2}\right)+\frac{\bm{\sigma}^{(i)}}{2}(\mu_V^+ + \mu_V^-\tau_3^{(i)}) | i \rangle ~ ,
\label{eq:M1_isospin}
\end{equation}
where $\mu_V^\pm = (\mu_p\pm \mu_n)$, with $\mu_p$ and $\mu_n$ the proton and neutron magnetic moments. Using the experimental values for these, it is clear that the isovector contribution is much more dominant than its isoscalar counterpart, leading to several quasi-selection rules \cite{MacDonald1955, Gell-Mann1953, Morpurgo1958}.

The use of the above equations depends on several approximations, with increasing importance:

\begin{enumerate}
\item \textbf{Long wavelength}: The expression for the $M1$ operator was derived for $kR \ll 1$, with $k$ the wave number of the emitted $\gamma$ ray. This is most certainly valid for the low energies coming from nuclear transitions, and is relevant only for energies upwards of 200\,MeV.

\item \textbf{Nuclear currents}: The orbital (convection) part of the nuclear current that interacts with the vector potential is typically constructed using $\bm{v} = \bm{p}/M$, whereas more precisely it should be $\bm{v} = i[H,\bm{r}]/\hbar$ where $H$ is the full Hamiltonian, including spin-orbit terms and exchange terms.

\item \textbf{Deformation}: The comparison of weak magnetism and the $M1$ matrix elements assumes the isobaric analog states are identical in all quantum numbers except for $T_3$. When introducing deformation, however, this picture can quickly turn around as often the analog state sits at high excitation energies and differing deformation. Furthermore, states can be parts of different rotational or vibrational bands, resulting in additional quenching of transition amplitudes.

\item \textbf{Isospin purity}: Typically we consider transitions for which $\Delta T = \pm 1$, meaning the isovector part is automatically the only contributing factor. This, however, assumes perfect isospin symmetry which is violated in practice so that isoscalar contributions can sneak in. 

Nevertheless, due to the smallness of the violation and the numerical weight of the isovector moment, this is typically negligible \cite{BlinStoyle1973}. However, in cases where the Gamow-Teller matrix element is strongly suppressed the dominant contribution can arise from the angular momentum part in Eq.~(\ref{eq:M1_isospin}) instead. The biggest culprit for isospin mixing is the isovector part of the Coulomb interaction, resulting in an isospin impurity proportional to $\sum_j [\langle f | V_C | j \rangle / (E_f-E_j)]^2$ \cite{BlinStoyle1973}. Due to a variety of mechanisms, the isospin purity remains rather low even for higher $Z$ nuclei, though individual variation can change significantly \cite{BlinStoyle1973}.

\item \textbf{Meson exchange}: An important distinction between Eqs. (\ref{eq:weak_magnetism_triplet}) and (\ref{eq:M1_isospin}) is the treatment of meson exchange currents. The strong CVC principle (Eq. (\ref{eq:cvc_strong})) relates weak interaction and electromagnetic \textit{form factors}, including all meson exchange effects, induced currents, et cetera. While Eq. (\ref{eq:weak_magnetism_triplet}) is always correct, the evaluation through Eq. (\ref{eq:M1_isospin}) is not and requires a remedy. In usual nuclear physics fashion this can be performed by introducing effective charges to both proton and neutron \cite{De-Shalit1990}. Recently much work was performed using ab initio methods \cite{Parzuchowski2017}.
\end{enumerate}

It is clear that many subtleties can underlie the proper evaluation of weak magnetism. Often, though, all of these approximations are sufficiently well-behaved. In case of discrepancies, a breakdown of one or more of these approximations can be investigated.

\paragraph{Evaluation of $g_A$}
\label{Evaluation-g_A}

In the neutron system, the value of $g_A/g_V$ is experimentally found to be 1.2756(13) \cite{PDG2020}. In more complex systems, however, typically lower (quenched) values of $g_A$ are required to reach agreement with experimental data when employing shell model or mean field theories \cite{Wilkinson1973, Wilkinson1973a, Wilkinson1974b}. The reason for this quenching lies in a failure to take into account the required degrees of freedom inside the nuclear medium meaning truncated valence spaces, three-body interactions and nucleon-nucleon correlations. The quenching is then an attempt at a parametrisation of theoretical deficiencies of commonly used theories. It speaks for itself then that no quenching is unique, and is inherently coupled to the details of the underlying theory with its effective interactions and basis states. Intensive research was performed in the second half of the twentieth century \cite{Ericson1973, Delorme1976, Towner1979}, and has recently become very actual again due to its relevance in neutrinoless double $\beta$ decay searches \cite{Kostensalo2017, Deppisch2016, Suhonen2017, Suhonen2017a}. Some explanation is required then, as we consider here both shell model and deformed single particle results.

Two effects are at play here. The first is traditionally called core polarization \cite{Fujita1964, Koshigiri1981}. One usually considers as starting point a single-particle Hamiltonian as $H_0 = T + U$, where $U$ is an average single particle potential. The true solutions of the system, however, are solutions of the true Hamiltonian, $H = H_0 + \mathcal{V}$, where $\mathcal{V} = V-U$ is additional influence from the true nucleon-nucleon potential. Further, solutions are constructed using single particle orbitals generated from only a subspace of finite dimension. The correction required using this approach is typically called \textit{core polarization}, and represents the lack of multi-particle correlations in the individual wave functions. This includes residual interactions between valence particles, core deformation, $\Delta(1236)$-isobar excitations and relativistic effects \cite{Towner1979}.

The second effect is that of meson exchange \cite{Barroso1975, Rho1978, Towner1987}. Typically, the proper nucleon-nucleon interaction can be reduced in an effective manner to correspond to the exchange of several mesons, principally the pion and heavier $\rho$ and $\omega$ excitations. Combined with possible $\Delta$-isobars, this leads to an impressive number of required calculations. A modern example can be found in Ref. \cite{Siiskonen2001}.

The choice for a quenching factor then depends on how well these effects are present in the models we employ \cite{Suhonen2017a}. In the case of the shell model, one often uses $g_A = 1.0$ (e.g.  \cite{Towner1987, Siiskonen2001}) to counter model-space truncations and meson exchange effects. The evaluation of the single particle results is slightly less straightforward. For low mass nuclei and transitions close to closed shells, meson exchange will be the dominating factor. Going to higher masses and more exotic transitions, deformation effects will start to play a role. This can be mitigated using a deformed potential (see Appendix). In mid-shell nuclei proton-neutron residual interactions can become non-trivial and give dominant contributions. It is clear and hardly unexpected that there is no easy solution. As a way of making do, one often uses the $g_A$ quenching factors extracted from early shell model calculations, i.e. those not containing extensive core polarization and meson exchange corrections, i.e. $g_A = 1.1$ for the $sd$ shell \cite{Brown1988a}, $g_A = 1.0$ for the $pf$ shell and onwards \cite{Martinez1996} (see \cite{Suhonen2017a} for a comprehensive overview and discussion of the renormalization of $g_A$).

As here experimental results for the $b$ and $c$ form factors and the Gamow-Teller and orbital current matrix elements, resp. $M_{GT}$ and $M_L$, will be compared to shell model calculations that used $g_A = 1$ (besides $g_V = 1$, $g_M = 4.706$, $g_P = -181.03$, and $g_S = g_T = 0$), we will also use $g_A = 1$ in dealing with the experimental quantities. Only for the the decay of tritium ($A =$ 3), where there is very little core polarization, we will use the free-nucleon value $g_A = 1.27$.

\subsection{Weak magnetism for the mirror $\beta$ transitions}
\label{wm-mirror-nuclei}

For the superallowed $\beta$~transitions between $T = 1/2$ isospin doublets in mirror nuclei ($^{A}_{Z+1}X_{N}\rightarrow^{A}_{Z}X'_{N+1}$) the vector form factor is well known, i.e. $a = g_V M_F = 1$ (Eq.~\ref{eq:MF_isospin}), while the weak magnetism form factor, $b$, and the axial-vector form factor, $c$, can be determined experimentally using the \textit{CVC} hypothesis (Eq.~\ref{eq:weak_magnetism_mirror}) and the experimental $ft$ value (Eq.~\ref{master}), respectively. Form factors $d$, $e$, $f$ and $j_2$ are set to zero as it was shown that the first-class contributions to these form factors that arise in the impulse approximation vanish for transitions between states of a common isospin multiplet \cite{Holstein1974}. Here we update the work of Ref.~\cite{Calaprice1976}, extending the range of mirror transitions for which experimental data on the weak magnetism term are available up to mass $A$~=~75.

\subsubsection{Magnetic moments of the $T=1/2$ mirror nuclei}
\label{sec:mirror moments Perez}

The experimental data required to calculate the weak magnetism form factor, $b$, from experimental data, i.e. the nuclear magnetic moments of the mother and the daughter isotopes (Eq.~\ref{eq:weak_magnetism_mirror}), are reviewed here.

Table~\ref{Magnetic-moments} lists the available experimental values for the magnetic moments of the mirror nuclei up to $A =$ 75, as published in the recent survey of Ref. \cite{Stone2014, Mertzimekis2016a} and updated up to April 2021 via the database of the Brookhaven National Nuclear Data Center \cite{ENSDF}. The publications containing the respective input data are again indicated with an alphanumeric code, with the reference key linking these codes with the actual reference numbers given in Table~\ref{reference-key-moments}. 

The same data-selection procedure that was used for the data leading to the $\mathcal{F} t$ values is used here for the analysis of the magnetic moments. Thus, data with an uncertainty that is a factor 10 or more larger than that of the most precise measurement, as well as values with other problems have again been rejected. These are listed and commented in Table~\ref{unusedref-moments}. The finally adopted values for the magnetic moments that were used in the further analysis are listed in column 4 of Table~\ref{Magnetic-moments}. When two or more input values are available and the error bars are of similar size, the weighted average is used with the uncertainty being increased by a factor $S = \sqrt{\chi^2/\nu}$ (Eq.~(\ref{eq:scale})) if $\chi^2/\nu > 1$. This scale factor is then listed in column 5. When two, often very precise measurements, differ by many standard deviations with no clear argument in favour of either one, the unweighted average is used as the adopted value. This is then indicated in the table.

For the heavier mirror nuclei, i.e. in the $fp$ shell, experimental magnetic moment values are often not available. For these nuclei, the magnetic moments were calculated from the strong linear correlations between the mirror pair magnetic moments and their $ft$ values obtained in Ref.~\cite{Perez2008}. For this, we used shell-model estimates for the $\it small$ quantities $S_e$ and $J_e$, which represent the contributions from the type of nucleon that is even in number to the $z$ components of the total spin $S$ and total angular momentum $J$ of the mirror pair \cite{Perez2008}. Because the quantities $S_e$ and $J_e$ are small \cite{Clement1979}, the ground state gyromagnetic ratios, $\gamma_{p/n} = \mu_{p/n}/J$, for the odd-proton and odd-neutron partners of the mirror nuclei could be related \cite{Perez2008} to their superallowed $\beta$-transition strengths, $\gamma_{\beta}$, defined as 
\begin{equation}
|\gamma_{\beta}| = \frac{1}{2} \sqrt{\left( \frac{6170} {\mathcal{F}t^\mathrm{mirror}} - 1 \right) \frac{1}{J(J+1)}}  ~  ,
\label{gamma-beta}
\end{equation}
\noindent via
\begin{equation}
(\gamma_p + \Delta \gamma_p) = g_p + \frac{G_p -g_p}{R} \gamma_{\beta}  ~ ,
\label{gamma-p}
\end{equation}
and
\begin{equation}
(\gamma_n + \Delta \gamma_n) = g_n + \frac{G_n -g_n}{R} \gamma_{\beta}  ~ .
\label{gamma-n}
\end{equation}
\noindent The sign of $\gamma_{\beta}$ is negative for isotopes with $j = l-1/2$ (with $j$ and $l$ being, respectively, the spin and orbital angular momentum of the subshell of the odd nucleon), i.e. for the $p_{1/2}$ shell (the mirror nuclei with $A =$ 13 and 15), the $d_{3/2}$ shell ($A =$ 33 to 39), and the $f_{5/2}$ shell ($A =$ 65 to 77). For all other mirror nuclei one has $j = l+1/2$ and a positive sign for $\gamma_{\beta}$.

The quantities $\Delta \gamma_p$ and $\Delta \gamma_n$ are small quantities representing the contributions to the total spin and total angular momentum of the mirror pair generated by the even type of nucleon in these odd-even nuclei. They were obtained from $0 \hbar \omega$ shell model calculations based on the Hamiltonians from Refs.~\cite{Cohen1965, Brown2006, Honma2002} and are listed in \cite{Perez2008}. As one is dealing here with nucleons in nuclei, the free-nucleon values of the $g$-factors, i.e. $g_p = 1.0$, $g_n = 0$, $G_p = 5.586$, and $G_n = -3.826$, as well as the ratio of the axial-vector to vector coupling constant $R = |C_A/C_V| \simeq 1.27$, were replaced by effective values denoted by $\widetilde{g}$, $\widetilde{G}$, and $\widetilde{R}$. The linear relations of Eqs.~(\ref{gamma-p}) and (\ref{gamma-n}) were used to determine values for $\widetilde{g}$, and $(\widetilde{G} - \widetilde{g})/\widetilde{R}$ for protons and neutrons from least-square fits to $(\gamma_{p,n} + \Delta_{\gamma_{p,n}})$ and $\gamma_{\beta}$, thereby using the experimental magnetic dipole moments and $\beta$-decay half-lives for the mirror nuclei with $A = 11$ to 43, yielding ('Fit (B)' in \cite{Perez2008}):
\begin{equation}
\widetilde{g}_p = 1.040 \pm 0.020 ~ ,
\end{equation}
\begin{equation}
\widetilde{g}_n = -0.011 \pm 0.016  ~ ,
\end{equation}
\begin{equation}
(\widetilde{G}_p - \widetilde{g}_p)/\widetilde{R} = 4.26 \pm 0.07  ~ ,
\end{equation}
\begin{equation}
(\widetilde{G}_n - \widetilde{g}_n)/\widetilde{R} = -3.70 \pm 0.05 ~ .
\end{equation}

Using the $\mathcal{F} t^\mathrm{mirror}$ values listed in Table~\ref{input-data-and-calculated-quantities} and Eqs.~(\ref{gamma-beta}) to (\ref{gamma-n}) then yield the calculated magnetic moment values listed in column 7 of Table~\ref{Magnetic-moments} for all mirror isotopes. For the mirror isotopes in the $fp$ shell with $A =$ 57 to 75, no $\Delta \gamma_{p,n}$ values were available. We therefore used the average values of these quantities for the isotopes with $A =$ 45 to 57 in the $fp$ shell listed in Table~V of Ref.~\cite{Perez2008} with a one standard deviation error so as to cover the full range of individual values listed there, i.e. $\Delta \gamma_p = +0.11(5)$ and $\Delta \gamma_n = -0.14(5)$. The errors due to the uncertainties on these estimated values have been added in quadrature to the other contributions to the error bars on the calculated magnetic moment values.

Good correspondence with the known experimental values is obtained as shows from the fact that from the 43 cases for which experimental values are available the difference between the experimental and calculated values, listed in column 10, is less than 0.1~$\mu_N$ for 27 nuclei, between 0.1~$\mu_N$ and 0.2~$\mu_N$ for another 11 nuclei, and larger than 0.2~$\mu_N$ for 5 nuclei (i.e. $^{17}$F, $^{41}$Sc, $^{41}$Ca, $^{45}$Ti, and $^{59}$Cu). Reasonably good confidence can thus also be given to the values calculated for those mirror nuclei for which experimental magnetic moment values are not available yet. For the heaviest isotopes, i.e. with mass $A = 61$ to 75, this is shown in Table~\ref{Magnetic-moments} by comparing the calculated moment values with those of neighbouring nuclei with the odd proton or neutron occupying the same shell model orbital. 




\newpage

{\scriptsize
\begin{longtable*}{  l  |  l l  |  c c  ||  c c c c}
\caption{Magnetic moments of the $T = 1/2$ mirror nuclei. Column 2 lists the values available in the literature~\cite{Stone2014, Mertzimekis2016a, ENSDF}, with adopted values in column 4. When an average was calculated the scale factor (Eq.~\ref{eq:scale}) is listed in column 5. Columns 6 to 9 relate to the values calculated (column 8) from the linear relations between the mirror pair magnetic moments and their $ft$ values of Ref.~\cite{Perez2008}. Columns 6 and 7 list the $\gamma_\beta$ values (calculated from the corrected $\mathcal{F}t^\mathrm{mirror}$ values in Table~\ref{input-data-and-calculated-quantities} using Eq.~(\ref{gamma-beta})), and the values for the quantities $\Delta \gamma_p$ and $\Delta \gamma_n$ to calculate the magnetic moments using Eqs.~(\ref{gamma-p}) and (\ref{gamma-n}), respectively. Calculated values for isotopes for which no experimental data are available are shown in italics. The last column lists the difference between the calculated value and the adopted experimental magnetic moment value. Note that when the sign for a magnetic moment value in column 2 is given, the sign has been explicitly measured.
The correlation between the alphabetical reference code used here and the actual reference numbers is listed in Table~\ref{reference-key-moments}.}
\label{Magnetic-moments}\\

\hline 

Parent  &  \multicolumn{2}{c|}{Measured magnetic moment(s)}  & Adopted value &  scale  &  $\gamma_{\beta}$  &  $\Delta_{\gamma(p,n)}$ &  Calculated &  $|exp - calc|$ \\
nucleus &     \multicolumn{2}{c|}{ ($\mu_N$) }               &    ($\mu_N$)  & $S$      &                   &                         &  value ($\mu_N$)  & ($\mu_N$)   \\


\hline
\endfirsthead

\multicolumn{9}{c}%
{ \tablename\ \thetable{}. (\textit{Continued})} \\
\hline 

Parent  &  \multicolumn{2}{c|}{Measured magnetic moment(s)} & Average value &  scale    &  $\gamma_{\beta}$  &  $\Delta_{\gamma(p,n)}$ &  Calculated     &  $|exp - calc|$   \\

nucleus &    \multicolumn{2}{c|}{ ($\mu_N$)  }             & ($\mu_N$)      & $S$     & or $sgn(\gamma_{\beta})$ &                   & value ($\mu_N$) &  ($\mu_N$)      \\


\hline
\endhead

\hline \hline
\endfoot

\hline \hline
\endlastfoot

%
%
   $^{11}$C &  $-0.964(1)$       & [Wo70] &     $-0.964(10)$      &      & $+0.19583(13)$ & $-0.1864$ & $-0.824(28)$  & $0.140$   \\

   $^{11}$B &  $+2.6886489(10)$  & [Mi39,Ep75] &     $+2.6886489(10)$  &      & $+0.19583(13)$ & $+0.1804$ & $+2.541(36)$  & $0.148$   \\

   $^{13}$N &  $0.3222(4)$       & [Be64] &     $-0.3222(4)$      &      & $-0.32558(71)$ & $+0.2646$ & $-0.306(15)$  & $0.016$   \\

   $^{13}$C &  $+0.7024118(14)$  & [Ro54] &     $+0.7024118(14)$  &      & $-0.32558(71)$ & $-0.2586$ & $+0.726(11)$  & $0.024$   \\

   $^{15}$O &  $0.7189(8)$       & [Co63] &     $+0.71950(12)$ &   $1$ & $-0.36585(86)$ & $0.0000$ & $+0.671(12)$   & $0.048$   \\
            &  $0.71951(12)$     & [Ta93] &                    &       &              &      &                  &           \\

   $^{15}$N &  $-0.28318884(5)$  & [Ba62] &     $-0.28318884(5)$  &      & $-0.36585(86)$ & $0.0000$ & $-0.259(16)$  & $0.024$   \\

   $^{17}$F &  $+4.7223(12)$     & [Su66] &     $+4.72136(29)$  &   $1$  & $+0.21993(14)$ & $0.0000$ & $+4.942(63)$  & $0.221$   \\
            &  $+4.7213(3)$    & [Mi93]    &                    &       &              &      &                  &           \\

   $^{17}$O &  $-1.89379(9)$     & [Al51a] &    $-1.89379(9)$     &      & $+0.21993(14)$ & $0.0000$ & $-2.062(49)$  & $0.168$      \\

   $^{19}$Ne & $-1.88542(8)$     & [Ma82] &     $-1.88542(8)$   &      & $+0.92810(37)$ & $+0.4032$ & $-1.924(25)$  & $0.039$   \\

   $^{19}$F &  $+2.628868(8)$ & [Li52,Ba64] &   $+2.628868(8)$    &      & $+0.92810(37)$ & $-0.3972$ & $+2.695(34)$  &  $0.067$     \\
   
   $^{21}$Na & $+2.38630(10)$    & [Am65] &     $+2.38630(10)$    &      & $+0.18527(25)$ & $+0.1657$ & $+2.495(36)$  &  $0.109$     \\

   $^{21}$Ne & $-0.661797(5)$    & [La57] &     $-0.661797(5)$    &      & $+0.18527(25)$ & $-0.1881$ & $-0.763(28)$ &  $0.101$     \\

   $^{23}$Mg & $0.5364(3)$       & [Fu93] &     $-0.5365(2)$    &  1  & $+0.14430(49)$ & $-0.2377$ & $-0.461(26)$ &  $0.076$     \\
             &  $-0.5366(3)$     &  [Yo17] &                    &       &              &      &                  &           \\

   $^{23}$Na & $+2.2176556(6)$ & [Wa54,Fu76] &  $+2.21759(7)$  &   $\rm{unw}$\footnote{Because the two (very precise) measurements differ by many standard deviations, the unweighted average is used.}  & $+0.14430(49)$ & $+0.2137$ & $+2.162(34)$  & $0.056$    \\
             &  $+2.217522(2)$   & [Be74]  &                &       &              &      &                  &           \\

   $^{25}$Al & $3.6455(12)$      & [Mi76a] &     $+3.6455(12)$     &      & $+0.13750(15)$ & $+0.1459$ & $+3.700(56)$  &  $0.054$     \\

   $^{25}$Mg & $-0.85545(8)$     & [Al51b] &     $-0.85545(8)$     &      & $+0.13750(15)$ & $-0.1678$ & $-0.880(44)$ &  $0.024$     \\

   $^{27}$Si & $(-)0.8652(4)$    & [Ma98]  &     $-0.86533(24)$   & $1$  & $+0.11851(12)$ & $-0.1692$ & $-0.701(43)$ &  $0.165$   \\
             &  $0.8654(3)$      & [Ma99] &               &       &              &      &                  &           \\

   $^{27}$Al & $+3.6415069(7)$   & [Ep68] &      $+3.6415069(7)$   &      & $+0.11851(12)$ & $+0.1459$ & $+3.497(54)$  &  $0.144$      \\

   $^{29}$P &  $1.2349(3)$       & [Su71] &      $+1.23475(21)$ & $1$ & $+0.3136(11)$ & $+0.0109$ & $+1.182(15)$ & $0.052$   \\
            &  $1.2346(3)$      & [Zh09] &                   &       &              &      &                  &           \\

   $^{29}$Si & $-0.55529(3)$     & [We53] &      $-0.55529(3)$     &      & $+0.3136(11)$ & $-0.0595$ & $-0.556(11)$ &  $0.001$     \\

   $^{31}$S &  $0.48793(8)$      & [Mi76b] &     $-0.48793(8)$     &      & $+0.30840(77)$ & $-0.3044$ & $-0.424(11)$ &  $0.064$     \\

   $^{31}$P &  $+1.13160(3)$     & [Wa54] &      $+1.13160(3)$     &      & $+0.30840(77)$ & $+0.2472$ & $+1.053(15)$  &  $0.078$    \\

   $^{33}$Cl & $+0.7523(16)$     & [Ro86] &      $+0.7548(5)$ & $1.6$ & $-0.08255(76)$ & $+0.1440$ & $+0.817(32)$ & $0.062$   \\
             &  $+0.7549(3)$     & [Ma04] &                   &       &              &      &                  &           \\

   $^{33}$S &  $+0.6438212(14)$ & [Dh51,Lu73] &  $+0.6438212(14)$  &      & $-0.08255(76)$ & $-0.1193$ & $+0.621(25)$  &  $0.023$     \\

   $^{35}$Ar & $+0.6322(2)$      & [Ma03] &      $+0.6322(2)$     &      & $-0.07459(51)$ & $-0.1424$ & $+0.611(25)$  & $0.021$   \\

   $^{35}$Cl & $+0.8218743(4)$   & [Bl72] &      $+0.8218743(4)$   &      & $-0.07459(51)$ & $+0.1681$ & $+0.831(31)$  & $0.009$      \\

   $^{37}$K &  $+0.20321(6)$     & [Vo71] &      $+0.20321(6)$     &      & $-0.15011(35)$ & $+0.2426$ & $+0.237(34)$  & $0.034$       \\

   $^{37}$Ar & $+1.145(5)$       & [Pi88] &      $+1.145(5)$       &      & $-0.15011(35)$ & $-0.2027$ & $+1.121(27)$  & $0.024$      \\

   $^{39}$Ca & $1.02168(12)$     & [Mi76c] &     $+1.02168(12)$    &      & $-0.17135(39)$ & $0.0000$ & $+0.934(27)$  & $0.087$      \\

   $^{39}$K &  $+0.39150731(12)$ & [Sa74a,b] &    $+0.391487(21)$ &  $\rm{unw}$\footnote{Because the two (very precise) measurements differ by many standard deviations, the unweighted average is used.}  & $-0.17135(39)$ & $0.0000$ & $+0.465(35)$  & $0.074$   \\
            &  $+0.3914662(3)$  & [Be50]  &                    &       &              &      &                  &           \\

   $^{41}$Sc & $+5.431(2)$       & [Mi90] &       $+5.431(2)$       &      & $+0.13602(49)$ & $0.0000$ & $+5.668(78)$  &  $0.237$     \\

   $^{41}$Ca & $-1.594781(9)$    & [Br62] &       $-1.594781(9)$ &       & $+0.13602(49)$ & $0.0000$ & $-1.800(61)$ & $0.205$   \\

   $^{43}$Ti & $0.85(2)$         & [Ma93] &       $-0.85(2)$        &      & $+0.1023(22)$ & $-0.1992$ & $-0.666(65)$ &  $0.184$      \\

   $^{43}$Sc & $+4.62(4)$        & [Co66a] &    $+4.533(22)$ & $2.2$ & $+0.1023(22)$ & $+0.1723$ & $+4.562(81)$ & $0.029$   \\
             &  $+4.528(10)$     & [Av11] &                   &       &              &         &                  &           \\

   $^{45}$V &                    &      &                     &       & $+0.0807(25)$  & $+0.1753$ & $\it +4.230(82)$     &    \\

   $^{45}$Ti & $0.095(2)$        & [Co66b] &   $(+)0.095(2)$\footnote{The sign was obtained from shell model calculations \cite{Buck2001}.}   &      & $+0.0807(25)$ & $-0.2139$ & $-0.335(66)$  & $0.430$       \\

   $^{47}$Cr &            &      &      &      & $+0.1511(43)$ & $-0.1360$ & $\it -0.651(36)$    &         \\

   $^{47}$V &             &      &      &      & $+0.1511(43)$ & $+0.1050$ & $\it +2.368(43)$     &          \\

   $^{49}$Mn &            &      &      &      & $+0.0916(57)$ & $+0.1222$ & $\it +3.270(80)$     &          \\

   $^{49}$Cr & $0.476(3)$ & [Jo70] &  $(-)0.476(3)$\footnote{The sign was obtained from shell model calculations \cite{Buck2001}.} &    & $+0.0916(57)$ & $-0.1595$ & $-0.476(67)$  &  $0.000$      \\

   $^{51}$Fe &           &      &      &      & $+0.0989(33)$ & $-0.1073$ & $\it -0.674(52)$    &         \\

   $^{51}$Mn & $3.5683(13)$  & [Jo71] &   $+3.5691(26)$ &  $2.1$    & $+0.0989(33)$ & $+0.0785$ & $+3.457(63)$  &  $0.111$      \\
             &  $+3.577(4)$  & [Ba15]  &               &           &              &         &               &           \\

   $^{53}$Co &            &       &      &      & $+0.0855(29)$ & $+0.1053$ & $\it +4.546(85)$   &          \\

   $^{53}$Fe & $-0.65(1)$ & [Mi17] &   $-0.65(1)$ &    & $+0.0855(29)$ & $-0.1431$ & $-0.645(69)$  &  $0.005$       \\

   $^{55}$Ni & $(-)0.976(26)$\footnote{The sign was obtained from shell model calculations \cite{Berryman2009}.}    & [Be09] &      $(-)0.976(26)$    &      & $+0.0855(32)$ & $-0.0708$ & $-0.898(71)$  & $0.078$      \\

   $^{55}$Co & $+4.822(3)$       & [Ca73] &    $+4.822(3)$       &      & $+0.0855(32)$ & $+0.0397$ & $+4.776(87)$   & $0.046$       \\

   $^{57}$Cu & $+2.582(7)$  & [Co10] &         $+2.582(7)$ &  & $+0.1462(30)$ & $+0.0607$ & $+2.403(39)$  & $0.179$          \\

   $^{57}$Ni & $-0.7975(4)$      & [Oh96] &    $-0.7975(14)$ &    & $+0.1462(30)$ & $-0.1061$ & $-0.669(31)$  & $0.129$         \\

   $^{59}$Zn &                   &      &                  &      & $+0.1263(57)$ & $-0.14(5)$ & $\it -0.51(8)$   &        \\

   $^{59}$Cu & $+1.891(9)$     & [Go04]  &  $+1.8919(29)$ &  $3.2$  & $+0.1263(57)$ & $+0.11(5)$ & $+2.20(8)$  &  $0.390$   \\
             & $+1.910(4)$     & [Co10]  &                &         &             &          &               &           \\
             & $+1.8910(9)$    &  [Vi11] &                &         &             &          &               &          \\

   $^{61}$Ga &            &      &      &      & $+0.1408(89)$ & $+0.11(5)$ & $\it +2.29(9)$     &        \\

   $^{61}$Zn &            &      &      &      & $+0.1408(89)$ & $-0.14(5)$ & $\it -0.59(8)$    &        \\

   $^{63}$Ga & $+1.469(5)$ & [Pr12] &   $+1.469(5)$ &        &    +     &        &           &           \\

   $^{67}$Se &            &      &      &      & $-0.067(12)$ & $-0.14(5)$ & $\it +0.94(13)$    &        \\

   $^{67}$As &            &      &      &      & $-0.067(12)$ & $+0.11(5)$ & $\it +1.61(14)$    &        \\

   $^{71}$Kr &            &      &      &      & $-0.078(17)$ & $-0.14(5)$ & $\it +1.04(14)$    &        \\

   $^{71}$Br &            &      &      &      & $-0.078(17)$ & $+0.11(5)$ & $\it +1.50(15)$    &        \\

   $^{75}$Sr &            &      &      &      & $-0.134(40)$ & $-0.14(5)$ & $\it +0.93(17)$    &        \\

   $^{75}$Rb &            &      &      &      & $-0.134(40)$ & $+0.11(5)$ & $\it +0.54(19)$     &        \\

   $^{77}$Sr & $-0.348(4)$ & [Li92] &   $-0.348(4)$ &        &    $-$     &        &           &           \\

\end{longtable*}
}


%


\begin{table}
\caption{References from which magnetic moment results have been rejected. The correlation between the alphabetical reference code used here and the actual reference numbers is listed in Table~\ref{reference-key-moments}.}
\label{unusedref-moments}
\begin{ruledtabular}

\begin{tabular} { l  |  l  l  |  p{4cm} }
Nucleus  &  Value  &  [Ref.]  &    Reason for rejection \\

 \hline

$^{27}$Si  &  $-0.8554(4)$  &  [Hu84]    & No correction for diamagnetism was applied. Result is strongly deviating from the two other more recent and mutually consistent results reported in [Ma89] and [Ma99].  \\
\hline
$^{19}$Ne  &  $-1.8846(8)$  &  [Ge05]    & Uncertainty is a factor 10 or more larger than that of the most precise measurement. \\
$^{35}$Ar  &  $+0.633(2)$   &  [Ca65]    & \\
$^{35}$Ar  &  $+0.633(7)$   &  [Kl96]    & \\
$^{37}$Ar  &  $+0.95(20)$   &  [Ro65]    & \\
$^{39}$K   &  $+0.39147(3)$ &  [Du93]    &     \\
$^{41}$Ca  &  $-1.61(2)$    &  [An82]    &     \\
           &  $-1.5942(7)$  &  [Ar83]    &     \\
$^{57}$Ni  &  $0.88(6)$     &  [Ro75]    &     \\
$^{59}$Cu  &  $+1.84(3)$    &  [St08]    &    \\
\hline
$^{57}$Cu  &  $2.00(5)$     &  [Mi06]    & Result is much less precise and differs by more than 10 standard deviations from the newer value from [Co10] which agrees well with $\mu \simeq 2.5 \mu_N$ as is calculated in the shell model \cite{Semon1996, Golovko2004, Honma2004}.    \\


\end{tabular}
\end{ruledtabular}

\end{table}



\newpage

\begin{table*}
\caption{Magnetic moments for the mirror nuclei with $A =$ 61, 67, 71 and 75, calculated with Eqs.~\ref{gamma-beta}-\ref{gamma-n} for the two possible signs of $\gamma_\beta$. The preferred values for the magnetic moments, listed in column 6 and corresponding to sgn$(\gamma_{\beta}) = +$ for the $j = l + 1/2$ $p_{3/2}$ subshell and sgn$(\gamma_{\beta}) = -$ for the $j = l - 1/2$ $f_{5/2}$ subshell, agree with the systematic of the magnetic moments for neighboring nuclei with the odd proton or neutron occupying the same shell model orbital, listed in the last column. Note that when the sign for a magnetic moment value in the last column is given, the sign has been explicitly measured. The correlation between the  alphabetical reference code used here and the actual reference numbers is listed in Table~\ref{reference-key-moments}.}
\label{moments-from-systematics}
\begin{ruledtabular}

\begin{tabular} { c c c | c | c | c | p{8cm} }
  
Isot.  & $J^\pi$ &     sub-    &   Magn. mom. ($\mu_N$)          &     Magn. mom. ($\mu_N$)       & Preferred       &  Reason, based on magnetic moments of neighbouring  \\
       &         &     shell   & for sgn$(\gamma_{\beta}) = -$   &  for sgn$(\gamma_{\beta}) = +$ & value ($\mu_N$) &  nuclei with similar single particle configuration  \\

 \hline

$^{61}$Ga & $3/2^{-}$ &   $p_{3/2}$  &  $+2.29(19)$  &    $+0.33(7)$  & $+2.29(19)$  &  similar to the value $\mu \approx$ +1.5 to +2.6 for the 3/2$^-$ ground states of $^{63-77}$Ga [Ch10] and $^{57-73}$Cu [Vi10]  \\

$^{61}$Zn & $3/2^{-}$ &              &  $-0.59(8)$   &   $+0.97(8)$   &  $-0.59(8)$  &  similar to the values $\mu = -0.28164(5)$ and $\mu = -0.8(2)$ for the 3/2$^-$ g.s. of $^{63}$Zn [La69], resp. the lowest-lying 3/2$^-$ state of $^{65}$Zn [We75]  \\

\hline

$^{67}$Se & $5/2^{-}$ &   $f_{5/2}$  &  $-0.30(13)$  &   $+0.94(13)$  &  $+0.94(13)$  &  similar to the values $\mu = 0.735(7)$ [Ol70], $\mu = +1.018(10)$ [Mo68], $\mu = +1.124(10)$ [Bl68] and $\mu = +1.12(3)$ [Za84] for the lowest lying 5/2$^-$ states in respectively $^{69}$Ge, $^{71}$Ge, $^{79}$Kr, and $^{77}$Se (see also text) \\

$^{67}$As & $5/2^{-}$ &              &  $+3.04(14)$  &   $+1.61(14)$  &  $+1.61(14)$  &  similar to the values $\mu = +1.6229(16)$ [Go05], $\mu = +1.674(2)$ [He76a,b], $\mu = +1.63(10)$ [Bo63], $\mu = 1.0(3)$ [Sp94] and $\mu = 1.6(5)$ [Ja96] for the lowest 5/2$^-$ states in respectively $^{69}$As, $^{71}$As, $^{73}$As, $^{79}$Br, and $^{81}$Br (see also text) \\

$^{71}$Kr & $5/2^{-}$ &              &  $-0.40(14)$  &   $+1.04(14)$  &  $+1.04(14)$  &  similar to the values $\mu = 0.735(7)$ [Ol70], $\mu = +1.018(10)$ [Mo68], $\mu = +1.124(10)$ [Bl68] and $\mu = +1.12(3)$ [Za84] for the lowest lying 5/2$^-$ states in respectively $^{69}$Ge, $^{71}$Ge, $^{79}$Kr, and $^{77}$Se  \\

$^{71}$Br & $5/2^{-}$ &              &  $+3.15(15)$  &   $+1.50(15)$  &  $+1.50(15)$  &  similar to the values $\mu = +1.6229(16)$ [Go05], $\mu = +1.674(2)$ [He76a,b], $\mu = +1.63(10)$ [Bo63], $\mu = 1.0(3)$ [Sp94] and $\mu = 1.6(5)$ [Ja96] for the lowest 5/2$^-$ states in respectively $^{69}$As, $^{71}$As, $^{73}$As, $^{79}$Br, and $^{81}$Br \\

$^{75}$Sr & $3/2^{-}$ &              &  $-0.55(17)$  &   $+0.93(17)$  &  $+0.93(17)$$^a$   &  (see $^{75}$Rb and text)   \\

$^{75}$Rb & $3/2^{-}$ &              &  $+2.25(19)$  &   $+0.54(19)$  &  $+0.54(19)$$^a$   &  similar to the value $\mu = +0.6544680(16)$ for the 3/2$^-$ g.s. state of $^{77}$Rb [Du86, Th81] (see also text)    \\


\end{tabular}
\footnotetext[1]{The calculated moments for sgn$(\gamma_{\beta}) = -1$ are preferred because of the agreement for $^{75}$Rb with the experimental value $\mu = +0.6544680(16)$ for the 3/2$^-$ g.s. state of the neighboring $^{77}$Rb [Du86, Th81], and because the 3/2$^-$ g.s. states of $^{79}$Sr [Bu90] and $^{81}$Rb [Th81] with, respectively, $\mu = -0.474(4)$ and $\mu = +2.0595(14)$, are farther away in mass and also situated above the N = 40 sub-shell closure.}
\end{ruledtabular}

\end{table*}

\begin{table*}
\caption{Reference key relating the reference codes used in Tables~\ref{Magnetic-moments} to \ref{moments-from-systematics} and to the actual reference numbers.}
\label{reference-key-moments}
\begin{ruledtabular}
\begin{tabular}{l c l c l c l c l c l c}


Table & Reference & Table & Reference & Table & Reference & Table & Reference & Table & Reference & Table & Reference \\
code  &   No.     & code  &    No.    & code  &   No.     & code  &    No.    & code  &    No.    & code  &    No.    \\

\hline

 & & & & & & & & & & & \\

Al51a & \cite{Alder1951a} &  Al51b & \cite{Alder1951b} &  Am65 & \cite{Ames1965} &  An82 & \cite{Andl1982} &  Ar83 & \cite{Arnold1983} &  Av11 & \cite{Avgoulea2011} \\
Ba15 & \cite{Babcock2015} &  Ba62 & \cite{Baldeschweiler1962} &  Ba64 & \cite{Baker1964} &  Be09 & \cite{Berryman2009} &  Be64 & \cite{Bernstein1964} &  Be74 & \cite{Beckmann1974} \\
Bl68 & \cite{Bleck1968} & Bl72 & \cite{Blaser1972} &  Bo63 & \cite{Bodenstedt1963} &  Br62 & \cite{Brun1962} &  Bu90 & \cite{Buchinger1990} &  Ca65 & \cite{Calaprice1965} \\
Ca73 & \cite{Callaghan1973} & Ch10 & \cite{Cheal2010} &  Co10 & \cite{Cocolios2010} &  Co63 & \cite{Commins1963} &  Co66a & \cite{Cornwell1966a} &  Co66b & \cite{Cornwell1966b} \\
Dh51 & \cite{Dharmatti1951} & Du86 & \cite{Duong1986} &  Du93 & \cite{Duong1993} &  Ep68 & \cite{Epperlein1968} &  Ep75 & \cite{Epperlein1975} &  Fu76 & \cite{Fuller1976} \\
Fu93 & \cite{Fukuda1993} & Ge05 & \cite{Geithner2005} &  Go04 & \cite{Golovko2004} &  Go05 & \cite{Golovko2005} &  He76a & \cite{Herzog1976a} &  He76b & \cite{Herzog1976b} \\
Hu84 & \cite{Hugg1984} & Ja96 & \cite{Jakob1996} &  Jo70 & \cite{Jonsson1970} &  Jo71 & \cite{Jonsson1971} &  Kl96 & \cite{Klein1996} &  La57 & \cite{LaTourrette1957} \\
La69 & \cite{Laulainen1969} & Li52 & \cite{Lindstrom1952} & Li92 & \cite{Lievens1992} & Lu73 & \cite{Lutz1973} &  Ma02 & \cite{Matsuta2002} &  Ma04 & \cite{Matsuta2004} \\
Ma82 & \cite{MacArthur1982} & Ma93 & \cite{Matsuta1993} & Ma98 & \cite{Matsuta1998} & Ma99 & \cite{Matsuta1999} &  Mi06 & \cite{Minamisono2006}  &  Mi17 & \cite{Miller2017} \\
Mi39 & \cite{Millman1939} & Mi76a & \cite{Minamisono1976a} & Mi76b & \cite{Minamisono1976b} & Mi76c & \cite{Minamisono1976c} & Mi90 & \cite{Minamisono1990} &  Mi93 & \cite{Minamisono1993}\\
Mo00 & \cite{Mohr2000} & Mo68 & \cite{Morgenstern1968} & Ne11 & \cite{Neronov2011} & Ne77 & \cite{Neronov1977} &  Oh96 & \cite{Ohtsubo1996} & Ol70 & \cite{Oluwole1970} \\
Pi88 & \cite{Pitt1988} & Pr12 & \cite{Procter2012} & Ro54 & \cite{Royden1954} & Ro65 & \cite{Robertson1965} & Ro75 & \cite{Rosenblum1975} & Ro86 & \cite{Rogers1986} \\
Sa74a & \cite{Sahm1974a} & Sa74b & \cite{Sahm1974b} & Sp94 & \cite{Speidel1994} &  St08 & \cite{Stone2008} & St14 & \cite{Stone2014} & Su66 & \cite{Sugimoto1966} \\
Su71 & \cite{Sugimoto1971} & Ta93 & \cite{Tanigaki1993} & Th81 & \cite{Thibault1981} &  Ul17 & \cite{Ulmer2017} & Vi10 & \cite{Vingerhoets2010} & Vi11 & \cite{Vingerhoets2011} \\
Vo71 & \cite{vonPlaten1971} &  Wa54 & \cite{Walchli1954} & We53 & \cite{Weaver1953} &  We75 & \cite{Wender1975} & Wo70 & \cite{Wolber1970} & Yo17 & \cite{Yordanov2017} \\
Za84 & \cite{Zamboni1984}  &  Zh09 & \cite{Zhou2009} 



\end{tabular}
\end{ruledtabular}

\end{table*}

\subsubsection{Experimental weak magnetism form factors $b$ and $b/Ac$}
\label{b and b/Ac for mirrors}

\begin{figure}
	\includegraphics[scale=0.28]{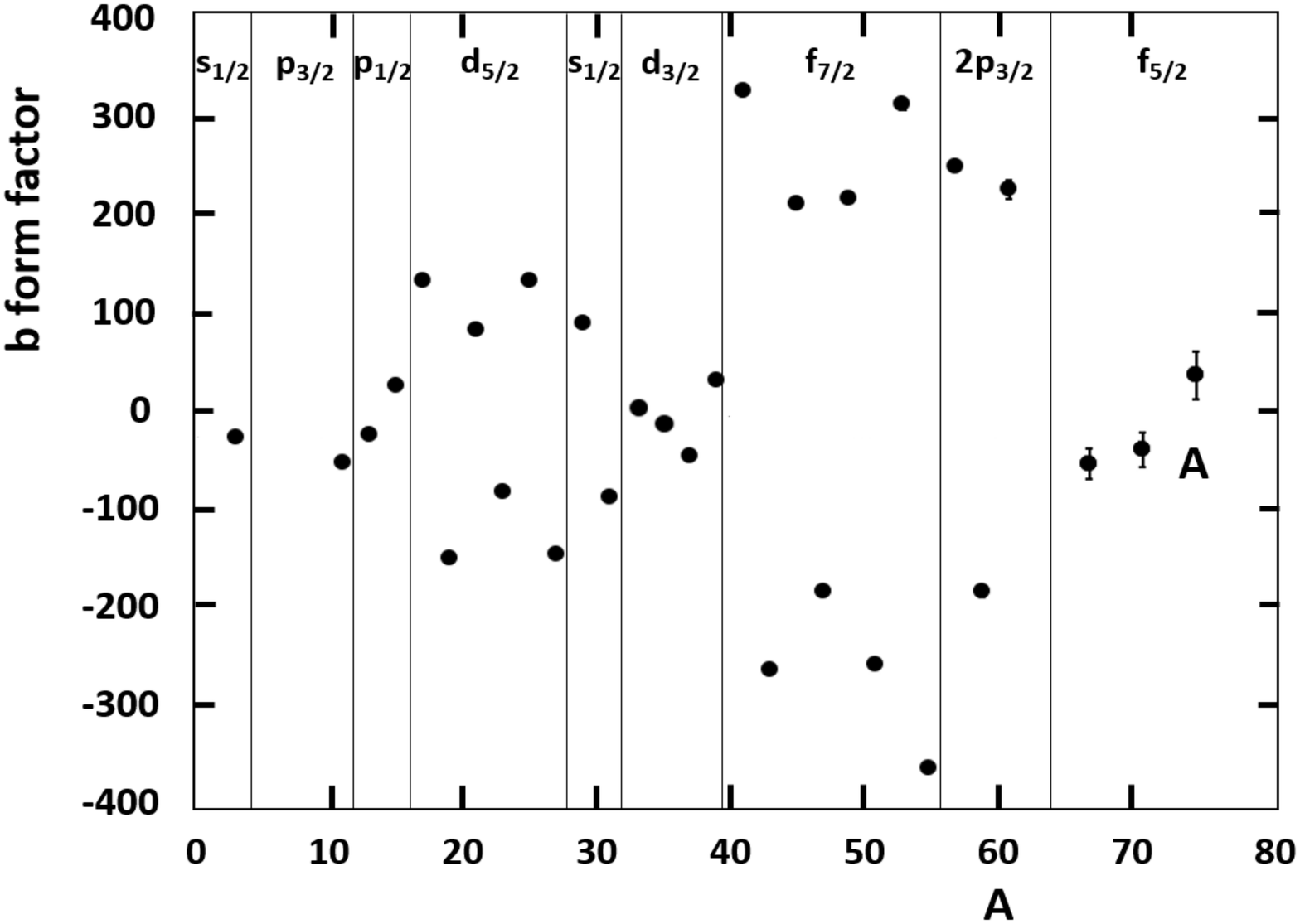}
	\caption{Weak magnetism form factor $b$ for the $T = 1/2$ mirror $\beta$ transitions from the experimental/calculated magnetic moments in column 8 of Table~\ref{table:spiegelkernen} and Eq.~(\ref{eq:weak_magnetism_mirror}). For $^{67}$Se, $^{71}$Kr, and $^{75}$Sr the values corresponding to the upper and lower limits of the BR values (see Tables~\ref{adopted-t12-BR-QEC} and \ref{input-data-and-calculated-quantities}) are shown. The thin vertical lines separate the shell model subshells, the labels of which are indicated. If not explicitly shown, error bars are smaller than the size of the symbol.}
		\label{fig:Mirror_b_vs_A}
\end{figure}

With the isovector magnetic moment for most of the $T = 1/2$ mirror nuclei pairs up to $A = 75$ determined, we can deduce the weak magnetism $b$ form factors using Eq.~(\ref{eq:weak_magnetism_mirror}) and combine these with the $c$ form factors deduced from the corrected $\mathcal{F}t^\mathrm{mirror}$ values (Table~\ref{input-data-and-calculated-quantities}) using Eq.~(\ref{master}), to obtain the normalized $b/Ac$ form factors ($A$ being the mass number). Results are listed in Table~\ref{table:spiegelkernen} and shown in Figure~\ref{fig:Mirror_b_vs_A}. As can be seen, large absolute values for $\mu_f - \mu_i$, and so for $b$, are observed for mirror pairs in subshells with $j = l+1/2$ ($j$ and $l$ being, respectively, the spin and orbital angular momentum of the subshell of the odd nucleon), i.e. in the $p_{3/2}$ ($A$ = 11), $d_{5/2}$ ($A$ = 17-27), $s_{1/2}$ ($A$ = 29-31), $f_{7/2}$ ($A$ = 41-55), and $2p_{3/2}$ ($A$ = 57-61) subshells. Small values for $\mu^{\mp}$ are observed for mirror pairs in subshells with $j = l-1/2$, i.e. the $p_{1/2}$ ($A$ = 13-15), $d_{3/2}$ ($A$ = 33-39), and $f_{5/2}$ ($A$ = 67-75) subshells. This is fully in line with the values for the difference of the odd-proton and odd-neutron Schmidt values, $\mu_p^{Sch} - \mu_n^{Sch}$, for the different nuclear spin values, which is large for $j = l+1/2$ (i.e. between 4.7 and 7.7 for the spins involved here) and relatively small for $j = l-1/2$ (i.e. ranging between -1.0 and +0.2) (see Table~\ref{table:schmidt}). As a consequence, the values for $b$ share the same systematic, as is clear from Fig.~\ref{fig:Mirror_b_vs_A} which shows large values for $b$ in the $d_{5/2}$, $f_{7/2}$, and $2p_{3/2}$ subshells ($j=l+1/2$), and significantly smaller values in the $p_{1/2}$, $d_{3/2}$, and $f_{5/2}$ subshells ($j=l-1/2$).
%
%
%
%

\begin{table*}

\begin{center}
\begin{ruledtabular}

{ \footnotesize
\caption{Input data and CVC results for the $b$ form factors and $b/Ac$ ratios for the mirror $\beta$ transitions up to $A = 39$.
	$\mu^{\mp}$ is the difference of the magnetic moment of the mother and daughter isotopes, where the values listed in Table~\ref{Magnetic-moments} were used. If several magnetic moments values are available, the weighted average was used. Values for $b$ were calculated from Eq.~(\ref{eq:weak_magnetism_mirror}) with $a(0) = 1$. Values for the Gamow-Teller form factor $c$ are from Table~\ref{input-data-and-calculated-quantities}.}

\begin{tabular}{c||c|c|c|c|c|c|c|c}

$\beta$ transition &             $A$ &        $J$ &  subshell  &  $\mu_f - \mu_i$  &    $b$\footnote{For $A =$ 45-53 and $A >$ 57, at least one of the mirror pair magnetic moments was not measured experimentally but calculated based on the experimental $\mathcal{F}$t value and the procedure outlined in Sec.~\ref{sec:mirror moments Perez} \cite{Perez2008}.}  &    $c$      & $(b/Ac)^{exp}$ & $(b/Ac)^{exp}$ \\
                   &                 &            &            &    [$\mu_N$]  &              &             &                &  subshell      \\
                   &                 &            &            &               &              &             &                &  average \footnote{As values and error bars sometimes vary significantly unweighted averages are given as well, in italics.}      \\

\hline

         H$\rightarrow$He &        3 &         1/2 &  $s_{1/2}$  & $-5.106460174(31)$ & $-26.53394540(16)$ &  $-2.1053(12)$ &   $4.2012(23)$ &  $4.2012(23)$\\
\hline
         C$\rightarrow$B &         11 &        3/2 &  $p_{3/2}$  &  $+3.6526(10)$ & $-51.871(14)$ &  $-0.75442(28)$ &   $6.2506(29)$ &   $6.2506(29)$    \\
\hline
         N$\rightarrow$C &         13 &        1/2 &  $p_{1/2}$  & $+1.02461(40)$ & $-23.0708(90)$ & $-0.55962(34)$ &   $3.1712(23)$ &   $2.93(21)$  \\

         O$\rightarrow$N &         15 &        1/2 &           &  $-1.00269(12)$ &  $26.0506(31)$ &   $0.63023(46)$ &  $2.7557(20)$ &     $\it{2.96(29)}$ \\
\hline
         F$\rightarrow$O &         17 &        5/2 &  $d_{5/2}$  &  $-6.61515(30)$ & $133.0616(61)$ &   $1.29555(65)$ &  $6.0416(30)$ &   $5.82(42)$  \\

        Ne$\rightarrow$F &         19 &        1/2 &           & $+4.51429(8)$ & $-148.5605(26)$ &  $-1.60203(65)$ &   $4.8807(20)$ &    $\it{6.18(95)}$      \\

        Na$\rightarrow$Ne &         21 &       3/2 &           &  $-3.04810(10)$ &  $82.6366(27)$ &   $0.71245(39)$ &    $5.5233(30)$  &          \\

        Mg$\rightarrow$Na &         23 &       3/2 &           &  $+2.75409(21)$ & $-81.7768(63)$ &  $-0.55413(47)$ &    $6.4164(55)$ &          \\

        Al$\rightarrow$Mg &         25 &       5/2 &           & $-4.5010(12)$ &  $133.140(36)$ &   $0.80844(42)$ &    $6.5875(38)$ &          \\

        Si$\rightarrow$Al &         27 &       5/2 &           & $+4.50684(24)$ & $-143.9792(77)$ &  $-0.69659(30)$ &    $7.6552(33)$ &   \\
\hline
         P$\rightarrow$Si &         29 &       1/2 &  $s_{1/2}$  &  $-1.79004(21)$ & $89.913(11)$ &   $0.53798(47)$ &    $5.7631(51)$ &   $5.43(21)$ \\

         S$\rightarrow$P &          31 &       1/2 &           & $+1.61953(9)$ & $-86.9584(46)$ &  $-0.52939(33)$ &    $5.2988(33)$ &    $\it{5.53(33)}$\\
\hline
        Cl$\rightarrow$S &         33 &        3/2 &  $d_{3/2}$  &   $-0.11098(50)$ &    $4.728(21)$ &  $-0.31416(28)$ & $-0.4561(21)$ &   $0.21(59)$  \\

        Ar$\rightarrow$Cl &         35 &       3/2 &           &  $+0.18967(20)$ &  $-8.5704(90)$ &   $0.28199(17)$ &  $-0.8684(11)$ &   $\it{0.5(14)}$      \\

         K$\rightarrow$Ar &         37 &       3/2 &           & $+0.9418(50)$ &    $-44.99(24)$ &  $-0.57789(39)$ &    $2.104(11)$ &          \\

        Ca$\rightarrow$K &         39 &        3/2 &           & $-0.63019(12)$ &  $31.7295(61)$ &   $0.66061(50)$ &   $1.2316(10)$ &    \\
\hline
         Sc$\rightarrow$Ca &        41 &       7/2 &  $f_{7/2}$  & $-7.0258(20)$ & $326.626(93)$ &   $1.0743(21)$ &   $7.415(14)$ &   $7.47(12)$  \\

         Ti$\rightarrow$Sc &        43 &       7/2 &           &  $+5.383(30)$ & $-262.5(14)$ &  $-0.8097(69)$ &    $7.539(77)$ &   $\it{8.05(97)}$       \\

         V$\rightarrow$Ti &         45 &       7/2 &           & $-4.135(82)$ & $211.0(42)$   & $0.6346(58)$ &   $7.39(16)$ &          \\

         Cr$\rightarrow$V &         47 &       3/2 &           &  $+3.012(57)$ &  $-182.8(34)$ &  $-0.5794(42)$ &  $6.71(14)$ &          \\

         Mn$\rightarrow$Cr &        49 &       5/2 &           &  $-3.746(80)$ & $217.2(46)$ &   $0.5373(75)$ &  $8.25(21)$ &          \\

        Fe$\rightarrow$Mn &         51 &       5/2 &           & $+4.243(52)$ & $-256.0(31)$ &  $-0.5811(49)$ &   $8.64(13)$ &          \\

        Co$\rightarrow$Fe &         53 &       7/2 &           &  $-5.196(86)$ &  $312.3(51)$ &   $0.6730(72)$ &    $8.75(17)$  &          \\

        Ni$\rightarrow$Co &         55 &       7/2 &           &  $+5.798(26)$ & $-361.6(16)$ &  $-0.6752(80)$ &    $9.74(12$) &   \\
\hline
        Cu$\rightarrow$Ni &         57 &       3/2 &  $p_{3/2}$  & $-3.3795(71)$ &  $248.69(53)$ &   $0.5639(28)$ &    $7.737(42)$ &   $7.67(20)$  \\

        Zn$\rightarrow$Cu &         59 &       3/2 &           & $+2.402(80)$ & $-182.9(61)$ &  $-0.4876(42)$ &    $6.36(30)$ &   $\it{6.98(70)}$       \\

         Ga$\rightarrow$Zn &        61 &       3/2 &           &  $-2.87(12)$ & $226.0(95)$ &   $0.5421(79)$ &    $6.84(30)$ &   \\
\hline
         Se$\rightarrow$As &        67 &       5/2 &  $f_{5/2}$  & $+0.67(19)$ & $-53(15)$ &  $-0.3873(89)$  &   $2.05(59)$\footnote[3]{Value for upper limit of BR; see also Table~\ref{adopted-t12-BR-QEC} and Sec.~\ref{Note on A>63}.} &     \\

                           &           &           &           &           &          &  $-0.2048(50)$  &   $3.9(11)$\footnote[4]{Value for lower limit of BR; see also Table~\ref{adopted-t12-BR-QEC} and Sec.~\ref{Note on A>63}.} &     \\
         
         Kr$\rightarrow$Br &        71 &       5/2 &           &  $0.46(21)$ & $-39(17)$ &  $0.454(16)$ &    $-1.20(54)$\footnotemark[3] &       \\

                          &           &           &           &           &          &   $0.1682(61)$  &   $-3.2(14)$\footnotemark[4] &     \\

        Sr$\rightarrow$Rb &         75 &       3/2 &           & $-0.38(25)$ & $37(25)$ &  $0.525(32)$ &   $0.93(63)$\footnotemark[3] &   \\

                          &           &           &           &           &          &  $0.367(22)$  &   $1.34(90)$\footnotemark[4] &     \\

\end{tabular}
\label{table:spiegelkernen}
}
\end{ruledtabular}
\end{center}
\end{table*}

\begin{figure}
\includegraphics[scale=0.30]{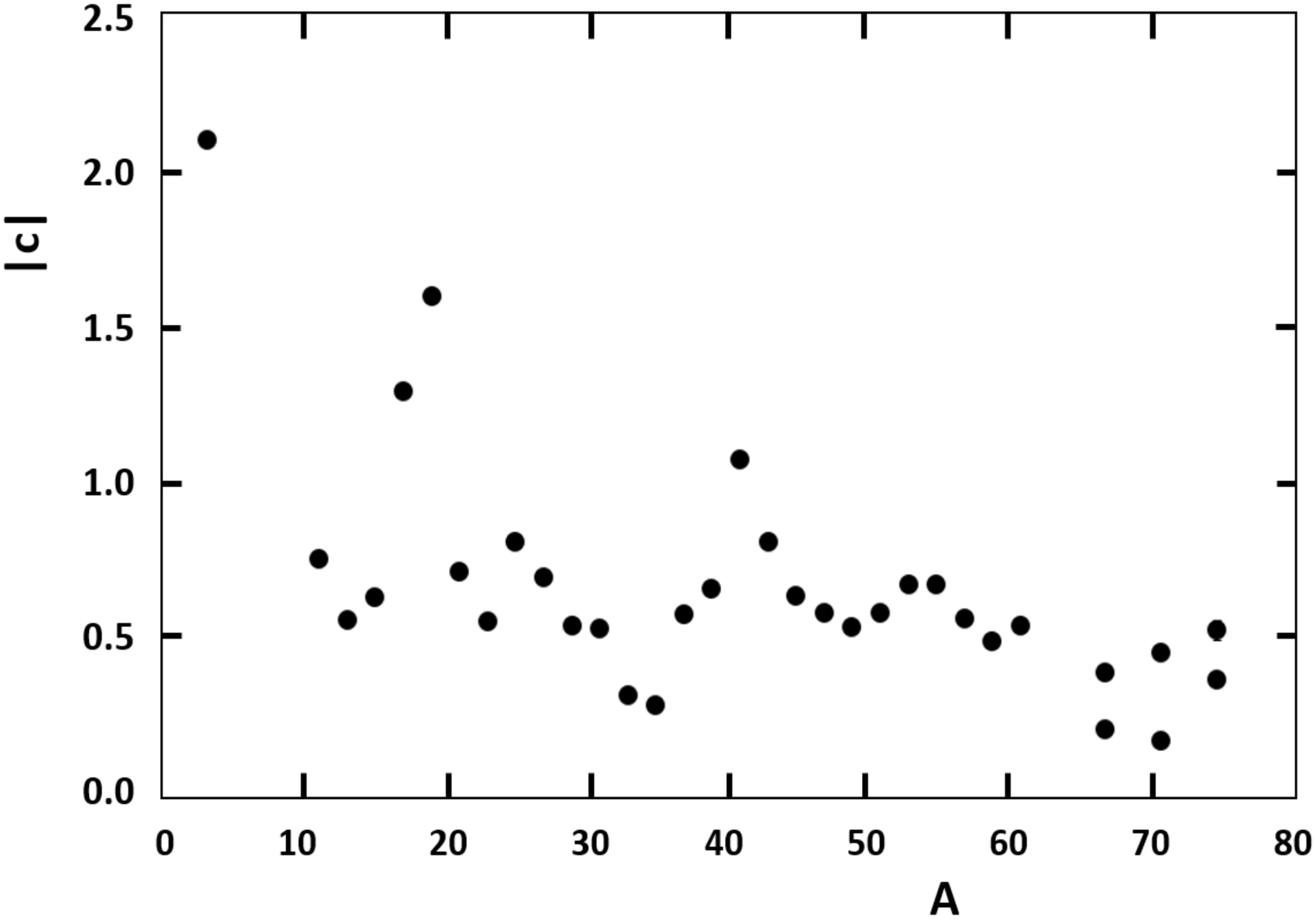}
\caption{Absolute value of the Gamow-Teller form factor, $c$, as a function of the mass number. A slight general downward trend with increasing mass is apparent, with clear shell closure effects around the doubly-magic $A=16$ and $A=40$ nuclei. For $^{67}$Se, $^{71}$Kr, and $^{75}$Sr the values corresponding to the upper and lower limits of the BR values (see Tables~\ref{adopted-t12-BR-QEC} and \ref{input-data-and-calculated-quantities}) are shown. If not explicitly shown, error bars are smaller than the size of the symbol.}
\label{fig:Mirror-c_vs_A}
\end{figure}

\begin{table}
\begin{ruledtabular}

{ \footnotesize
\caption{Differences, $\mu_p^{Sch} - \mu_n^{Sch}$, of the Schmidt values for magnetic moments of odd-proton and odd-neutron nuclei with spin $j = 1/2$ to $j = 7/2$, for $j = l+1/2$ and $j = l-1/2$, and for both $g_S^{free}$ and $0.6 \times g_S^{free}$. Large values are obtained for $j = l+1/2$ and much smaller values for $j = l-1/2$.}
\label{table:schmidt}

\centering
\begin{tabular}{c | c | c | c | c} 

     &    \multicolumn{4}{c}{$\mu_p^{Sch} - \mu_n^{Sch}$}   \\
\hline
     &   \multicolumn{2}{c|}{$j = l+1/2$}    &    \multicolumn{2}{c}{$j = l-1/2$}    \\
\hline
$j$ & $g_S^{free}$ & $0.6 \times g_S^{free}$ & $g_S^{free}$ & $0.6 \times g_S^{free}$ \\
\hline

$1/2$ & $4.71$ & $2.82$ & $-0.90$ & $-0.27$ \\
$3/2$ & $5.71$ & $3.82$ & $-1.02$ & $+0.11$ \\
$5/2$ & $6.71$ & $4.82$ & $-0.50$ & $+0.84$ \\
$7/2$ & $7.71$ & $5.82$ & $+0.23$ & $+1.69$ \\

\end{tabular}
}
\end{ruledtabular}
\end{table}
%
%
%
%
%
Figure~\ref{fig:Mirror-c_vs_A} shows the evolution of the Gamow-Teller form factor, $|c|$, (the sign of $c$ is straightforwardly obtained from shell model calculations) as a function of the mass number. While a clear local increase is visible when crossing the doubly-magic nuclei at $A=16$, respectively $A=40$, a general overall, slow decrease can be observed. This can be understood intuitively as the Fermi surfaces of proton and neutron become increasingly distanced for higher masses.

\begin{figure}
\includegraphics[width=0.45\textwidth]{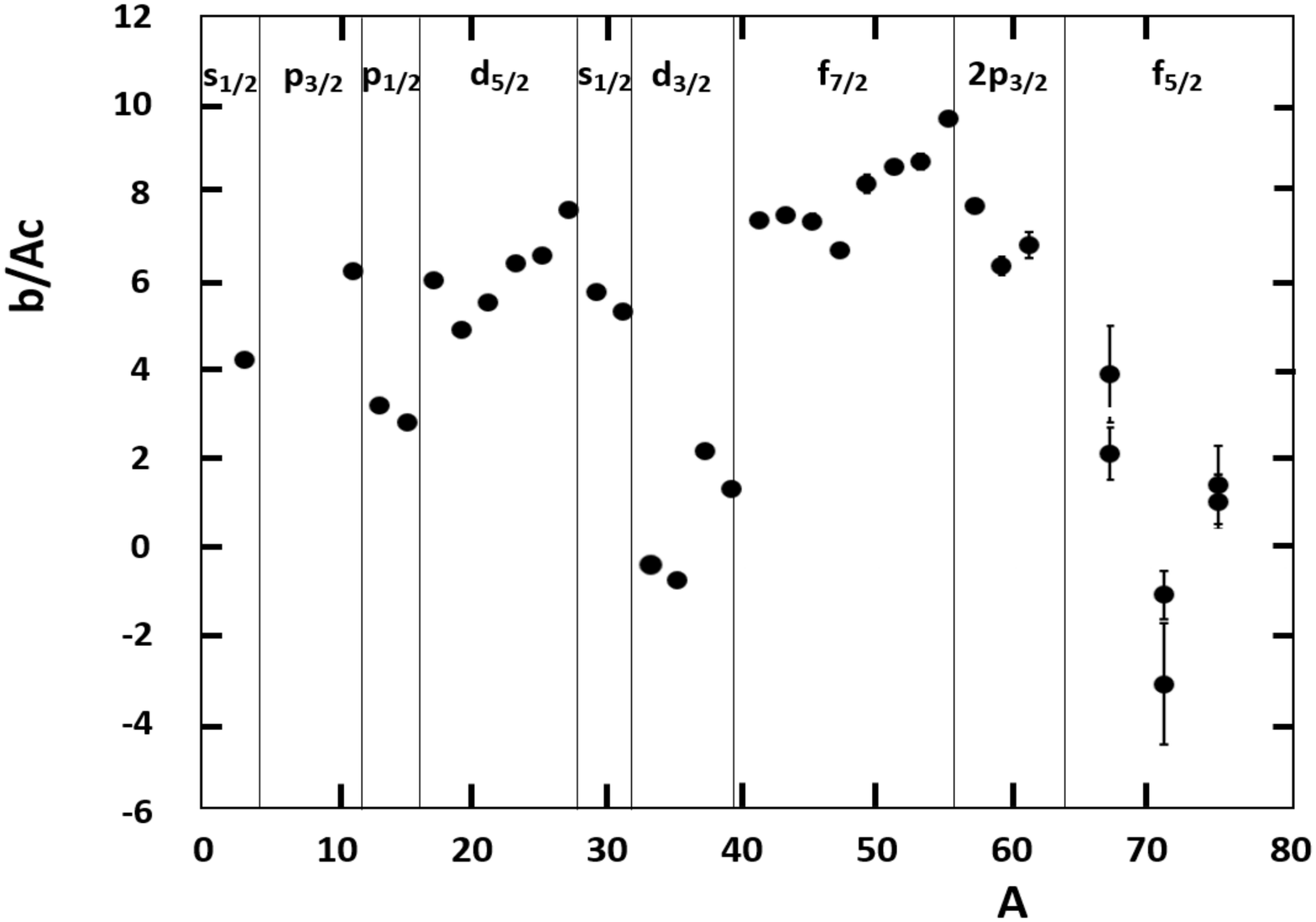}
\caption{Experimental values of $b/Ac$ for the $T = 1/2$ mirror $\beta$ transitions (Table~\ref{table:spiegelkernen}). The different shell model subshells are indicated. For $^{67}$Se, $^{71}$Kr, and $^{75}$Sr the values corresponding to the upper and lower limits of the BR values are shown. See also Tables~\ref{adopted-t12-BR-QEC}, \ref{input-data-and-calculated-quantities} and \ref{table:spiegelkernen}, and text. If not explicitly shown, error bars are smaller than the size of the symbol.
}
\label{fig:Mirror-bAc_vs_A}
\end{figure}

This decrease persists in the ratio $b/Ac$ as is seen in Fig.~\ref{fig:Mirror-bAc_vs_A}, where again a distinction is made between $j=l\pm 1/2$ transitions. Despite significant variation in the absolute value of $b$ (Figure~\ref{fig:Mirror_b_vs_A} and Table~\ref{table:spiegelkernen}), subshell behaviour of $b/Ac$ is reasonably uniform. The already large values in the large $d_{5/2}$ and $f_{7/2}$ subshells are further accentuated by the decrease of the $|c|$ values throughout these subshells (Fig.~\ref{fig:Mirror-c_vs_A}). The behavior of $b/Ac$ versus $A$ for the mirror decays is thus understood in terms of the single particle Schmidt values for the magnetic moments and the slight decrease of the size of the Gamow-Teller form factor $c$ with increasing $A$.
\\

%
%
%
%
%
%


\subsubsection{Theoretical form factor comparison}
\label{sec:Theoretical form factor comparison}

While some substructure is clearly visible in the experimental results of Fig. \ref{fig:Mirror-bAc_vs_A}, a variation of a full order of magnitude is observed. In the face of ever-increasing experimental precision, it is paramount that theory can meet this challenge. Here we will compare experimental results to theoretical predictions, using both single-particle estimates and the nuclear shell model.

\paragraph{\bf Single particle calculations}

As noted in the previous sections, significant variation occurs in the values for $b/Ac$ for different spin-orbit orbitals. This can be understood intuitively from a single particle view of the transition. Using Eq. (\ref{eq:bAc_impulse_approx}) this means we are interested in the ratio of matrix elements $R = M_L/M_{GT}$. Using simple harmonic oscillator wave functions this ratio is
\begin{align}
\frac{M_L}{M_{GT}} &= \frac{\langle n_f l_f j_f | \bm{l} | n_i l_i j_ \rangle}{\langle n_f l_f j_f | \bm{\sigma} | n_i l_i j_ \rangle} = (-1)^{j_i-j_f} \nonumber \\
&\times\frac{\left\{\begin{array}{ccc}
1/2 & l & j_i \\
1 & j_f & l
\end{array} \right\}}{\left\{\begin{array}{ccc}
l & 1/2 & j_i \\
1 & j_f & 1/2
\end{array} \right\}}
\frac{\sqrt{l(l+1)(2l+1)}}{\sqrt{6}} ~ .
\end{align}
\noindent Using the four possibilities for transitions with $j_{i,f} = l \pm 1/2$, this results in
\begin{align}
    R^{--} &= -(l+1) \qquad R^{-+} = -1/2 \\
    R^{+-} &= -1/2 \qquad R^{++} = l.
    \label{eq:ratio_ML_MGT}
\end{align}

\begin{figure}
    \centering
    \includegraphics[width=0.5\textwidth]{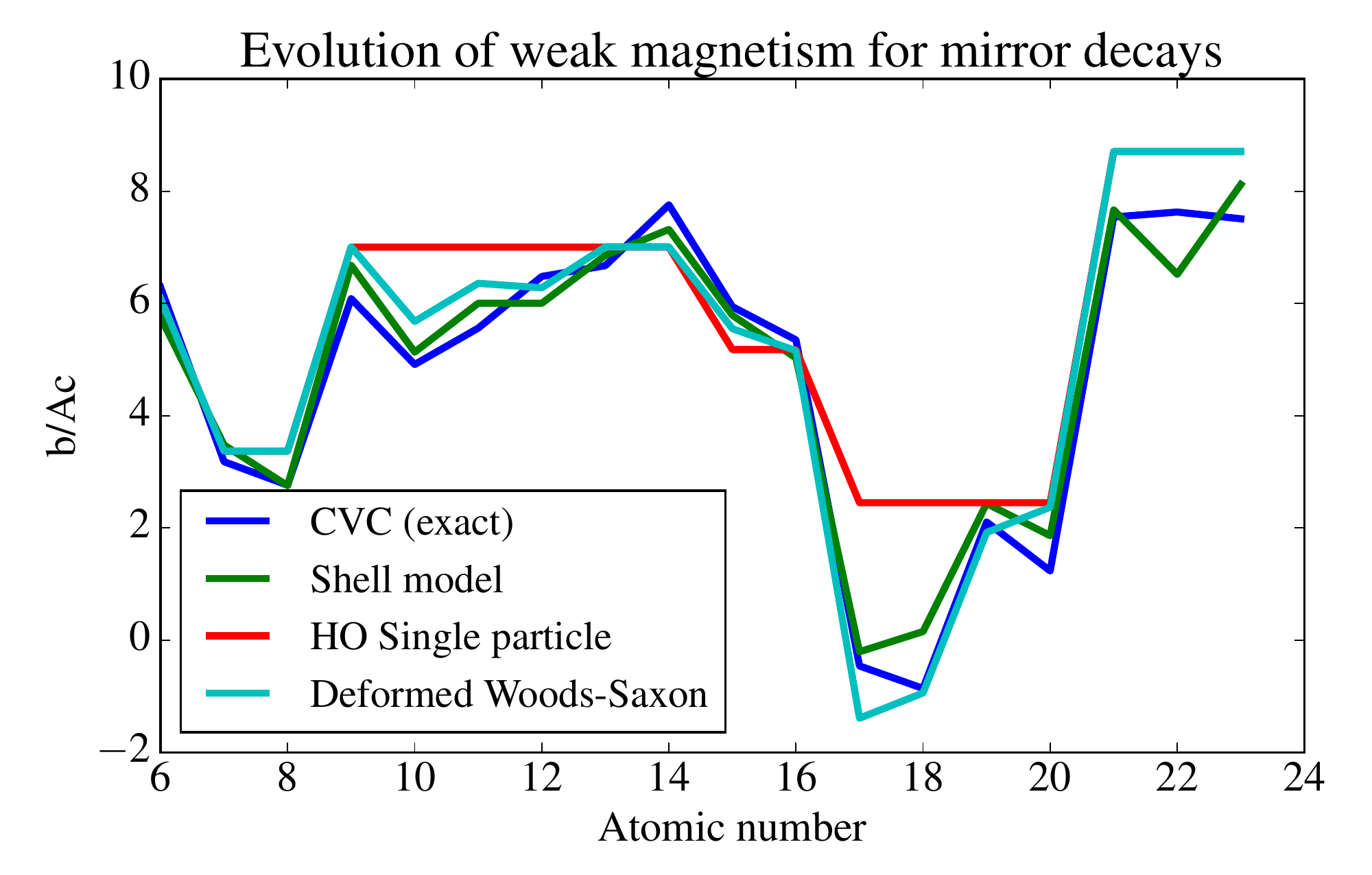}
    \caption{Comparison of different methods of calculating the weak magnetism contribution $b/Ac$ to experimental data using CVC, for $Z =$6 ($^{11}$C) to $Z =$23 ($^{45}$V). The simple spherical harmonic oscillator results are summarized in Eq.~(\ref{eq:ratio_ML_MGT}). Clearly, single particle results using the deformed Woods-Saxon potential give excellent agreement across the board, including the outliers $^{33}$Cl and $^{35}$Ar at atomic numbers 17 and 18.}
    \label{fig:weak_magnetism_SP}
\end{figure}

In the case of mirror decays, the signs for initial and final states are the same. As such, the expected behaviour of $b/Ac$ is opposite for spin-orbit partners. This split can clearly be seen in the data previously presented. Despite this insight, there is still significant variation in the experimental values within the same subshell (see Table~\ref{table:spiegelkernen} and Fig~\ref{fig:Mirror-bAc_vs_A}). Additional insight can be gleaned by incorporating deformation into the single particle potential. As discussed in the appendix, we move on to an axially deformed Woods-Saxon potential. Thanks to mean-field results \cite{Moller2016} we have theoretical deformations available for all participating nuclei. Wave functions are computed numerically \cite{Hayen2019c} and projected on a spherical harmonic oscillator basis, for which we can use the analytical results of Eq. (\ref{eq:ratio_ML_MGT}). Performing the calculation, we find the results shown in Fig.~\ref{fig:weak_magnetism_SP}.

As was noted, the general trends are replicated by the simple spherical harmonic oscillator results of Eq. (\ref{eq:ratio_ML_MGT}). When taking into account nuclear deformation through the single particle potential of Eq. (\ref{eq:hamiltonian_deformed_WS}), deviations are introduced at an individual level. It is remarkable to note that the agreement with experimental data now becomes excellent for all transitions considered (Fig.~\ref{fig:weak_magnetism_SP}). Of particular interest is the agreement with cases of large deviations from traditional estimates for $A = 33$ and $A=35$. In contrast to most of the other nuclei, both initial and final states are oblate deformed resulting in large cancellation effects. The case of $^{33}$Cl has been discussed in more detail in \cite{Hayen2019c}, including the dependence on the input deformation parameters (see also Sec.~\ref{Mirrors-Impulse approximation}).

A final feature to note is the convergence to the simple harmonic oscillator results as subshells are filled. This can be understood using well-known Nilsson orbitals \cite{Nilsson1955}. In the case of prolate deformation, projections with low spin relative to the symmetry axis are pushed down in energy and vice versa for high spin projections. When a subshell is filled, valence nucleons reside in low spin projections. Due to the proximity of other low spin states, significant mixing occurs and the wave function of Eq.(\ref{eq:wf_deformed_exp}) becomes non-trivial. All possible combinations in Eq. (\ref{eq:ratio_ML_MGT}) occur, and lower (higher) values occur for $j=l+1/2$ ($j=l-1/2$) orbitals. As the subshell is filled, nearby equally high spin states become sparse and the resultant mixing is strongly reduced. The results then converge to the spherical harmonic oscillator estimates.

\paragraph{\bf Nuclear shell model}
\label{shell model-mirrors}

For the mirror nuclei up to mass 45 the $b$ and $c$ form factors, which depend on the nuclear matrix elements $M_{GT}$ and M$_L$, were also calculated in the nuclear shell model. The choice of an effective interaction for these calculations in light nuclei, whose principal configurations involve several valence nucleons away from major shell closures, is easily made. There are well established interactions that give excellent fits to spectra.  For $p$-shell nuclei, we use the Cohen-Kurath \cite{Cohen1965} interaction, i.e. (8 - 16)POT, for $s,d$-shell nuclei the universal $s,d$-interaction USD of Wildenthal \cite{Wildenthal1984}, and for $p,f$-shell nuclei the Kuo-Brown $G$-matrix \cite{KB66} as modified by Poves and Zuker \cite{PZ81} and denoted KB3.  Close to major shell closures the choice of a model space and effective interaction is more problematic.  Our approach is to construct a hybrid interaction comprising the Millener-Kurath \cite{Millener1975} interaction for the cross-shell matrix elements, and the (8-16)POT, USD, or KB3 interactions for the in-shell matrix elements. The calculations further used $g_A = g_V = 1$, $g_M = 4.706$, $g_P = -181.03$, and $g_S = g_T = 0$.

\begin{figure}
\centering
\includegraphics[width=0.48\textwidth]{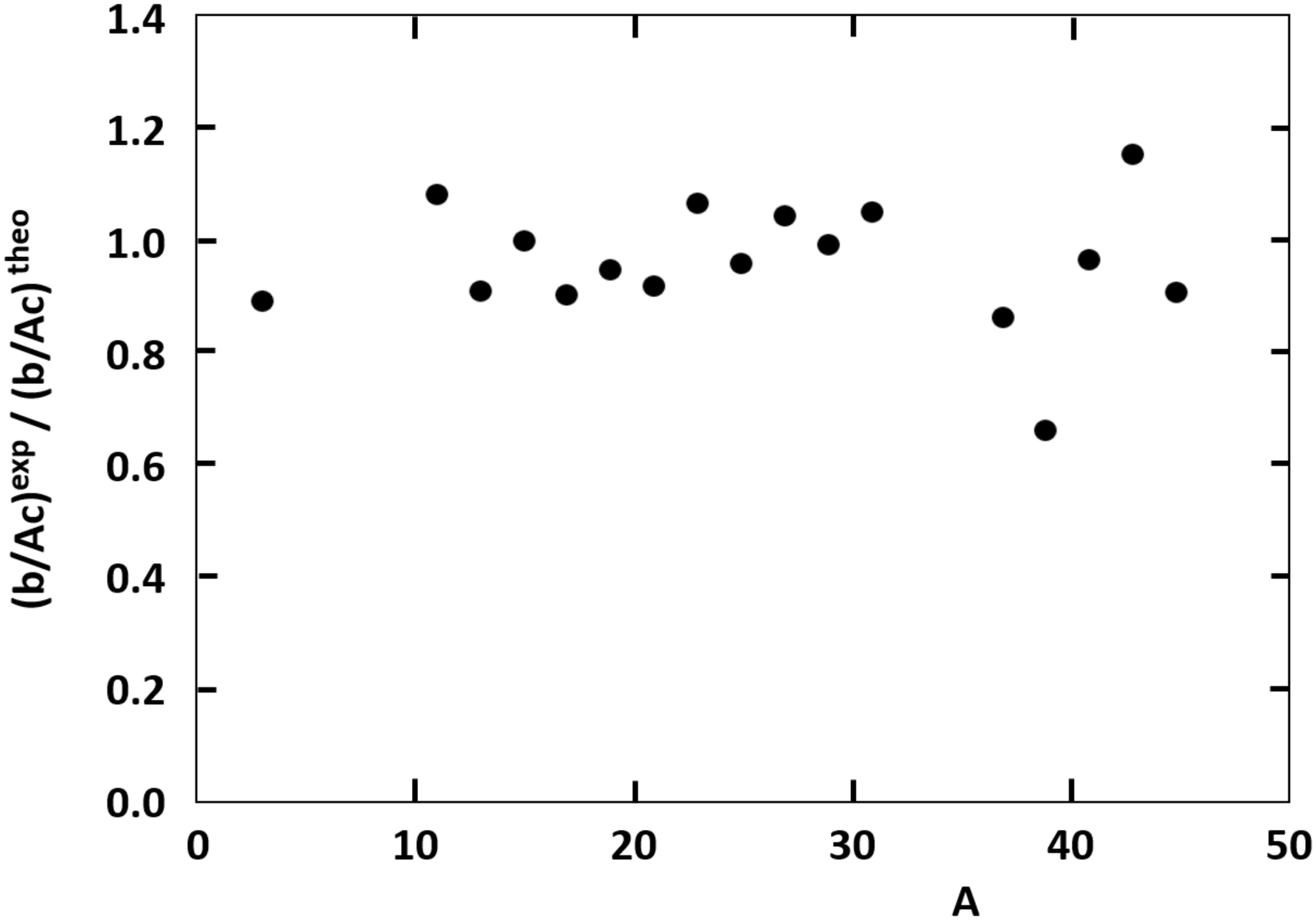}
\caption{Ratio of experimental values (Table~\ref{input-data-and-calculated-quantities}) and nuclear shell model calculated values (Sec.~\ref{shell model-mirrors}) for the b/Ac form factor ratio for the T = 1/2 mirror beta transitions up to A = 45. The values for $A =$ 33 and 35 have been omitted (see text). Error bars (based only on the experimental $b/Ac$ values) are smaller than the size of the symbols.}
\label{fig:bAc-exp-vs-theo}
\end{figure}

Results thus obtained for the form factor ratio $b/Ac$ up to $A=45$ are listed in column 4 of Table~\ref{matrixelementenspiegelkernen-bAc-vs-theo}. For higher masses the required truncation of the shell model space, due to calculation power limitations, was too important to yield reliable values. The ratio of the values for $b/Ac$ extracted from experiment (column 3 in Table~\ref{matrixelementenspiegelkernen-bAc-vs-theo}) and these shell model calculated values are listed in the last column of Table~\ref{matrixelementenspiegelkernen-bAc-vs-theo} and displayed in Fig.~\ref{fig:bAc-exp-vs-theo}. The average for the ratios $(b/Ac)^{exp}/(b/Ac)^{theo}$ for the mass range up to $A = 45$, omitting the values for $A =$ 33 and 35, which both deviate significantly from unity, is $0.96(11)$. The cases of $A =$ 33 and 35 will be discussed in more detail in the next paragraph.


\subsubsection{Impulse approximation: access to $M_{GT}$ and $M_L$ and comparison with shell-model calculations}
\label{Mirrors-Impulse approximation}
%

%
%
Using the impulse approximation, the ratio $b/Ac_1$ was written in Eq.~(\ref{eq:bAc_impulse_approx}) in terms of the Gamow-Teller and orbital current matrix elements, $M_{GT}$ and $M_L$, respectively. The first is directly obtained from the Gamow-Teller form factor, $c \cong g_A M_{GT}$ that was extracted from the experimental ${\mathcal F}t$ values (cf. Table~\ref{input-data-and-calculated-quantities}). Using then this value for $M_{GT}$, Eq.~(\ref{eq:bAc_impulse_approx}) allows extracting $M_L$. Note that we use here the sign convention of Holstein, which uses a positive value for $g_A$, so that $M_{GT}$ and $c$ have the same sign. The resulting values for $M_{GT}^{exp}$ and $M_L^{exp}$ (using $g_A = +1.27$ for $A =$ 3 and $g_A = +1.00$ for all other cases (Sec.~\ref{Evaluation-g_A})) are listed in columns 4 and 8 of Table~\ref{matrixelementenspiegelkernenTHEO}. The corresponding shell-model calculated values are listed in columns 5 and 9.

\begin{figure}
	\centering
	\begin{tabular}{c}
		\includegraphics[width=0.45\textwidth]{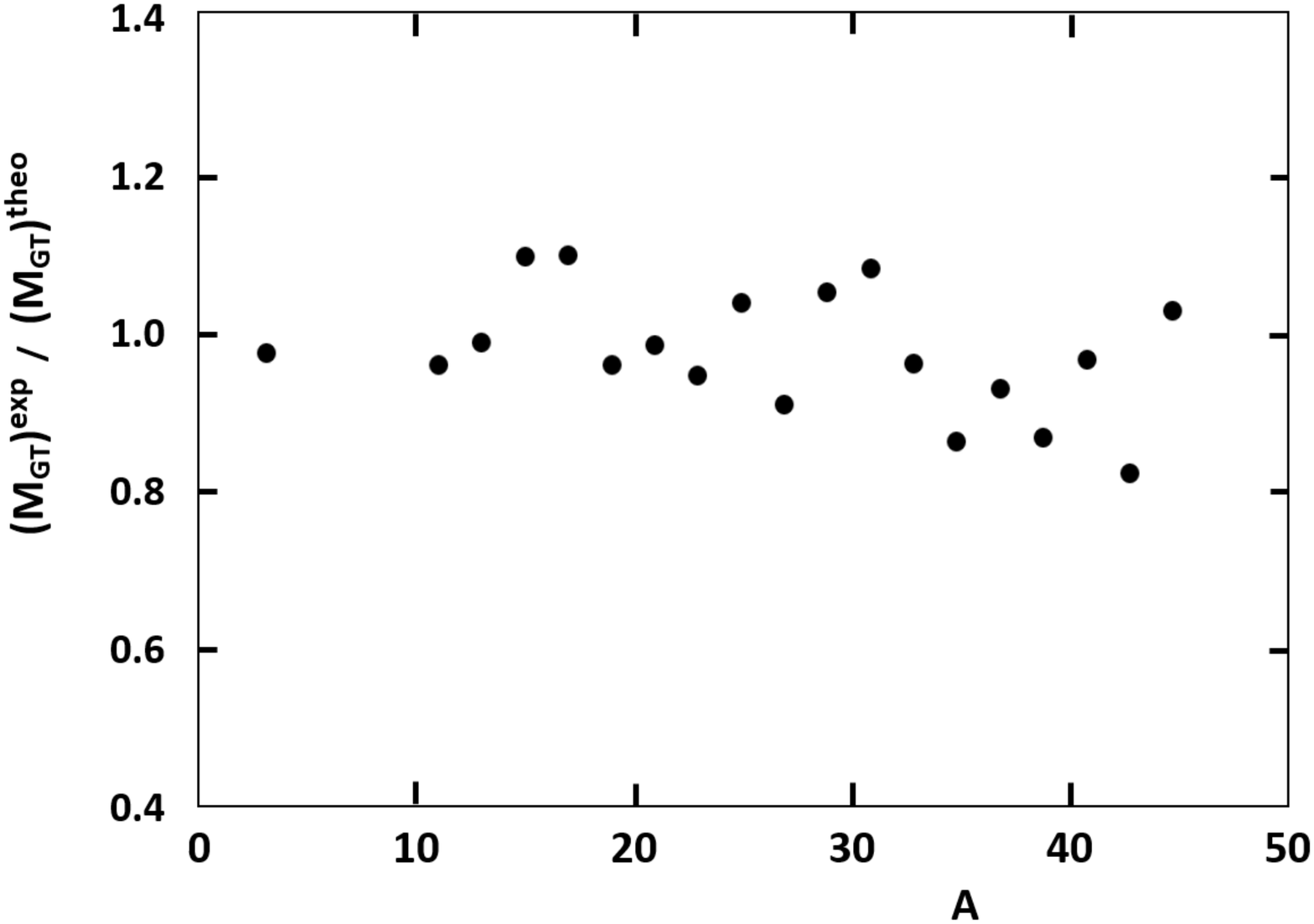}  \\[0.2cm]
		\includegraphics[width=0.45\textwidth]{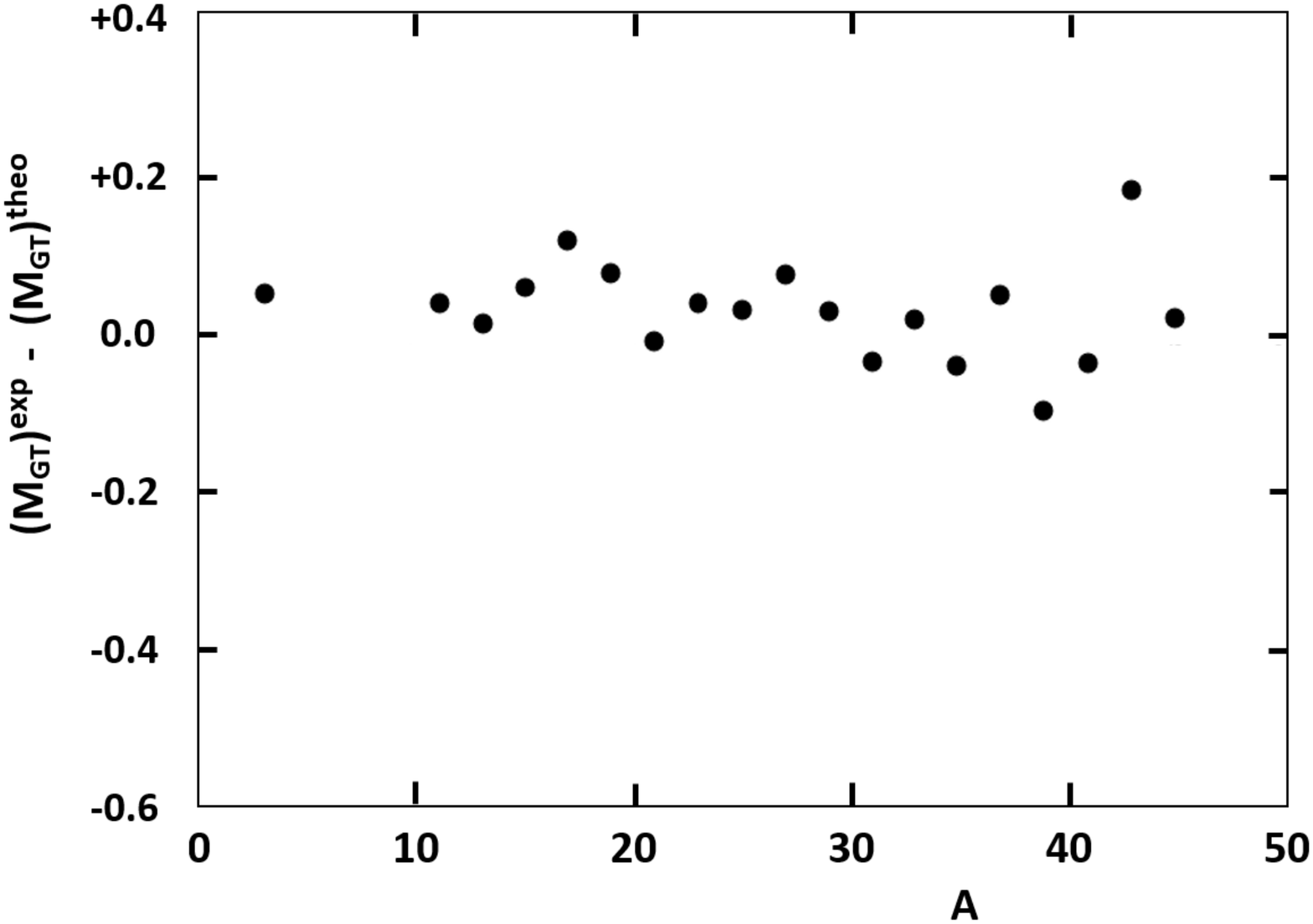}  \\
	\end{tabular}
	\caption{Ratio (top) and difference (bottom) of experimental and nuclear shell model (Sec.~\ref{shell model-mirrors}) calculated values of the $M_{GT}$ matrix elements for the $T = 1/2$ mirror $\beta$ transitions with masses $A = 3$ to 45. See also Table~\ref{matrixelementenspiegelkernenTHEO}. Error bars (based only on the experimental $M_{GT}$ values) are smaller than the size of the symbols.}
	\label{fig:MGT-ratio-diff-exp-theo}
\end{figure}

The ratio and the difference of the experimental and shell-model calculated values for $M_{GT}$ is shown in Fig.~\ref{fig:MGT-ratio-diff-exp-theo}. It is seen that the shell model is very capable in calculating $M_{GT}$, with the ratio differing 10\% to 20\% from unity and the difference being typically limited to about 0.1. Taking an unweighted average, we find the ratio of experimental to theoretical values to be $M_{GT}^{exp}/M_{GT}^{theo} = 0.97(8)$.

\begin{figure}
	\centering
	\begin{tabular}{c}
		\includegraphics[width=0.45\textwidth]{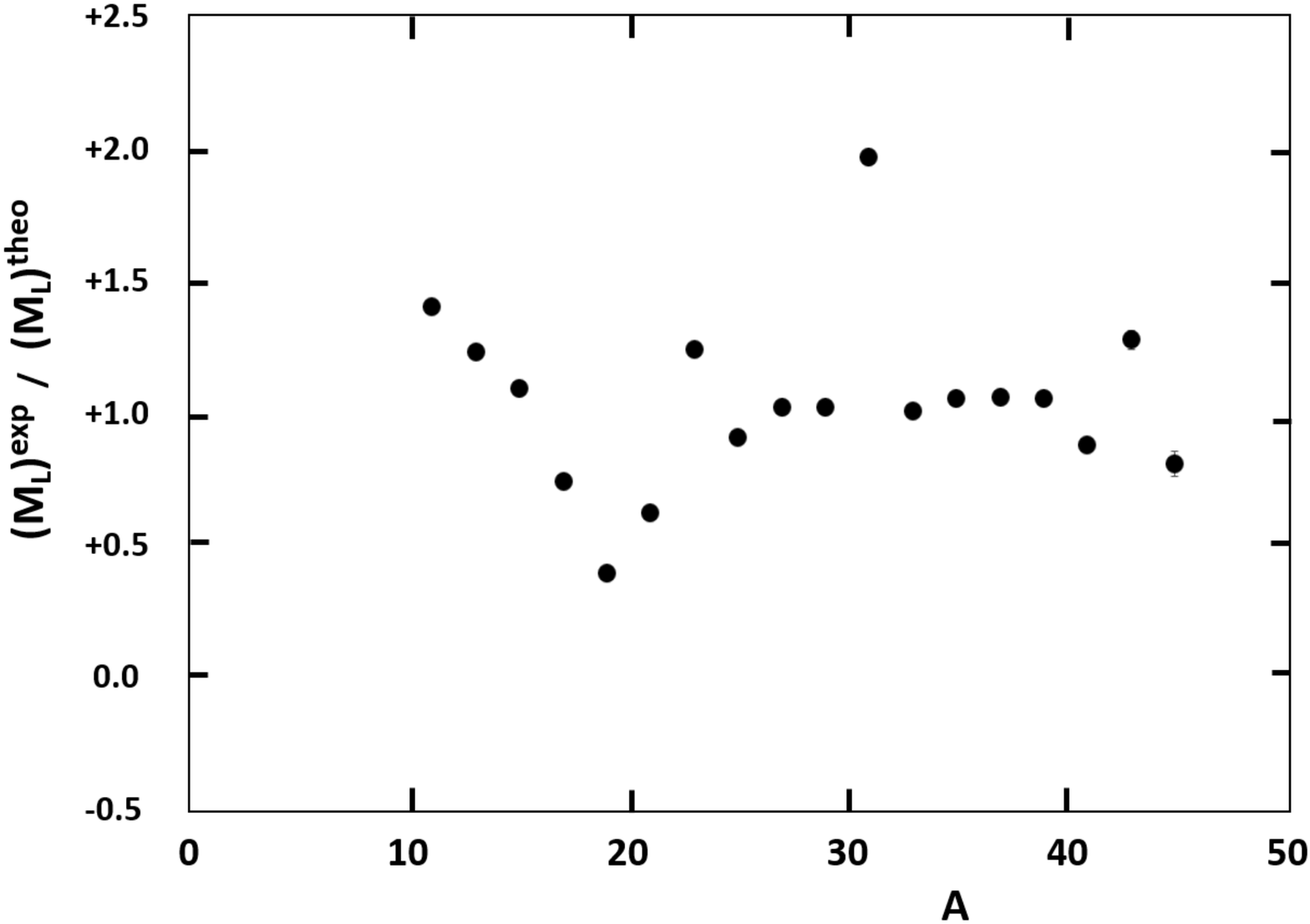}  \\[0.2cm]
		\includegraphics[width=0.45\textwidth]{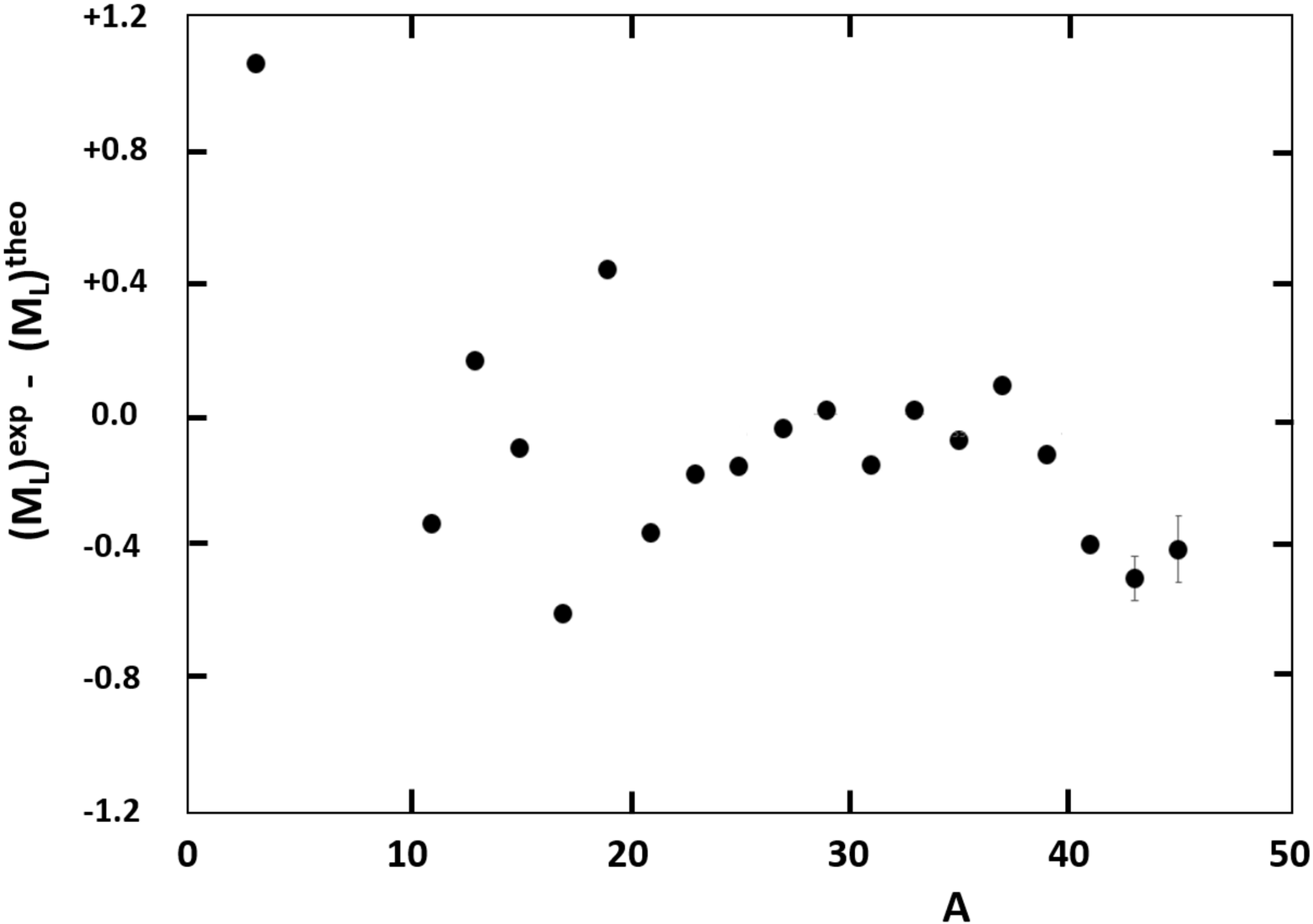}  \\
	\end{tabular}
	\caption{Ratio (top) and difference (bottom) of experimental and nuclear shell model (Sec.~\ref{shell model-mirrors}) calculated values of the $M_L$ matrix elements for the $T = 1/2$ mirror $\beta$ transitions with masses $A = 3$ to 45. The ratio for $A = 3$ is not shown as $M_L^{theo} = 0$ in this case. See also Table~\ref{matrixelementenspiegelkernenTHEO}. If not explicitly shown, error bars (based on the experimental $M_L$ values only) are smaller than the size of the symbols.}
\label{fig:ML-ratio-diff-exp-theo}
\end{figure}

As a further check for the quality of the shell model calculations, Fig.~\ref{fig:ML-ratio-diff-exp-theo} shows the ratio and the difference between the experimental and theoretical values for the orbital current matrix element, $M_L$, for the mirror transitions up to mass 45 (listed in Table~\ref{matrixelementenspiegelkernenTHEO}). Again, reasonably good agreement between theory and experiment is observed. A small tendency of the shell-model calculation to slightly overestimate the value of $M_{L}$ by up to 0.5 $\it{fm}$ is seen. The average value of $M_L^{exp}/M_L^{theo} = 0.99(35)$ for $A = 11-$29 and 33$-$45 is close to unity, however. The masses 3 and 31 are left out when constructing this average because their shell model calculations give $M_L =$ 0 and $M_L =$ -0.159, respectively. Both being close to zero, due to the dominant $s$-state configuration in their wave function. Not omitting the case of $A = 31$ slightly modifies the ratio to $1.04(34)$. It is unclear why the experimentally deduced $M_L$ value for these two transitions, i.e. -1.06 for $^3$He and -0.31 for $^{31}$S, is not smaller than it is. Contributions from meson exchange which are incorrectly estimated in the quenching of $g_A$ could partially explain this discrepancy.

As can be seen from Table~\ref{table:spiegelkernen} and Figure~\ref{fig:Mirror-bAc_vs_A}, most $(b/Ac)^{exp}$ values are positive, except for the mirror $\beta$ decays of $^{33}$Cl, $^{35}$Ar, and $^{71}$Kr, for which slightly negative values are found. This is what one could expect from Eq.~(\ref{eq:bAc_impulse_approx}), in the sign convention of Holstein \cite{Holstein1974}. Indeed, since the first term in Eq.~(\ref{eq:bAc_impulse_approx}) is equal to $g_M/g_A = 4.706$ (for $g_A = +1$), a negative value for $b/Ac$ can only occur if $M_L$ and $M_{GT}$ have opposite signs and $|M_L| > 4.706 |M_{GT}|$ in absolute value. 
As can be seen from the last two columns in Table~\ref{matrixelementenspiegelkernenTHEO}, the ratio $M_L/M_{GT}$ is found (both from experiment and theory) to be positive in subshells with $j = l+1/2$ (i.e. the $s_{1/2}$, $p_{3/2}$, $d_{5/2}$, and $f_{7/2}$ subshells), leading to large values for $b/Ac$ (Eq.~(\ref{eq:bAc_impulse_approx})), and negative in subshells with $j = l-1/2$ (the $p_{1/2}$ and $d_{3/2}$ subshells, i.e. the masses $A =$ 13-15, and $A = $ 33-39), resulting in much smaller values for $b/Ac$. This is in line with the CVC-based relation Eq.~(\ref{eq:weak_magnetism_mirror}) and the values for $j = l \pm 1/2$ states in Table~\ref{table:schmidt} (see also Sec.~\ref{b and b/Ac for mirrors}), combined with small values for $c = M_{GT}$. For the cases of $^{33}$Cl, $^{35}$Ar, and $^{71}$Kr, the values of $M_{GT}$ are indeed the smallest among all mirror $\beta$ transitions, as is reflected by their large $\mathcal{F}t$ values (Table~\ref{Ft-values}). For all three cases the ratios $(M_L/M_{GT})^{exp} = (b/Ac)^{exp} - 4.7$ (see Eq.~\ref{eq:bAc_impulse_approx}) are $\approx -5$ thereby over-compensating the value of 4.7 of the first term in Eq.~(\ref{eq:bAc_impulse_approx}) and causing $b/Ac$ to be small and negative (cf. the last but one column in Table~\ref{input-data-and-calculated-quantities}). This, together with the fact that $^{33}$Cl, $^{35}$Ar, and $^{71}$Kr and their respective daughter isotopes are rare cases of oblate deformation (with deformation parameters $\beta_2 \approx -0.23$ for $^{33}$Cl and $^{35}$Ar, and $\approx -0.36$ for $^{71}$Kr \cite{Moller2016}) then poses a challenge to theoretical calculations. However, with $M_L/M_{GT}$ being a ratio of matrix elements of the same order in spherical tensor formulation, such that complex many-body couplings drop out when neglecting core polarisation and meson exchange, one is left with a ratio of single-particle matrix elements such that one can expect the extreme single-particle to capture most of the required dynamic \cite{Hayen2019c}. The challenge then remains to pick a suitable single-particle state based on an underlying potential. The case of $^{71}$Kr obviously is too complex for this. For $^{33}$Cl and $^{35}$Ar Table~\ref{table:bAc-33Cl-35Ar} compares the values for $b/Ac$ obtained in Ref.~\cite{Hayen2019c} using the spherical harmonic oscillator, the Woods-Saxon potential, and the deformed Woods-Saxon model obtained in Ref. [119] with the here-performed shell model calculations and the experimental results. As can be seen, only the deformed Woods-Saxon potential can reproduce the experimental values within a factor of about two and with the correct sign.

The good agreement with single particle estimates in a deformed potential - particularly $^{33}$Cl and $^{35}$Ar, which turned out to be difficult for the shell model - stresses the importance of an appropriate basis choice. While the shell model reaches for most cases similar accuracy, it of course comes as a result of many higher-order particle-hole excitations to effectively reproduce nuclear deformation. Additionally, results are obtained using different effective interactions with tuned parameters, effective charges and renormalization of coupling constants. When the transition permits it, results of similar accuracy can be obtained for the mirror nuclei in a much more simple fashion in the extreme single particle approximation. Even so, while reassuring, the possibilities for a just application of single-particle results are slim and limited by transitions dominated by single particle states. Inspection of $g$-factors can be a useful tool here, so that single-particle results are an interesting cross-check when applicable.


%
%
%

\begin{table}
\caption{Comparison of experimental and theoretical values for the ratio $(b/Ac)$ for the mirror $\beta$ transitions up to $^{45}$V. Theoretical values were calculated using the shell model. Values for $(b/Ac)^{exp}$ are obtained from the experimental ${\mathcal F}t$ values and the magnetic moments of the mirror nuclei pairs. See text for more details.}
\label{matrixelementenspiegelkernen-bAc-vs-theo}
\begin{ruledtabular}

\begin{tabular}{c|c||c|c|c}

         $\beta$ decay  & $A$ &  $(b/Ac)^{exp}$ &  $(b/Ac)^{theo}$  &  $\frac{(b/Ac)^{exp}}{(b/Ac)^{theo}}$  \\
         \hline

         H $\rightarrow$ He &  3 & $4.2012(23)$ &  $4.706$  &  $0.893$ \\

         C $\rightarrow$  B & 11 & $6.2506(29)$  &  $5.759$  &  $1.085$  \\

         N $\rightarrow$  C & 13 & $3.1712(23)$ &  $3.479$  &  $0.912$  \\

         O $\rightarrow$  N & 15 & $2.7556(20)$  &  $2.753$  &  $1.001$  \\

         F $\rightarrow$  O & 17 & $6.0416(30)$ &  $6.682$  &  $0.904$  \\

        Ne $\rightarrow$  F & 19 & $4.8807(20)$ &  $5.134$  &  $0.951$   \\

        Na $\rightarrow$ Ne & 21 & $5.5233(30)$ &  $6.005$  &  $0.920$   \\

        Mg $\rightarrow$ Na & 23 & $6.4164(55)$  &  $6.004$  &  $1.069$ \\

        Al $\rightarrow$ Mg & 25 & $6.5875(38)$  &  $6.858$  &  $0.961$   \\

        Si $\rightarrow$ Al & 27 & $7.6552(33)$  &  $7.320$  &  $1.046$  \\

         P $\rightarrow$ Si & 29 & $5.7631(51)$  &  $5.790$  &  $0.995$  \\

         S $\rightarrow$  P & 31 & $5.2988(33)$  &  $5.030$  &  $1.053$   \\

        Cl $\rightarrow$  S & 33 & $-0.4561(21)$ &  $-0.206$ &  $2.219$   \\

        Ar $\rightarrow$ Cl & 35 & $-0.8684(11)$ &  $0.154$  &  $-5.633$   \\

         K $\rightarrow$ Ar & 37 & $2.104(11)$  &  $2.437$  &  $0.863$   \\

        Ca $\rightarrow$  K & 39 & $1.2316(10)$ &  $1.863$  &  $0.661$  \\

        Sc $\rightarrow$ Ca & 41 & $7.415(14)$  &  $7.669$  &  $0.967$  \\

        Ti $\rightarrow$ Sc & 43 & $7.539(77)$   &  $6.520$  &  $1.156$   \\

        V $\rightarrow$  Ti & 45 & $7.39(16)$   &  $8.121$ &  $0.910$   \\
\end{tabular}
\end{ruledtabular}
\end{table}

\begin{table*}
\caption{Comparison of experimental and theoretical values for the $M_{GT}$ and $M_L$ matrix elements (in $fm$ units) for the mirror $\beta$ transitions up to $^{45}$V. Theoretical values were calculated using the shell model (see text for details). Values for $M_{GT}^{exp}$ are obtained from the Gamow-Teller form factors, $c$, listed in Table~\ref{input-data-and-calculated-quantities}. Values for $M_{L}^{exp}$ were calculated from Eq.~(\ref{eq:bAc_impulse_approx}) using the $(b/Ac)^{exp}$ values listed in Table~\ref{matrixelementenspiegelkernen-bAc-vs-theo}. In both cases $g_A = 1.27$ was used for $A =$ 3, and $g_A = 1.00$ for all other cases (see Sec.~\ref{Evaluation-g_A}).}
\label{matrixelementenspiegelkernenTHEO}
\begin{ruledtabular}

\begin{tabular}{c|c|c||c|c|c|c||c|c|c|c||c|c}

         $\beta$ decay           &      $A$ &  shell &  $M_{GT}^{exp}$ & $M_{GT}^{theo}$ & $\frac{M_{GT}^{exp}}{M_{GT}^{theo}}$ & $M_{GT}^{(exp - theo)}$ & $M_{L}^{exp}$ &  $M_{L}^{theo}$ &  $\frac{M_{L}^{exp}}{M_{L}^{theo}}$ & $M_{L}^{(exp - theo)}$ & $\frac{M_{L}^{theo}}{M_{GT}^{theo}}$ & $\frac{M_{L}^{exp}}{M_{GT}^{exp}}$\\
         \hline

         H $\rightarrow$ He &  3 & $s_{1/2}$  &   $-1.6577(9)$  & $-1.706$  &  $0.972$   & $+0.048$  &  $-1.0436(48)$ &    $0$ &  $-$       & $-1.044$   &  $0.000$ & $+0.627$\\

         C $\rightarrow$  B & 11 & $p_{3/2}$  &  $-0.75442(28)$ & $-0.789$ &  $0.956$   & $+0.035$   &  $-1.1653(22)$ & $-0.831$ &  $1.402$ &  $-0.334$ &   $+1.053$ & $+1.545$   \\

         N $\rightarrow$  C & 13 & $p_{1/2}$ &   $-0.55962(34)$ & $-0.568$ &  $0.985$  & $+0.008$   &  $+0.8589(14)$  &  $+0.697$ &  $1.232$  & $+0.162$ & $-1.227$ & $-1.535$  \\

         O $\rightarrow$  N & 15 &           &   $+0.63023(46)$ & $+0.576$  &  $1.094$  &  $+0.054$  &  $-1.2291(15)$ & $-1.125$ &  $1.093$ & $-0.104$  & $-1.953$ & $-1.950$    \\

         F $\rightarrow$  O & 17 & $d_{5/2}$ &    $+1.29555(65$)& $+1.182$  &   $1.096$ & $+0.114$   &  $+1.7303(40)$  &  $+2.336$ &  $0.741$   &  $-0.606$ & $+1.976$ & $+1.336$  \\

        Ne $\rightarrow$  F & 19 &           &   $-1.60203(65)$& $-1.676$ &  $0.956$  & $+0.074$   &  $-0.2799(32)$ & $-0.717$ &  $0.390$   &  $+0.437$ & $+0.428$  & $+0.175$ \\

        Na $\rightarrow$ Ne & 21 &           &    $+0.71245(39)$ & $+0.726$  &  $0.981$ & $-0.014$   &  $+0.5823(22)$  &  $+0.943$ &  $0.617$  & $-0.361$  & $+1.299$ & $+0.817$ \\

        Mg $\rightarrow$ Na & 23 &           &  $-0.55413(47)$ & $-0.588$ &  $0.942$  & $+0.034$   &  $-0.9478(32)$ & $-0.763$ &  $1.242$   & $-0.185$  & $+1.298$ & $+1.710$\\

        Al $\rightarrow$ Mg & 25 &           &   $+0.80844(42)$ &  $+0.781$ &   $1.035$ & $+0.027$   &  $+1.5211(32)$ &  $+1.681$ &  $0.905$  & $-0.160$  & $+2.152$ & $+1.881$    \\

        Si $\rightarrow$ Al & 27 &           &  $-0.69659(30)$ & $-0.769$ &  $0.906$  & $+0.072$   &  $-2.0544(25)$ & $-2.010$ &  $1.022$  &  $-0.044$ & $+2.614$ &  $+2.949$   \\

         P $\rightarrow$ Si & 29 & $s_{1/2}$  &   $+0.53798(47)$ &  $+0.513$ &  $1.049$ & $+0.025$   &  $+0.5687(28)$ &  $+0.556$ &  $1.023$  & $+0.013$  & $+1.084$ & $+1.057$   \\

         S $\rightarrow$  P & 31 &           &   $-0.52939(33)$ & $-0.490$ &   $1.080$ & $-0.039$   &  $-0.3138(18)$ & $-0.159$ &  $1.974$ &  $-0.155$ & $+0.324$ &  $+0.593$   \\

        Cl $\rightarrow$  S & 33 & $d_{3/2}$ &   $-0.31416(28)$ & $-0.328$ &   $0.958$ & $+0.014$   &  $+1.6217(16)$ &  $+1.611$ &  $1.007$   & $+0.011$  & $-4.912$ &  $-5.162$ \\

        Ar $\rightarrow$ Cl & 35 &           &   $+0.28199(17)$ &  $+0.328$ &  $0.860$  & $-0.046$   &  $-1.5719(10)$ & $-1.493$ &  $1.053$   & $-0.079$  & $-4.552$  &  $-5.574$ \\

         K $\rightarrow$ Ar & 37 &           &   $-0.57789(39)$ & $-0.624$ &  $0.926$  & $+0.046$   &   $+1.5037(64)$ &  $+1.416$ &  $1.062$   & $+0.088$  & $-2.269$  &  $-2.602$ \\

        Ca $\rightarrow$  K & 39 &           &   $+0.66061(50)$ &  $+0.764$ &  $0.865$  & $-0.103$   &  $-2.2952(19)$ & $-2.172$ &  $1.057$   & $-0.123$  & $-2.843$ &  $-3.474$ \\
        
        Sc $\rightarrow$ Ca & 41 & $f_{7/2}$ &   $+1.0743(21)$ &  $+1.116$ &  $0.963$  & $-0.042$    &  $+2.910(16)$ &  $+3.307$ &  $0.880$     & $-0.397$  & $+2.963$ & $+2.709$  \\

        Ti $\rightarrow$ Sc & 43 &           &   $-0.8097(69)$ & $-0.989$ &  $0.819$  & $+0.179$   &  $-2.294(65)$ & $-1.794$ &  $1.279$  & $-0.500$  & $+1.814$ &  $+2.833$   \\

        V $\rightarrow$  Ti & 45 &           &   $+0.6346(58)$ &  $+0.619$ &  $1.025$  & $+0.016$   &   $+1.70(10)$ &  $+2.114$ &  $0.806$     & $-0.411$  & $+3.415$ &  $+2.684$   \\
\end{tabular}
\end{ruledtabular}
\end{table*}

%
%
%


%
%
\begin{table}
\caption{Extreme single-particle values of $b/Ac$ for the mirror $\beta$ decays of $^{33}$Cl and $^{35}$Ar calculated with three different possibilities for the nuclear potential, i.e. spherical harmonic oscillator (SHO), spherical Woods–Saxon (WS) and deformed Woods–Saxon (DWS), and within the nuclear shell model (NSM) using the universal $s, d$-interaction USD of Wildenthal \cite{Wildenthal1984}. The experimental values are listed in the last column.}
\label{table:bAc-33Cl-35Ar}

\begin{ruledtabular}

\begin{tabular}{c|c c c c|c}

         decay     & SHO      &  WS     &  DWS     &  NSM      &  exp.       \\
         \hline

         $^{33}$Cl &  $2.46$  &  $2.46$ & $-1.13$  &  $-0.21$  &  $-0.456$   \\
 
         $^{35}$Ar &  $2.46$  &  $2.46$ & $-0.94$  &  $+0.15$  &  $-0.868$   \\

\end{tabular}
\end{ruledtabular}
\end{table}

\subsubsection{Discussion}

Good agreement between theoretical and experimental values for the Gamow-Teller matrix elements, $M_{GT}$, the orbital current matrix elements, $M_L$, and the form factor ratio $b/Ac$ for the mirror $\beta$ transitions considered here is obtained. Differences reveal no significant trend in $A$ or in the shell model states occupied, except for a small downward slope towards higher $A$ values related to the required truncation of the shell-model space. The observed deviation in $b/Ac$ for the cases of $^{33}$Cl and $^{35}$Ar was understood in terms of the smallness of these $b/Ac$ values, due to the large ratio of $M_L/M_{GT}$ in the $d_{3/2}$ subshell ($j = l - 1/2$), and the large oblate deformation of these isotopes. 

The good correspondence between experimental and theoretical values for the $M_L$ matrix elements, with the exception of the case of $^{31}$S with a strong $s$-state configuration in the wave function, gives confidence in the reliability of the shell-model calculations performed and thus perhaps also in the calculations of the matrix elements determining yet other, although usually less important form factors (e.g. the tensor form factor, $d$, and the induced pseudoscalar form factor, $h$, \cite{Holstein1974}). We therefore list in Table~\ref{table:mirror-form-factors} the full set of form factors from the shell model calculations discussed above (Sec.~\ref{shell model-mirrors}). These can be used in the analysis of experiments determining the $\beta-\nu$ correlation or the $\beta$ asymmetry parameter for the mirror $\beta$ transitions, that are being planned at different laboratories (e.g. \cite{Brodeur2016, Lienard2015, Fenker2018}). Additional information on the effect of recoil terms, radiative corrections and specific experimental conditions on such measurements are also discussed in Refs.~\cite{Hayen2021, Vanlangendonck2021}.

\begin{table*}
	\centering
	\begin{ruledtabular}
		
		{ \footnotesize
		\caption{Form factors calculated in the shell model for the $T = 1/2$ mirror $\beta$ transitions up to mass $A = 45$. Details on the calculations and the interactions used can be found in Sec.~\ref{shell model-mirrors}. Note that the first-class part of the $d$ form factor, i.e. $d \approx A g_A M_{\sigma L}$ (see Table~\ref{table:summary}), is zero because these are transitions within an isospin multiplet. Note that the form factor $j_2$ is in absolute value smaller than $8 \times 10^{-2}$ and so negligible for all transitions listed here.} \label{table:mirror-form-factors}
		
		\begin{tabular}{c|c||c|c|r|l|l|c|l|l}
				
$\beta$ decay & Shell model & $c = c_1$ & $c_2$ & $b/Ac$ & $f/Ac$ & $g/A^{2}c$ &  $j_1/A^{2}c$ & $j_3/A^{2}c$ & $h/A^{2}c$ \\
              & interaction &           &       &        &        &           &              &             &           \\
				
\hline
$^{3}$He      & MSDI3     & $-1.706$ & $-0.971$ & $4.709$ & $0.0$  & $0.0$  & $0.0$ & $0.0$     & $+181.2$ \\

$^{11}$C      & CK816POT  & $-0.789$ & $-1.092$ & $5.761$ & $0.064$ & $-66.83$ & $-70.39$ & $+185.4$  & $+114.2$ \\

$^{13}$N      & CK816POT  & $-0.568$ & $-1.133$ & $3.480$ & $0.0$ & $0.0$  & $-232.3$ & $0.0$    & $-39.38$ \\

$^{15}$O      & MK        & $+0.576$ &  $+1.267$ & $2.755$ & $0.0$ & $0.0$  & $-277.0$ & $0.0$     & $-81.79$ \\

$^{17}$F      & MK        &  $+1.182$ &  $+2.319$ & $6.684$ & $-0.188$ & $+127.9$  & $-57.38$ & $+172.1$  & $+126.5$ \\

$^{19}$Ne     & USD       & $-1.676$ & $-3.044$ & $5.134$ & $0.0$ & $0.0$  & $-1.311$ & $0.0$     & $+180.2$ \\

$^{21}$Na     & USD       &  $+0.726$ &  $+1.397$ & $6.008$ & $0.098$ & $-49.66$  & $-19.12$ & $-206.8$  & $+163.0$ \\

$^{23}$Mg     & USD       & $-0.588$ & $-1.219$ & $6.004$ & $-0.059$ & $+25.49$  & $-49.19$ & $-244.3$  & $+134.4$ \\

$^{25}$Al     & USD       &  $+0.781$ &  $+1.767$ & $6.858$ & $-0.161$ & $+65.56$  & $-91.98$ & $+41.79$  & $+93.83$   \\

$^{27}$Si     & USD       & $-0.769$ & $-1.871$ & $7.321$ & $0.419$ & $-149.3$ & $-130.2$ & $+130.2$  & $+57.44$ \\

$^{29}$P      & USD       &  $+0.513$ &  $+1.293$ & $5.794$ & $0.0$ & $0.0$  & $-143.9$ & $0.0$    & $+44.73$   \\

$^{31}$S      & USD       & $-0.490$ & $-1.260$ & $5.030$ & $0.0$ & $0.0$  & $-149.3$ & $0.0$  & $+39.29$ \\

$^{33}$Cl     & USD       & $-0.328$ & $-1.133$ & $-0.212$ & $0.898$ & $-271.3$ & $-393.9$ & $+166.3$ & $-191.8$ \\

$^{35}$Ar     & USD       &  $+0.328$ &  $+1.143$ & $0.148$ & $-0.092$ & $+25.88$  & $-390.7$ & $+161.0$ & $-190.6$ \\

$^{37}$K      & USD       & $-0.624$ & $-1.873$ & $2.434$ & $0.147$ & $-40.03$  & $-244.7$ & $+49.63$ & $-50.92$ \\

$^{39}$Ca     & MK        &  $+0.764$ &  $+2.336$ & $1.859$ & $-0.856$ & $+217.7$  & $-247.0$ & $+67.30$ & $-53.35$ \\

$^{41}$Sc     & MK        &  $+1.116$ &  $+3.591$ & $7.667$ & $-0.848$ & $+217.5$ & $-109.3$ & $+253.7$  & $+77.29$ \\

$^{43}$Ti     & KB3       & $-0.989$ & $-3.095$ & $6.523$ & $0.437$ & $-105.5$ & $-71.64$ & $+141.1$ & $+113.2$ \\

$^{45}$V      & KB3       &  $+0.619$ &  $+2.218$ & $8.120$ & $-0.689$ & $+159.6$ & $-149.2$ & $+308.7$ & $+392.5$ \\

			\end{tabular}
		}
		
	\end{ruledtabular}
	
\end{table*}



\subsection{Isospin $T =$ 1, 3/2, and 2 multiplet decays}
\label{triplet-states}

We have also investigated weak magnetism for $\beta$ decays from states in isospin $T = 1, T = 3/2$, and $T = 2$ isobar multiplets for which experimental input data
required to calculate the $b$ and $c$ form factors are available. Weak magnetism studies for $\beta$ decays of isobaric $T = 1$ triplet states have been performed before \cite{Calaprice1976, Huber2011}. Here we update and significantly extend these works, yielding results for as many as 68 (previously only 14 \cite{Calaprice1976, Huber2011}) $\beta$ transitions from isospin multiplet states and the corresponding analog $\gamma$ transitions.

\begin{figure}
\centering
\includegraphics[width=0.45\textwidth]{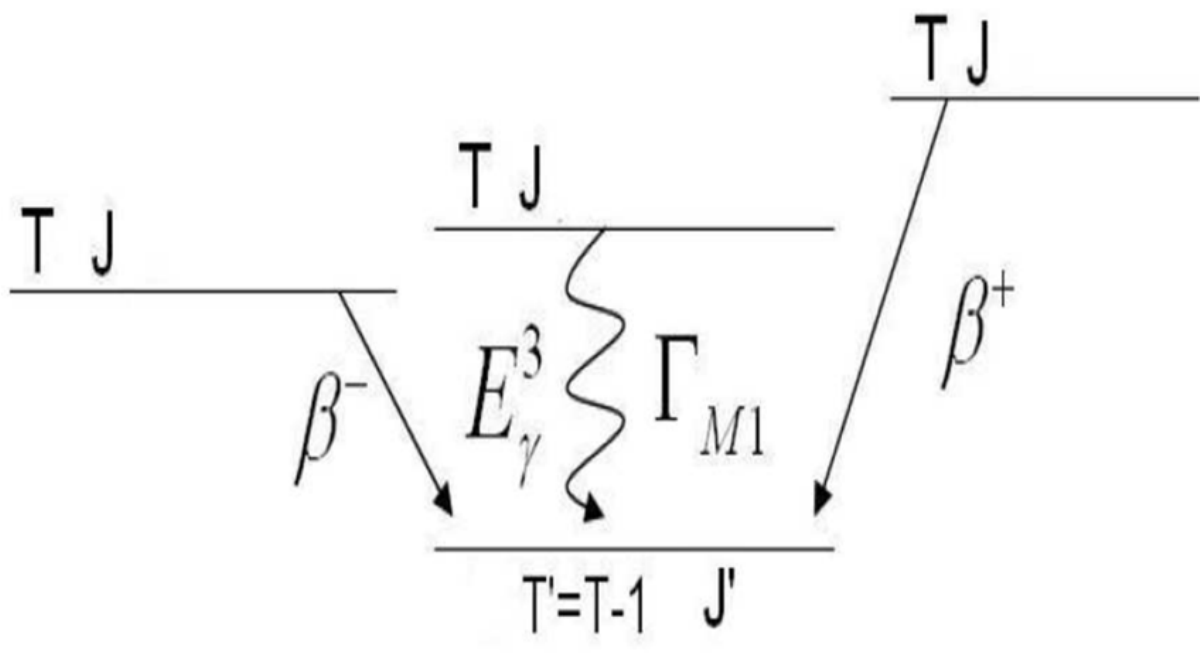}
\caption{Generic scheme showing the $\beta$ and $\gamma$ decays from the isobaric analog states in a $T = 1$ triplet to a common $T^\prime = 0$ daughter state. The weak magnetism form factor of the $\beta$ transitions can be obtained from the energy, $E_\gamma$, and the M1-decay width, $\Gamma_{M1}$, of the corresponding isovector transition from the isobaric $T = 1$ analog state.}
\label{fig:triplet-scheme}
\end{figure}

Figure~\ref{fig:triplet-scheme} displays the generic decay scheme for the $\beta$ and $\gamma$ decays of isobaric analog states in a $T = 1$ triplet to a common $T^\prime = 0$ final state. For the $\Delta T = 1, \Delta T = \pm 1$ transitions considered here the weak magnetism form factor for the $\beta$ transitions is now given by Eq.~(\ref{eq:weak_magnetism_triplet}). The decay width, $\Gamma$, of the $\gamma$-decaying analog state is related to its half-life $t_{1/2}$ (in seconds) by
\begin{equation}
\Gamma = \frac{4.56238 \times 10^{-16}} {t_{1/2} } ~ \rm{eV} ~ .
\label{eq:Gamma-halflife}
\end{equation}
\noindent When several $\gamma$ rays are deexciting the analog state the partial decay width for each transition is obtained by multiplying $\Gamma$ with the transition's fractional intensity. In tables, sometimes the $B(M1)$ transition strength for the specific $\gamma$ ray is listed. The $\Gamma_{M1}$ width is then obtained as
\begin{equation}
\Gamma_{M1} = 20.723 \times 10^{-3} ~ E_{\gamma}^3 ~ B(M1) ~ \rm{eV}  ~ ,
\label{eq:Gamma-BM1}
\end{equation}
with $E_{\gamma}$ in MeV and $B(M1)$ in Weiskopff units.

%
%


Using Eqs.~(\ref{eq:ft_general}) and (\ref{Fermi}), the partial half-life, $t$, for a general mixed Fermi/Gamow-Teller (F/GT) transition Eq.~(\ref{eq:ft_general}) can be rewritten as
\begin{align}
t = \frac{2 ~ \mathcal{F} t^{0^+ \rightarrow 0^+} (1 + \Delta_R^V)} {(1 + \delta_R^{\prime}) ~ \left[ f_V ~ a^2 + f_A ~ c^2 \right]} .
\label{F-GT-full}
\end{align}
%
%
All $\beta$ transitions considered here are either $J^\pi \rightarrow J^\pi$, $\Delta T = 1$ transitions or $J^\pi \rightarrow J^\pi \pm 1$, $\Delta T = 1$ transitions. The latter are of pure Gamow-Teller type while in the former isospin-symmetry breaking due to the electromagnetic interaction may induce a small isospin-forbidden Fermi component into the otherwise pure Gamow-Teller decay. In the latter case the Fermi matrix element in the form factor $a  = g_V M_F$ can be written as $M_F = \alpha \sqrt{2T}$ with $\alpha$ the amplitude of the admixture of the analog state to the final state of the $\beta$ decay into the initial state \cite{Raman1975}. An extensive survey of isospin-forbidden $J^\pi \rightarrow J^\pi$ Fermi transtions \cite{Raman1975} has shown $\alpha$ to be of the order of $10^{-3}$ and lower (see also \cite{Schuurmans2000, Severijns2005, Soti2014}). At the level of precision we are dealing with here (which is determined by the uncertainty on the values of $\Gamma_{M1}$ or $B_{M1}$ used to extract the weak magnetism form factor, $b$), the term $a^2$ in Eq.~(\ref{F-GT-full}) can be neglected with respect to $c^2$. 
%
%
%
%
One then has, for all transitions considered here:
\begin{align}
f_A t = \frac{2 ~ \mathcal{F} t^{0^+ \rightarrow 0^+} (1 + \Delta_R^V)} {(1 + \delta_R^{\prime}) ~ c^2 } .
\label{f_GT}
\end{align}
The inner radiative correction, $\Delta_R^V$, has recently been subject of new, more complete calculations \cite{Czarnecki2019, Seng2020, Hayen2021}. We use here the value $\Delta_R^V = 0.02473(27)$ of Ref.\cite{Hayen2021}, which is in agreement with the value of \cite{Seng2020} (both calculations being also slightly more encompassing than Ref.~\cite{Czarnecki2019}) and used a slightly more conservative approach in determining the uncertainty. The values for $\delta_{R^\prime}$ are listed in the last column of Tables~\ref{table:multiplets-input-1} and \ref{table:multiplets-input-2}.

Note that Eq.~(\ref{f_GT}) only holds for 'normal' allowed $\beta$ transitions. For strongly hindered transitions, with large log~$ft$ values (i.e. typically larger than about 6.7) the beta spectrum is unlikely to have an allowed shape, and the connection between log$ft$ and $c$ is in principle lost because the electromagnetic interaction is very much amplified in this case \cite{Calaprice1976}. For similar reasons Huber \cite{Huber2011} recently suggested that a detailed study of the breakdown of the impulse approximation in nuclei with large log~$ft$ transitions would be useful.

Combining the weak magnetism form factor, $b$, obtained from the energy and decay width of the analog M1 $\gamma$ transition via Eq.~(\ref{eq:weak_magnetism_triplet}), and the Gamow-Teller form factor, $c$, obtained from the $\beta$ decay $ft$ value via Eq.~(\ref{f_GT}), the $b/Ac$ value for the $\beta$ transitions in $T \geq 1$ isospin multiplets can be obtained.
%
%
%
The new analysis presented here significantly extends the available data set. Thereafter, experimental values for the $M_L$ matrix elements will again be deduced from these experimental $b/Ac$ ratios and Eq.~(\ref{eq:bAc_impulse_approx}). Finally, the experimental results can again be compared to those from shell model calculations that have been performed for several $\beta$ transitions from the $T = 1$ multiplets, as will be discussed below.

\subsubsection{Experimental values for Gamow-Teller and weak magnetism form factors}
\label{experimental-values-multiplets}

Searching the Brookhaven National Nuclear Data Base (https://www.nndc.bnl.gov/ensdf/) files, input data were found for 52 pairs of $\beta$ and $\gamma$ transitions from analog states
in isospin $T = 1$ isobaric triplets with masses ranging from $A$~=~6 to 42, for 15 pairs of transitions from analog states in $T = 3/2$ quartets with $A$~=~9 to 23 and $A =$ 53, and a single pair of transitions from analog states in the $T = 2$ multiplet with $A$ = 32.

Detailed information and input data related to the $\beta$ and $\gamma$ transitions considered here, as well as the results obtained in the analysis, are listed in the Tables~\ref{table:multiplets-input-1} to \ref{table:higher-multiplets-bAc} that are discussed below. To clearly indicate which transitions are considered here, Figs.~\ref{fig12:triplets-6-14} to \ref{fig16:triplets-38-42} show the partial decay scheme for all pairs of analog $\beta$ and $\gamma$ transitions from states in the $T = 1$ triplets and the $T = 2$ multiplet, indicating the log$ft$ value for the $\beta$ transitions and the energy and the $\Gamma_{M1}$ or $B(M1)$ value (Eq.~(\ref{eq:Gamma-BM1})) for the $\gamma$ transitions. Figs.~\ref{fig17:T3/2-9-13} and \ref{fig18:T3/2-15-53} provide the same information for the pairs of analog $\beta$ and $\gamma$ transitions from states in the $T = 3/2$ multiplets with mass $A$~=~9 to 23 and $A =$ 53.

\begin{widetext}
\begin{figure*}
	\centering
	\includegraphics[width=1.00\textwidth]{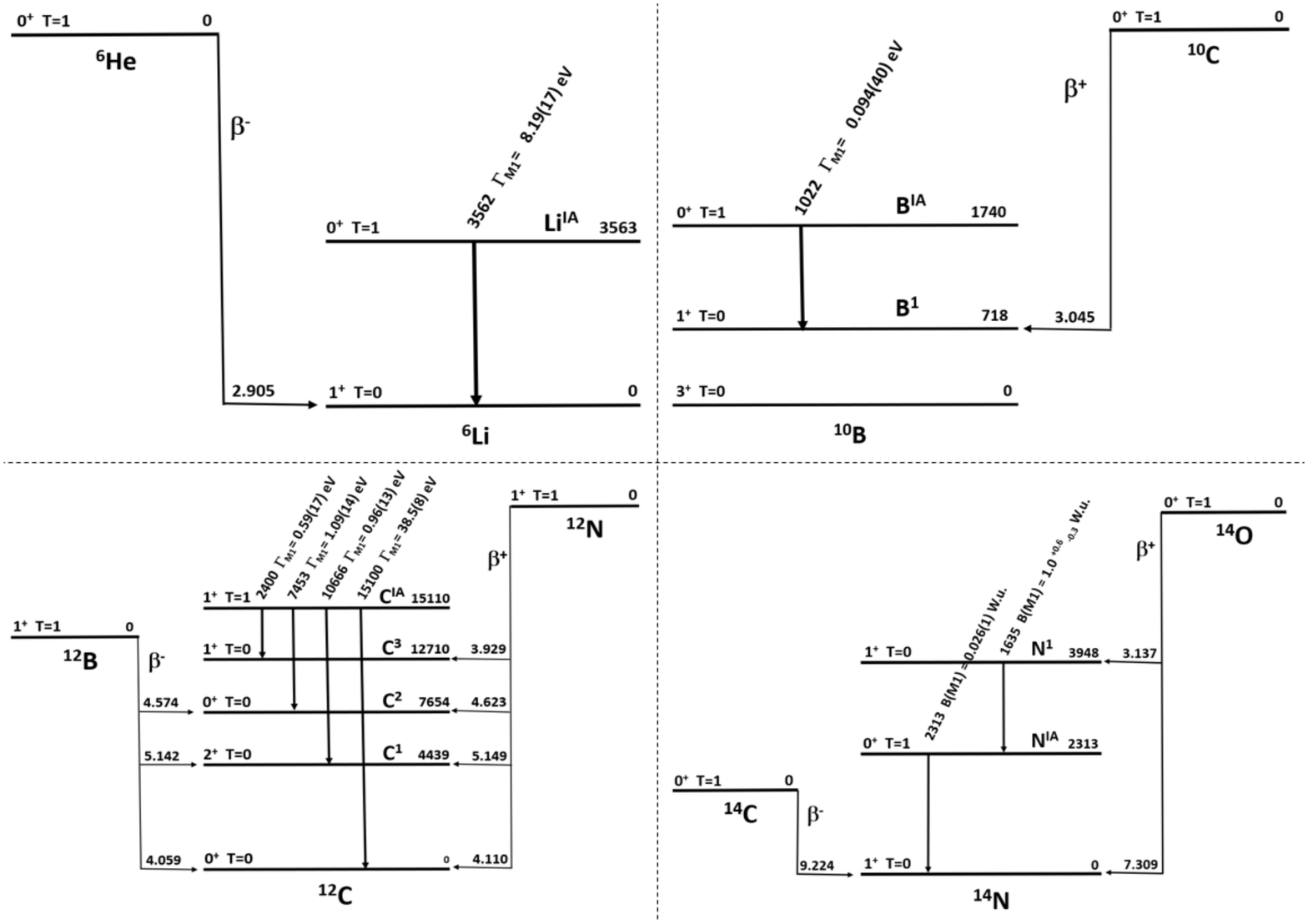}
	\caption{Partial decay schemes (not to scale) showing the $\beta$ transitions and their analog $\gamma$ transitions in $T = 1$ multiplets with $A =$ 6 to 14 for which the weak magnetism form factor is determined, with their respective log$ft$ value ($\beta$ transitions), or the transition energy and $\Gamma_{M1}$ or $B(M1)$ value ($\gamma$ transitions). For each ground state and excited state the spin-parity, isospin, and energy (in keV) is given. The excited state labels (e.g. Li$^{IA}$ or C$^1$) are also used in the first column of Tables~\ref{table:multiplets-input-1} to \ref{table:higher-multiplets-bAc} for easy reference.}
	\label{fig12:triplets-6-14}
\end{figure*}
\begin{figure*}
	\centering
	\includegraphics[width=1.00\textwidth]{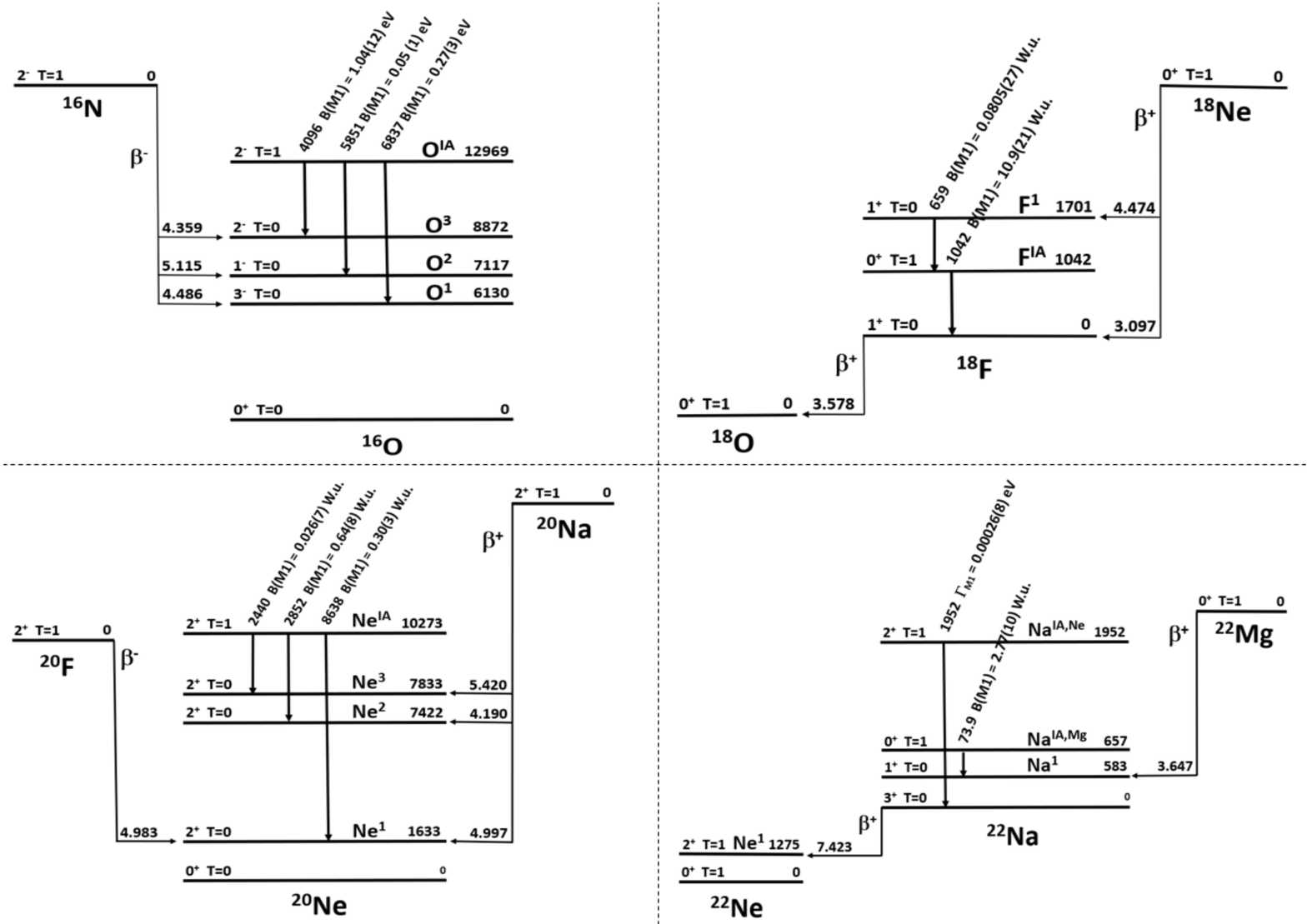}
	\caption{Partial decay schemes (not to scale) showing the $\beta$ transitions and their analog $\gamma$ transitions in $T = 1$ multiplets with $A =$ 16 to 22 for which the weak magnetism form factor is determined, with their respective log$ft$ value ($\beta$ transitions), or the transition energy and $\Gamma_{M1}$ or $B(M1)$ value ($\gamma$ transitions). For each ground state and excited state the spin-parity, isospin, and energy (in keV) is given. The excited state labels (e.g. Li$^{IA}$ or C$^1$) are also used in the first column of Tables~\ref{table:multiplets-input-1} to \ref{table:higher-multiplets-bAc} for easy reference.}
	\label{fig13:triplets-16-22}
\end{figure*}
\begin{figure*}
	\centering
	\includegraphics[width=1.00\textwidth]{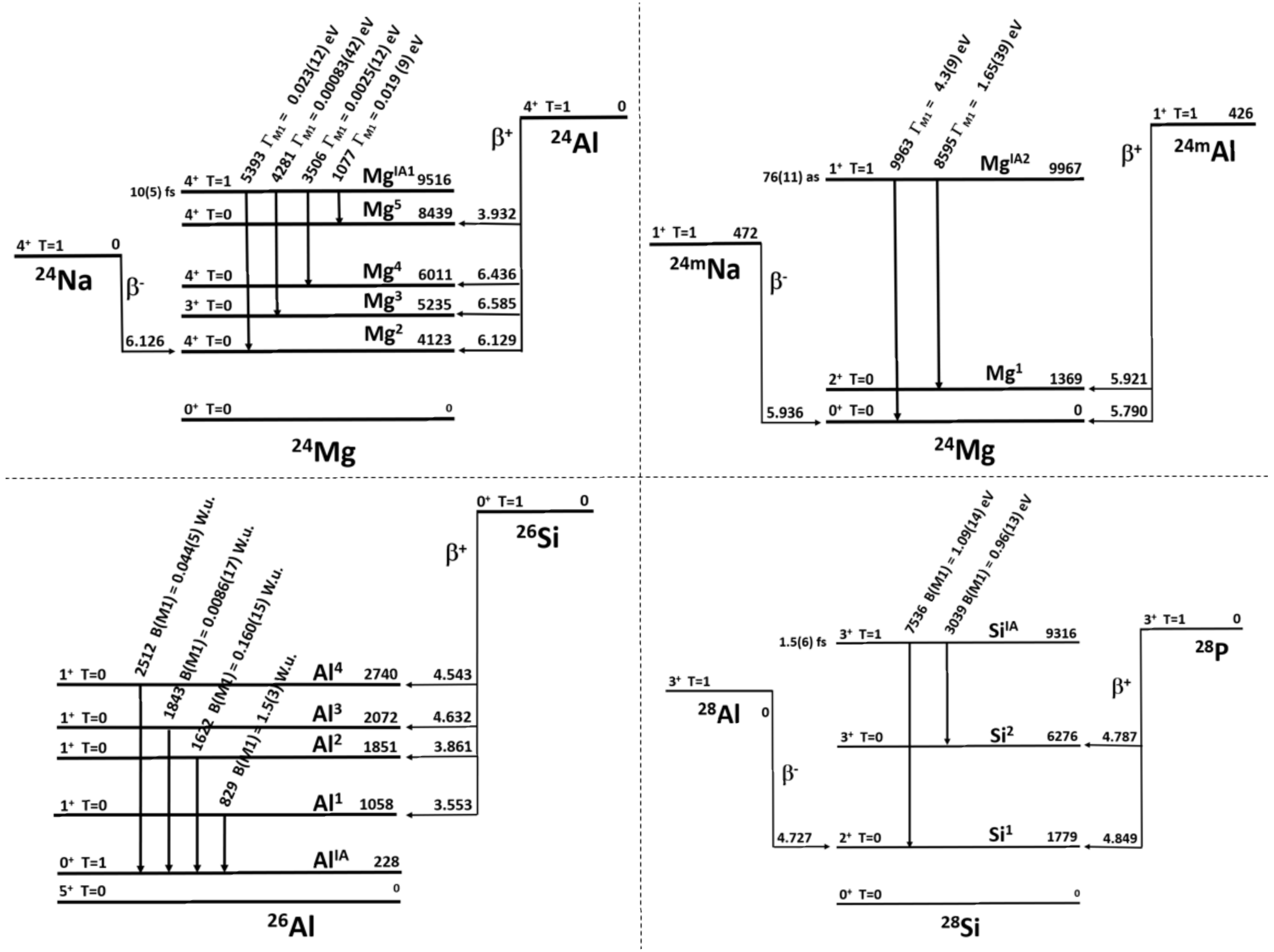}
	\caption{Partial decay schemes (not to scale) showing the $\beta$ transitions and their analog $\gamma$ transitions in $T = 1$ multiplets with $A =$ 24 to 28 for which the weak magnetism form factor is determined, with their respective log$ft$ value ($\beta$ transitions), or the transition energy and $\Gamma_{M1}$ or $B(M1)$ value ($\gamma$ transitions). For each ground state and excited state the spin-parity, isospin, and energy (in keV) is given. The excited state labels (e.g. Li$^{IA}$ or C$^1$) are also used in the first column of Tables~\ref{table:multiplets-input-1} to \ref{table:higher-multiplets-bAc} for easy reference.}
	\label{fig14:triplets-24-28}
\end{figure*}
\begin{figure*}
	\centering
	\includegraphics[width=1.00\textwidth]{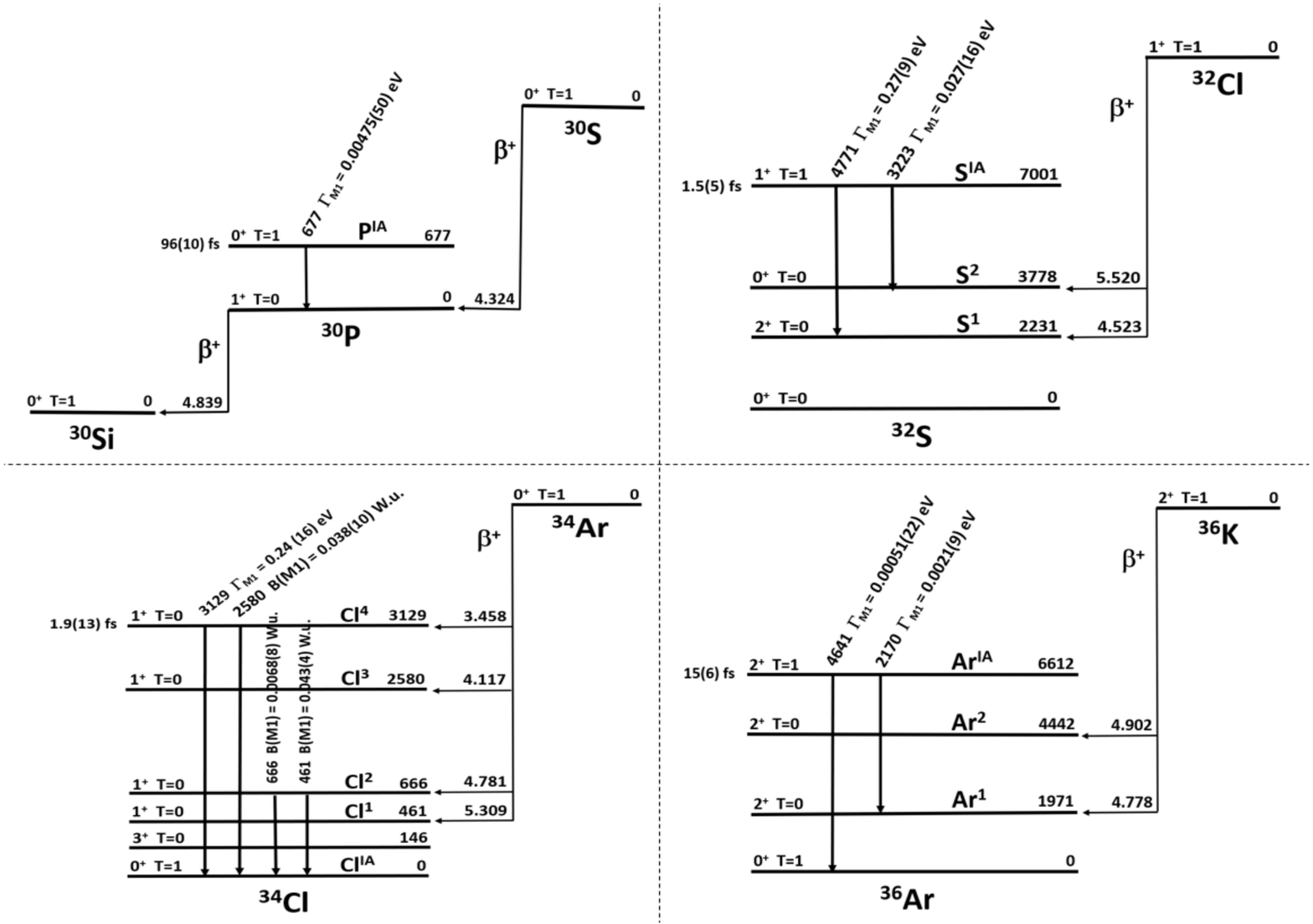}
	\caption{Partial decay schemes (not to scale) showing the $\beta$ transitions and their analog $\gamma$ transitions in $T = 1$ multiplets with $A =$ 30 to 36 for which the weak magnetism form factor is determined, with their respective log$ft$ value ($\beta$ transitions), or the transition energy and $\Gamma_{M1}$ or $B(M1)$ value ($\gamma$ transitions). For each ground state and excited state the spin-parity, isospin, and energy (in keV) is given. The excited state labels (e.g. Li$^{IA}$ or C$^1$) are also used in the first column of Tables~\ref{table:multiplets-input-1} to \ref{table:higher-multiplets-bAc} for easy reference.}
	\label{fig15:triplets-30-36}
\end{figure*}
\begin{figure*}
	\centering
	\includegraphics[width=1.00\textwidth]{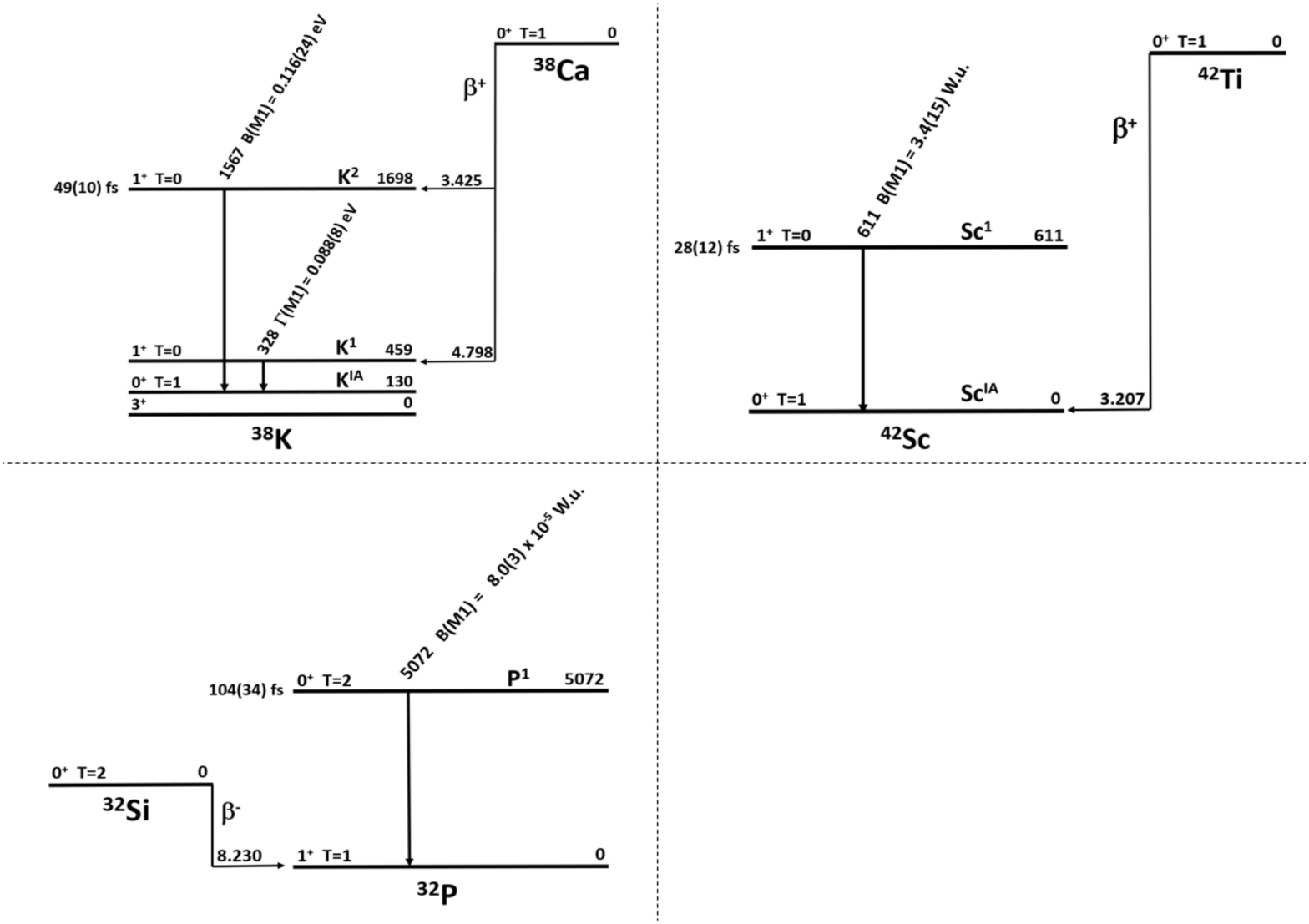}
	\caption{Partial decay schemes (not to scale) showing the $\beta$ transitions and their analog $\gamma$ transitions in $T = 1$ multiplets with $A =$ 38 to 42, and a single $T = 2$ multiplet with $A =$ 32 for which the weak magnetism form factor is determined, with their respective log$ft$ value ($\beta$ transitions), or the transition energy and $\Gamma_{M1}$ or $B(M1)$ value ($\gamma$ transitions). For each ground state and excited state the spin-parity, isospin, and energy (in keV) is given. The excited state labels (e.g. Li$^{IA}$ or C$^1$) are also used in the first column of Tables~\ref{table:multiplets-input-1} to \ref{table:higher-multiplets-bAc} for easy reference.}
	\label{fig16:triplets-38-42}
\end{figure*}
\begin{figure*}
	\centering
	\includegraphics[width=1.00\textwidth]{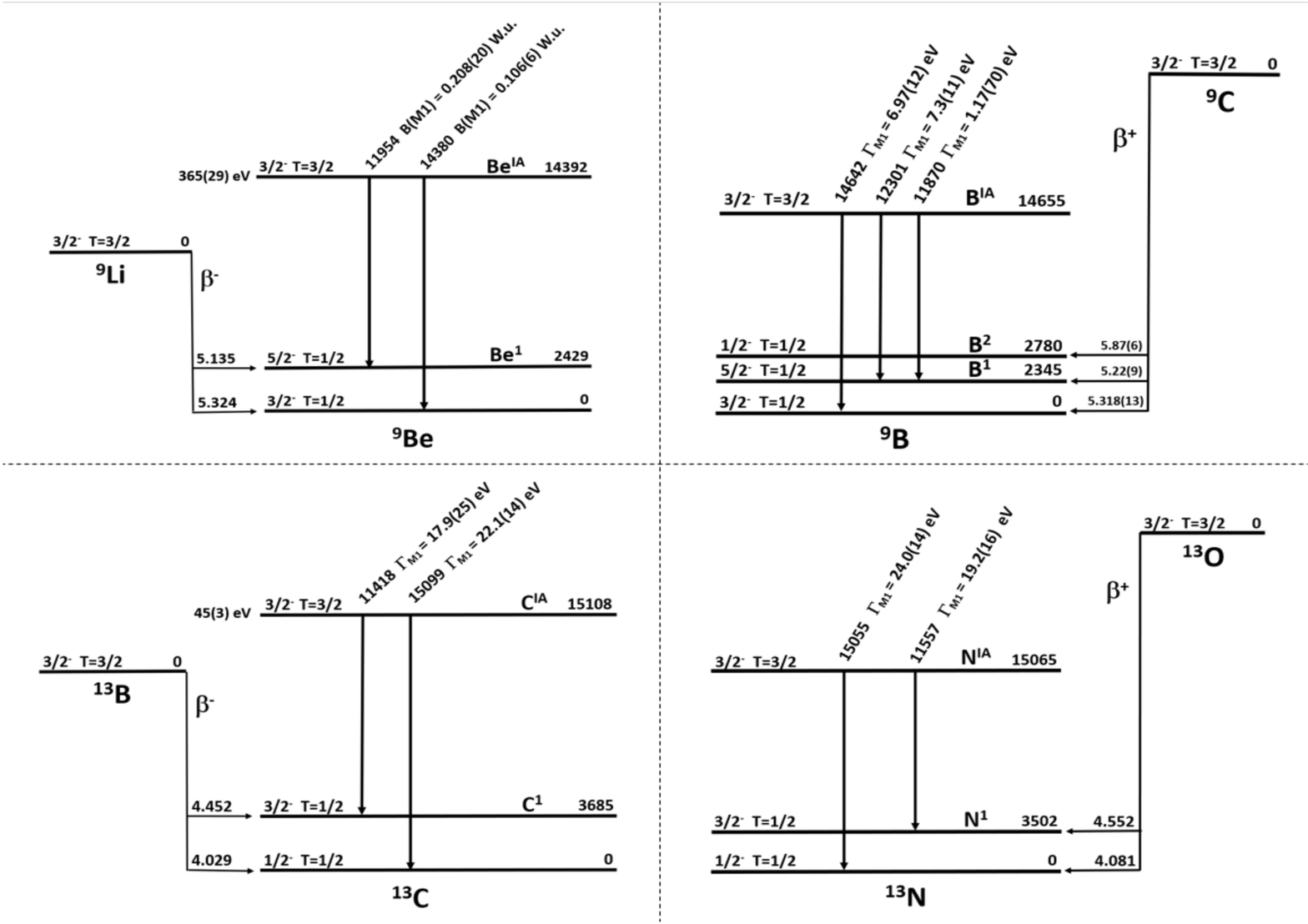}
	\caption{Partial decay schemes (not to scale) showing the $\beta$ transitions and their analog $\gamma$ transitions in $T = 3/2$ multiplets with $A =$ 9 to 13 for which the weak magnetism form factor is determined, with their respective log$ft$ value ($\beta$ transitions), or the transition energy and $\Gamma_{M1}$ or $B(M1)$ value ($\gamma$ transitions). For each ground state and excited state the spin-parity, isospin, and energy (in keV) is given. The excited state labels (e.g. Li$^{IA}$ or C$^1$) are also used in the first column of Tables~\ref{table:multiplets-input-1} to \ref{table:higher-multiplets-bAc} for easy reference.}
	\label{fig17:T3/2-9-13}
\end{figure*}
\begin{figure*}
	\centering
	\includegraphics[width=1.00\textwidth]{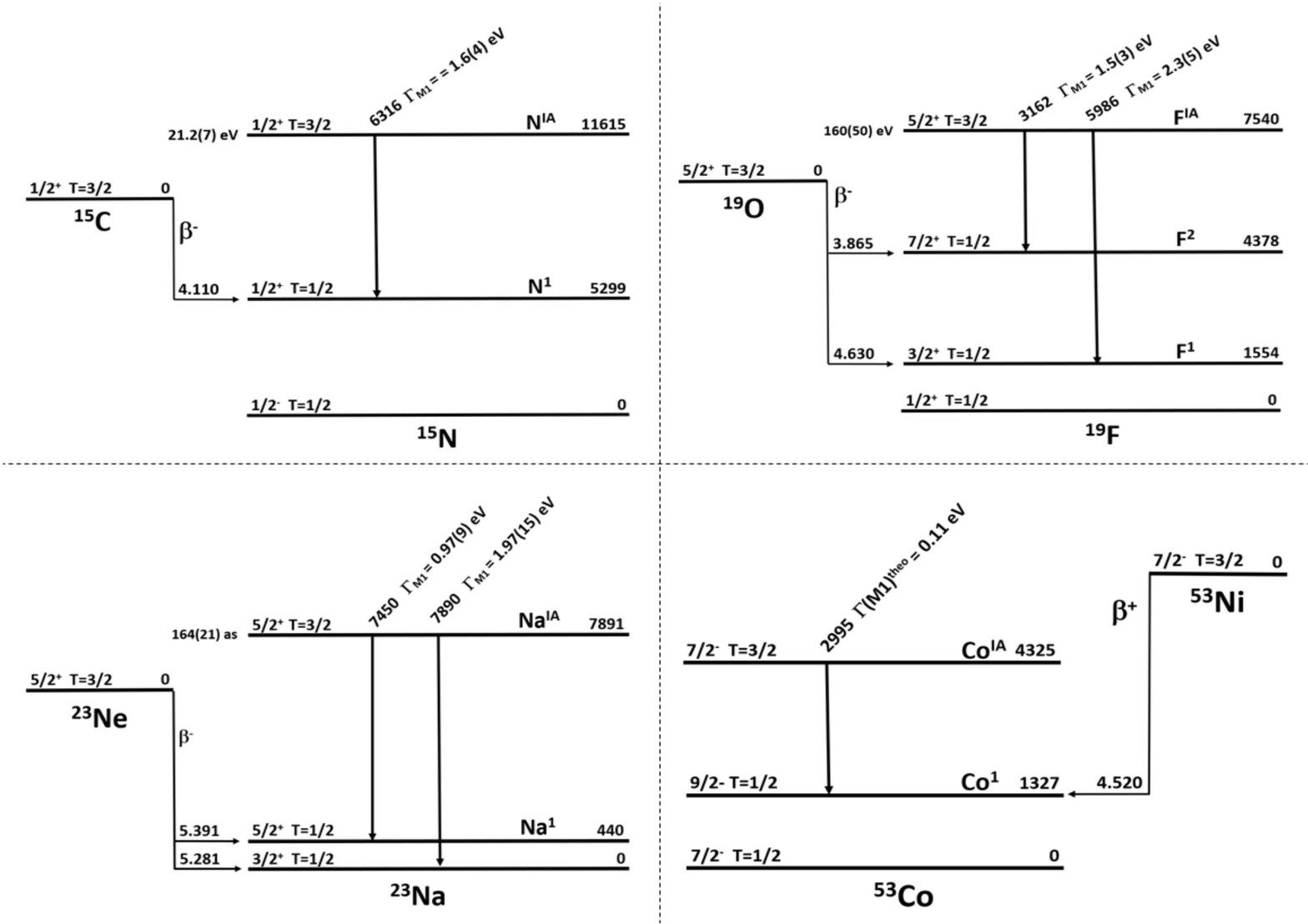}
	\caption{Partial decay schemes (not to scale) showing the $\beta$ transitions and their analog $\gamma$ transitions in $T = 3/2$ multiplets with $A =$ 15 to 23, and 53 for which the weak magnetism form factor is determined, with their respective log$ft$ value ($\beta$ transitions), or the transition energy and $\Gamma_{M1}$ or $B(M1)$ value ($\gamma$ transitions). For each ground state and excited state the spin-parity, isospin, and energy (in keV) is given. The excited state labels (e.g. Li$^{IA}$ or C$^1$) are also used in the first column of Tables~\ref{table:multiplets-input-1} to \ref{table:higher-multiplets-bAc} for easy reference.}
	\label{fig18:T3/2-15-53}
\end{figure*}
\end{widetext}

Tables~\ref{table:multiplets-input-1} and \ref{table:multiplets-input-2} provide, for each $\beta$ transition considered here, 
the input data leading to the $ft$ value, as well as the transition-dependent radiative correction, $\delta^\prime_R$, 
required to extract the Gamow-Teller form factor, $c$, from this $ft$ value via Eq.~(\ref{f_GT}). Columns 4 and 5 thus list, 
respectively, the energy of the final state of the $\beta$ transition, $E_{level}(J_f)$, 
and the $\beta$ decay transition energy, $Q_{\beta^-}$ or $Q_{EC}$ (obtained by combining $E_{level}(J_f)$ 
with the ground state-to-ground state $Q$ values listed in \cite{AME2020-2}). 
The resulting $f_{A}$ value is listed in column 6 and was calculated from Eq.~(\ref{eq:f_def_integral}) using the 
beta spectrum generator code described in Ref.~\cite{Hayen2019c}. The latter is based on the recent high-precision analytical 
description of the allowed $\ensuremath{\beta}$ spectrum shape \cite{Hayen2018}. 

Columns 7 to 10 list the half-life, $t_{1/2}$, of the $\beta$-decaying states, the branching ratio, $BR$ and the electron-capture fraction, $P_{EC}$, of the respective $\beta$ transitions, and the partial half-life, $t$, resulting from these input values (Eq.~(\ref{eq:partial-t})). Most half-lives and all branching ratio values were obtained from the Brookhaven National Nuclear Data Base (https://www.nndc.bnl.gov/ensdf/). For isotopes with a superallowed pure Fermi transition, i.e. $^{10}$C, $^{14}$O, $^{18}$Ne, $^{22}$Mg, $^{26}$Si, $^{30}$S, $^{34}$Ar, $^{38}$Ca, and $^{42}$Ti, the half-lives were taken from the detailed analysis presented in Ref.~\cite{Hardy2020}. For $^{20}$F, with varying values being reported in the literature, the weighted average of the values listed in Table V of \cite{Hughes2018} was used with the error bar increased by a factor $\sqrt(\chi^2/\nu)$. The electron-capture fractions were calculated using the same procedure as described in Sec.~\ref{adopted input for mirror Ft} for the mirror $\beta$ transitions.



For all 68 pairs of analog $\beta$ and $\gamma$ transitions considered, the log$ft$ values are given in column 4 in Tables~\ref{table:tripletkernen-bAc-1} to \ref{table:higher-multiplets-bAc}. These values result from the values for $f_A$ and $t$ listed in Tables~\ref{table:multiplets-input-1} and \ref{table:multiplets-input-2}. The $\beta$  transitions are identified by the information in columns 1 to 3. Column 5 to 8 list for the analog $\gamma$ transitions the energies of the initial state, the transition energies, the M1 decay widths and the resulting weak magnetism values, $b_{\gamma}$, respectively. Combining the latter with the Gamow-Teller form factors, $|c|^{exp}$, obtained from the log$ft$ values using Eq.~(\ref{f_GT}) and listed in column 9, the $|b/Ac|^{exp}$ values listed in column 10 are obtained for the different $\beta$ transitions.

From Eqs.~(\ref{eq:weak_magnetism_triplet}) and (\ref{f_GT}) it follows that the signs of $b$ and $c$ cannot be 
determined experimentally. However, as the ratio $b/Ac$ was found to be positive (in the convention of Holstein 
\cite{Holstein1974}, where the sign of $g_A$ is taken to be positive) for nearly all mirror $\beta$ transitions 
(see Table~\ref{table:spiegelkernen} and Sec.~\ref{Mirrors-Impulse approximation}), it was chosen to be positive 
for the transitions considered here as well. Shell model calculations for a subset of the transitions considered
here indeed showed the ratio $b/Ac$ to be positive when using $g_A = +1$ (see Sec.~\ref{matrix-elements-multiplets}, and Sec.~\ref{Mirrors-Impulse approximation}).
As it turns out, only one value out of 68 is found to be close to zero (i.e. $b/Ac = 0.189(42)$ for the transition
$^{36}$K$\rightarrow ^{36}$Ar$^1$ (Table~\ref{table:tripletkernen-bAc-3})) and could thus possibly have a negative sign
as was seen for the transitions of $^{33}$Cl, $^{35}$Ar, and $^{71}$Kr (Table~\ref{table:spiegelkernen} and Sec.~\ref{Mirrors-Impulse approximation})).

In the mass $A = 14$ triplet the $\beta$ decays of the ground states of $^{14}$C and $^{14}$O to 
the ground state of $^{14}$N yield values for $b/Ac$ that are very large ({\it i.e.} $|b/Ac|^{exp} =$
324(7) and 35.6(8), respectively) compared to all other $T$~=~1 decays listed in 
Tables~\ref{table:tripletkernen-bAc-1} to ~\ref{table:tripletkernen-bAc-3}.
These are clear cases of strongly hindered transitions (with log~$ft$ values of 9.225(2) and 7.309(7)) 
for which the beta spectrum is unlikely to have an allowed shape and the connection between 
log $ft$ and $c$ (Eq.~(\ref{f_GT})) is lost (see also \cite{Calaprice1976}). When nevertheless trying to use 
Eq.~(\ref{f_GT}) for such cases very small values for $c$ are obtained, often leading to large values 
for the ratio $b/Ac$. However, other hindered transitions, i.e. $^{22}$Na (with log$ft = 7.4227(6)$;
Table~\ref{table:tripletkernen-bAc-2}) and $^{32}$Si (log$ft = 8.23(55)$; Table~\ref{table:higher-multiplets-bAc}),
do not lead to similarly large $b/Ac$ values, but rather yield values that are in line with the non-hindered transitions. The reason for this is not clear.

The $|b/Ac|$ values for the beta transitions from the $T = 1$ and $T = 3/2$ states, except for the transitions from 
the $A = 14$ triplet, are shown graphically in Fig.~\ref{fig:bAc-T=1-T=3/2}. 
As can be seen, for almost all $\beta$ transitions the $|b/Ac|$ values range
between 0 and 10, except for the transitions from the $A = 24$ triplet consisting of the isomeric states 
$^{24m}$Na and $^{24m}$Al and the excited state at 9967 keV in $^{24}$Mg (Fig.~\ref{fig14:triplets-24-28}). 
Even though the three $\beta$ transitions in this triplet have low log$ft$ values (i.e. between 5.79(13) 
and 5.936(88)), they show large $|b/Ac|$ values ranging from 15.8(21) to 20.8(30). Other $\beta$ transitions
with similar or even larger log$ft$ values, such as e.g. the $\beta$ transitions from $^9$C ($T = 3/2$) or
from the ground states of $^{24}$Na and $^{24}$Al $(T = 1)$ (with log$ft$ values from 5.876(45) to 6.585(40)), 
and even the strongly hindered transitions from $^{22}$Na (log$ft = 7.4227(6)$) and $^{32}$Si (log$ft = 8.23(55)$), 
all show values for $|b/Ac|$ ranging from 2.2(6) to 7.3(11). The large values of $b/Ac$ for the $A = 24$ triplet 
of the isomeric states $^{24m}$Na and $^{24m}$Al and the excited state at 9967 keV in $^{24}$Mg are, however, not
so much due to the small value for the Gamow-Teller form factor, $c$, but mainly to the large values of the 
M1 $\gamma$-decay width, $\Gamma_{M1}$ of the analog $\gamma$ transitions, i.e. $\Gamma_{M1}$ = 4.3(9) eV and 
1.65(39) eV (Table~\ref{table:tripletkernen-bAc-2}). The value of $|b/Ac|$ = 17.5(32) for the transition from 
$^{24m}$Al to the ground state of $^{24}$Mg was also addressed in shell model calculations (to be discussed in the 
next section), with the theoretical result being in agreement with experiment within error bars.

Fig.~\ref{fig:bAc-T=1-T=3/2} does not show evidence for the single particle-related structure 
that was noticed for the $\beta$ transitions of the $T = 1/2$ mirror nuclei (Fig.~\ref{fig:Mirror-bAc_vs_A}). This 
is not surprising as for most of the $T = 1$, 3/2 and 2 transitions the final states are in fact excited states that usually have 
a more complicated structure. Nevertheless, some lower values of $b/Ac$ are clearly present in the $p_{1/2}$ (around 
$A =$ 13 to 15) and $d_{3/2}$ (region from $A =$ 33 to 40) subshells with $j = l - 1/2$ (see Sec.~\ref{b and b/Ac for mirrors}). 

The ratios $|b/Ac|^{exp}$ for the $\beta$ transitions from $T = 1$ triplet states listed in Tables~\ref{table:tripletkernen-bAc-1} 
to \ref{table:tripletkernen-bAc-3}, except for the strongly hindered transitions from the ground states of $^{14}$O and $^{14}$C 
to the ground state of $^{14}$N, are plotted in Fig.~\ref{fig:bAc-T=1-T=3/2}. The unweighted average (an unweighted average is used 
because individual error bars differ a lot) is $|b/Ac|^{exp} = 4.9 \pm 3.5$, with the error bar indicating one standard deviation. 
Similarly, for the transitions from $T = 3/2$ states listed in Table~\ref{table:higher-multiplets-bAc}  
and also plotted in Fig.~\ref{fig:bAc-T=1-T=3/2}, an unweighted average value $|b/Ac|^{exp} = 6.3 \pm 2.4$ is found. 

%
%
%
\begin{table*}
	
	\begin{center}
		\begin{ruledtabular}
			
			{ \footnotesize
			\caption{Input data to the log$f_A t$ values and values for the radiative correction $\delta^\prime_R$ (last column) for the $\beta$ transitions from the $A =$ 6 to $A =$ 26, $T = 1$ triplet states (see Fig.~\ref{fig12:triplets-6-14} to \ref{fig14:triplets-24-28}) for which $|b/Ac|^{exp}$ can be obtained from the $\Gamma_{M1}$ transition strength of the corresponding analog $\gamma$ transitions. $Q_{\beta^{-}/EC}$ values were obtained from the ground state-to-ground state $Q$ values from the 2016 Atomic Mass Evaluation \cite{AME2016-2} corrected for the energy of the final state, $E_{level}(J_f)$, as listed in the Brookhaven National Nuclear Data Base \cite{ENSDF}, from which also most values for the half-life, $t_{1/2}$ (see text), and all branching ratios, $BR$, were obtained. Values for the phase space factor $f_A$ were calculated with the beta spectrum generator code described in Ref.~\cite{Hayen2019c} (see text for more details). Element labels $X$ without superscripts denote nuclear ground states. Superspcripts "1" to "4" indicate excited states (see Fig.~\ref{fig12:triplets-6-14} to \ref{fig14:triplets-24-28}). For the energies of the $^{24m}$Al and $^{24m}$Na isomers the values 472.2074(8) keV and 425.8(1) keV were used \cite{ENSDF}.} \label{table:multiplets-input-1}
				
				\begin{tabular}{c|c||c|c|c|c|c|c|c|c|c}
	
	&    &                               &          &           &          &          &         &          &        \\
	Transition                         & $A$ & $J_{i} \xrightarrow{\beta} J_{f}$     & $E_{level}(J_f)$ & $Q_{\beta^{-}/EC}$ & $f_A$   &  $t_{1/2}$  & $BR$ & $P_{EC}$ &  $t$  & $\delta^\prime_R$ \\
	
	&    &                               & (keV)    & (keV)     &          &  (s)     & (\%)    & (\%)     &  (s)     &   (\%) \\
	\hline
	\hline
					
He $\xrightarrow{\beta^{-}}$ Li        & 6  & 0$^{+}$ $\rightarrow$ 1$^{+}$ & 0           &3505.215(53)& 997.795(68) &0.8067(1)&100      &   -    & 0.8067(10) & 1.203   \\
					
C $\xrightarrow{\beta^{+}}$ B$^1$      & 10 & 0$^{+}$ $\rightarrow$ 1$^{+}$ & 718.380(11) &2929.682(70)& 56.6306(90) &19.3016(24)&98.53(2)& 0.0285  & 19.583(13) & 1.454      \\
					
B $\xrightarrow{\beta^{-}}$ C          & 12 &1$^{+}$ $\rightarrow$ 0$^{+}$  & 0           &13369.4(13) & 557180(260) &0.02020(2)&98.216(28)& -    & 0.020567(21) & 0.740      \\
					
N $\xrightarrow{\beta^{+}}$ C          &    &1$^{+}$ $\rightarrow$ 0$^{+}$  & 0           &17338.1(10) & 1128063(335) &0.011000(16)&96.17(5)& 0.0000817  & 0.011438(18) & 0.700     \\
					
B $\xrightarrow{\beta{-}}$ C$^1$      & 12 &1$^{+}$ $\rightarrow$ 2$^{+}$   & 4439.82(21) &8929.6(13)& 81196(57) &0.02020(2)&1.182(19)& -   & 1.709(28) & 0.911      \\
					
N $\xrightarrow{\beta^{+}}$ C$^1$      &    &1$^{+}$ $\rightarrow$ 2$^{+}$  & 4439.82(21) &12898.3(60) & 243533(591) &0.011000(16)&1.898(32)& 0.000209 & 0.5796(98) & 0.814      \\
					
B $\xrightarrow{\beta^{-}}$ C$^2$      & 12 &1$^{+}$ $\rightarrow$ 0$^{+}$  & 7654.07(19) &5715.3(13)& 10016(11) &0.02020(2)&0.54(2) &  -  & 3.74(14) & 1.093      \\
					
N $\xrightarrow{\beta^{+}}$ C$^2$      &    &1$^{+}$ $\rightarrow$ 0$^{+}$  & 7654.07(19) &9684.0(60) & 53912(177) &0.011000(16)&1.41(3)& 0.000533  & 0.780(17) & 0.934      \\
					
N $\xrightarrow{\beta^{+}}$ C$^3$      & 12 &1$^{+}$ $\rightarrow$ 1$^{+}$  & 12710(6)    &4628.1(61) & 925.5(99) &0.011000(16)&0.120(3)& 0.00708 & 9.17(23) & 1.265     \\

O $\xrightarrow{\beta^{+}}$ N$^1$      & 14 &0$^{+}$ $\rightarrow$ 1$^{+}$  & 3948.10(20) &1196.26(20) & 0.003800(24) &70.619(11)&0.054(2) & 176  & 360(13)$\times 10^3$ & 2.000      \\
					
C $\xrightarrow{\beta^{-}}$ N          & 14 &0$^{+}$ $\rightarrow$ 1$^{+}$  & 0           &156.476(4) & 0.00933138(82) &5700(30) y&100   & -   & 179.88(95)$\times 10^9$ & 2.033      \\
					
O $\xrightarrow{\beta^{+}}$ N          &    &0$^{+}$ $\rightarrow$ 1$^{+}$  & 0           &5144.364(25)& 1760.522(49) &70.619(11)&0.61(1)& 0.00701   & 11576(190) & 1.268      \\
					
N $\xrightarrow{\beta^{-}}$ O$^{1}$    & 16 & 2$^{-}$ $\rightarrow$ 3$^{-}$ & 6129.89(4)  &4291.0(13)& 2847.3(39) &7.13(2)&66.2(6)  & -   & 10.77(10) & 1.267      \\
					
N $\xrightarrow{\beta^{-}}$ O$^{2}$    & 16 & 2$^{-}$ $\rightarrow$ 1$^{-}$ & 7116.85(14) &3304.1(23)& 876.9(27) &7.13(2)&4.8(4)   & -   & 149(12)  & 1.371      \\
					
N $\xrightarrow{\beta^{-}}$ O$^{3}$    & 16 & 2$^{-}$ $\rightarrow$ 2$^{-}$ & 8871.9(5)   &1549.0(24)& 33.94(22) &7.13(2)&1.06(7)  & -  & 673(44) & 1.643     \\
					
Ne $\xrightarrow{\beta^{+}}$ F         & 18 &0$^{+}$ $\rightarrow$ 1$^{+}$  & 0           &4444.50(59)& 689.42(54) &1.66422(47)&92.11(21)& 0.0270  & 1.8157(96) & 1.368      \\
					
F $\xrightarrow{\beta^{+}}$ O          &    &1$^{+}$ $\rightarrow$ 0$^{+}$  & 0           &1655.93(46)& 0.5564(16) &6586.2(30)&100 & 3.33   & 6805.6(31) & 1.851  \\
					
Ne $\xrightarrow{\beta^{+}}$ F$^{1}$   & 18 &0$^{+}$ $\rightarrow$ 1$^{+}$  & 1700.81(18) &2743.69(62)& 33.411(54) &1.66422(47)&0.188(6)& 0.212   & 891(29) & 1.611 \\
					
F $\xrightarrow{\beta^{-}}$ Ne$^{1}$   & 20 & 2$^{+}$ $\rightarrow$ 2$^{+}$ &1633.674(15) &5390.795(34)& 8626.32(27) &11.096(36)\footnote{Weighted average of the values listed in Table V of \cite{Hughes2018}.} &99.9913(8)& -   & 11.097(36) & 1.213      \\

Na $\xrightarrow{\beta^{+}}$ Ne$^{1}$  &    & 2$^{+}$ $\rightarrow$ 2$^{+}$ &1633.674(15) &12258.7(11)& 175906(83) &0.4479(23)&79.3(11)& 0.00109   & 0.5648(84) & 0.929      \\

Na $\xrightarrow{\beta^{+}}$ Ne$^{2}$  & 20 & 2$^{+}$ $\rightarrow$ 2$^{+}$ & 7421.9(12)  &6470.5(16)& 5675.1(97) &0.4479(23)&16.4(13)& 0.00943   & 2.73(22) & 1.208      \\

Na $\xrightarrow{\beta^{+}}$ Ne$^{3}$  & 20 & 2$^{+}$ $\rightarrow$ 2$^{+}$ & 7833.4(15)  &6059.0(19)& 3933.2(87) &0.4479(23)&0.67(6) & 0.0119   & 66.9(60) & 1.238 \\

Mg $\xrightarrow{\beta^{+}}$ Na$^{1}$  & 22 & 0$^{+}$ $\rightarrow$ 1$^{+}$ & 583.05(10)  &4198.36(18)& 473.65(14) &3.87445(69)&41.33(20)& 0.0626   & 9.383(45) & 1.449     \\

Na $\xrightarrow{\beta^{+}}$ Ne$^{1}$  & 22 & 3$^{+}$ $\rightarrow$ 2$^{+}$ & 1274.537(7) &1568.78(15)& 0.29144(32) &82108(69)$\times 10^3$& 99.944(14)& 10.8  & 91016(78)$\times 10^3$ & 1.950      \\

Na $\xrightarrow{\beta^{-}}$ Mg$^{2}$  & 24 & 4$^{+}$ $\rightarrow$ 4$^{+}$ &4122.889(12) &1392.79(24)& 24.726(17) &53989(43)&99.855(5)& -  & 54067(43) & 1.815  \\

Al $\xrightarrow{\beta^{+}}$ Mg$^{2}$  &    & 4$^{+}$ $\rightarrow$ 4$^{+}$ &4122.889(12) &9761.88(23)& 50529.1(63) &2.053(4)&7.7(10)& 0.00413   & 26.7(35) & 1.054 \\

Al $\xrightarrow{\beta^{+}}$ Mg$^{3}$  & 24 & 4$^{+}$ $\rightarrow$ 3$^{+}$ & 5235.12(4)  &8649.65(23)& 26246.4(38) &2.053(4)&1.40(13)& 0.00624   & 147(14) & 1.105 \\

Al $\xrightarrow{\beta^{+}}$ Mg$^{4}$  & 24 & 4$^{+}$ $\rightarrow$ 4$^{+}$ & 6010.84(4)  &7873.93(23)& 15951.0(26) &2.053(4)&1.2(1) & 0.00851  & 171(14) & 1.157  \\

Al $\xrightarrow{\beta^{+}}$ Mg$^{5}$  & 24 & 4$^{+}$ $\rightarrow$ 4$^{+}$ & 8439.36(4)  &5445.41(23)& 2081.61(50) &2.053(4)&50.0(20)& 0.0312  & 4.11(16) & 1.340 \\

$^m$Na $\xrightarrow{\beta^{-}}$ Mg     & 24 & 1$^{+}$ $\rightarrow$ 0$^{+}$ & 0           &5987.884(21)& 21350.49(33) &0.02020(7)&0.05(1)& -   & 40.4(81) & 1.177  \\

$^m$Al $\xrightarrow{\beta^{+}}$ Mg     &    & 1$^{+}$ $\rightarrow$ 0$^{+}$ & 0           &14310.6(25)& 466680(410) &0.1313(25)&10(3)& 0.000961  & 1.31(39) & 0.941 \\

$^m$Al$\xrightarrow{\beta^{+}}$Mg$^{1}$& 24 & 1$^{+}$ $\rightarrow$ 2$^{+}$ & 1368.672(5) &12941.83(23)& 279859(25) &0.1313(25)&4.4(5)& 0.00131  & 2.98(34) & 0.968 \\

Si $\xrightarrow{\beta^{+}}$ Al$^{1}$  & 26 & 0$^{+}$ $\rightarrow$ 1$^{+}$ &1057.739(12) &4011.397(86)& 348.764(45) &2.2453(7)&21.9(4)& 0.128  & 10.27(19) & 1.529\\

Si $\xrightarrow{\beta^{+}}$ Al$^{2}$  & 26 & 0$^{+}$ $\rightarrow$ 1$^{+}$ & 1850.62(3)  &3218.52(9) & 87.844(17) &2.2453(7)&2.73(7)& 0.328   & 82.5(21) & 1.646 \\

Si $\xrightarrow{\beta^{+}}$ Al$^{3}$  & 26 & 0$^{+}$ $\rightarrow$ 1$^{+}$ & 2071.64(4)  &2997.50(9)& 55.087(12) &2.2453(7)&0.290(11)& 0.454   & 778(30) & 1.685 \\

Si $\xrightarrow{\beta^{+}}$ Al$^{4}$  & 26 & 0$^{+}$ $\rightarrow$ 1$^{+}$ & 2740.03(3)  &2329.11(9)& 9.4555(29) &2.2453(7)&0.0618(25)& 1.60   & 3691(149) & 1.826 \\

\end{tabular}

			}
		\end{ruledtabular}
	\end{center}
\end{table*}
%
%
%

%
%
%

\begin{table*}
	
	\begin{center}
		\begin{ruledtabular}
			
			{ \footnotesize
			\caption{Input data to the log$f_A t$ values and values for the radiative correction $\delta^\prime_R$ (last column) for the $\beta$ transitions from the $A =$ 28 to $A =$ 53, $T =$ 1, 3/2 and 2 multiplet states (see Fig.~\ref{fig14:triplets-24-28} to \ref{fig18:T3/2-15-53}) for which $|b/Ac|^{exp}$ can be obtained from the $\Gamma_{M1}$ transition strength of the corresponding analog $\gamma$ transitions. $Q_{\beta^{-}/EC}$ values were obtained from the ground state-to-ground state $Q$ values from the 2020 Atomic Mass Evaluation \cite{AME2020-2} corrected for the energy of the final state, $E_{level}(J_f)$, as listed in the Brookhaven National Nuclear Data Base \cite{ENSDF}, from which most values for the half-life, $t_{1/2}$ (see text), and all branching ratios, $BR$, were obtained. Values for the phase space factor $f_A$ were calculated with the beta spectrum generator code described in Ref.~\cite{Hayen2019c}. See text for more details. Element labels $X$ without superscripts denote nuclear ground states. Superscripts "1" to "4" indicate excited states. For the $^{53}$Ni$\rightarrow ^{53}$Co transition the half-life is from \cite{Dossat2007} and the branching ratio from \cite{Su2016}.} \label{table:multiplets-input-2}
				
				\begin{tabular}{c|c||c|c|c|c|c|c|c|c|c}
	
	&    &                               &          &           &          &          &         &          &      \\
	
	Transition                         & $A$& $J_{i} \xrightarrow{\beta} J_{f}$     & $E_{level}(J_f)$ & $Q_{\beta^{-}/EC}$ & $f_A$   &  $t_{1/2}$  & $BR$ & $P_{EC}$ &  $t$  & $\delta^\prime_R$ \\
	
	&    &                               & (keV)    & (keV)     &          &  (s)     & (\%)    & (\%)     &  (s)  &  (\%) \\
	\hline
	\hline
					
Al $\xrightarrow{\beta^{-}}$ Si$^{1}$  & 28 & 3$^{+}$ $\rightarrow$ 2$^{+}$  &1779.030(11) &2663.048(48)& 396.2(35) &134.70(12)&99.99(1)& -   & 134.71(12) & 1.640 \\
					
P $\xrightarrow{\beta^{+}}$ Si$^{1}$   &    & 3$^{+}$ $\rightarrow$ 2$^{+}$  &1779.030(11) &12566.1(11) & 180803(1554) &0.2703(5)&69.1(7)& 0.00302   & 0.3912(40) & 0.944  \\
					
P $\xrightarrow{\beta^{+}}$ Si$^{2}$   & 28 & 3$^{+}$ $\rightarrow$ 3$^{+}$  & 6276.20(7)  &8065.9(11)& 17215(331) &0.2730(5)&7.6(4)& 0.0129  & 3.56(19) & 1.177 \\
					
S $\xrightarrow{\beta^{+}}$ P        & 30 & 0$^{+}$ $\rightarrow$ 1$^{+}$ & 0           & 6141.60(20)& 3804(104) &1.17977(77)&21.3(5)& 0.0415   & 5.53(13) & 1.344  \\
					
P$\xrightarrow{\beta^{+}}$ Si        &    &1$^{+}$ $\rightarrow$ 0$^{+}$       & 0           &4232.106(61)& 459(12) &149.88(24)&99.939(3)& 0.132   & 150.17(24) & 1.522  \\
					
Cl $\xrightarrow{\beta^{+}}$ S$^{1}$   & 32 & 1$^{+}$ $\rightarrow$ 2$^{+}$  & 2230.57(15) &10450.26(58)& 67076(20) &0.298(1)&60(4)& 0.00855  & 0.497(33) & 1.086 \\
					
Cl $\xrightarrow{\beta^{+}}$ S$^{2}$   & 32 & 1$^{+}$ $\rightarrow$ 0$^{+}$  & 3778.4(10)  &8902.4(11)& 28917(26) &0.298(1)&2.6(8)& 0.0144  & 11.5(35) & 1.178  \\
					
Ar $\xrightarrow{\beta^{+}}$ Cl$^{1}$  & 34 & 0$^{+}$ $\rightarrow$ 1$^{+}$  & 461.00(4)   &5600.793(75)& 2189(84) &0.84644(35)&0.91(10)& 0.0874   & 93(10) & 1.439 \\
					
Ar $\xrightarrow{\beta^{+}}$ Cl$^{2}$  & 34 & 0$^{+}$ $\rightarrow$ 1$^{+}$  & 665.56(4)   &5396.233(75)& 1774(15) &0.84644(35)&2.49(11)& 0.103   & 33.9(15) & 1.460 \\
					
Ar $\xrightarrow{\beta^{+}}$ Cl$^{3}$  & 34 & 0$^{+}$ $\rightarrow$ 1$^{+}$  & 2589.5(14)  &3482.3(14)& 132.3(13) &0.84644(35)&0.86(5)& 0.585   & 98.7(57) & 1.711 \\
					
Ar $\xrightarrow{\beta^{+}}$ Cl$^{4}$  & 34 & 0$^{+}$ $\rightarrow$ 1$^{+}$  & 3129.2(10)  &2932.5(10)& 43.59(60) &0.84644(35)&1.30(7)& 1.23   & 65.7(35) & 1.812  \\
					
K $\xrightarrow{\beta^{+}}$ Ar$^{1}$   & 36 & 1$^{+}$ $\rightarrow$ 2$^{+}$  & 1970.38(5)  &10883.98(33)& 79044(77) &0.342(2)&44(4)& 0.0114  & 0.777(71) & 1.084 \\
					
K $\xrightarrow{\beta^{+}}$ Ar$^{2}$   & 36 & 1$^{+}$ $\rightarrow$ 0$^{+}$  & 4440.11(19) &8374.25(38)& 19589(137) &0.342(2)&8.4(10)& 0.0263   & 4.07(49) & 1.234 \\
					
Ca $\xrightarrow{\beta^{+}}$ K$^{1}$  & 38 & 0$^{+}$ $\rightarrow$ 1$^{+}$   & 458.53(16)  &6283.73(17)& 4019(134) &0.44370(25)&2.84(6)& 0.0856  & 15.64(33) & 1.421\\
					
Ca $\xrightarrow{\beta^{+}}$ K$^{2}$  & 38 & 0$^{+}$ $\rightarrow$ 1$^{+}$   & 1697.65(25) &5044.61(26)& 1166.0(92) &0.44370(25)&19.48(13)& 0.195   & 2.282(15) & 1.549 \\
					
Ti $\xrightarrow{\beta^{+}}$ Sc$^{1}$ & 42 & 0$^{+}$ $\rightarrow$ 1$^{+}$   & 611.051(6)  &6405.60(22)& 4318(152) &0.20833(80)&55.9(36)& 0.114   & 0.374(24) & 1.452  \\
								
\hline

Li $\xrightarrow{\beta^{-}}$ Be        &  9 &3/2$^{-}$ $\rightarrow$ 3/2$^{-}$& 0           &13606.45(20)& 582680(41) &0.1783(4)&49.2(9)& -   & 0.3624(67) & 0.676 \\

Li $\xrightarrow{\beta^{-}}$ Be$^{1}$  &  9 &3/2$^{-}$ $\rightarrow$ 5/2$^{-}$& 2429.4(13)  &11177.1(13)& 227139(179) &0.1783(4)&29.7(30)& -   & 0.600(61) & 0.759 \\

C $\xrightarrow{\beta^{+}}$ B          &  9 &3/2$^{-}$ $\rightarrow$ 3/2$^{-}$& 0           &16494.5(23)& 895467(644) &0.1265(9)&54.1(15)& 0.0000571   & 0.2338(67) & 0.735 \\

C $\xrightarrow{\beta^{+}}$ B$^{1}$    &  9 &3/2$^{-}$ $\rightarrow$ 5/2$^{-}$& 2345(11)    &14150(11)& 405910(2317) &0.1265(9)&30.4(58)& 0.0000927   & 0.416(79) & 0.796\\

C $\xrightarrow{\beta^{+}}$ B$^{2}$    &  9 &3/2$^{-}$ $\rightarrow$ 1/2$^{-}$& 2780(16)    &13715(16)& 344584(2957) &0.1265(9)&5.8(6)& 0.000103   & 2.18(23) & 0.802   \\

B $\xrightarrow{\beta^{-}}$ C          & 13 &3/2$^{-}$ $\rightarrow$ 1/2$^{-}$& 0           &13436.9(10)& 569259(203) &0.01736(16)&92.1(18)& -   & 0.01885(41) & 0.744 \\

B $\xrightarrow{\beta^{-}}$ C$^{1}$    & 13 &3/2$^{-}$ $\rightarrow$ 3/2$^{-}$&3684.507(19) &9752.4(10)& 123890(60) &0.01736(16)&7.6(8)& -   & 0.228(24) & 0.863 \\

O $\xrightarrow{\beta^{+}}$ N          & 13 &3/2$^{-}$ $\rightarrow$ 1/2$^{-}$& 0           &17770.0(95)& 1253437(3444) &0.00858(5)&89.2(22)& 0.000118   & 0.00962(24) & 0.693   \\

O $\xrightarrow{\beta^{+}}$ N$^{1}$    & 13 &3/2$^{-}$ $\rightarrow$ 3/2$^{-}$& 3502(2)     &14268(10)& 406858(1509) &0.00858(5)&9.8(20)& 0.000234   & 0.088(18) & 0.809 \\

C $\xrightarrow{\beta^{-}}$ N$^{1}$    & 15 &1/2$^{+}$ $\rightarrow$ 1/2$^{+}$&5298.822(14) &4472.89(80)& 3325.0(27) &2.449(5)&63.2(8)& -   & 3.875(50) & 1.213  \\

O $\xrightarrow{\beta^{-}}$ F$^{1}$    & 19 &5/2$^{+}$ $\rightarrow$ 3/2$^{+}$& 1554.038(9) &3266.3(26)& 863.9(31) &26.88(5)&54.4(12)& -   & 49.4(11) & 1.399   \\

O $\xrightarrow{\beta^{-}}$ F$^{2}$    & 19 &5/2$^{+}$ $\rightarrow$ 7/2$^{+}$&4377.700(42) &442.6(26)& 0.2684(58) &26.88(5)&0.0984(30)& -   & 27317(834) & 1.991  \\

Ne $\xrightarrow{\beta^{-}}$ Na        & 23 &5/2$^{+}$ $\rightarrow$ 3/2$^{+}$& 0           &4375.80(10)& 3431.23(36) &37.24(12)&66.9(13)& -   & 55.7(11) & 1.331  \\

Ne $\xrightarrow{\beta^{-}}$ Na$^{1}$  & 23 &5/2$^{+}$ $\rightarrow$ 5/2$^{+}$& 439.990(9)  &3935.82(10)& 2115.26(24) &37.24(12)&32.0(13)& -   & 116.4(47) & 1.379 \\

Ni $\xrightarrow{\beta^{+}}$ Co$^{1}$  & 53 &7/2$^{-}$ $\rightarrow$ 9/2$^{-}$& 1327.0(9)   &11702(25)& 101177(1135) &0.0552(7)&17(8)  & 0.0389   & 0.32(15) & 1.201 \\

\hline

Si $\xrightarrow{\beta^{-}}$ P         & 32 & 0$^{+}$ $\rightarrow$ 1$^{+}$  & 0           &227.19(30)& 0.03376(86) &4828(600)$\times 10^6$ & 100 & 4.09  & 5.03(62)$\times 10^3$ & 2.329     \\

\end{tabular}
			}
		\end{ruledtabular}
	\end{center}
\end{table*}
%
%
%

%
%
%
%
\begin{table*}

\begin{center}
\begin{ruledtabular}

{ \footnotesize
\caption{Data for the $\beta$ transitions between $A =$ 6 to 18, $T = 1$ triplet states and their corresponding analog $\gamma$ transitions (see Fig.~\ref{fig12:triplets-6-14} and \ref{fig13:triplets-16-22}), leading to the form factor ratio $|b/Ac|^{exp}$. Log$(f_{A}t)$ values are from Table~\ref{table:multiplets-input-1}. Level energies and $E_{\gamma}$ and $\Gamma_{M1}$ values are from Ref.~\cite{ENSDF}. Element labels $X$ in column 1 without superscripts denote nuclear ground states, while superscripts "1" to "3" indicate excited states, and "IA" indicates the analogue state to the $\beta$ decaying state(s) (see Fig.~\ref{fig12:triplets-6-14} and \ref{fig13:triplets-16-22}). For the $|b/Ac|^{exp}$ ratios flagged with a dagger ($^{\dag}$) the value of $|b|^{exp}$ is divided (multiplied) by $\sqrt{3}$ because the gamma transition that yields $b$ is from a $T = 1$ ($T = 0$) to a $T = 0$ ($T =1$) state while the $\beta$ transition that yields $|c|^{exp}$ is in the reverse direction (see Eq.~\eqref{eq:weak_magnetism_triplet}). Note that the sign of $M_L$ is relative to the sign of $M_{GT}$ since $M_L$ is extracted using the absolute value of the $c$ form factor.} \label{table:tripletkernen-bAc-1}

\begin{tabular}{c|c||c|c|c|c|c|c|c|c|c}

     $T = 1$                           &    &                                &           &            &             &           &         &          &      &      \\

$\xrightarrow{\beta,\gamma}$           &    &                                &           & level of $\gamma$ & E$_{\gamma}$ & $\Gamma_{M1}$ &   &    &            &         \\

      $T=0$                            & $A$& $J_{i}(T_i) \rightarrow J_{f}(T_f)$      & $\log ft$ &     (keV)  &   (keV)     &    (eV)   & $|b|_{\gamma}^{exp}$ & $|c|^{exp}$ &  $|b/Ac|^{exp}$  & $M_L^{exp}$ \\\hline

He $\xrightarrow{\beta^{-}}$ Li        & 6  & 0$^{+}(1)$ $\rightarrow$ 1$^{+}(0)$ & 2.905753(61) &            &             &           &         &2.7802(19)&    4.088(42) & $-1.72(12)$\\

Li$^{IA}$ $\xrightarrow{\gamma}$ Li    &    & 0$^{+}(1)$ $\rightarrow$ 1$^{+}(0)$ &           & 3562.88(10)& 3561.75(10) & 8.19(17)  & 68.2(7) &          &            &         \\

                                       &    &                               &           &            &             &           &         &          &            &         \\

C $\xrightarrow{\beta^{+}}$ B$^1$      & 10 & 0$^{+}(1)$ $\rightarrow$ 1$^{+}(0)$ & 3.04520(13) &            &             &           &         & 2.3649(19) & 3.34(72) & $-3.2(17)$ \\

B$^{IA}$ $\xrightarrow{\gamma}$ B$^1$  &    & 0$^{+}(1)$ $\rightarrow$ 1$^{+}(0)$ &           & 1740.05(4) &1021.646(14) & 0.094(40) &  79(17) &          &            &         \\

                                       &    &                               &           &            &             &           &         &          &            &         \\

B $\xrightarrow{\beta^{-}}$ C          & 12 &1$^{+}(1)$ $\rightarrow$ 0$^{+}(0)$  & 4.05917(49)  &            &             &           &         &0.73852(55)&  3.825(45) & $-0.650(33)$ \\

N $\xrightarrow{\beta^{+}}$ C          &    &1$^{+}(1)$ $\rightarrow$ 0$^{+}(0)$  & 4.11069(68)  &            &             &           &         &0.69613(64)&  4.058(48) & $-0.451(33)$ \\

C$^{IA}$ $\xrightarrow{\gamma}$ C      &    &1$^{+}(1)$ $\rightarrow$ 0$^{+}(0)$  &           &   15110(3) &    15100(3) &  38.5(8)  & 33.9(4) &          &            &         \\

                                       &    &                               &           &            &             &           &         &          &            &         \\

B $\xrightarrow{\beta^{-}}$ C$^1$      & 12 &1$^{+}(1)$ $\rightarrow$ 2$^{+}(0)$  & 5.1423(70)  &            &             &           &         &0.2121(17)&  3.54(24)  & $-0.248(50)$ \\

N $\xrightarrow{\beta^{+}}$ C$^1$      &    &1$^{+}(1)$ $\rightarrow$ 2$^{+}(0)$  & 5.1497(74)  &            &             &           &         &0.2104(18)&  3.57(24)  & $-0.240(50)$ \\

C$^{IA}$ $\xrightarrow{\gamma}$ C$^1$  &    &1$^{+}(1)$ $\rightarrow$ 2$^{+}(0)$  &           &   15110(3) &    10666(3) &  0.96(13)  & 9.0(6) &          &            &         \\

                                       &    &                               &           &            &             &           &         &          &            &         \\

B $\xrightarrow{\beta^{-}}$ C$^2$      & 12 &1$^{+}(1)$ $\rightarrow$ 0$^{+}(0)$  & 4.574(16)   &            &             &           &         &0.4077(76) &  3.35(23)  & $-0.552(96)$ \\

N $\xrightarrow{\beta^{+}}$ C$^2$      &    &1$^{+}(1)$ $\rightarrow$ 0$^{+}(0)$  & 4.6239(94)   &            &             &           &         &0.3851(42) &  3.55(24)  & $-0.446(93)$ \\

C$^{IA}$ $\xrightarrow{\gamma}$ C$^2$  &    &1$^{+}(1)$ $\rightarrow$ 0$^{+}(0)$  &           &   15110(3) &    7453(3)  &  1.09(14) & 16.4(11)&          &            &         \\

                                       &    &                               &           &            &             &           &         &          &            &         \\

N $\xrightarrow{\beta^{+}}$ C$^3$      & 12 &1$^{+}(1)$ $\rightarrow$ 1$^{+}(0)$  & 3.929(12)  &            &             &           &         & 0.856(12) &  6.44(93)  & 1.49(80) \\

C$^{IA}$ $\xrightarrow{\gamma}$ C$^3$  &    &1$^{+}(1)$ $\rightarrow$ 1$^{+}(0)$  &           &   15110(3) &    2400(7)  &  0.59(17) & 66.2(95)&          &            &         \\

                                       &    &                               &           &            &             &           &         &          &            &         \\

O $\xrightarrow{\beta^{+}}$ N$^1$      & 14 &0$^{+}(1)$ $\rightarrow$ 1$^{+}(0)$  & 3.137(16) &            &             &           &         & 2.123(40)&  3.15(70)$^{\dag}$  & $-3.3(15)$ \\

N$^1$ $\xrightarrow{\gamma}$ N$^{IA}$  &    &1$^{+}(0)$ $\rightarrow$ 0$^{+}(1)$  &           &3948.10(20) &  1635.2(2)  & 0.091(40) & 54(12)  &          &            &         \\

                                       &    &                               &           &            &             &           &         &          &            &         \\

C $\xrightarrow{\beta^{-}}$ N          & 14 &0$^{+}(1)$ $\rightarrow$ 1$^{+}(0)$  & 9.2249(23)   &            &             &           &         &0.0019174(54)&  324.1(71)   & 0.612(14) \\

O $\xrightarrow{\beta^{+}}$ N          &    &0$^{+}(1)$ $\rightarrow$ 1$^{+}(0)$  & 7.3093(71)  &            &             &           &         &0.01747(14)&  35.58(83)  & 0.539(15) \\

N$^{IA}$ $\xrightarrow{\gamma}$ N      &    &0$^{+}(1)$ $\rightarrow$ 1$^{+}(0)$  &           &2312.798(11)& 2312.593(11)& 0.0067(3) & 8.70(19)&          &            &         \\

                                       &    &                               &           &            &             &           &         &          &            &     \\
                                       
N $\xrightarrow{\beta^{-}}$ O$^{1}$    & 16 & 2$^{-}(1)$ $\rightarrow$ 3$^{-}(0)$ & 4.4867(42)   &            &             &           &         & 0.4503(22)&  4.46(25)  & $-0.11(11)$  \\

O$^{IA}$ $\xrightarrow{\gamma}$ O$^{1}$&    & 2$^{-}(1)$ $\rightarrow$ 3$^{-}(0)$ &           & 12968.6(4) &   6837.1(4) & 1.8(2)    & 32.1(18)&          &            &         \\

                                       &    &                               &           &            &             &           &         &          &            &         \\

N $\xrightarrow{\beta^{-}}$ O$^{2}$    & 16 & 2$^{-}(1)$ $\rightarrow$ 1$^{-}(0)$ & 5.115(36)   &            &             &           &         & 0.2184(91)&   3.95(41) & $-0.165(89)$ \\

O$^{IA}$ $\xrightarrow{\gamma}$ O$^{2}$&    & 2$^{-}(1)$ $\rightarrow$ 1$^{-}(0)$ &           & 12968.6(4) &   5850.7(5) &  0.21(4)  & 13.8(13) &         &            &         \\

                                       &    &                               &           &            &             &           &         &          &            &         \\

N $\xrightarrow{\beta^{-}}$ O$^{3}$    & 16 & 2$^{-}(1)$ $\rightarrow$ 2$^{-}(0)$ & 4.359(29)  &            &             &           &         &0.521(17)&   7.57(56)   & 1.49(30)\\

O$^{IA}$ $\xrightarrow{\gamma}$ O$^{3}$&    & 2$^{-}(1)$ $\rightarrow$ 2$^{-}(0)$ &           & 12968.6(4) &   4096.1(7) & 1.5(2)    & 63.1(42)&          &            &         \\

                                       &    &                               &           &            &             &           &         &          &            &         \\

Ne $\xrightarrow{\beta^{+}}$ F         & 18 &0$^{+}(1)$ $\rightarrow$ 1$^{+}(0)$  &  3.0955(11) &            &             &           &         & 2.2328(32)&   5.72(55) & 2.3(12) \\

F $\xrightarrow{\beta^{+}}$ O          &    &1$^{+}(0)$ $\rightarrow$ 0$^{+}(1)$  &3.5783(13) &            &             &           &         &1.2778(22)&5.77(55)$^{\dag}$ & 1.36(71)\\

F$^{IA}$ $\xrightarrow{\gamma}$ F      &    &0$^{+}(1)$ $\rightarrow$ 1$^{+}(0)$  &           & 1041.55(8) &  1041.55(8) & 0.26(5)   & 230(22) &          &            &          \\

                                       &    &                               &           &            &             &           &         &          &            &          \\

Ne $\xrightarrow{\beta^{+}}$ F$^{1}$   & 18 &0$^{+}(1)$ $\rightarrow$ 1$^{+}(0)$  & 4.472(14) &            &             &           &         &0.4572(73)& 4.13(10)$^{\dag}$ & $-0.262(44)$ \\

F$^{1}$ $\xrightarrow{\gamma}$ F$^{IA}$&    &1$^{+}(0)$ $\rightarrow$ 0$^{+}(1)$  &           & 1700.81(18)& 659.25(20) &0.000478(16)&19.64(33)&          &            &          \\

                                       &    &                               &           &            &             &           &         &          &            &          \\

\end{tabular}
}
\end{ruledtabular}
\end{center}
\end{table*}
%
%
%

%
%

\begin{table*}
	
\begin{center}
\begin{ruledtabular}
			
{ \footnotesize
\caption{Data for the $\beta$ transitions between $A =$ 20 to 26, $T = 1$ triplet states and their corresponding analog $\gamma$ transitions (see Fig.~\ref{fig13:triplets-16-22} and \ref{fig14:triplets-24-28}), leading to the form factor ratio $|b/Ac|^{exp}$. Log$(f_{A}t)$ values are from Table~\ref{table:multiplets-input-1}. Level energies and $E_{\gamma}$ and $\Gamma_{M1}$ values are from Ref.~\cite{ENSDF}. The $\Gamma_{M1}$ value for the transition from $^{22}$Na $\rightarrow ^{22}$Ne$^1$ is from \cite{Triambak2017a, Triambak2017b}. Element labels $X$ in column 1 without superscripts denote nuclear ground states, while superscripts "1" to "4" indicate excited states, and "IA" indicates the analogue state to the $\beta$ decaying state(s) (see Fig.~\ref{fig13:triplets-16-22} and \ref{fig14:triplets-24-28}). For the $|b/Ac|^{exp}$ ratios flagged with a dagger ($^{\dag}$) the value of $|b|^{exp}$ is divided (multiplied) by $\sqrt{3}$ because the gamma transition that yields $b$ is from a $T = 1$ ($T = 0$) to a $T = 0$ ($T =1$) state while the $\beta$ transition that yields $|c|^{exp}$ is in the reverse direction (see Eq.~(\ref{eq:weak_magnetism_triplet})). Note that the sign of $M_L$ is relative to the sign of $M_{GT}$ since $M_L$ is extracted using the absolute value of the $c$ form factor.} \label{table:tripletkernen-bAc-2}
				
\begin{tabular}{c|c||c|c|c|c|c|c|c|c|c}
					
     $T = 1$                           &    &                                &           &            &             &           &         &          &            &         \\
					
$\xrightarrow{\beta,\gamma}$           &    &                                &           & level of $\gamma$ & E$_{\gamma}$ & $\Gamma_{M1}$ &   &    &            &         \\
					
      $T=0$                            & $A$& $J_{i}(T_i) \rightarrow J_{f}(T_f)$      & $\log ft$ &     (keV)  &   (keV)     &    (eV)   & $|b|_{\gamma}^{exp}$ & $|c|^{exp}$ &  $|b/Ac|^{exp}$  & $M_L^{exp}$ \\\hline
					
F $\xrightarrow{\beta^{-}}$ Ne$^{1}$   & 20 & 2$^{+}(1)$ $\rightarrow$ 2$^{+}(0)$  &4.9810(14) &            &             &           &         &0.25493(45)& 8.26(41)  & 0.91(11) \\

Na $\xrightarrow{\beta^{+}}$ Ne$^{1}$  &    & 2$^{+}(1)$ $\rightarrow$ 2$^{+}(0)$  &  4.9972(64) &            &             &           &         &0.2506(19) &  8.40(42)  & 0.93(11) \\

Ne$^{IA}\xrightarrow{\gamma}$Ne$^{1}$&    & 2$^{+}(1)$ $\rightarrow$ 2$^{+}(0)$  &           & 10273.2(19)&  8638(3)    & 4.0(4)    &42.1(21) &          &            &          \\

                                       &    &                                &           &            &             &           &         &          &            &          \\

Na $\xrightarrow{\beta^{+}}$ Ne$^{2}$  & 20 & 2$^{+}(1)$ $\rightarrow$ 2$^{+}(0)$  &  4.190(35)  &            &             &           &         & 0.634(25)&   4.88(37) & 0.11(23) \\

Ne$^{IA}\xrightarrow{\gamma}$Ne$^{2}$&    & 2$^{+}(1)$ $\rightarrow$ 2$^{+}(0)$  &           & 10273.2(19)&  2852(4)    & 0.31(4)   & 61.8(40)&          &            &          \\

                                       &    &                                &           &            &             &           &         &          &            &          \\

Na $\xrightarrow{\beta^{+}}$ Ne$^{3}$  & 20 & 2$^{+}(1)$ $\rightarrow$ 2$^{+}(0)$  &  5.420(39)  &            &             &           &         &0.1538(69)&   4.03(58) & $-0.104(90)$ \\

Ne$^{IA}\xrightarrow{\gamma}$Ne$^{3}$&    & 2$^{+}(1)$ $\rightarrow$ 2$^{+}(0)$  &           & 10273.2(19)&  2440.4(33) & 0.0078(21)& 12.4(17)&          &            &          \\

                                       &    &                                &           &            &             &           &         &          &            &          \\

Mg $\xrightarrow{\beta^{+}}$ Na$^{1}$  & 22 & 0$^{+}(1)$ $\rightarrow$ 1$^{+}(0)$  & 3.6477(21)  &            &             &           &         &1.1819(30)&   5.419(93)  & 0.84(11) \\

Na$^{IA(Mg)}\xrightarrow{\gamma}$Na$^{1}$ &   & 0$^{+}(1)$ $\rightarrow$ 1$^{+}(0)$ &           & 657.00(14) &  73.9(1)    &0.0000232(8)&140.9(24)&         &            &          \\

                                       &    &                                &           &            &             &           &         &          &            &          \\

Na $\xrightarrow{\beta^{+}}$ Ne$^{1}$  & 22 & 3$^{+}(0)$ $\rightarrow$ 2$^{+}(1)$  & 7.42366(60)  &            &             &           &         &0.015259(19)&  7.5(12)$^{\dag}$ & 0.043(19) \\

Na$^{IA(Ne)}\xrightarrow{\gamma}$Na  &    & 2$^{+}(1)$ $\rightarrow$ 3$^{+}(0)$  &           & 1951.8(3)  & 1951.8(3)   &0.000269(88) & 2.99(49)&        &            &          \\

                                       &    &                                &           &            &             &           &         &          &            &          \\

Na $\xrightarrow{\beta^{-}}$ Mg$^{2}$  & 24 & 4$^{+}(1)$ $\rightarrow$ 4$^{+}(0)$ &  6.12609(46)   &            &             &           &         &0.068014(74)&   4.8(12) & 0.005(83)   \\

Al $\xrightarrow{\beta^{+}}$ Mg$^{2}$  &    & 4$^{+}(1)$ $\rightarrow$ 4$^{+}(0)$ &  6.129(57)   &            &             &           &         &0.0680(44)&    4.8(13) & 0.005(86)   \\

Mg$^{IA1}\xrightarrow{\gamma}$Mg$^{2}$&    & 4$^{+}(1)$ $\rightarrow$ 4$^{+}(0)$ &            & 9516.28(4) &  5392.68(9) & 0.023(12) & 7.8(20) &          &            &          \\

                                       &    &                               &            &            &             &           &         &          &            &          \\

Al $\xrightarrow{\beta^{+}}$ Mg$^{3}$  & 24 & 4$^{+}(1)$ $\rightarrow$ 3$^{+}(0)$ &  6.585(40)   &            &             &           &         &0.0402(19)&   2.16(56) & $-0.102(23)$   \\

Mg$^{IA1}\xrightarrow{\gamma}$Mg$^{3}$&    & 4$^{+}(1)$ $\rightarrow$ 3$^{+}(0)$ &            & 9516.28(4) & 4280.62(13) &0.00083(42)& 2.09(53)&          &            &          \\

                                       &    &                               &            &            &             &           &         &          &            &          \\

Al $\xrightarrow{\beta^{+}}$ Mg$^{4}$  & 24 & 4$^{+}(1)$ $\rightarrow$ 4$^{+}(0)$ &  6.436(36)   &            &             &           &         &0.0478(20)&   4.28(11)  & $-0.021(51)$   \\

Mg$^{IA1}\xrightarrow{\gamma}$Mg$^{4}$&    & 4$^{+}(1)$ $\rightarrow$ 4$^{+}(0)$ &            & 9516.28(4) & 3505.61(9)  & 0.0025(12)& 4.9(12) &          &            &          \\

                                       &    &                               &            &            &             &           &         &          &            &          \\

Al $\xrightarrow{\beta^{+}}$ Mg$^{5}$  & 24 & 4$^{+}(1)$ $\rightarrow$ 4$^{+}(0)$ &  3.932(17)   &            &             &           &         &0.852(17) &   3.86(93) & $-0.72(79)$   \\

Mg$^{IA1}\xrightarrow{\gamma}$Mg$^{5}$&    & 4$^{+}(1)$ $\rightarrow$ 4$^{+}(0)$ &            & 9516.28(4) & 1076.86(4)  & 0.019(9)  & 79(19)  &          &            &          \\

                                       &    &                               &            &            &             &           &         &          &            &          \\

Na(m) $\xrightarrow{\beta^{-}}$ Mg     & 24 & 1$^{+}(1)$ $\rightarrow$ 0$^{+}(0)$ &  5.936(88)    &            &             &           &         &0.0849(85) &   20.7(30) & 1.36(29)   \\

Al(m) $\xrightarrow{\beta^{+}}$ Mg     &    & 1$^{+}(1)$ $\rightarrow$ 0$^{+}(0)$ &  5.79(13)  &            &             &           &         &0.101(15) &   17.5(32) & 1.29(38)   \\

Mg$^{IA2}$ $\xrightarrow{\gamma}$ Mg   &    & 1$^{+}(1)$ $\rightarrow$ 0$^{+}(0)$ &            & 9967.19(22)&  9963.0(15) &  4.3(9)   & 42.3(44)&          &            &          \\

                                       &    &                               &            &            &             &           &         &          &            &          \\

Al(m)$\xrightarrow{\beta^{+}}$Mg$^{1}$& 24 & 1$^{+}(1)$ $\rightarrow$ 2$^{+}(0)$ &  5.921(50)   &            &             &           &         &0.0860(50) &   15.8(21) & 0.96(19)   \\

Mg$^{IA2}\xrightarrow{\gamma}$Mg$^{1}$&  & 1$^{+}(1)$ $\rightarrow$ 2$^{+}(0)$ &            & 9967.19(22)&  8595.1(15) & 1.65(39)  & 32.7(39)&          &            &          \\

                                       &    &                               &            &            &             &           &         &          &            &          \\

Si $\xrightarrow{\beta^{+}}$ Al$^{1}$  & 26 & 0$^{+}(1)$ $\rightarrow$ 1$^{+}(0)$ &  3.5539(79)   &            &             &           &         &1.316(12) &   6.18(61)$^{\dag}$ & 1.93(80)   \\

Al$^{1}\xrightarrow{\gamma}$Al$^{IA}$&    & 1$^{+}(0)$ $\rightarrow$ 0$^{+}(1)$ &            &1057.739(12)&  829.3(4)   & 0.0177(35)& 122(12) &          &            &          \\

                                       &    &                               &            &            &             &           &         &          &            &          \\

Si $\xrightarrow{\beta^{+}}$ Al$^{2}$  & 26 & 0$^{+}(1)$ $\rightarrow$ 1$^{+}(0)$ &  3.861(11)   &            &             &           &         &0.924(12) &   2.88(13)$^{\dag}$ & $-1.69(13)$   \\

Al$^{2}\xrightarrow{\gamma}$Al$^{IA}$&    & 1$^{+}(0)$ $\rightarrow$ 0$^{+}(1)$ &            & 1850.62(3) &  1622(7)    & 0.0141(13)& 39.9(18)&          &            &          \\

                                       &    &                               &            &            &             &           &         &          &            &          \\

Si $\xrightarrow{\beta^{+}}$ Al$^{3}$  & 26 & 0$^{+}(1)$ $\rightarrow$ 1$^{+}(0)$ &  4.632(16)   &            &             &           &         &0.3802(72)&   1.63(16)$^{\dag}$ & $-1.170(66)$   \\

Al$^{3}\xrightarrow{\gamma}$Al$^{IA}$&    & 1$^{+}(0)$ $\rightarrow$ 0$^{+}(1)$ &            & 2071.64(4) & 1842.8(7)   &0.00112(22)& 9.29(91)&          &            &          \\

                                       &    &                               &            &            &             &           &         &          &            &          \\

Si $\xrightarrow{\beta^{+}}$ Al$^{4}$  & 26 & 0$^{+}(1)$ $\rightarrow$ 1$^{+}(0)$ &  4.543(18)   &            &             &           &         &0.4209(85) &  3.31(20)$^{\dag}$ & $-0.588(86)$   \\

Al$^{4}\xrightarrow{\gamma}$Al$^{IA}$&    & 1$^{+}(0)$ $\rightarrow$ 0$^{+}(1)$ &            & 2740.03(3) & 2511.59(10) & 0.0144(16) &20.9(12)&          &            &          \\

\end{tabular}
}
\end{ruledtabular}
\end{center}
\end{table*}
%
%
%
%
%

\begin{table*}
	
\begin{center}
\begin{ruledtabular}
			
{ \footnotesize
\caption{Data for the $\beta$ transitions between $A =$ 28 to 42, $T = 1$ triplet states and their corresponding analog $\gamma$ transitions (see figs.~\ref{fig14:triplets-24-28} to \ref{fig16:triplets-38-42}), leading to the form factor ratio $|b/Ac|^{exp}$. Log$(f_{A}t)$ values are from Table~\ref{table:multiplets-input-2}. Level energies and $E_{\gamma}$ and $\Gamma_{M1}$ values are from Ref.~\cite{ENSDF}. Element labels $X$ in column 1 without superscripts denote nuclear ground states, while superscripts "1" to "4" indicate excited states, and "IA" indicates the analogue state to the $\beta$ decaying state(s) (see figs.~\ref{fig14:triplets-24-28} to \ref{fig16:triplets-38-42}). For the $|b/Ac|^{exp}$ ratios flagged with a dagger ($^{\dag}$) the value of $|b|^{exp}$ is divided (multiplied) by $\sqrt{3}$ because the gamma transition that yields $b$ is from a $T = 1$ ($T = 0$) to a $T = 0$ ($T =1$) state while the $\beta$ transition that yields $|c|^{exp}$ is in the reverse direction (see Eq.~(\ref{eq:weak_magnetism_triplet})). Note that the sign of $M_L$ is relative to the sign of $M_{GT}$ since $M_L$ is extracted using the absolute value of the $c$ form factor.} \label{table:tripletkernen-bAc-3}
				
\begin{tabular}{c|c||c|c|c|c|c|c|c|c|c}
					
	$T = 1$                            &    &                                &           &            &             &           &         &          &            &         \\
					
	$\xrightarrow{\beta,\gamma}$       &    &                                &           & level of $\gamma$ & E$_{\gamma}$ & $\Gamma_{M1}$ &   &    &            &         \\
					
	$T=0$                              & $A$& $J_{i}(T_i) \rightarrow J_{f}(T_f)$      & $\log ft$ &     (keV)  &   (keV)     &    (eV)   & $|b|_{\gamma}^{exp}$ & $|c|^{exp}$ &  $|b/Ac|^{exp}$  & $M_L^{exp}$ \\\hline

Al $\xrightarrow{\beta^{-}}$ Si$^{1}$  & 28 & 3$^{+}(1)$ $\rightarrow$ 2$^{+}(0)$  &  4.7273(38)&            &             &           &         &0.3407(15) &  1.74(38) & $-1.01(13)$   \\

P $\xrightarrow{\beta^{+}}$ Si$^{1}$   &    & 3$^{+}(1)$ $\rightarrow$ 2$^{+}(0)$  &  4.8496(58) &            &             &           &         & 0.2970(20) &  2.00(43) & $-0.80(13)$   \\

Si$^{IA}\xrightarrow{\gamma}$Si$^{1}$&    & 3$^{+}(1)$ $\rightarrow$ 2$^{+}(0)$  &           & 9315.92(10)&  7535.7(4)  &  0.21(9)  & 16.6(36)&            &           &          \\

                                       &    &                                &            &            &             &           &         &           &           &          \\

P $\xrightarrow{\beta^{+}}$ Si$^{2}$   & 28 & 3$^{+}91)$ $\rightarrow$ 3$^{+}(0)$  &  4.787(24) &            &             &           &         &0.3188(89) &  4.47(85) & $-0.08(27)$   \\

Si$^{IA}\xrightarrow{\gamma}$Si$^{2}$&    & 3$^{+}(1)$ $\rightarrow$ 3$^{+}(0)$  &            & 9315.92(10)&  3039.16(17)& 0.08(3)   & 39.9(75)&           &            &          \\

                                       &    &                                &            &            &             &           &         &           &            &          \\

S $\xrightarrow{\beta^{+}}$ P          & 30 &0$^{+}(1)\rightarrow$1$^{+}(0)$& 4.324(16) &            &             &           &         & 0.5429(98)&  6.08(34)  & 0.75(18) \\

P$\xrightarrow{\beta^{+}}$ Si          &    &1$^{+}(0)\rightarrow$0$^{+}(1)$& 4.839(12) &            &             &           &         &0.2998(41)& 6.36(34)$^{\dag}$ & 0.50(10)\\

P$^{IA}$ $\xrightarrow{\gamma}$ P      &    &0$^{+}(1)\rightarrow$1$^{+}(0)$     &            & 677.01(3)  &   677.01(3) & 0.0048(5) & 99.1(52)&           &            &           \\

                                       &    &                                &            &            &             &           &         &           &            &           \\

Cl $\xrightarrow{\beta^{+}}$ S$^{1}$   & 32 & 1$^{+}(1)$ $\rightarrow$ 2$^{+}(0)$  &  4.523(29)   &            &             &           &         & 0.432(14)  & 3.08(52)  & $-0.70(23)$   \\

S$^{IA}\xrightarrow{\gamma}$S$^{1}$  &    & 1$^{+}(1)$ $\rightarrow$ 2$^{+}(0)$  &            & 7001.4(4)  &  4770.5(3)  & 0.27(9)   & 42.6(71)&           &            &          \\

                                       &    &                                &            &            &             &           &         &           &            &          \\

Cl $\xrightarrow{\beta^{+}}$ S$^{2}$   & 32 & 1$^{+}(1)$ $\rightarrow$ 0$^{+}(0)$  &  5.52(14)  &            &             &           &         & 0.137(21) &  5.5(19)   & 0.11(25)   \\

S$^{IA}\xrightarrow{\gamma}$S$^{2}$  &    & 1$^{+}(1)$ $\rightarrow$ 0$^{+}(0)$  &            & 7001.4(4)  &  3223.4(10) & 0.027(16) & 24.3(72)&           &            &          \\

                                       &    &                                &            &            &             &           &         &           &            &          \\

Ar $\xrightarrow{\beta^{+}}$ Cl$^{1}$  & 34 & 0$^{+}(1)$ $\rightarrow$ 1$^{+}(0)$  &  5.309(51)   &            &             &           &         & 0.175(10) & 7.91(58)$^{\dag}$  & 0.56(11)   \\

Cl$^{1}\xrightarrow{\gamma}$Cl$^{IA}$&    & 1$^{+}(0)$ $\rightarrow$ 0$^{+}(1)$  &            & 461.00(4)  &  461.00(4)  &0.000087(8)& 27.1(12)&           &            &          \\

                                       &    &                                &            &            &             &           &         &           &            &          \\

Ar $\xrightarrow{\beta^{+}}$ Cl$^{2}$  & 34 & 0$^{+}(1)$ $\rightarrow$ 1$^{+}(0)$  &  4.781(20)   &            &             &           &         & 0.3206(72)& 1.72(11)$^{\dag}$  & $-0.957(41)$   \\

Cl$^{2}\xrightarrow{\gamma}$Cl$^{IA}$&    & 1$^{+}(0)$ $\rightarrow$ 0$^{+}(1)$  &            & 665.56(4)  &  665.55(5)  &0.000042(5)&10.84(65)&           &            &          \\

                                       &    &                                &            &            &             &           &         &           &            &          \\

Ar $\xrightarrow{\beta^{+}}$ Cl$^{3}$  & 34 & 0$^{+}(1)$ $\rightarrow$ 1$^{+}(0)$  &  4.117(26)   &            &             &           &         & 0.688(20) & 1.89(26)$^{\dag}$  & $-1.94(19)$   \\

Cl$^{3}\xrightarrow{\gamma}$Cl$^{IA}$&    & 1$^{+}(0)$ $\rightarrow$ 0$^{+}(1)$  &            & 2580.4(2)  & 2579.4(14)  & 0.0135(36)& 25.5(34)&           &            &          \\

                                       &    &                                &            &            &             &           &         &           &            &          \\

Ar $\xrightarrow{\beta^{+}}$ Cl$^{4}$  & 34 & 0$^{+}(1)$ $\rightarrow$ 1$^{+}(0)$  &  3.458(24) &            &             &           &         & 1.467(41) & 2.78(94)$^{\dag}$  & $-2.8(14)$   \\

Cl$^{4}\xrightarrow{\gamma}$Cl$^{IA}$&    & 1$^{+}(0)$ $\rightarrow$ 0$^{+}(1)$  &            & 3129.13(12)& 3129(10)    & 0.24(16)  & 80(27)  &           &            &          \\

                                       &    &                                &            &            &             &           &         &           &            &          \\

K $\xrightarrow{\beta^{+}}$ Ar$^{1}$   & 36 & 1$^{+}(1)$ $\rightarrow$ 2$^{+}(0)$  &  4.778(40)   &            &             &           &         & 0.318(15) &  0.189(42)   & $-1.438(67)$   \\

Ar$^{IA}\xrightarrow{\gamma}$Ar$^{1}$&    & 1$^{+}(1)$ $\rightarrow$ 2$^{+}(0)$  &            & 6612.12(20)&  4641.0(5)  &0.00051(22)& 2.17(47)&           &            &          \\

                                       &    &                                &            &            &             &           &         &           &            &          \\

K $\xrightarrow{\beta^{+}}$ Ar$^{2}$   & 36 & 1$^{+}(1)$ $\rightarrow$ 0$^{+}(0)$  &  4.902(52)   &            &             &           &         & 0.279(17) &  1.37(31)  & $-0.93(10)$   \\

Ar$^{IA}\xrightarrow{\gamma}$Ar$^{2}$&    & 1$^{+}(1)$ $\rightarrow$ 0$^{+}(0)$  &            & 6612.12(20)& 2170.29(20) & 0.0021(9) & 13.8(30)&           &            &          \\

                                       &    &                                &            &            &             &           &         &           &            &          \\

Ca $\xrightarrow{\beta^{+}}$ K$^{1}$  & 38 & 0$^{+}(1)$ $\rightarrow$ 1$^{+}(0)$   &  4.798(17)   &            &             &           &         & 0.3143(62) & 6.26(32)$^{\dag}$  & 0.49(10)   \\

K$^{1}\xrightarrow{\gamma}$K$^{IA}$ &    & 1$^{+}(0)$ $\rightarrow$ 0$^{+}(1)$   &            & 458.4(4)   &  327.9(2)   &0.000064(6)& 43.2(20)&           &            &          \\

                                      &    &                                 &            &            &             &           &         &           &            &          \\

Ca $\xrightarrow{\beta^{+}}$ K$^{2}$  & 38 & 0$^{+}(1)$ $\rightarrow$ 1$^{+}(0)$   &  3.4251(45) &            &             &           &         & 1.5262(80) & 1.49(15)$^{\dag}$  & $-4.91(23)$   \\

K $^{2}\xrightarrow{\gamma}$ K$^{IA}$ &  & 1$^{+}(0)$ $\rightarrow$ 0$^{+}(1)$   &            & 1697.84(12)& 1567.39(12) &0.0093(19) & 49.9(51)&           &            &          \\

                                      &    &                                 &            &            &             &           &         &           &            &          \\

Ti $\xrightarrow{\beta^{+}}$ Sc$^{1}$ & 42 & 0$^{+}(1)$ $\rightarrow$ 1$^{+}(0)$   &  3.207(32)   &            &             &           &         & 1.963(72) & 6.2(14)$^{\dag}$   & 3.0(27)   \\

Sc$^{1}\xrightarrow{\gamma}$Sc$^{IA}$&   & 1$^{+}(0)$ $\rightarrow$ 0$^{+}(1)$   &            & 611.051(6) &  611.046(6) & 0.016(7)  & 297(65) &           &            &          \\

                                      &    &                                 &            &            &             &           &         &           &            &          \\



\end{tabular}
}
\end{ruledtabular}
\end{center}
\end{table*}
%
%
%

%
%

\begin{table*}
\begin{center}
\begin{ruledtabular}
			
{ \footnotesize
\caption{Data for the $\beta$ transitions between $A =$ 9 to 53, $T = 3/2$ quartet states and their corresponding analog $\gamma$ transitions (see Fig.~\ref{fig17:T3/2-9-13} and \ref{fig18:T3/2-15-53}), leading to the form factor ratio $|b/Ac|^{exp}$. At the bottom of the table the single $\beta$ transition from a $T = 2$ quintet state and its analog $\gamma$ transition are included as well. Log$(f_{A}t)$ values are from Table~\ref{table:multiplets-input-2}. Level energies and $E_{\gamma}$ and $\Gamma_{M1}$ values are from Ref.~\cite{ENSDF}. The data for the transition $^{53}$Ni $\rightarrow ^{53}$Co$^1$ are from \cite{Su2016}. The $\Gamma_{M1}$ value for this case is from theory \cite{Su2016}, with a 20\% error being assumed. Element labels $X$ in column 1 without superscripts denote nuclear ground states, while superscripts "1" to "2" indicate excited states, and "IA" indicates the analogue state to the $\beta$ decaying state(s) (see Fig.~\ref{fig17:T3/2-9-13} and \ref{fig18:T3/2-15-53}). Note that the sign of $M_L$ is relative to the sign of $M_{GT}$ since $M_L$ is extracted using the absolute value of the $c$ form factor.} \label{table:higher-multiplets-bAc}
\begin{tabular}{c|c||c|c|c|c|c|c|c|c|c}

	$T = 3/2$       &    &                                &           & level of $\gamma$ & E$_{\gamma}$ & $\Gamma_{M1}$ &   &    &             &         \\
					
	$decays$                            & $A$& $J_{i}(T_i) \rightarrow J_{f}(T_f)$      & $\log ft$ &     (keV)   &   (keV)     &    (eV)    & $|b|_{\gamma}^{exp}$ & $|c|^{exp}$ &  $|b/Ac|^{exp}$  & $M_L^{exp}$ \\ \hline

Li $\xrightarrow{\beta^{-}}$ Be        &  9 &3/2$^{-}(3/2)$ $\rightarrow$ 3/2$^{-}(1/2)$& 5.3246(80)  &             &             &           &          &0.1721(16)&  7.31(23)  & 0.448(40)   \\

Be$^{IA}$ $\xrightarrow{\gamma}$Be      &    & 3/2$^{-}(3/2)$ $\rightarrow$ 3/2$^{-}(1/2)$ &           & 14392.2(18) & 14380.0(18) &  6.6(4)   & 11.32(34)&          &            &          \\

                                       &    &                                &           &             &             &           &          &          &            &          \\

Li $\xrightarrow{\beta^{-}}$ Be$^{1}$  &  9 &3/2$^{-}(3/2)$ $\rightarrow$ 5/2$^{-}(1/2)$ &  5.135(44)  &             &             &           &          &0.214(11) &   8.26(42) & 0.760(98)   \\

Be$^{IA}$ $\xrightarrow{\gamma}$Be$^{1}$&    &3/2$^{-}(3/2)$ $\rightarrow$ 5/2$^{-}(1/2)$&           & 14392.2(18) & 11954.3(22) & 7.48(7)   &15.907(75)&          &            &          \\

                                       &    &                                &           &             &             &           &          &          &            &          \\
                                       
C $\xrightarrow{\beta^{+}}$ B          &  9 &3/2$^{-}(3/2)$ $\rightarrow$ 3/2$^{-}(1/2)$& 5.320(12) &             &             &           &          &0.1728(25)&   7.42(25) & 0.469(44)   \\

B$^{IA}$ $\xrightarrow{\gamma}$B        &    &3/2$^{-}(3/2)$ $\rightarrow$ 3/2$^{-}(1/2)$&           & 14655.0(25) & 14462.2(25) &  6.97(42) & 11.54(35)&          &            &          \\

                                       &    &                                &           &             &             &           &          &          &            &          \\

C $\xrightarrow{\beta^{+}}$ B$^{1}$    &  9 &3/2$^{-}(3/2)$ $\rightarrow$ 5/2$^{-}(1/2)$&  5.228(84)  &             &             &           &          &0.192(18) &   8.7(10)  & 0.77(21)   \\

B$^{IA}$ $\xrightarrow{\gamma}$B$^{1}$  &    &3/2$^{-}(3/2)$ $\rightarrow$ 5/2$^{-}(1/2)$&           & 14655.0(25) & 12301(11)   & 7.3(11)   & 15.1(11) &          &            &          \\

                                       &    &                                &           &             &             &           &          &          &            &          \\

C $\xrightarrow{\beta^{+}}$ B$^{2}$    &  9 &3/2$^{-}(3/2)$ $\rightarrow$ 1/2$^{-}(1/2)$&  5.876(45)  &             &             &           &          &0.0912(47)&   7.8(24)  & 0.28(21)   \\

B$^{IA}$ $\xrightarrow{\gamma}$B$^{2}$  &    &3/2$^{-}(3/2)$ $\rightarrow$ 1/2$^{-}(1/2)$&           & 14655.0(25) & 11870(160)  & 1.17(70)  & 6.4(19)  &          &            &          \\

                                       &    &                                &           &             &             &           &          &          &            &          \\

B $\xrightarrow{\beta^{-}}$ C          & 13 &3/2$^{-}(3/2)$ $\rightarrow$ 1/2$^{-}(1/2)$& 4.0294(94)  &             &             &           &          &0.7632(83)&   2.804(94)  & $-1.452(73)$   \\

C$^{IA}$ $\xrightarrow{\gamma}$ C       &    &3/2$^{-}(3/2)$ $\rightarrow$ 1/2$^{-}(1/2)$&           & 15108.2(12) & 15098.8(12) &  22.1(14) & 27.82(88)&          &            &          \\

                                       &    &                                &           &             &             &           &          &          &            &          \\

B $\xrightarrow{\beta^{-}}$ C$^{1}$    & 13 &3/2$^{-}(3/2)$ $\rightarrow$ 3/2$^{-}(1/2)$&  4.452(46)  &             &             &           &          &0.470(25) &   6.24(55) & 0.72(26)   \\

C$^{IA}$ $\xrightarrow{\gamma}$C$^{1}$  &    &3/2$^{-}(3/2)$ $\rightarrow$ 3/2$^{-}(1/2)$&           & 15108.2(12) & 11418.2(12) & 17.9(25)  & 38.1(27) &          &            &          \\

                                       &    &                                &           &             &             &           &          &          &            &          \\

O $\xrightarrow{\beta^{+}}$ N          & 13 &3/2$^{-}(3/2)$ $\rightarrow$ 1/2$^{-}(1/2)$& 4.081(11) &             &             &           &          &0.7202(92)&   3.11(10) & $-1.149(73)$   \\

N$^{IA}$ $\xrightarrow{\gamma}$ N       &    &3/2$^{-}(3/2)$ $\rightarrow$ 1/2$^{-}(1/2)$&           & 15064.6(4)  & 15055.2(4)  &  24.0(14) & 29.12(85)&          &            &          \\

                                       &    &                                &           &             &             &           &          &          &            &          \\

O $\xrightarrow{\beta^{+}}$ N$^{1}$    & 13 &3/2$^{-}(3/2)$ $\rightarrow$ 3/2$^{-}(1/2)$&  4.552(90)  &             &             &           &          &0.419(43) &   7.11(78) & 1.01(34)   \\

N$^{IA}$ $\xrightarrow{\gamma}$N$^{1}$  &    &3/2$^{-}(3/2)$ $\rightarrow$ 3/2$^{-}(1/2)$&           & 15064.6(4)  & 11557(2)    & 19.2(16)  & 38.7(16) &          &            &          \\

                                       &    &                                &           &             &             &           &          &          &            &          \\

C $\xrightarrow{\beta^{-}}$ N$^{1}$    & 15 &1/2$^{+}(3/2)$ $\rightarrow$ 1/2$^{+}(1/2)$&  4.1101(56) &             &             &           &          &0.6949(45)&   3.06(38) & $-1.14(27)$   \\

N$^{IA}$ $\xrightarrow{\gamma}$N$^{1}$  &    &1/2$^{+}(3/2)$ $\rightarrow$ 1/2$^{+}(1/2)$&           & 11615(4)    &  6316(4)    & 1.6(4)    & 31.9(40) &          &            &          \\

                                       &    &                                &           &             &             &           &          &          &            &          \\

O $\xrightarrow{\beta^{-}}$ F$^{1}$    & 19 &5/2$^{+}(3/2)$ $\rightarrow$ 3/2$^{+}(1/2)$& 4.6303(97) &             &             &           &          &0.3814(43)&   7.26(79) & 0.97(30)   \\

F$^{IA}$ $\xrightarrow{\gamma}$F$^{1}$  &    &5/2$^{+}(3/2)$ $\rightarrow$ 3/2$^{+}(1/2)$&           &  7539.6(9)  &  5986(9)   & 2.3(5)    & 52.6(57) &          &            &          \\

                                       &    &                                &           &             &             &           &          &          &            &          \\

O $\xrightarrow{\beta^{-}}$ F$^{2}$    & 19 &5/2$^{+}(3/2)$ $\rightarrow$ 7/2$^{+}(1/2)$& 3.865(16) &             &             &           &          &0.918(17) &   6.31(64) & 1.47(59)   \\

F$^{IA}$ $\xrightarrow{\gamma}$F$^{2}$  &    &5/2$^{+}(3/2)$ $\rightarrow$ 7/2$^{+}(1/2)$&           & 7539.6(9)   &  3161.9(9)  & 1.5(3)    & 110(11)  &          &            &          \\

                                       &    &                                &           &             &             &           &          &          &            &          \\

Ne $\xrightarrow{\beta^{-}}$ Na        & 23 &5/2$^{+}(3/2)$ $\rightarrow$ 3/2$^{+}(1/2)$&  5.2810(86)  &             &             &           &          &0.1804(18)&   9.38(37) & 0.842(68)   \\

Na$^{IA}$ $\xrightarrow{\gamma}$ Na     &    &5/2$^{+}(3/2)$ $\rightarrow$ 3/2$^{+}(1/2)$&           & 7891.19(25) & 7889.7(3)  &  1.97(15)  & 38.9(15) &          &            &          \\

                                       &    &                                &           &             &             &           &          &          &            &          \\

Ne $\xrightarrow{\beta^{-}}$ Na$^{1}$  & 23 &5/2$^{+}(3/2)$ $\rightarrow$ 5/2$^{+}(1/2)$&  5.391(18)  &             &             &           &          &0.1588(32)&   8.16(42) & 0.548(67)   \\

Na$^{IA}$ $\xrightarrow{\gamma}$Na$^{1}$&    &5/2$^{+}(3/2)$ $\rightarrow$ 5/2$^{+}(1/2)$&           & 7891.19(25) & 7449.9(3)   & 0.97(9)   & 29.8(14) &          &            &          \\

                                       &    &                                &           &             &             &           &          &          &            &          \\

Ni $\xrightarrow{\beta^{+}}$ Co$^{1}$  & 53 &7/2$^{-}(3/2)$ $\rightarrow$ 9/2$^{-}(1/2)$&  4.52(22) &             &             &           &          & 0.44(10) &   2.26(57) & $-1.06(35)$   \\

Co$^{IA}$ $\xrightarrow{\gamma}$Co$^{1}$&   &7/2$^{-}(3/2)$ $\rightarrow$ 9/2$^{-}(1/2)$&           & 4325(2)     &  4325(2)    & 0.11(2)   & 52.2(47) &          &            &          \\

\hline
\hline

	$T = 2$                            &    &                                &           &             &             &           &          &          &            &         \\

	$\xrightarrow{\beta,\gamma}$       &    &                                &           & level of $\gamma$ & E$_{\gamma}$ & $\Gamma_{M1}$ &   &    &            &         \\

	$T=1$                              & $A$& $J_{i}(T_i) \rightarrow J_{f}(T_f)$      & $\log ft$ &     (keV)   &   (keV)     &    (eV)   & $|b|_{\gamma}^{exp}$ & $|c|^{exp}$ &  $|b/Ac|^{exp}$  & $M_L^{exp}$ \\\hline

Si $\xrightarrow{\beta^{+}}$ P         & 32 & 0$^{+}(2)$ $\rightarrow$ 1$^{+}(1)$  & 8.230(55)   &             &             &           &          &0.00602(38) &  5.76(39)  & 0.0063(24) \\

P$^{1}$ $^{IA}$ $\xrightarrow{\gamma}$P  &    & 0$^{+}(2)$ $\rightarrow$ 1$^{+}(1)$  &           & 5072.44(6)  & 5072.00(6)  & 0.00022(1)&1.110(25) &           &           &          \\

\end{tabular}
}
\end{ruledtabular}
\end{center}
\end{table*}



%
%

\subsubsection{Nuclear matrix elements $M_{GT}$ and $M_L$}
\label{matrix-elements-multiplets}

\paragraph{Experimental values for $M_{GT}$ and $M_L$}

Using the experimental values for the $b$ and $c$ form factors in Tables~\ref{table:tripletkernen-bAc-1} to 
\ref{table:higher-multiplets-bAc}, and assuming the ratio $b/Ac$ to be positive (Sec.~\ref{experimental-values-multiplets}), 
the matrix element $M_L$ can be derived using Eq.~(\ref{eq:bAc_impulse_approx}). 
As we use for this the absolute values of the $c$ form factor the sign obtained for $M_L$ is relative to that of 
$M_{GT}$ (Eq.~(\ref{eq:bAc_impulse_approx})). The resulting values for $M_L$ are listed in the last columns of 
Tables~\ref{table:tripletkernen-bAc-1} to \ref{table:higher-multiplets-bAc}.
%
%
%

\paragraph{Shell model values for $M_{GT}$ and $M_L$}

For a number of $\beta$ transitions in the $T = 1$ multiplets shell-model calculations of the matrix elements relevant 
for the different form factors have again been performed. These transitions correspond to the ones considered in an 
early version of this paper dating from before one of the authors retired and before the data set was significantly 
extended by a thorough inspection of the Nuclear Data Base \cite{ENSDF}. For each case, several calculations were 
performed with different effective shell-model interactions (Table~\ref{table:triplet-vs-Exp-Towner}).
In the p-shell ($p_{3/2}$ and $p_{1/2}$ orbits), three interactions of Cohen-Kurath (i.e. CK6162BME, CK8162BME, 
and CK816POT, and labelled as such in Table~\ref{table:triplet-vs-Exp-Towner}) 
\cite{Cohen1965}, and a more recent one from Brown-Warburton (PWBT) \cite{Warburton1992} were used. In the sd-shell 
($d_{5/2}$, $s_{1/2}$, and $d_{3/2}$ orbits), we used three versions of the universal sd-shell interaction, i.e. USD 
\cite{Wildenthal1984}, and two more recent ones, USD-A and USD-B \cite{Brown2006}). For A=16 and possibly A=18 a 
cross-shell interaction is needed. Here, the Millener-Kurath (MK) interaction \cite{Millener1975} that was designed 
to give 1p-1h matrix elements (relative to a closed-shell $^{16}$O) and including the $p_{3/2}$, $p_{1/2}$, $d_{5/2}$,
$s_{1/2}$, and $d_{3/2}$ orbits was used, together with USD for $sd$-shell interactions and one of Cohen-Kurath for $p$-shell 
interactions. That left the 2p-2h interaction to be determined. This was computed from the Millener-Kurath potential, 
but then adjusting its strength so the excited $0^+$ state in $A = 18$ (4$p$-2$h$ configuration) came at about the right 
energy. The accuracy of this interaction for A=16 and A=18 might therefore be inferior to the others.

\paragraph{Comparison between theoretical and experimental matrix elements}

Results from the shell model calculations of the $M_{GT}$ and $M_L$ matrix elements and the corresponding values for the $b$ form factor and the $b/Ac$ ratio (the latter being obtained from Eq.~(\ref{eq:bAc_impulse_approx})) for the fourteen $\beta$ transitions considered, are listed in Table~\ref{table:triplet-vs-Exp-Towner}. Comparing the theoretically calculated and experimentally obtained values for the Gamow-Teller matrix element, the Cohen-Kurath interaction seems to perform best for $A =$ 6 (indicated by the boldface font in the fifth column), with there not being a preferred choice for $A =$ 10, and with the PWBT interaction yielding results closest to experiment for $A =$ 12. As can be seen by comparing the last two columns in Table~\ref{table:triplet-vs-Exp-Towner}, for all these cases the theoretically predicted value for $|b/Ac|$ (listed in the last column) is significantly larger than the experimental one (last but one column). This is due to the systematically too low theoretical values for $|M_L|$ as is clear from a comparison of columns 6 and 7. The reason for this is not clear. For $A =$ 16 and 18, the MK interaction seems to perform quite well, reproducing the value for $|b/Ac|^{exp}$ always within about one standard deviation. For $A =$ 18 the USD interaction is further found to provide similar results to the MK interaction. For $A =$ 20 and 24 the USD-A interaction seems to reproduce the experimental Gamow-Teller matrix elements slightly better than the USD and USD-B interactions, whereas for $A =$ 30 the USD interaction provides best agreement. Note that even for the case of $^{24m}$Al, the large value of $|b/Ac|^{exp} = 17.5(32)$, being due to the combination of a very small Gamow-Teller matrix element and a relatively large M1 $\gamma$-decay width, $\Gamma_{M1}$, is reproduced quite well. 

\begin{figure}
	\centering
	\includegraphics[width=0.49\textwidth]{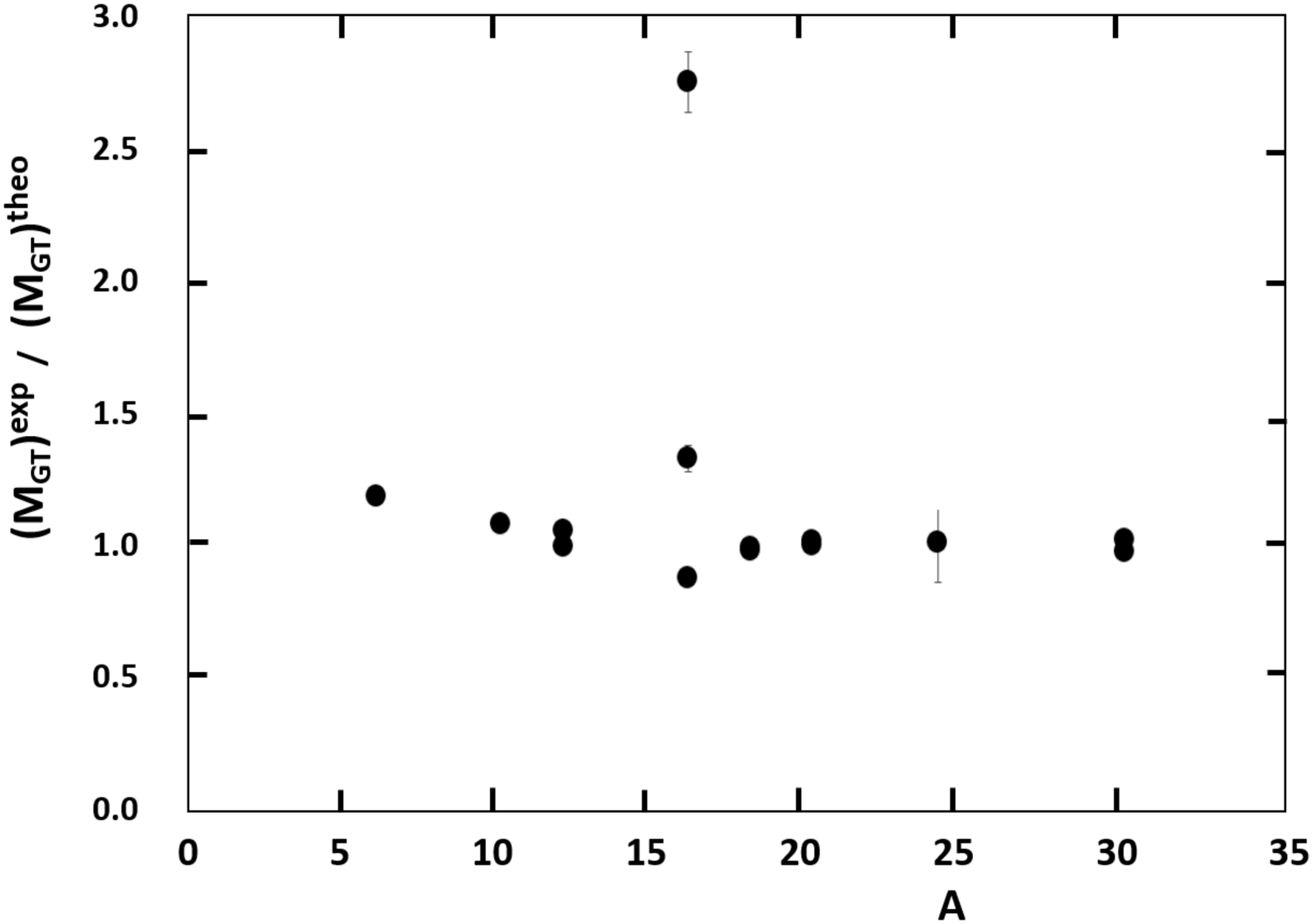}  
	\includegraphics[width=0.49\textwidth]{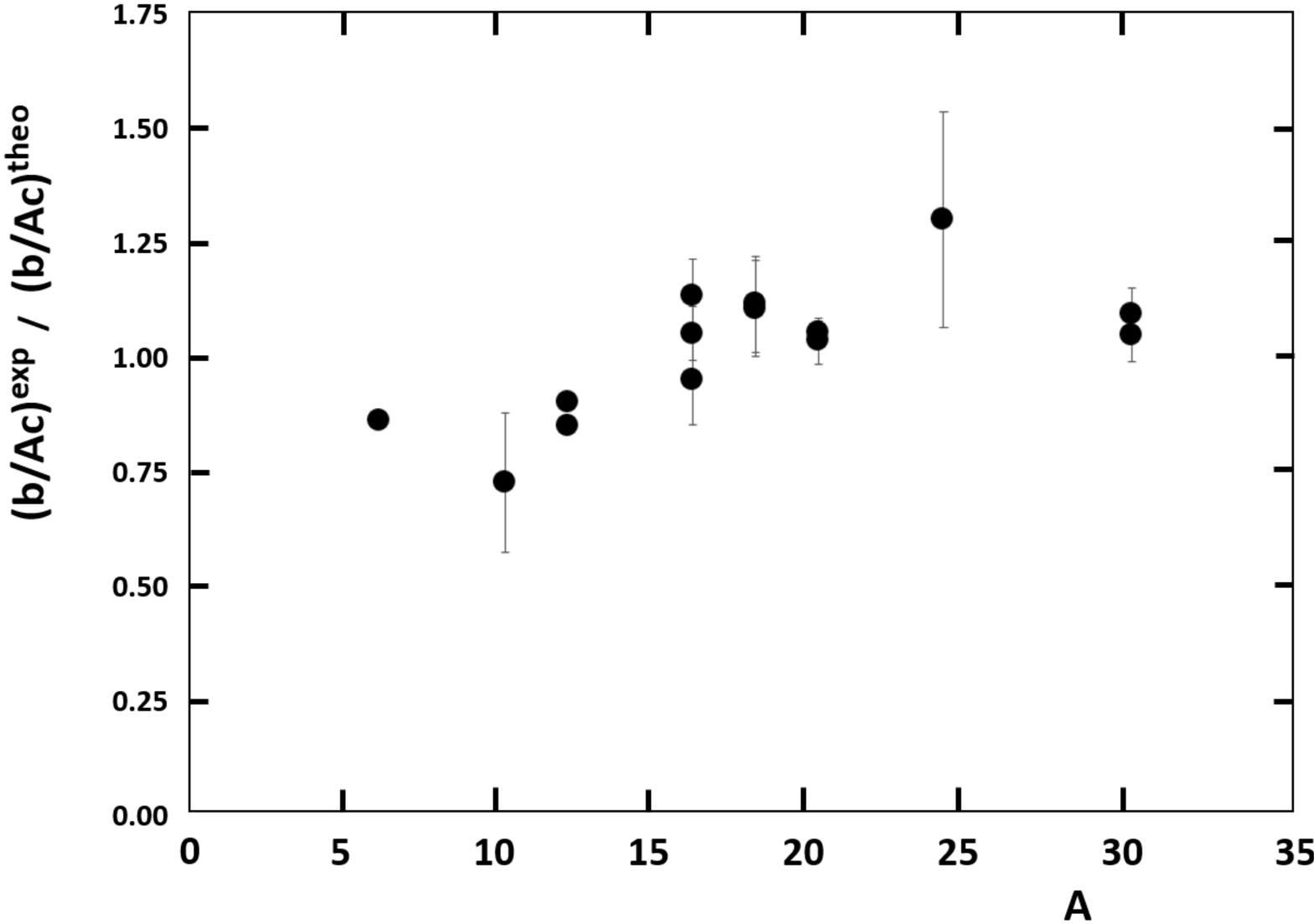}  
	\caption{Ratio of theoretically calculated and experimental Gamow-Teller matrix elements (top) and $b/Ac$ values (bottom) for the triplet $\beta$ transitions in Table~\ref{table:triplet-vs-Exp-Towner}. If not explicitly shown, error bars (based on the experimental $M_{GT}$ and $b/Ac$ values only) are smaller than the size of the symbols.} 
	\label{fig:T1-bAc-MGT-exp-theo}
\end{figure}

Fig.~\ref{fig:T1-bAc-MGT-exp-theo} shows the theoretical and calculated Gamow-Teller matrix elements to agree within about 20\% (see also columns 4 and 5 in Table~\ref{table:triplet-vs-Exp-Towner}). The unweighted average of their ratio yields $|M_{GT}|^{exp}/|M_{GT}|^{theo} = 1.16(47)$ (or 1.04(11) when omitting the $^{16}$N $(1^-) \rightarrow ^{16}$O $(2^-)$ transition for which the calculated Gamow-Teller matrix element is about a factor three lower than the experimental one). This average value being in agreement with unity indicates that also for these $\beta$ decays from $T =$ 1 states, with in general a more complex nuclear structure than the $T =$ 1/2 mirror nuclei, the shell model reproduces the Gamow-Teller matrix elements on average quite well. The unweighted average of the $|b/Ac|$ values for all 14 transitions (obtained from columns 10 and 11 in Table~\ref{table:triplet-vs-Exp-Towner} and using the $(b/Ac)^{theo}$ values corresponding to the $M_{GT}^{theo}$ values listed in boldface in column 5) is found to be $(b/Ac)^{exp}/(b/Ac)^{theo} = 1.01(15)$, again showing good overall agreement (within typically about 20\%) between the shell model calculations and experiment (Fig.~\ref{fig:T1-bAc-MGT-exp-theo}). However, as was mentioned already, the $M_L$ values calculated for the transitions with $A =$ 6 to 12 are systematically too low, while for the higher masses they do agree with experiment within about one standard deviation (see Table~\ref{table:triplet-vs-Exp-Towner}). Evidently the weak magnetism form factor, $b$, which depends on both the spin matrix element, $M_{GT}$, and the orbital current matrix element, $M_L$, (see Eq.~(\ref{eq:bAc_impulse_approx})) is nonetheless dominated by the spin terms in these isovector transitions.

\begin{table*}
	\centering
	\begin{ruledtabular}
		
		{ \footnotesize
		\caption{Comparison of the experimental values for the Gamow-Teller matrix element, $M_{GT}$ (the $c$ form factor), the orbital current matrix element, $M_L$, the form factor $|b|$, and the ratio $|b/Ac|$, with theoretical values calculated in the shell model using the interactions mentioned in the text (Sec.~\ref{matrix-elements-multiplets}), for fourteen $\beta$ transitions from $T = 1$ triplet states in Tables~\ref{table:tripletkernen-bAc-1} to ~\ref{table:higher-multiplets-bAc}. The theoretically calculated values of $|M_{GT}|^{theo}$ in column 5 that agree best with the experimental results, $|M_{GT}|^{exp}$, listed in column 4 are indicated in boldface. For the transition from $^{10}$C to $^{10}$B no preference can be given. Values for $|b/Ac|^{theo}$ were obtained from the values of $|M_{GT}|^{theo}$ and $|M_L|^{theo}$ using Eq.~(\ref{eq:bAc_impulse_approx}). The value $g_A$~=~1 was used throughout.} \label{table:triplet-vs-Exp-Towner}
			\begin{tabular}{c|c|c||c|c||c|c||c|c||c|c}
				
				$\beta$ decay & $J_{i}\rightarrow J_{f}$ & shell model & $|M_{GT}|^{exp}$ & $|M_{GT}|^{theo}$ & $|M_L|^{exp}$ & $|M_L|^{theo}$ &  $|b|^{exp}$ & $|b|^{theo}$ & $|b/Ac|^{exp}$ & $|b/Ac|^{theo}$ \\
				              &                          & interaction &                  &                   &               &                &              &              &                & \\
				
				\hline
				$^{6}$He $\xrightarrow{\beta^{-}}$ $^{6}$Li     &0$^{+}$ $\rightarrow$ 1$^{+}$& CK6162BME & 2.7802(20)& \bf 2.348 & 1.72(12) & 0.17 & 68.2(7) & 67.3  & 4.088(42) & \bf 4.78  \\
												             	&							  & PWBT      &           & 1.819     & 		& 0.003 &         & 51.4  &           & 4.71  \\[1mm]
				$^{10}$C $\xrightarrow{\beta^{+}}$ $^{10}$B$^1$   &0$^{+}$ $\rightarrow$ 1$^{+}$& CK6162BME & 2.3649(19)& 2.210     & 3.2(17) & 0.138 & 79(17)  & 102.6 & 3.34(72)  & 4.64  \\
				   								                &							  & CK8162BME &           & 2.141     & 		& 0.288 &         & 97.9  &           & 4.57  \\
												            	&							  & CK816POT  &           & 2.213     & 		& 0.241 &         & 101.7 &           & 4.60  \\
												            	&							  & PWBT      &           & 2.208     & 		& 0.238 &         & 106.3 &           & 4.81  \\[1mm]
				$^{12}$B $\xrightarrow{\beta^{-}}$ $^{12}$C     &1$^{+}$ $\rightarrow$ 0$^{+}$& CK6162BME &0.73852(55)& 0.558     &0.650(33)& 0.012 & 33.9(4) & 31.4  & 3.825(45) & 4.69  \\
								   				                &							  & CK8162BME &           & 0.573     & 		& 0.050 &         & 31.8  &           & 4.62  \\
												            	&							  & CK816POT  &           & 0.554     & 		& 0.065 &         & 30.5  &           & 4.59  \\
												            	&							  & PWBT      &           & \bf 0.699 & 		& 0.123 &         & 38.0  &           & \bf 4.53  \\[1mm]
				$^{12}$N $\xrightarrow{\beta^{+}}$ $^{12}$C     &1$^{+}$ $\rightarrow$ 0$^{+}$& CK6162BME &0.69613(64)& 0.558     &0.451(33)& 0.012 & 33.9(4) & 31.4  & 4.058(48) & 4.69  \\
								   				                &							  & CK8162BME &           & 0.573     & 		& 0.050 &         & 31.8  &           & 4.62  \\
												            	&							  & CK816POT  &           & 0.554     & 		& 0.065 &         & 30.5  &           & 4.59  \\
												            	&							  & PWBT      &           & \bf 0.699 & 		& 0.123 &         & 38.0  &           & \bf 4.53  \\
\hline				
                $^{16}$N $\xrightarrow{\beta^{-}}$ $^{16}$O$^1$   &2$^{-}$ $\rightarrow$ 3$^{-}$& MK        &0.4503(22) & \bf 0.338 & 0.11(11)& 0.155 & 32.1(18)& 23.0  & 4.46(25)  & \bf 4.25  \\[1mm]
				$^{16}$N $\xrightarrow{\beta^{-}}$ $^{16}$O$^2$   &2$^{-}$ $\rightarrow$ 1$^{-}$& MK        & 0.2184(91)&\bf 0.079 &0.165(89)& 0.041 & 13.8(13)& 5.3   & 3.95(41)  & \bf 4.18  \\[1mm]
				$^{16}$N $\xrightarrow{\beta^{-}}$ $^{16}$O$^3$   &2$^{-}$ $\rightarrow$ 2$^{-}$& MK        & 0.521(17) & \bf 0.595 & 1.49(30)& 1.181 &63.1(42) & 63.7  & 7.57(56)  & \bf 6.69  \\[1mm]
				$^{18}$Ne $\xrightarrow{\beta^{+}}$ $^{18}$F    &0$^{+}$ $\rightarrow$ 1$^{+}$& MK        & 2.2328(32)&\bf 2.252 & 2.3(12) & 1.060 & 230(22) & 209.8 & 5.72(55)  & \bf 5.18  \\
										            			&							  & USD       &           & 2.250     & 		& 1.128 &         & 210.9 &           & 5.21  \\[1mm]
				$^{18}$F $\xrightarrow{\beta^{+}}$ $^{18}$O     &1$^{+}$ $\rightarrow$ 0$^{+}$& MK        & 1.2778(22)&\bf 1.300 &1.36(71) & 0.612 & 230(22) & 121.2 & 5.77(55)  & \bf 5.18  \\
											            		&							  & USD       &           & 1.299     & 		& 0.651 &         & 121.8 &           & 5.21  \\
\hline
				$^{20}$F $\xrightarrow{\beta^{-}}$ $^{20}$Ne$^1$  &2$^{+}$ $\rightarrow$ 2$^{+}$& USD     &0.25493(45)& 0.246     & 0.91(11)& 0.851 & 42.1(21)& 40.2  & 8.26(41) & 8.16  \\
											            		&							  & USD-A     &           & \bf 0.251 & 		& 0.827 &         & 40.2  &         & \bf 8.00  \\
												            	&							  & USD-B     &           & 0.244     & 		& 0.868 &         & 40.3  &         & 8.27  \\[1mm]
				$^{20}$Na $\xrightarrow{\beta^{+}}$ $^{20}$Ne$^1$ &2$^{+}$ $\rightarrow$ 2$^{+}$& USD 	  &0.2506(19) & 0.246     & 0.93(11)& 0.851 & 42.1(21)& 40.2  & 8.40(22) & 8.16  \\
												            	&							  & USD-A     &           & \bf 0.251 & 		& 0.827 &         & 40.2  &         & \bf 8.00  \\
												            	&							  & USD-B     &           & 0.244     & 		& 0.868 &         & 40.3  &         & 8.27  \\[1mm]
				$^{24m}$Al $\xrightarrow{\beta^{+}}$ $^{24}$Mg  &1$^{+}$ $\rightarrow$ 0$^{+}$& USD	      & 0.101(15) & 0.139     & 1.29(38)  & 1.020 & 42.3(44)& 40.2 & 17.5(32)& 12.05 \\
												            	&							  & USD-A     &           & \bf 0.100 & 		& 0.873 &         & 32.2  &         & \bf 13.44  \\
												            	&							  & USD-B     &           & 0.199     & 		& 1.009 &         & 46.7  &         & 9.78 \\[1mm]
				$^{30}$S $\xrightarrow{\beta^{+}}$ $^{30}$P     &0$^{+}$ $\rightarrow$ 1$^{+}$& USD       &0.5429(98) & \bf 0.532 & 0.75(18)& 0.593 & 99.1(52)& 92.9  & 6.08(34) & \bf 5.82  \\
												            	&							  & USD-A     &           & 0.354     & 		& 0.637 &         & 69.0  &         & 6.51  \\
												            	&							  & USD-B     &           & 0.473     & 		& 0.606 &         & 84.9  &         & 5.99  \\[1mm]
				$^{30}$P $\xrightarrow{\beta^{+}}$ $^{30}$Si    &1$^{+}$ $\rightarrow$ 0$^{+}$& USD       &0.2998(41) & \bf 0.307 & 0.50(10)& 0.342 & 99.1(52)& 53.6  & 6.36(34)  & \bf 5.82  \\
												             	&							  & USD-A     &           & 0.204     & 		& 0.368 &         & 39.9  &         & 6.51  \\
												            	&							  & USD-B     &           & 0.273     & 		& 0.350 &         & 49.0  &         & 5.99  \\

			\end{tabular}%
		}
	\end{ruledtabular}

\end{table*}

\subsection{General discussion of the weak magnetism results}
\label{general discussion}

\subsubsection{Relation to nuclear structure}

\paragraph{$\beta$ transitions from the $T = 1/2$ mirror nuclei}

For the $T = 1/2$ mirror $\beta$ transitions for which CVC relates the weak magnetism form factor $b$ to the difference between the mirror pair magnetic moments, a clear sub-shell behaviour related to the single particle Schmidt values of the magnetic moments is observed, despite significant variation in the absolute values of $b$. This behaviour is further accentuated in the values of $b/Ac$, in part due to the slight decrease of the size of the Gamow-Teller form factor, $c$, with increasing mass (Fig.~\ref{fig:Mirror-bAc_vs_A} and Sec.~\ref{b and b/Ac for mirrors}). 

For the mirror nuclei up to mass $A = 45$ the $b$ and $c$ form factors, which are related to the $M_{GT}$ and $M_L$ matrix elements, were also calculated in the nuclear shell model (Sections~\ref{shell model-mirrors} and~\ref{Mirrors-Impulse approximation}). For this, well-established interactions that give excellent fits to experimental spectra were used. Calculations were limited to mass $A = 45$ as for higher masses the truncation of the shell model space required, due to calculation power limitations, was too important to yield reliable results. Good agreement between the experimental and theoretically calculated values for the form factor ratio $b/Ac$ was obtained (Fig.~\ref{fig:bAc-exp-vs-theo}). Further, the shell model calculated values for the $M_{GT}$ and $M_L$ matrix elements also show good correspondence with the experimental values extracted from the experimental $b/Ac$ ratio when using the impulse approximation (Figs.~\ref{fig:MGT-ratio-diff-exp-theo} and \ref{fig:ML-ratio-diff-exp-theo}). This is also the case for the special cases of $^{33}$Cl and $^{35}$Ar with large oblate deformation, although the shell model could not reproduce the sign of $b/Ac$ for $^{35}$Ar. However, an extreme single particle calculation in a deformed basis could reproduce the sign as well as the magnitude of $b/Ac$ for both cases within a factor of about 2.

\paragraph{$\beta$ transitions from $T = 1$ and $T = 3/2$ states}

\begin{figure} \centering
\includegraphics[width=0.49\textwidth]{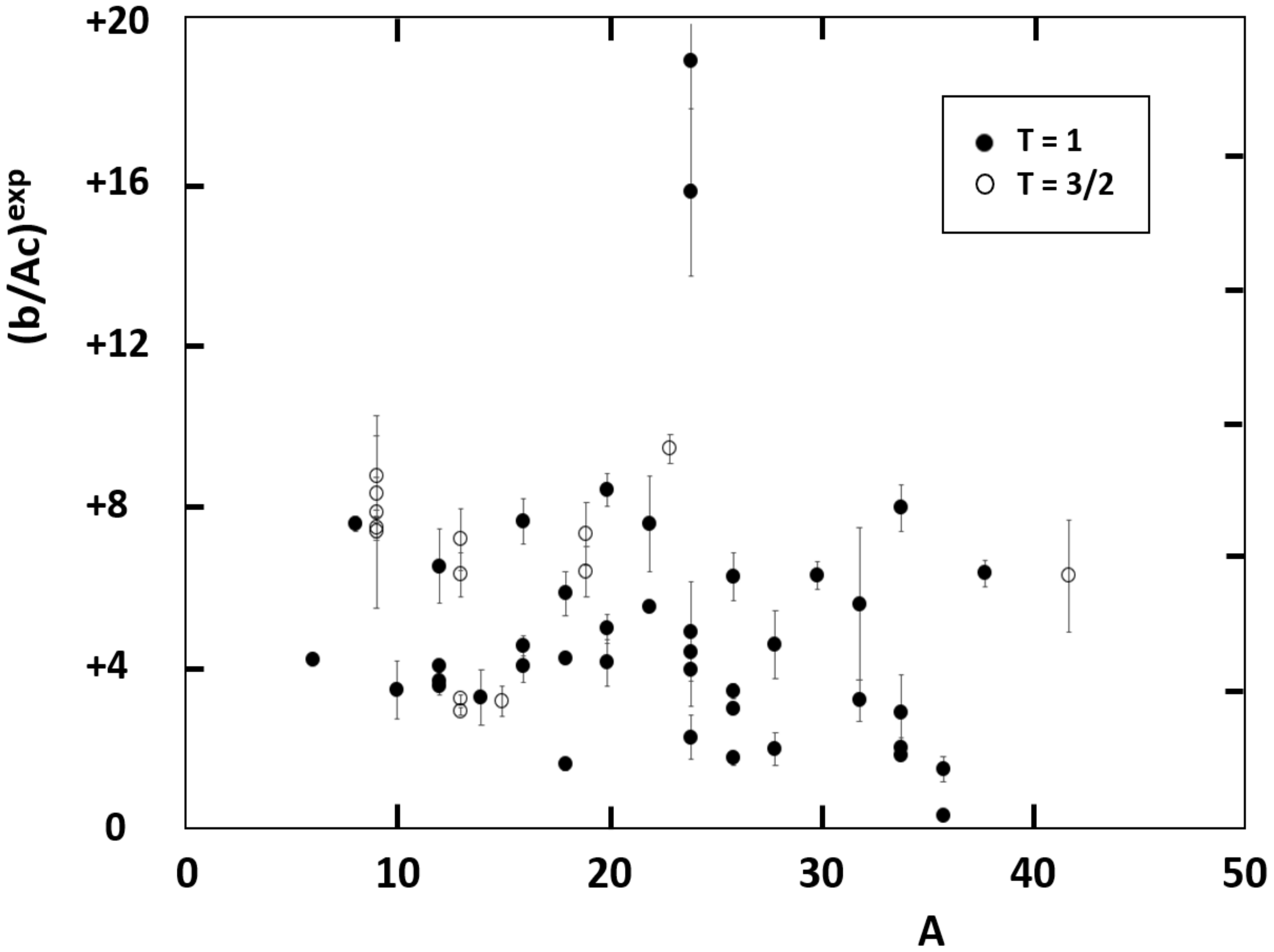}
\caption{Experimental $b/Ac$ values for the beta transitions from $T =$ 1  and $T =$ 3/2 states (Tables~\ref{table:tripletkernen-bAc-1} to \ref{table:higher-multiplets-bAc}).}
\label{fig:bAc-T=1-T=3/2}
\end{figure}

The $b/Ac$ values for the $\beta$ transitions from $T = 1$ and $T = 3/2$ states (Fig.~\ref{fig:bAc-T=1-T=3/2}) do not exhibit the clear single-particle related systematic that is obeyed by the $T = 1/2$ mirror transitions. Of course, many of the $T = 1$ and $T = 3/2$ transitions involve excited states with often strongly mixed configurations, implying the involvement of many shell-model orbitals, each with a fractional occupation. One may nevertheless hope that the transitions to ground states of even or even-even nuclei, and transitions to high-lying first-excited $2^+$ states, which in general all exhibit a rather 'simple' nuclear structure, do show some simple shell structure related systematic as well. Table~\ref{table:overview ground states} therefore lists all $T =$ 1/2, 1 and 3/2 transitions to ground states or high-lying first-excited $2^+$ states. As can be seen (but of course statistics is rather limited) the $T = 1$ and $T = 3/2$ transitions in this table indeed turn out to follow the trend of $|b/Ac|$ being typically large for the subshells with $j = l+1/2$ and smaller for $j = l-1/2$. The (unweighted) averages of $(b/Ac)^{exp}$ including all transitions in a given sub-shell turn out not to differ significantly from the values calculated for only the mirror $\beta$ transitions. Note, however, that many transitions from the higher isospin multiplets discussed here cannot be included in this comparison as they involve odd-odd nuclei with the odd proton in a $j = l+1/2$ orbital and the odd neutron in a $j = l-1/2$ orbital, or vice versa.

%

\begin{table*}
	
\begin{center}
\begin{ruledtabular}

{ \footnotesize
\caption{Overview of the values for $(b/Ac)^{exp}$ for all ground state-to-ground state and ground state-to-first-excited 2$^+$ state transitions with $A < 39$ discussed here ('mir' = mirror $\beta$ transition). Values of $(b/Ac)^{exp}$ for $\beta$ transitions from $T =$ 1 and $3/2$ states are in italics. The last column lists the average values for all transitions in a subshell. Because of widely differing error bars unweighted averages were calculated. When for a given mass value the $(b/Ac)^{exp}$ value is available for both the $\beta^-$ and $\beta^+$ transitions in a $T =$ 1 multiplet, sharing the same analog $\gamma$ transition (i.e. for $A =$ 18, 20, 24, 24, and 30), the average of both $(b/Ac)^{exp}$ values is used to calculate the unweighted average for the respective sub-shell. Averages in square brackets indicate the averages for the mirror transitions separately (from Table~\ref{table:spiegelkernen}). Note that the $f_{7/2}$ sub-shell is not shown here, but that the value $(b/Ac)^{exp} = 6.2(14)$ for the $^{42}$Sc $\rightarrow ^{42}$Sc$^1$ transition from a $T =$ 1 state (see Table~\ref{table:tripletkernen-bAc-3}) agrees with the $f_{7/2}$ sub-shell average of 8.1(10) for the mirror transitions (Table~\ref{table:spiegelkernen}).}
\label{table:overview ground states}

\begin{tabular}{c|c||c|c|c|c|c|c|c}
					
$\beta$ transition &    $A$ & $J_{i}^\pi\rightarrow J_{f}^\pi$ &  sub-shell  &  $j$  &    type  &    log~$ft$   & $(b/Ac)^{exp}$ & $(b/Ac)^{exp}$ \\
                   &        &                                  &            &       &          &               &                &  sub-shell avg.  \\
	               &        &                                  &            &       &          &               &                &  [only mir]     \\
					
\hline
					
H$\xrightarrow{\beta^{-}}$He &     3 &1/2$^{+}$ $\rightarrow$ 1/2$^{+}$&  $s_{1/2}$  & $l+1/2$ &   mir    & 3.0532(38)    &   4.2012(23) &  4.145(80)  \\
He$\xrightarrow{\beta^{-}}$Li &    6 &0$^{+}$ $\rightarrow$ 1$^{+}$    &             &         &  $T=1$   & 2.905753(61)  &   4.088(42)    & [4.2012(23)]  \\
\hline
Li$\xrightarrow{\beta^{-}}$Be\footnote[1]{The $A =$8 multiplet was not included in the analysis performed here since error bars on the log $ft$ values are difficult to estimate because broad levels of $^8$Be participate in the $\beta$ decay \cite{ENSDF, Barker1989}. The value for $(b/Ac)^{exp}$ listed here was obtained from $\beta$-ray angular distribution measurements \cite{Sumikama2011}.}  &    8  &2$^{+}$ $\rightarrow$ 2$^{+}$    &  $p_{3/2}$  & $l+1/2$ & $T=1$   &  5.72         & \it{7.5(2)}   &   7.04(69)    \\
B$\xrightarrow{\beta^{+}}$Be\footnotemark[1]  &    8  &2$^{+}$ $\rightarrow$ 2$^{+}$    &             &         & $T=1$   &  5.77         & \it{7.5(2)}   &  [6.2506(29)] \\
Li$\xrightarrow{\beta^{-}}$Be &    9  &3/2$^{-}$ $\rightarrow$ 3/2$^{-}$&             &         & $T=3/2$ &  5.3246(80)   & \it{7.31(23)} &               \\
C$\xrightarrow{\beta^{+}}$B &      9  &3/2$^{-}$ $\rightarrow$ 3/2$^{-}$&             &         & $T=3/2$ &  5.320(12)    & \it{7.42(25)} &             \\					
C$\xrightarrow{\beta^{+}}$B &      11 &3/2$^{+}$ $\rightarrow$ 3/2$^{+}$&             &         &   mir   & 3.59294(21)   & 6.2506(29)    &               \\   			
\hline
N$\xrightarrow{\beta^{+}}$C &      13 &1/2$^{+}$ $\rightarrow$ 1/2$^{+}$&  $p_{1/2}$  & $l-1/2$ &   mir   & 3.67037(45)   &  3.1712(23) &   [2.93(21)]  \\
O$\xrightarrow{\beta^{+}}$N &      15 &1/2$^{+}$ $\rightarrow$ 1/2$^{+}$&             &         &   mir   & 3.64368(58)   & 2.7557(20) &             \\					
\hline
F$\xrightarrow{\beta^{+}}$O &      17 &5/2$^{+}$ $\rightarrow$ 5/2$^{+}$&  $d_{5/2}$  & $l+1/2$ &   mir   & 3.36006(36)   & 6.0416(30) &   6.5(15)\footnote[2]{The values for the transitions from the $^{24m}$Na and $^{24m}$Al isomeric states were not included.}  \\
Ne$\xrightarrow{\beta^{+}}$F &     18 &0$^{+}$ $\rightarrow$ 1$^{+}$    &             &         &  $T=1$  &  3.0955(11)   & \it{5.72(55)}   &  [5.82(42)] \\
F$\xrightarrow{\beta^{+}}$O &      18 &1$^{+}$ $\rightarrow$ 0$^{+}$    &             &         &  $T=1$  &  3.5783(23)   & \it{5.77(55)}   &             \\					
Ne$\xrightarrow{\beta^{+}}$F &     19 &1/2$^{+}$ $\rightarrow$ 1/2$^{+}$&             &         &   mir   & 3.23591(25)   & 4.8807(20)  &          \\
F$\xrightarrow{\beta^{-}}$Ne$^1$ &   20 &2$^{+}$ $\rightarrow$ 2$^{+}$    &             &         &  $T=1$  &  4.98364(31) &  \it{8.28(41)}   &             \\					
Na$\xrightarrow{\beta^{+}}$Ne$^1$ &  20 &2$^{+}$ $\rightarrow$ 2$^{+}$    &             &         &  $T=1$  &  4.9972(64)) &  \it{8.40(42)}   &             \\					
			
Na$\xrightarrow{\beta^{+}}$Ne &    21 &3/2$^{+}$ $\rightarrow$ 3/2$^{+}$&             &         &   mir   & 3.60991(40)   & 5.5233(30) &          \\
					
Mg$\xrightarrow{\beta^{+}}$Na &    23 &3/2$^{+}$ $\rightarrow$ 3/2$^{+}$&             &         &   mir   & 3.67225(70)   & 6.4164(55) &          \\
Ne$\xrightarrow{\beta^{-}}$Na &    23 &5/2$^{+}$ $\rightarrow$ 3/2$^{+}$&             &         & $T=3/2$ &  5.2810(86)   &   9.38(37) &          \\					
Na$\xrightarrow{\beta^{-}}$Mg$^2$ &  24 &4$^{+}$ $\rightarrow$ 4$^{+}$    &             &         &  $T=1$  &  6.12609(46)  &  \it{4.8(12)}  &          \\
Al$\xrightarrow{\beta^{+}}$Mg$^2$ &  24 &4$^{+}$ $\rightarrow$ 4$^{+}$    &             &         &  $T=1$  &  6.129(57)    &  \it{4.8(13)}   &         \\					
Na(m)$\xrightarrow{\beta^{-}}$Mg & 24 &1$^{+}$ $\rightarrow$ 0$^{+}$    &             &         &  $T=1$  &  5.936(88)    &  \it{20.7(30)}   &          \\
Al(m)$\xrightarrow{\beta^{+}}$Mg & 24 &1$^{+}$ $\rightarrow$ 0$^{+}$    &             &         &  $T=1$  &  5.79(13)     &  \it{17.5(32)}   &         \\					
Al(m)$\xrightarrow{\beta^{+}}$Mg$^{1}$ & 24 &1$^{+}$ $\rightarrow$ 0$^{+}$ &           &         &  $T=1$  &  5.921(50)     &  15.8(21)   &         \\					
					
Al$\xrightarrow{\beta^{+}}$Mg &    25 &5/2$^{+}$ $\rightarrow$ 5/2$^{+}$&             &        &   mir    & 3.56972(37)   &  6.5875(38) &          \\
					
Si$\xrightarrow{\beta^{+}}$Al &    27 &5/2$^{+}$ $\rightarrow$ 5/2$^{+}$&             &        &   mir    & 3.61665(28)   & 7.6551(33) &          \\
\hline
P$\xrightarrow{\beta^{+}}$Si &     29 &1/2$^{+}$ $\rightarrow$ 1/2$^{+}$&  $s_{1/2}$ & $l+1/2$  &   mir   & 3.67802(72)   & 5.7631(51) &   5.76(46) \\
S$\xrightarrow{\beta^{+}}$P  &     30 &0$^{+}$ $\rightarrow$ 1$^{+}$    &             &         &  $T=1$  & 4.324(16)     & \it{6.08(34)}   & [5.43(21)]  \\
P$\xrightarrow{\beta^{+}}$Si &     30 &1$^{+}$ $\rightarrow$ 0$^{+}$    &             &         &  $T=1$  & 4.839(12)     & \it{6.36(34)}   &             \\					
					
S$\xrightarrow{\beta^{+}}$P &      31 &1/2$^{+}$ $\rightarrow$ 1/2$^{+}$&           &           &   mir   & 3.68127(48)   & 5.2988(33) &            \\
\hline

Cl$\xrightarrow{\beta^{+}}$S &     33 &3/2$^{+}$ $\rightarrow$ 3/2$^{+}$&  $d_{3/2}$ & $l-1/2$ &   mir   &  3.74802(74) & -0.4561(21) &   0.6(12)  \\
Ar$\xrightarrow{\beta^{+}}$Cl &    35 &3/2$^{+}$ $\rightarrow$ 3/2$^{+}$&           &          &   mir   &  3.75548(46) & -0.8684(11) &   [0.5(14)]  \\
K$\xrightarrow{\beta^{+}}$Ar$^1$ &   36 &1$^{+}$ $\rightarrow$ 2$^{+}$&           &             &   $T=1$  &  4.778(40)   &  0.189(42)  &          \\
K$\xrightarrow{\beta^{+}}$Ar$^2$ &   36 &1$^{+}$ $\rightarrow$ 0$^{+}$&           &             &   $T=1$  &  4.902(52)   &  1.37(31)   &          \\
K$\xrightarrow{\beta^{+}}$Ar &     37 &3/2$^{+}$ $\rightarrow$ 3/2$^{+}$&           &           &   mir  &  3.66383(52) &  2.104(10)  &          \\
Ca$\xrightarrow{\beta^{+}}$K &     39 &3/2$^{+}$ $\rightarrow$ 3/2$^{+}$&           &          &   mir   &  3.63180(61) &  1.2316(10) &          \\

\end{tabular}
}

\end{ruledtabular}
\end{center}
\end{table*}

\subsubsection{Averages and their role for $\beta$ decay experiments and the reactor neutrino anomaly}

\paragraph{$\beta$ decay experiments}

\begin{figure}[h!] \centering
	\includegraphics[width=0.49\textwidth]{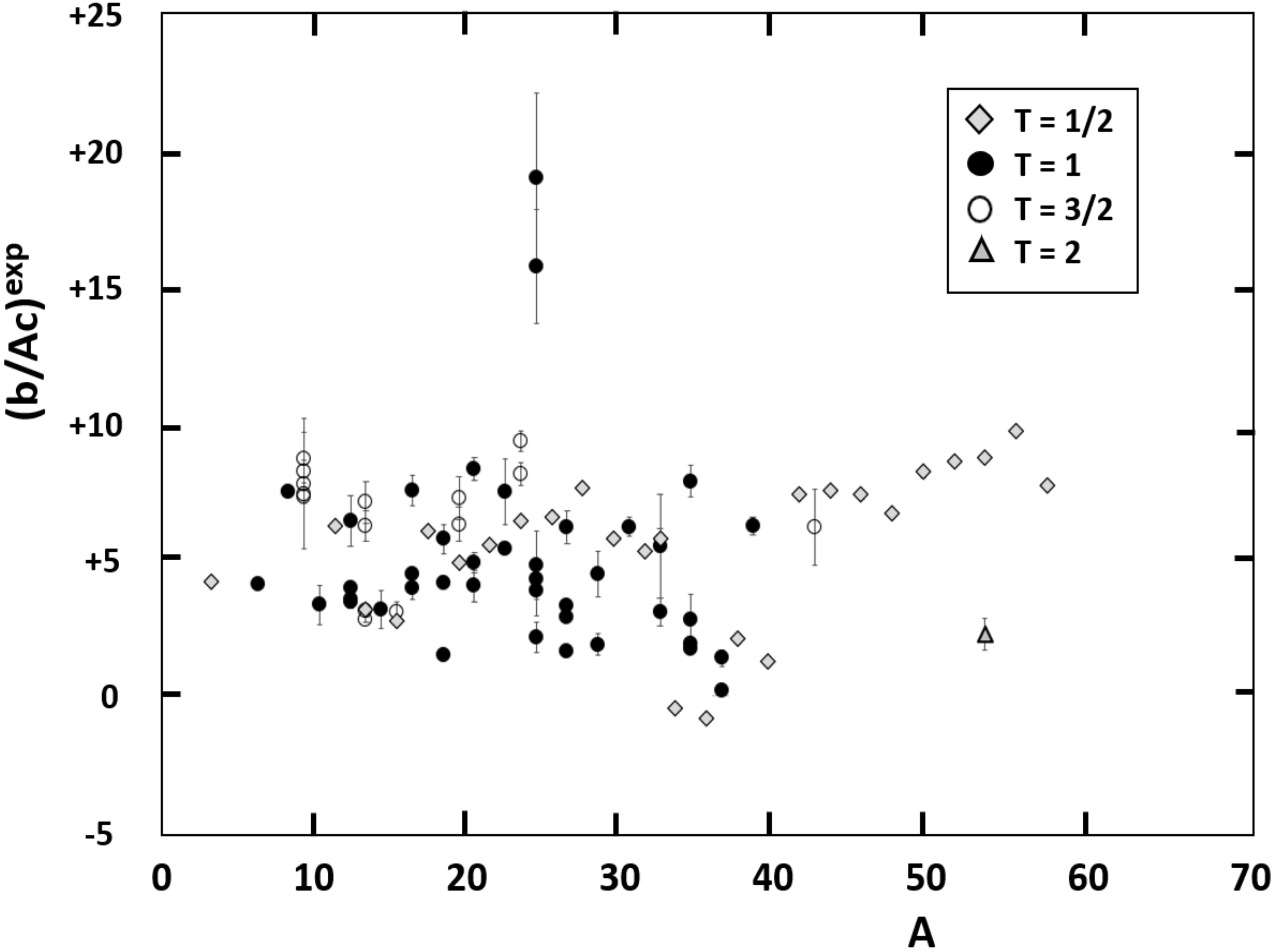}  
	\caption{Overview of the experimental $b/Ac$ values for all 85 $\beta$ transitions from isotopes with $A =$ 3 to 61 considered here (Tables~\ref{input-data-and-calculated-quantities} and \ref{table:tripletkernen-bAc-1} to \ref{table:higher-multiplets-bAc}). The unweighted average of all values shown is $b/Ac = 5.4 \pm 3.1$.} \label{fig:bAc-All-85-transitions}
\end{figure}

In Fig.~\ref{fig:bAc-All-85-transitions} the experimental values for $b/Ac$ for all $\beta$ transitions up to mass $A = 61$ considered here are shown on a single plot. For the $\beta$ transitions from $T =$ 1, 3/2 and 2 states the sign of $b/Ac$ is assumed positive (see Sec.~\ref{triplet-states}). As can be seen, all values are in the same range, with a maximum of about +10.0 and a minimum value of about -1.0. The transitions from $^{14}$O and $^{14}$C are again excluded, while for the nine cases where both the $\beta^-$ and $\beta^+$ transition in a $T = 1$ triplet yield a value for $|b/Ac|$ only the average of both values is used. The unweighted average of the 27 $T =$ 1/2 transitions with mass $A =$ 3 to 61, showing a clear subshell dependence related to their single-particle structure (Table~\ref{input-data-and-calculated-quantities} and Fig.~\ref{fig:Mirror-bAc_vs_A}), is $(b/Ac)^{exp} = 5.6 \pm 2.7$. This is in good agreement with the overall unweighted average of the 57 transitions from $T =$1, 3/2 and 2 states (Fig.~\ref{fig:bAc-T=1-T=3/2}), i.e. $|b/Ac|^{exp} = 5.3 \pm 3.2$. The average for the entire set of 84 $\beta$ transitions discussed here is $b/Ac = 5.4 \pm 3.1$. In all cases the one standard deviation error is rather large due to the spread in the individual values.
An overview of all averages and spreads of the weak magnetism term, $b/Ac$, and the matching matrix elements, $M_{GT}$ and $M_L$, mentioned in this text is given in Table \ref{table:overview_averages}.


%
\begin{table*}

  \centering
    \begin{ruledtabular}

    { \footnotesize
  \caption{Overview of $b/Ac$ averages and ratios of experimental and shell model calculated values for $M_{GT}$, $M_L$ and $b/Ac$ mentioned in the text (the relevant sections and figures are mentioned in the last column).}\label{table:overview_averages}
\begin{tabular}{c||c|c|c|c}

            	& quantity 		  				& mass region or isospin value 	& average  		& Section / Figure \\
\hline
$T = 1/2$   	& $(b/Ac)^{exp}$				&   $A =$ 3, 11-61; 27 transitions	&  5.6 $\pm$ 2.7 & Sec.~\ref{general discussion}       / Fig.~\ref{fig:Mirror-bAc_vs_A} \\

 mirror     	& $(b/Ac)^{exp}/(b/Ac)^{theo}$ 	&	$A =$ 3, 11-31, 37-45 		    &  0.96(11)\footnote{The values for $^{33}$Cl and $^{35}$Ar (i.e. resp. 2.22 and -5.63), which deviate significantly from all other values (see Table~\ref{matrixelementenspiegelkernen-bAc-vs-theo}), were not included.} 	& Sec.~\ref{shell model-mirrors} / Fig.~\ref{fig:bAc-exp-vs-theo} \\
 
transitions 	& $M_{GT}^{exp}/M_{GT}^{theo}$  &   $A =$ 3, 11-45 			    	&  0.97(8) 	& Sec.~\ref{Mirrors-Impulse approximation} / Fig.~\ref{fig:MGT-ratio-diff-exp-theo} \\

            	& $M_{L}^{exp}/M_{L}^{theo}$ 	&   $A =$ 11-29, 33-45	     	    &  1.04(34) 	& Sec.~\ref{Mirrors-Impulse approximation} / Fig.~\ref{fig:ML-ratio-diff-exp-theo} \\
\hline

$T = 1, 3/2, 2$ & $(b/Ac)^{exp}$, $T = 1$    	&   $T = 1$; $A =$ 6-42; 42 transitions		&  4.9 $\pm$ 3.5 	  & Sec.~\ref{experimental-values-multiplets} / Fig.~\ref{fig:bAc-T=1-T=3/2}\\

transitions     & $(b/Ac)^{exp}$, $T = 3/2$		&   $T = 3/2$; $A =$ 9-53; 15 transitions	&  6.3 $\pm$ 2.4 	  & Sec.~\ref{experimental-values-multiplets} / Fig.~\ref{fig:bAc-T=1-T=3/2}\\

            	& $(b/Ac)^{exp}$, $T = 1, 3/2, 2$ & $T =$ 1, 3/2, and 2; $A =$ 6-53; 58 transitions&  5.3 $\pm$ 3.2 	& Sec.~\ref{general discussion} / Fig.~\ref{fig:bAc-T=1-T=3/2} \\
            	
            	& $(b/Ac)^{exp}/(b/Ac)^{theo}$	& 	$T = 1$; $A =$ 6-30; 14 transitions		&  1.01(15)		& Sec.~\ref{matrix-elements-multiplets} / Fig.~\ref{fig:T1-bAc-MGT-exp-theo}  \\
            	
            	& $M_{GT}^{exp}/M_{GT}^{theo}$  &   $T = 1$; $A =$ 6-30; 13 transitions		&  1.04(11)\footnote{Including the very large value of 2.76 for the 2$^{-}$ $\rightarrow$ 1$^{-}$ transition $^{16}$N $\xrightarrow{\beta^{-}}$ $^{16}$O$^2$ (see Table~\ref{table:triplet-vs-Exp-Towner}), which was omitted here, shifts the average to 1.16(47).}  & Sec.~\ref{matrix-elements-multiplets} / Fig.~\ref{fig:T1-bAc-MGT-exp-theo} \\
\hline

all         	& $(b/Ac)^{exp}$				&$T =$ 1/2, 1, 3/2, and 2; $A =$ 3-61; 85 transitions&  5.4 $\pm$ 3.1	& Sec.~\ref{general discussion} /  Fig.~\ref{fig:bAc-All-85-transitions} \\

\end{tabular}

   }
    \end{ruledtabular}
\end{table*}
%
%
%


The weak-magnetism form factor provides in most cases the major recoil correction to the $\beta$-spectrum shape and $\beta$-correlation coefficients for allowed transitions with 'normal' strength (\textit{i.e} non-retarded transitions with $\log ft$ values less than about 6.7). This is true for both analog transitions, i.e. between members of an isospin multiplet such as the $\beta$ transitions within the $T = 1/2$ isospin doublets discussed here, and non-analog transitions, as e.g. the ones from states with isospin $T =$1, 3/2 and 2. The above analysis of weak magnetism form factors for the $T = 1/2$ mirror $\beta$ transitions allows taking this form factor explicitly into account in the analysis of experimental data, as was done in the determination of the $\beta \nu$-correlation with $^{21}$Na \cite{Vetter2008} and of the $\beta$-asymmetry parameter for $^{37}$K \cite{Fenker2018}. In this way higher precision and sensitivity can be reached when searching for new physics or when determining the $V_{ud}$ quark mixing matrix element in experiments in $\beta$-decay correlation measurements with the mirror nuclei. 

The observation that for non-analog allowed $\beta$ transitions the $b/Ac$ value is approximately nucleus independent, within certain limits, enables estimating its effect on $\beta$ decay correlation results for allowed transitions for which no experimental nor theoretical data leading to $b/Ac$ are readily available. This was done already in the analysis of e.g. the $\beta$-asymmetry parameters in the decay of $^{114}$In \cite{Wauters2009a} and the recoil-asymmetry parameter in the decay of $^{80}$Rb \cite{Pitcairn2009}. The present analysis now offers for non-analog transitions, for which one cannot rely on CVC, the value of $b/Ac = 5.4 ~ \pm ~ 3.1$ obtained as an average over a large number of $\beta$ transitions for nuclei with masses ranging from $A =$ 3 to 61 (Table~\ref{table:overview_averages}).

\paragraph{Reactor neutrino anomaly}

So far, an average value for the weak magnetism contribution has been used in detailed calculations of the cumulative electron and antineutrino spectra emerging from a nuclear reactor. In one of the original works \cite{Huber2011}, it was noted that the variation in $b/Ac$ values for all known transitions causes substantial uncertainty in these spectra. Instead, a subset all known transitions was used with a fairly constrained average value of \cite{Huber2011} ($M_n$ is the nucleon mass)
\begin{equation}
b/Ac = \frac{3}{4} M_n (0.5 \pm 0.5)\% MeV^{-1} = 3.5 \pm 3.5~ .
\end{equation}
\noindent The downside to this approach is twofold, as firstly it is not clear whether the chosen subset is a reasonable representation of the behaviour of nuclear transitions at fission-fragment masses, while secondly each of those transitions has an unequal weighting. Specifically, transitions with a large $b/Ac$ value might have a large production rate whereas one with a low $b/Ac$ value might only be barely populated, or vice versa, thereby invalidating a straight average. Finally, it has been noted that the inclusion of recoil-order corrections to the allowed $\beta$ spectrum shape has been oversimplified \cite{Hayen2019, Hayen2019b}, with the traditional calculations lacking additional energy-dependent terms related to weak magnetism and an absence of induced tensor currents. As these transitions are non-analog, the latter is non-zero also in the Standard Model without second-class currents. Our results presented here serve both as as a demonstration of the spread in $b/Ac$ values much higher in mass than previously available, and as a motivation for experimental measurement of its value in the fission-fragment mass region.

\subsubsection{Higher-order form factors $d, f, g$, etc.}

In contrast to the transitions between analog states discussed in the previous two paragraphs, for hindered Gamow-Teller transitions (\textit{i.e.} with log$ft$ values of about 6.7 and larger) between non-analog states, matrix elements of rank-1 spherical tensor operators with an $M1$ character \cite{Calaprice1977,Holstein1974}, {\it i.e.} the $b$, $c$ and $d$ form factors, are usually suppressed, whereas matrix elements of rank-2 spherical tensor operators with $E2$ character \cite{Calaprice1977} (as in e.g. the $f$ and $g$ form factors) are not. As a consequence, the $f/Ac$ and $g/A^2c$ recoil term contributions \cite{Holstein1974} to $\beta$-decay observables can become typically as large as or even larger than the $b/Ac$ contribution from the weak-magnetism term (see e.g. the cases of $^{60}$Co \cite{Wauters2010} and $^{67}$Cu \cite{Soti2014}). It was shown in Ref. \cite{Wauters2010} that whereas for such hindered $\beta$ transitions theoretical calculations often have difficulties to reproduce the experimental value for the Gamow-Teller matrix element, combining the value for this matrix element (and so the $c$ form factor) obtained from the experimental $ft$ value (Eq.~\eqref{f_GT}) with theoretically calculated matrix elements for the other form factors and using the thus obtained ratios for e.g $f/Ac$ and $g/A^2 c$, leads to satisfactory results.

Finally, given the reasonably good performance of the shell model calculations as to the $b$ form factor (i.e. the $M_{GT}$ and $M_L$ matrix elements) for the 14 $\beta$ transitions from $T =$ 1 states  discussed here in some detail, it may be of interest for experiments involving these (ground state-to-ground state) transitions to also know the values calculated for the other form factors. These are therefore listed in Table \ref{table:form-factors-T=1}. Of special interest is the tensor form factor ratio $d/Ac$ (assuming no second-class currents), which is non-zero for Gamow-Teller transitions and is second in line, after the weak magnetism form factor, as to its effect in $\beta$ spectrum shape measurements and angular correlation measurements. The calculated values for $d/Ac$ (listed in column 7 of Table~\ref{table:form-factors-T=1}) turn out to be a factor of about two or more smaller than the corresponding values of $b/Ac$, except for the already mentioned transition $^{16}$N $(1^-) \rightarrow ^{16}$O $(2^-)$ for which the calculated Gamow-Teller matrix element is about a factor of three lower than experiment. 

For the $A = 12$ and $A  = 20$ transitions the calculated values can be compared to experimental results from the alignment correlation terms in the $\beta$-ray angular distributions of the purely spin aligned mirror pairs $^{12}$B-$^{12}$N \cite{Minamisono2001} and $^{20}$F-$^{20}$Na \cite{Minamisono2011}. While very good agreement was obtained for the $b/Ac$ form factor (Table~\ref{table:triplet-vs-Exp-Towner}), the values calculated here for $d/Ac$, $g/A^2c$, $j_2/A^2c$, and $j_3/A^2c$ for the $A =$ 12 and $A =$ 20 transitions turn out to reproduce the experimentally obtained values within a factor of two to three, as can be seen in Table~\ref{table:form-factors-A=12and20}. 

Finally it is of interest to note here the renewed interest in precision $\beta$-spectrum shape measurements for nuclear decays, using a variety of techniques \cite{Knutson2011, Mougeot2012, Huyan2016, Huyan2018, Huyan2018a, Lojek2009, Lojek2015, Broussard2019, Perkowski2018}, and which include two of the cases considered here, i.e. $^6$He $\rightarrow ^6$Li \cite{Huyan2018, Huyan2018a} and $^{20}$F $\rightarrow ^{20}$Ne$^1$ (Fig.~\ref{fig13:triplets-16-22}) \cite{Naviliat2021}. Such measurements could possibly contribute to extending the experimental knowledge on the $d$ form factor as well.

\begin{table*}
	\centering
	\begin{ruledtabular}
	
		{ \footnotesize
			\caption{Form factors calculated in the shell model for the 14 $T = 1$ $\beta$ transitions in Table~\ref{table:triplet-vs-Exp-Towner}. The shell model interactions that performed best in reproducing the experimentally obtained Gamow-Teller matrix elements (see Table~\ref{table:triplet-vs-Exp-Towner}) are indicated in boldface. Note that the shell model sign for the individual form factors is arbitrary but that this arbitrariness disappears in the form factor ratios.} \label{table:form-factors-T=1}	
			
			\begin{tabular}{c c c|c c c c c c c c c c}
				
$\beta$ decay & $J_{i}\rightarrow J_{f}$ & shell model & $c = c_1$ & $c_2$ & $b/Ac$ & $d/Ac$ & $f/Ac$ & $g/A^{2}c$ &  $j_1/A^{2}c$ & $j_2/A^{2}c$ & $j_3/A^{2}c$ & $h/A^{2}c$ \\
      &         & interaction &           &       &        &        &        &        &       &         &       &           \\
				
				\hline
$^{6}$He $\xrightarrow{\beta^{-}}$ $^{6}$Li  &0$^{+}$ $\rightarrow$ 1$^{+}$& \bf CK6162BME & $-2.348$ & $-2.241$ & 4.777 & $-0.028$ & 0.0 & 0.0 & 14.43 & 0.0  & 0.0  & 195.2 \\
						            	     &							   &     PWBT      & $-1.819$ & $-1.482$ & 4.710 & $0.330$  & 0.0 & 0.0 & 54.52 & 0.0  & 0.0  & 232.1 \\
$^{10}$C $\xrightarrow{\beta^{+}}$ $^{10}B^1$&0$^{+}$ $\rightarrow$ 1$^{+}$&     CK6162BME & $+2.210$  & $+2.346$  & 4.643 & $-0.498$ & 0.0 & 0.0 & 15.29 & 0.0  & 0.0  & 195.5 \\
							                 &							   &     CK8162BME & $+2.141$  & $+2.191$  & 4.573 & $-0.444$ & 0.0 & 0.0 & 26.30 & 0.0  & 0.0  & 206.0   \\
							             	 &							   &     CK816POT  & $+2.213$  & $+2.306$  & 4.596 & $-0.443$ & 0.0 & 0.0 & 20.74 & 0.0  & 0.0  & 200.6   \\
								             &							   &     PWBT     & $+2.208$  & $+2.322$  & 4.814 & $-0.385$ & 0.0 & 0.0 & 18.16 & 0.0  & 0.0  & 198.4    \\
$^{12}$B $\xrightarrow{\beta^{-}}$ $^{12}$C &1$^{+}$ $\rightarrow$ 0$^{+}$ &     CK6162BME & $-0.558$ & $-0.569$ & 4.689 & $3.554$ & 33.9(4) & 31.4 & 40.07 & 0.0  & 0.0  & 219.0  \\
							            	 &							   &     CK8162BME & $+0.573$  & $+0.589$  & 4.625 & $3.549$ & 0.0 & 0.0  & 38.42 & 0.0 & 0.0   & 218.1    \\
								             &							   &     CK816POT  & $-0.554$ & $-0.574$ & 4.588 & $3.715$ & 0.0 & 0.0  & 35.73 & 0.0 & 0.0   & 215.6  \\
								             &							   & \bf PWBT      & $+0.699$  & $+0.704$  & 4.530 & $2.861$ & 0.0 & 0.0  & 43.81 & 0.0 & 0.0   & 222.5    \\ 
$^{12}$N $\xrightarrow{\beta^{+}}$ $^{12}$C  &1$^{+}$ $\rightarrow$ 0$^{+}$&     CK6162BME & $+0.558$  & $+0.569$  & 4.689 & $3.554$ & 0.0 & 0.0  & 40.07 & 0.0 & 0.0   & 219.0    \\
								             &							   &     CK8162BME & $-0.573$ & $-0.589$ & 4.625 & $3.549$ & 0.0 & 0.0  & 38.42 & 0.0 & 0.0   & 218.2\\
								             &							   &     CK816POT  & $+0.554$ & $+0.574$  & 4.588 & $3.715$ & 0.0 & 0.0  & 35.73 & 0.0 & 0.0   & 215.6  \\
								             &							   & \bf PWBT      & $-0.699$ & $-0.704$ & 4.530 & $2.861$ & 0.0 & 0.0  & 43.81 & 0.0 & 0.0   & 222.5    \\
\hline
$^{16}$N $\xrightarrow{\beta^{-}}$ $^{16}$O$^1$&2$^{-}$ $\rightarrow$ 3$^{-}$ & \bf MK     & $-0.338$ & $-0.904$ & 4.253  & $1.313$  & $-0.617$ & 196.5  & $-317.8$ & $46.62$  & $257.7$ & $-120.2$ \\[1mm]
$^{16}$N $\xrightarrow{\beta^{-}}$ $^{16}$O$^2$&2$^{-}$ $\rightarrow$ 1$^{-}$ & \bf MK     & $-0.079$ & $-0.251$ & 4.193  & $-6.013$ & $-0.142$ & 57.36  & $100.9$  & $-204.2$ & $786.2$ & $275.9$ \\[1mm]
$^{16}$N $\xrightarrow{\beta^{-}}$ $^{16}$O$^3$&2$^{-}$ $\rightarrow$ 2$^{-}$ & \bf MK     & $-0.595$ & $-1.464$ & 6.691  & $0.063$  & $-0.193$ & 143.8  & $-160.2$ & $-14.18$ & $188.4$ & $29.08$ \\[1mm]

$^{18}$Ne $\xrightarrow{\beta^{+}}$ $^{18}$F &0$^{+}$ $\rightarrow$ 1$^{+}$  & \bf MK      & $-2.252$ & $-3.941$ & 5.176  & $-0.521$ & 0.0  & 0.0  & 11.44  & 0.0  & 0.0 & 191.9  \\
								        	 &							     &     USD     & $-2.250$ & $-3.952$ & 5.207  & $-0.509$ & 0.0  & 0.0  & 9.026  & 0.0  & 0.0 & 189.3 \\ 
$^{18}$F $\xrightarrow{\beta^{+}}$ $^{18}$O  &1$^{+}$ $\rightarrow$ 0$^{+}$  & \bf MK      & $-1.300$ & $-2.276$ & 5.179  & $0.521$  & 0.0  & 0.0  & 11.44  & 0.0  & 0.0 & 191.8 \\
									         &							     &     USD     & $-1.299$ & $-2.282$ & 5.209  & $0.509$  & 0.0  & 0.0  & 9.029  & 0.0  & 0.0 & 189.6    \\
\hline
$^{20}$F $\xrightarrow{\beta^{-}}$ $^{20}$Ne$^1$ &2$^{+}$ $\rightarrow$ 2$^{+}$ & USD       & $+0.246$  & $+0.602$  & 8.171  & $3.211$  & 0.035  & $-9.187$ & $-174.8$ & 310.0  & $-605.7$ & $15.35$ \\
									         &							     & \bf USD-A    & $-0.251$ & $-0.607$ & 8.008  & $3.068$  & 0.124  & $-32.07$ & $-166.3$ & 296.8  & $-603.6$ & $23.51$ \\
									         &							     &    USD-B     & $-0.244$ & $-0.595$ & 8.258  & $3.238$  & 0.091  & $-23.67$ & $-174.2$ & 302.3  & $-616.8$ & $15.98$ \\
$^{20}$Na $\xrightarrow{\beta^{+}}$ $^{20}$Ne$^1$ &2$^{+}$ $\rightarrow$ 2$^{+}$ & USD      & $-0.246$ & $-0.602$ & 8.171  & $3.211$  & 0.070  & $-9.187$ & $-174.8$ & 310.0  & $-605.7$ & $15.34$ \\
									         &							     & \bf USD-A    & $+0.251$  & $+0.607$  & 8.008  & $3.068$  & 0.246  & $-32.07$ & $-166.3$ & 296.8  & $-603.6$ & $23.51$ \\
									         &							     &     USD-B    & $+0.244$  & $+0.595$  & 8.258  & $3.238$  & 0.181  & $-23.67$ & $-174.2$ & 302.3  & $-616.8$ & $15.98$ \\
$^{24m}$Al $\xrightarrow{\beta^{+}}$ $^{24}$Mg &1$^{+}$ $\rightarrow$ 0$^{+}$&    USD	    & $+0.139$  & $+0.401$  & 12.05  & $6.385$  & 42.3(44)& $40.2$  & $-276.0$ & 0.0  & $0.0$  & $-81.31$ \\
									         &							     & \bf USD-A    & $-0.100$ & $-0.332$ & 13.42  & $8.500$  & 0.0  & $0.0$  & $-401.0$ & 0.0  & $0.0$  & $-199.7$    \\
									         &							     &     USD-B    & $+0.199$  & $+0.504$  & 9.778  & $5.465$  & 0.0  & $0.0$  & $-175.4$ & 0.0  & $0.0$  & $14.57$ \\
$^{30}$S $\xrightarrow{\beta^{+}}$ $^{30}$P  &0$^{+}$ $\rightarrow$ 1$^{+}$  & \bf USD      & $+0.532$  & $+0.919$  & 5.821  & $-1.065$ & 0.0  & 0.0  & 87.72  & 0.0  & 0.0  & 263.2 \\
									         &							     &     USD-A    & $-0.354$ & $-0.580$ & 6.497  & $-1.243$ & 0.0  & 0.0  & 113.0  & 0.0  & 0.0  & 287.8 \\
									         &							     &     USD-B    & $-0.473$ & $-0.768$ & 5.983  & $-1.029$ & 0.0  & 0.0  & 117.7  & 0.0  & 0.0  & 293.6  \\
$^{30}$P $\xrightarrow{\beta^{+}}$ $^{30}$Si &1$^{+}$ $\rightarrow$ 0$^{+}$   & \bf USD     & $+0.307$  & $+0.530$  & 5.820  & $1.064$  & 0.0  & 0.0  & 87.59  & 0.0  & 0.0  & 264.2 \\
									         &							     &      USD-A   & $-0.204$ & $-0.335$ & 6.520  & $1.242$  & 0.0  & 0.0  & 113.3  & 0.0  & 0.0  & 288.7 \\
									         &							     &      USD-B   & $-0.273$ & $-0.443$ & 5.983  & $1.038$  & 0.0  & 0.0  & 117.6  & 0.0  & 0.0  & 292.6 \\
				
	\end{tabular}
}
	\end{ruledtabular}
	
\end{table*}

\begin{table*}
	\centering
	\begin{ruledtabular}
	
		{ \footnotesize
			\caption{Form factors calculated in the shell model for the $T = 1$ $\beta$ transitions of $^{12}$B/$^{12}$N and $^{20}$F/$^{20}$Na (from Table~\ref{table:triplet-vs-Exp-Towner}), compared with experimental results from $\beta$-ray angular distribution measurements \cite{Minamisono2001, Minamisono2011}. For $A =$ 12 the CK816POT and PWBT shell model interaction values are listed. For $A =$ 20 the USD, USD-A and USD-B interactions all give similar results (see Table~\ref{table:triplet-vs-Exp-Towner}) so only results for USD-A are listed here. For $A =$ 20 the form factors were also calculated in Ref.~\cite{Calaprice1977} (labeled CCW), which we show here for comparison.}
			\label{table:form-factors-A=12and20}	
			
			\begin{tabular}{c c c|c c c c c}
				
$\beta$ decay & $J_{i}\rightarrow J_{f}$ & exp. value or & $d/Ac$ & $f/Ac$ &     $g/A^{2}c$ &   $j_2/A^{2}c$ & $j_3/A^{2}c$ \\
              &                          & shell model   &        &        &                &                &             \\
              &                          & interaction   &        &        &                &                &              \\
				
				\hline
$^{12}$B/$^{12}$N $\xrightarrow{\beta^{-}/\beta^{+}}$ $^{12}$C &  1$^{+}$ $\rightarrow$ 0$^{+}$ &  exp\footnote{From \cite{Minamisono1998, Minamisono2001}.}   & $+4.9(10)$  & &  &  &   \\
                                                       &                                        & CK816POT & $3.715$  &  &  &   &   \\ 
                                                       &                                        &  PWBT    & $2.861$  &  &  &   &   \\ 
\hline
$^{20}$F/$^{20}$Na $\xrightarrow{\beta^{-}/\beta^{+}}$ $^{20}$Ne$^1$ & 2$^{+}$ $\rightarrow$ 2$^{+}$  & exp\footnote{From \cite{Minamisono2011} ($d, j_2$ and $j_3$) and \cite{Elmbt1987} ($f$ and $g$).}
& $8.00(73)$ & $+0.31(14)$ &  $-54(25)$ & $-21(130)$  & $-1273(211)$  \\
                                                       &                                        & USD-A  & $3.068$    & $+0.124$ ($^{20}$F)  & $-32.1$   & $296.8$     & $-603.6$  \\
                                                       &                                        &        &            & $+0.246$ ($^{20}$Na) &           &             &         \\
	                                                   &                                        & CCW    & $3.51 $    & $-0.080$ ($^{20}$F)  & $-20.6$   & $333.0$     & $-663.5$  \\
	                                                   &                                        &        &            & $+0.160$ ($^{20}$Na)  &           &             &         \\
	\end{tabular}
}
	\end{ruledtabular}
	
\end{table*}


\section{Conclusion}

In the first section of this paper the input data for the corrected $\mathcal{F}$t values of the isospin $T = 1/2$ mirror $\beta$ transitions up to $A = 75$ were updated and combined with new and/or extended calculations of the required correction factors, finally leading to new and usually also more precise $\mathcal{F}$t values. The latter have been used already to extract, in combination with existing results from $\beta$-correlation measurements for several of the mirror $\beta$ transitions, a new value for the $V_{ud}$ quark-mixing matrix element for the $T = 1/2$ mirror $\beta$ decays \cite{Hayen2021}. here we have extracted from these $\mathcal{F}$t values the Gamow-Teller weak form factor, $c$. Combining this with an analysis of the magnetic moments of the $T = 1/2$ mirror nuclei, gave access to the weak magnetism form factor, $b$, and allowed determining the 'normalized' weak magnetism form factor ratios, $b/Ac$, for the mirror $\beta$ transitions. Taking into account the effect of weak magnetism in correlation measurements performed with these mirror nuclei, will enhance the sensitivity of such measurements to probe possible new physics phenomena, such as e.g. searches for scalar or tensor type weak currents or tests of parity or time-reversal violation. 

In order to get a broader picture on the size of the weak magnetism form factor, e.g. in view of correlation measurements involving $\beta$ transitions between non-analog states or for inclusion in calculations related to the reactor neutrino anomaly, a much broader survey was performed as well. This survey used existing experimental data for analog $\beta$ and $\gamma$ transitions from common isobaric multiplets. This resulted in values for the Gamow-Teller and weak magnetism form factors for almost 70 $\beta$ transitions from states with isospin $T =$ 1, 3/2 and 2, and for masses up to $A~=~53$. 


Experimental results were also compared with theoretical calculations performed in the shell model. Thus, the Gamow-Teller matrix element, $M_{GT}$, and orbital current matrix element, $M_L$, deduced from the experimental Gamow-Teller and weak magnetism form factors, $c$ and $b$ respectively, using the impulse approximation, were compared to theoretical values calculated in the shell model for the mirror $\beta$ decays up to $A$~=~45 as well as for 14 $\beta$ decays from $T = 1$  states with $A =$ 6 to 30. Good agreement between theory and experiment was found for $M_{GT}$ and to a lesser extent also for $M_L$, with good overall correspondence between experiment and theory being found for the ratio $b/Ac$. This gives some confidence in the theoretical calculations so that the weak magnetism recoil correction can also reliably be addressed theoretically for other allowed $\beta$ transitions in such light nuclei. Note, however, that for nuclei heavier than the ones dealt with here the shell-model calculations necessarily have to be performed in a truncated model space leading to uncertainties in the calculated matrix elements and form factors. More extended calculations then have to be performed or other theoretical approaches used.

Should detailed theoretical calculations not be available for the interpretation in terms of new physics of $\beta$-correlation measurements with a specific isotope in the mass range $A =$ 3 to 61, one could alternatively use the overall average value $b/Ac$~=~5.4~$\pm$~3.1 (Table~\ref{table:overview_averages}) that was obtained for the set of 84 $\beta$ transitions considered here. Of course direct measurements of the weak magnetism term, $b$, would be even more beneficial for this and could also help to further improve theory. At present, a series of dedicated $\beta$-spectrum shape measurements are ongoing and planned, using a wide variety of detection systems. All focus on extracting the weak magnetism term and/or the so-called Fierz interference term (which is sensitive to the presence of possible scalar or tensor type contributions to the weak interaction \cite{Jackson1957}) from the spectrum shape. Indeed, the $\beta$ spectrum depends on the weak magnetism, $b/Ac$, via a term that is linear in the $\beta$-particle energy, and on the Fierz interference term via a term that is inversely proportional to the energy \cite{Gonzalez2016}. Very recently, the electron-spectrum shape in the decay of the free neutron was investigated to produce the first precise results for the Fierz interference term in neutron decay \cite{Sun2020, Saul2020}. At the nuclear side, large-volume NaI and CsI scintillators stopping all radiation so as to minimize systematic errors have been used for the transitions $^6$He $\rightarrow ^6$Li \cite{Huyan2016, Huyan2018, Huyan2018a} and $^{20}$F $\rightarrow ^{20}$Ne$^1$ (Fig.~\ref{fig13:triplets-16-22}) \cite{Naviliat2021} (using the setup described in \cite{Hughes2018}). Other authors used a superconducting spectrometer to observe the $\beta$ spectrum of $^{68}$Ga \cite{Knutson2011} or metallic magnetic calorimeters to study the $\beta$ spectra of $^{63}$Ni and $^{241}$Po \cite{Mougeot2012}. More recently, the $\beta$ spectrum of $^{45}$Ca was measured using the UCNA spectrometer \cite{Plaster2012} with the Nab/UCNB prototype detection system (consisting of 1.5 mm thick, highly segmented silicon detectors, with active area diameter of 11.5 cm, and thin front dead layer), installed at each end \cite{Broussard2019}. Further, at the LIRAT facility at GANIL (Caen) a $^6$He source was sandwiched between two Yttrium Aluminum Perovskite (YAP:Ce) scintillators providing a 4$\pi$ geometry \cite{Naviliat2021}. A 4$\pi$ solid angle was also achieved by installing two plastic scintillators in a strong magnetic field for measuring the $\beta$-spectrum shape of the pure Gamow-Teller decay of $^{114}$In \cite{Vanlangendonck2021a}. The same $\beta$ decay is presently also investigated by a combination of a plastic scintillator, serving as a trigger device, and a hexagonally structured multi-wire drift chamber (MWDC), filled with a mixture of helium and isobutane gas \cite{Lojek2009, Lojek2015, DeKeukeleere2021}. The determination of the weak magnetism form factor for the $1^+ \rightarrow 0^+$ the pure Gamow-Teller $\beta$ decay of $^{114}$In in those two experiments would be the first direct determination of weak magnetism in the mass range of fission fragments and would be of special interest for further theoretical work on the reactor neutrino problem. 

A better control of the induced form factors would allow future correlation coefficients measurements in nuclear $\beta$ decay to better take into account the recoil correction when interpreting results in terms of new physics or when determining the $V_{ud}$ quark-mixing matrix element. When a precision of the order of 1\% or better is obtained in such measurements, radiative corrections have, in addition, to be considered as well. For measurements of the $\beta \nu$-correlation and the $\beta$-asymmetry parameter with the mirror nuclei these are discussed in detail in Refs.~\cite{Hayen2021, Vanlangendonck2021}.

\appendix
\section{Single particle evaluation of weak magnetism}

\subsection{Overview}
The evaluation of the weak magnetism form factor using the impulse approximation (Eq. (\ref{eq:M1_isospin})) requires the evaluation of both the $M_{GT}$ and $M_L$ matrix elements. Expressed in the usual way, we find
\begin{equation}
\frac{b}{Ac} = \frac{1}{g_A}\left(g_M + g_V \frac{M_L}{M_{GT}}\right) ~ .
\end{equation}
For a general operator $\mathcal{O}_i$ we can project onto a basis state in second quantization to find
\begin{equation}
\langle f | \mathcal{O}_i \tau^{\pm} | i \rangle = \sum_{\alpha, \beta} \langle \alpha | \mathcal{O}_i | \beta \rangle \langle f | a^\dagger_\alpha a_\beta | i \rangle ~ ,
\label{eq:operator_decomp}
\end{equation}
where $\alpha$ and $\beta$ are single particle proton (neutron) and neutron (proton) states for $\beta^-$ ($\beta^+$) decay. The quantities $\langle f | a^\dagger_\alpha a_\beta | i \rangle$ are called one body transition densities, and can be calculated using both the shell model and mean field techniques.

The simplest way of evaluating Eq. (\ref{eq:operator_decomp}) assumes a single particle in a spherical harmonic oscillator potential, thereby reducing the sum to a single element, $\gamma$, with a trivial $\langle f | a^\dagger_\alpha a_\beta | i \rangle = \delta_{\alpha \beta}\delta_{\alpha \gamma}$. The single particle state is chosen based on regular $jj$-coupling, assuming the same radial functions for $j = l \pm 1/2$. In this case the ratio trivially reduces to
\begin{align}
\frac{M_L}{M_{GT}} &= \frac{\langle n_f l_f j_f | \bm{l} | n_i l_i j_ \rangle}{\langle n_f l_f j_f | \bm{\sigma} | n_i l_i j_ \rangle} = (-1)^{j_i-j_f} \nonumber \\
&\times\frac{\left\{\begin{array}{ccc}
1/2 & l & j_i \\
1 & j_f & l
\end{array} \right\}}{\left\{\begin{array}{ccc}
l & 1/2 & j_i \\
1 & j_f & 1/2
\end{array} \right\}}
\frac{\sqrt{l(l+1)(2l+1)}}{\sqrt{6}} ~ ,
\label{eq:ML_MGT_ESP}
\end{align}
where $J_i$ and $J_f$ are initial and final nuclear spins. In case $j_i = j_f = l + 1/2$, Eq. (\ref{eq:ML_MGT_ESP}) reduces to $l$, while for $j_i=j_f=l-1/2$ it reduces to $-(l+1)$, and $-1/2$ otherwise. For transitions where $j_i=j_f$, the orbital component can constitute a large part of the total $b/Ac$ value. This scenario is typically limited to lower nuclear masses, in particular mirror nuclei and transitions from isospin multiplets, precisely the two cases we have discussed in Sec. \ref{sec:cvc}.

The simple spherical harmonic oscillator potential, while clearly of use, paints an overly simplified picture of the nuclear environment. We can then move forwards to a more realistic potential, such as a Woods-Saxon form. In this case we construct the Hamiltonian as
\begin{equation}
\mathcal{H} = -\frac{\hbar^2}{2m}\nabla^2 - V_0f(r) - V_s\left(\frac{\hbar}{m_\pi c}\right)^2\frac{1}{r}\frac{df}{dr}\bm{l}\cdot \bm{s} ~ ,
\label{eq:hamiltonian_spherical_WS}
\end{equation}
where $f(r)$ has the typical Woods-Saxon form
\begin{equation}
f(r) = \frac{1}{1+\exp{(r-R)/a_0}} ~ ,
\end{equation}
with $R$ the nuclear radius, and
\begin{equation}
V_0 = V\left(1\pm \chi \frac{N-Z}{N+Z} \right).
\end{equation}
Here $V$ and $\chi$ are free parameters based on the approach by Refs. \cite{Hird1973, Dudek1980}. These are typically put to $49.6$\,MeV and $0.86$, respectively. As the spherical harmonic oscillator states form a good basis for our improved wave functions and analytical results are available, we write the new nuclear state as
\begin{equation}
| \nu j \rangle = \sum_{nl} C^\nu_{nlj} | nl j \rangle ~ ,
\label{eq:wf_spherical_exp}
\end{equation}
where $| nlj \rangle$ are the spherical harmonic oscillator basis states as before. As $j$ remains a good quantum number in a spherical potential, the Hamiltonian of Eq. (\ref{eq:hamiltonian_spherical_WS}) serves to mix different radial quantum numbers. In keeping with the extreme single particle approximation, the result of Eq. (\ref{eq:ML_MGT_ESP}) is then trivially extended to
\begin{equation}
\frac{M_L}{M_{GT}} = \frac{\sum_{kl}C^{\nu_k*}_{n_kl_kj_k}C^{\nu_l}_{n_ll_lj_l}\langle n_k l_kj_k | \bm{l} | n_ll_lj_l \rangle}{\sum_{kl}C^{\nu_k*}_{n_kl_kj_k}C^{\nu_l}_{n_ll_lj_l}\langle n_k l_kj_k | \bm{\sigma} | n_ll_lj_l \rangle} ~ .
\end{equation}
This result is strictly only valid for a state with only one single particle in the final state responsible for the nuclear spin, i.e. odd-$A$ nuclei. In even-$A$ nuclei, even though we only consider one active nucleon in the decay process, we have to consider at least two nucleons coupling to the correct total spin. It is shown \cite{Behrens1982} that this can be established by multiplying our previous result with a factor $C(K)$ which depends only on the spherical tensor operator rank, $K$, i.e. $\langle f| a^\dagger_\alpha a_\beta | i \rangle = C(K)$. As both the Gamow-Teller and orbital matrix element are rank 1 operators, this factor drops out in our simple approximation we have made here. As this is in general not true for states described by multiple nucleon configurations - as is done for instance in the shell model - we consider here the simple case of an even-even to odd-odd transition with two valence nucleons in initial and final states. In this case valence particle 1 transforms into particle 2 and we write for $\beta^\pm$ decay
\begin{align}
C(K) &= \sqrt{\frac{\hat{J_i}\hat{J_f}\hat{T_i}\hat{T_f}}{1+\delta_{j_1j_2}}}(-1)^{T_f-T_{3f}}\left(
\begin{array}{ccc}
T_f & 1 & T_i \\
-T_{3f} & \pm 1 & T_{3i}
\end{array}
\right) \nonumber \\
&\times\left\{
\begin{array}{ccc}
\frac{1}{2} & T_f & \frac{1}{2}(T_f+T_i) \\
T_i & \frac{1}{2} & 1
\end{array}
\right\} \sqrt{\frac{3}{2}} (-1)^K \nonumber \\
& \times   2 [\delta_{j_1j_1}-(-1)^{j_1+j_2}]
\left\{
\begin{array}{ccc}
j_2 & J_f & j_1 \\
J_i & j_1 & K
\end{array} 
\right\} ~ ,
\end{align}
where we introduced the hat notation $\hat{j} = 2j+1$. An equivalent formula can be written down for odd-odd to even-even decays using the results from Ref. \cite{Rose1954}.

\subsection{Axially deformed potentials}
While the generalization to a Woods-Saxon potential described in the previous section most certainly helps in the correct determination of the valence particle, its corresponding state is typically still dominated by a single harmonic oscillator component. The reason for this is the large distance between nuclear levels with equal $j$ and parity, particularly visible in the lower-$Z$ nuclei. As many of the cases studied in this work lie close to the $N=Z$ line, large (mainly quadrupole) deformations are found even for light nuclei, with also higher-order deformations being non-negligible in the low to medium-$Z$ region of interest here \cite{Moller2016}.


It is of interest then to extend the potential of Eq. (\ref{eq:hamiltonian_spherical_WS}) to include axial deformations
\begin{equation}
\mathcal{H}' = \mathcal{H} + V_0R\frac{df}{dr}\sum_{n=1} \beta_{2n}Y^0_{2n} ~ , 
\label{eq:hamiltonian_deformed_WS}
\end{equation}
with $Y_l^m$ the standard spherical harmonic function. In this case $j$ is not any more a good quantum number, and we must instead to resort to the projection of $\bm{j}$ along the symmetry axis of the nucleus written as $\Omega$. Writing the angular momentum of a rotating deformed core as $\bm{R}$, we denote by $K$ the projection along the symmetry axis of the sum $\bm{j}+\bm{R}$. It follows then directly that in the rotational ground state, we have $K=\Omega$ for odd-$A$ nuclei. In even-even nuclei the valence particles couple to $K=0$, while in odd-odd nuclei $K$ can be $|\Omega_p\pm\Omega_n|$. The new single particle wave function will be of the Nilsson type and be a combination of wave functions of the type of Eq. (\ref{eq:wf_spherical_exp}) with the appropriate spin projections such that
\begin{equation}
| \mu \Omega \rangle = \sum_{nl} \sum_{\nu j} C^\mu_{\nu \Omega} C^\nu_{nlj} | nl j \rangle \equiv \sum_{j} C_{j\Omega} | N l j \Omega \rangle ~ ,
\label{eq:wf_deformed_exp}
\end{equation}
where in the last term we have written our solution using the notation by Davidson\footnote{The difference with the treatment of Davidson lies in the spherical potential used. Unless this is exactly a modified spherical oscillator wave function as per Nilsson \cite{Nilsson1955}, more than one $C^\nu_{Nj}$ will be non-zero. In the limit of zero deformation, Eq. (\ref{eq:wf_spherical_exp}) will generally contain more than one term.} \cite{Davidson1968}. The matrix element for odd-$A$ decays is then written as \cite{Behrens1982}

\begin{align}
&\langle \phi(J_fK_f;\Omega_f) || O_{KLs} \tau^\pm || \phi(J_iK_i;\Omega_i) \rangle  \nonumber \\
&= \sqrt{\frac{\hat{J_i}\hat{J_f}}{(1+\delta_{K_f0})(1+\delta_{K_i0})}} \sum_{j_2j_1} C_{j_2\Omega_2}C_{j_1\Omega_1} \nonumber \\
&\times \left\{(-1)^{J_2-K_2+j_2-\Omega_2} \left( \begin{array}{ccc}
J_f & K & J_i \\
-K_f & \Omega_2-\Omega_1 & K_i
\end{array} \right) \right. \nonumber \\
&\times\left( \begin{array}{ccc}
j_2 & K & j_1 \\
-\Omega_2 & \Omega_2-\Omega_1 & \Omega_1
\end{array}\right) + \left( \begin{array}{ccc}
J_f & K & J_i \\
K_f & -\Omega_2-\Omega_1 & K_i
\end{array}\right)\nonumber \\
&\times\left. \left(\begin{array}{ccc}
j_2 & K & j_1 \\
\Omega_2 & -\Omega_2-\Omega_1 & \Omega_1
\end{array} \right) \right\}\langle j_2 || O_{KLs} || j_1 \rangle ~ . 
\label{eq:deformed_matrix_element_odd}
\end{align}
In the case of even-$A$ decays initial and final states are approximated by a coupling of two valence particles to the correct total angular momentum. Using the results of Ref. \cite{Berthier1966}, the spin-reduced matrix element for even-even to odd-odd decays is found to be
\begin{align}
&\langle \phi(J_fK_f;\Omega_f) || \sum_{n=1,2} \{O_{KLs}\tau^\pm_n\} || \phi(J_iK_i=0;\Omega_i=0) \rangle \nonumber \\
&= \sqrt{\frac{\hat{J_i}\hat{J_f}}{2(1+\delta_{K_f0})}} \left(\begin{array}{ccc}
J_f & K & J_i \\
-K_f & K_f & 0
\end{array} \right) [1+(-1)^{J_i}]\nonumber \\
&\times \sum_{j_2j_1} C_{j_2-\Omega_2}C_{j_1\Omega_1} (-1)^{j_2-\Omega_2}  \left(\begin{array}{ccc}
j_2 & K & j_1 \\
-\Omega_2 & K_f & -\Omega_1
\end{array} \right)\nonumber \\
&\times \langle j_2 || O_{KLs} || j_1 \rangle ~ ,
\label{eq:deformed_matrix_element_even}
\end{align}
while the reverse case can be found in several publications \cite{Berthier1966, Behrens1982}. Here $\langle j_2 || O_{KLs} || j_1 \rangle$ are simple spin-reduced single-particle matrix elements for the harmonic oscillator wave functions.

The final task is then to calculate the different $C_{\Omega j}$, a task which is performed by the code described in Ref.~\cite{Hayen2019c}. This code automatically uses these results to calculate the relevant factors such as $b/Ac$ and $d/Ac$. It enables coupling to results from more advanced calculations through the single particle density matrix elements $\rho_{\alpha \beta} = \langle f | a^\dagger_\alpha a_\beta | i \rangle$. Further, it calculates the spectrum shape and integrated $f$ value using these nuclear structure inputs based on the spectral shape description from Ref.~\cite{Hayen2018}.

\bibliography{bibliografie_IAN_10-09-2014}

\end{document}